%% file: main.tex
\documentclass[a4paper,fleqn]{cas-sc}

\input{00_preamble/preamble.tex}

\begin{document}

\shorttitle{Variational quantum eigensolver: review and best practices}

\shortauthors{\phantom{}} 

\title[mode=title]{The Variational Quantum Eigensolver: a review of methods and best practices}

\author[1,2]{Jules Tilly}[type=editor,
                        orcid=0000-0001-5034-474X]
\cortext[cor1]{Corresponding author}
\ead{jules.tilly.18@ucl.ac.uk}
\author[2,3]{Hongxiang Chen}[
    type=editor, orcid=0000-0001-6792-7394]
\author[2,4]{Shuxiang Cao}[
    type=editor, orcid=0000-0001-7178-4250]
\author[1,2]{Dario Picozzi}[orcid=0000-0002-6517-2458]
\author[5]{Kanav Setia}[orcid=0000-0003-2032-3993]
\author[6]{Ying Li}[orcid=0000-0002-1705-2494]
\author[2,3]{Edward Grant}[orcid=0000-0003-0657-1915]
\author[2,3]{Leonard Wossnig}[orcid=0000-0002-0861-9540]
\author[7]{Ivan Rungger}[type=editor,
                        orcid=0000-0002-9222-9317]
\author[8]{George H. Booth}[orcid=0000-0003-2503-4904]
\author[1]{Jonathan Tennyson}[orcid=0000-0002-4994-5238]

\address[1]{University College London, Department of Physics and Astronomy, WC1E 6BT London, United  Kingdom}
\address[2]{Odyssey Therapeutics, WC1X 8BB London,  United  Kingdom}
\address[3]{University College London, Department of Computer Science, WC1E 6BT London, United  Kingdom}
\address[4]{University of Oxford, Department of Physics, Clarendon Laboratory, OX1 3PU, United  Kingdom}
\address[5]{qBraid, 5235, Harper Court, Chicago, IL}
\address[6]{Graduate School of China Academy of Engineering Physics, Beijing 100193, China}
\address[7]{National Physical Laboratory, TW11 0LW Teddington, United Kingdom}
\address[8]{King's College London, Department of Physics, Strand, London, WC2R 2LS, United  Kingdom}

\begin{abstract}
The variational quantum eigensolver (or VQE), first developed by Peruzzo \textit{et al.} (Nature Comm.5, 4213 (2014)), has received significant attention from the research community in recent years. It uses the variational principle to compute the ground state energy of a Hamiltonian, a problem that is central to quantum chemistry and condensed matter physics. Conventional computing methods are constrained in their accuracy due to the computational limits facing exact modeling of the exponentially growing electronic wavefunction for these many-electron systems. The VQE may be used to model these complex wavefunctions in polynomial time, making it one of the most promising near-term applications for quantum computing. One important advantage is that variational algorithms have been shown to present some degree of resilience to the noise in the quantum hardware. Finding a path to navigate the relevant literature has rapidly become an overwhelming task, with many methods promising to improve different parts of the algorithm, but without clear descriptions of how the diverse parts fit together. The potential practical advantages of the algorithm are also widely discussed in the literature, but with varying conclusions. Despite strong theoretical underpinnings suggesting excellent scaling of individual VQE components, studies have pointed out that their various pre-factors could be too large to reach a quantum computing advantage over conventional methods.

This review aims at disentangling the relevant literature to provide a comprehensive overview of the progress that has been made on the different parts of the algorithm, and to discuss future areas of research that are fundamental for the VQE to deliver on its promises. All the different components of the algorithm are reviewed in detail. These include the representation of Hamiltonians and wavefunctions on a quantum computer, the optimization process to find ground state energies, the post processing mitigation of quantum errors, and suggested best practices. We identify four main areas of future research:(1) optimal measurement schemes for reduction of circuit repetitions required; (2) large scale parallelization across many quantum computers;  (3) ways to overcome the potential appearance of vanishing gradients in the optimization process for large systems, and how the number of iterations required for the optimization scales with system size; (4) the extent to which VQE suffers for quantum noise, and whether this noise can be mitigated in a tractable manner. The answers to these open research questions will determine the routes for the VQE to achieve quantum advantage as the quantum computing hardware scales up and as the noise levels are reduced.
\end{abstract}


\maketitle
\tableofcontents 

\input{\ProjectRoot/01_Introduction.tex}
\input{\ProjectRoot/02_Overview_of_VQE.tex}

\input{\ProjectRoot/03_hamiltonian_main.tex}

\input{\ProjectRoot/04_encoding_main.tex}

\input{\ProjectRoot/05_group_main.tex}
\input{\ProjectRoot/06_ansatz_main.tex}

\input{\ProjectRoot/07_optimization_main.tex}

%
%
%
%
%
%
%
%
%
%
%

\input{\ProjectRoot/08_error-mit.tex}

\input{\ProjectRoot/09_VQE_extensions.tex}

\section{Conclusion and outlook}

The VQE is among the most promising near-term applications for quantum computers. While it is supported by strong theoretical arguments, there are still numerous open questions about the future applicability of the method. In this review, we outline the most relevant research that has been produced on the different components of the VQE, and how these are connected. To begin with, the choice of the Hamiltonian has direct implications for the number of qubits required, and for the capacity of VQE to achieve accurate results. A canonical orbital basis in second quantization has been the dominant choice in the community due to limited number of qubits required, although alternatives such as plane wave bases \cite{Babbush2018, Barkoutsos2018} and frozen natural orbitals basis \cite{Gonthier2020} have been proposed. Once the problem Hamiltonian has been constructed (and provided one decides to use second quantization), it is critical to decide on a mapping from fermionic operators to spin operators that best suits the system studied. While the Jordan-Wigner mapping \cite{Jordan1928} has been most commonly used due to its convenient implementation, low-weight mappings (in particular the ternary tree mapping \cite{Jiang2020} for molecular Hamiltonians, and the generalized superfast encoding \cite{Setia2019} for lattice models) have been shown to bring significant benefits with respect to circuit depth \cite{Tranter2015} for fermionic-based ans\"atze and resilience to the barren plateau problem \cite{Cerezo2021_BP, Uvarov2020, Uvarov2020_frustrated}. 

One of the most discussed drawbacks of VQE is the large number of measurements required to perform the optimization process \cite{Wecker2015, Elfving2020, Gonthier2020}. Some savings can be accomplished by managing the distribution of measurements, by weighting shots allocated to each operators by their respective weights in the Hamiltonian \cite{Wecker2015, Rubin2018, Arrasmith2020} or by truncating terms with significant impact on the energy \cite{mccleanTheoryVariationalHybrid2015}. However, to significantly reduce the scaling of VQE one should turn towards methods which jointly measure Hamiltonian terms. At present for molecular Hamiltonian this can most efficiently be done using a decomposed interactions method \cite{Huggins2021, Yen2021_Cartan}. For lattice models, however, because low-weight encoding are more easily achieved, a simple qubit-wise commutativity grouping scheme may the most cost effective \cite{mccleanTheoryVariationalHybrid2015, Kandala2017, Hempel2018, Rubin2018, Kokail2019, Izmaylov2019, Nam2020, Verteletskyi2020, Hamamura2020, Gokhale2019_long}. 

The core of the VQE remains the ansatz, or parametrized quantum circuit, chosen to model to the trial wavefunction. Among fixed structured ans\"atze the k-Unitary paired Generalized Coupled Cluster (k-UpCCGSD) \cite{Lee2019} and its extension \cite{Kottmann2021_1} stand out due to their ability to achieve excellent accuracy (evidenced numerous times \cite{Grimsley2019_UCC_Review, Lee2019, GreeneDiniz2020, Kottmann2021_1, Sokolov2020}), while maintaining a scaling linear in the number of qubits. For lattice models, while Unitary Coupled Cluster ans\"atze have been tested (for instance \cite{Sokolov2020}), one may prefer to use an ansatz for which the physically-motivated construction is more adapted. One such ansatz is the Hamiltonian Variational Ansatz \cite{Wecker2015, Wiersema2020}, and its extensions \cite{Babbush2018, Choquette2021}, which have been shown to be a promising methods for the study of many-body physics through VQE \cite{Wiersema2020, Babbush2018, Choquette2021}. Adaptive methods (all stemming from ADAPT-VQE \cite{Grimsley2019}), are promising avenues with demonstrated benefits \cite{Claudino2020}, but their overall cost and scaling remains unclear. 

The choice of optimizer directly impacts the ability of the VQE to converge and the overall cost of the method. Comparative studies remain sparse, and it is challenging to infer the ability of specific optimizer to navigate the complex optimization landspace of VQE \cite{Bittel2021}. However, two optimizers have received at least some level of numerical supports regarding their convergence rate and ability to avoid optimization pitfalls: the quantum natural gradient \cite{amari_natural_1998,martens_new_2020,wierichs_avoiding_2020}, and RotoSolve \cite{nakanishi_sequential_2020,ostaszewskiStructureOptimizationParameterized2021}. Finally, appropriate mitigation of quantum noise is likely required for VQE. Variational algorithms have been shown to exhibit at least some degree of ability to learn the biases created by quantum noise \cite{mccleanTheoryVariationalHybrid2015, Enrico2021EvaluatingNoiseResilience,Enrico2020NoiseInducedBreakingOfSymmetries}. However, further methods, such as symmetry verification and extrapolation based methods, despite coming at a significant computational cost, could be required to achieve the target accuracy with VQE. 

Despite numerical simulations being presented in the vast majority of papers proposing new methods, comprehensive, comparative and applied studies remain rare. This is most likely due to the rapid development of the field and the amount of different interacting components the VQE requires. Some notable exceptions (in particular \cite{Gonthier2020}) have attempted to focus on best practices to study applicability of VQE to practical electronic structure problems. Further numerical studies of these aspects can undoubtedly help guiding future research in the field. We identify four broad areas of suggested further research: (1) optimal measurement strategies to reduce the substantial number of shots required to estimate the expectation value of a Hamiltonian using a VQE ansatz; (2) the real cost and benefits of the massive parallelization potential of VQE, which is likely required for the method to ever be relevant; (3) the extent to which the method is subject to the barren plateau problem and whether current mitigation techniques are sufficient for VQE to scale, this would also involve further studying the complexity of the VQE optimization landscape, the expected convergence rate of the method and the relevance of various optimizers in avoiding local minima; and (4) the extent of the noise resilience of VQE, and the extent to which noise can be mitigated in a tractable manner. 

One interesting aspect of VQE that our review has not covered is that numerous algorithms have been inspired by it. For example, the Full Quantum Eigensolver \cite{Wei2020} (FQE) includes a quantum gradient descent to perform an optimization similar to VQE, but directly on the quantum computer to speed up convergence; the Projective Quantum Eigensolver \cite{Stair2021} optimizes energy using residuals from projection of the Schr\"odinger equation, rather than gradients. The Quantum Assisted Simulator (QAS) \cite{Bharti2020, Haug2020} and its extensions \cite{Bharti2020_IQAE} aims to avoid the barren plateau problem by replacing the usual optimization loop of VQE by a three-step process: selection of an ansatz, measurement of overlap matrices on a quantum computer, conventional computing post-processing based on the data from the second step. The Quantum Sampling Regression \cite{Rivero2020} also bypasses the VQE optimization loop and focuses on sampling the Fourier basis of the energy landscape using an ansatz to perform post-processing on a conventional computer, thereby bypassing the quantum-classical communication costs such as latency. The Variational Quantum State Eigensolver (VQSE) \cite{Cerezo2020_VQSE} adapts the VQE to cases in which one searches eigenvalues and eigenvectors of a quantum state density matrix. The Permutation VQE (PermVQE) \cite{Tkachenko2021} nests the VQE in an additional optimization loop with the objective to permute the order of the qubits at each iteration and minimize long-range correlations in ground state energy. The development of such algorithms is based on a deep knowledge of both the advantages and limitations of the VQE. The Filtering VQE \cite{Amaro2022} uses filtering operators in a VQE process to find ground states more efficiently and accurately, showing the method outperforms standard VQE and Quantum Approximate Optimization Algorithm in the case of a quadratic unconstrained binary optimization Hamiltonian \cite{Kochenberger2014}. Deep VQE \cite{Fujii2020} can be used to break down a given Hamiltonian into sub-units which can be solved separately to reduce the qubit requirements on very large systems. This methods was also extended to excited states computation by Mizuta \textit{et al.} \cite{Mizuta2021}. Meitei \textit{et al.} \cite{Meitei2021} presented a method, later optimized by Asthana \textit{et al.} \cite{Asthana2022}, to perform VQE directly at pulse level (ctrl-VQE), removing the need for a gate based ansatz. Finally, variational methods have also been used to model real and imaginary time evolution of quantum systems for ground and excited states discovery \cite{Yuan2019, Gustafson2021, Lin2021, Mansuroglu2021, McArdle2019, Nishi2021,Benedetti2021, Gomes2021}.

To assess the future worth of the VQE as an electronic structure method, one must also consider how it can be paired with other algorithms, such as those described in the paragraph above or with QPE, to help accelerate computation where needed. The algorithm is no doubt among the most promising methods for near-term applications of quantum computers. As  is the case for many emerging technologies, it faces daunting obstacles that will need to be overcome before it can be used in practical calculations, but the VQE also promises to deliver outstanding applications in fields such as drug discovery, chemical engineering, or material sciences.

\input{\ProjectRoot/acknowledgements.tex}

\appendix

\input{\ProjectRoot/appendices.tex}

\clearpage
\bibliographystyle{bibstyle/model3-num-names} 


\end{document}

%% file: 00_preamble/preamble.tex
\usepackage[utf8]{inputenc}
\usepackage[T1]{fontenc}
\usepackage[english]{babel}
\usepackage{graphicx}
\usepackage{dcolumn}
\usepackage{physics}
\usepackage{braket}
\usepackage{amsmath}
\usepackage{dsfont} 
\newcommand{\unit}{\mathds{1}}  
\usepackage{appendix}
\usepackage[dvipsnames]{xcolor}
\usepackage{pgf}
\usepackage{hyperref}
\hypersetup{ 
    colorlinks=true,
    linkcolor=blue,
    filecolor=blue,      
    urlcolor=blue,
}
\usepackage{blochsphere}
\usepackage[colorinlistoftodos]{todonotes}
\usepackage{qcircuit} 
\usepackage{blkarray}
\usepackage{booktabs}
\usepackage{tabularx} 
\usepackage{multirow} 
\usepackage{longtable} 
\usepackage{mathtools}
\usepackage[ruled]{algorithm2e} 
\usepackage[numbers, sort&compress]{natbib} 
\usepackage{subfig}

\DeclareMathOperator*{\argmin}{\arg\!\min}
\DeclareMathOperator{\arctantwo}{arctan2}
\newcommand*{\ProjectRoot}{.}
\renewcommand{\citet}[1]{Ref.~\cite{#1}}

\DeclareMathOperator{\Had}{Had}
\DeclareMathOperator{\QFT}{QFT}
\newcommand{\nEES}{K}

%% file: 01_Introduction.tex
\section{Introduction}

Quantum computing has undergone rapid development over recent years: from first conceptualization in the 1980s \cite{Benioff1980, Feynman1982}, and early proof of principles for hardware in the 2000s \cite{Chuang1998, Jones1999, Leung2000, Vandersypen2001, Haffner2005, Negrevergne2006, Plantenberg2007, Hanneke2009, Monz2011, Devitt2013, Devitt2016, Monz2016}, quantum computers can now be built with hundreds of qubits \cite{Jurcevic2021, Pino2021, Ebadi2021}. While the technology remains in its infancy, the fast progress of quantum hardware and the massive financial investments all over the world have led many to assert that so-called Noisy-Intermediate Scale Quantum (NISQ) devices \cite{preskillQuantumComputingNISQ2018, Brooks2019} could outperform conventional computers in the near future \cite{Arute2019, Zhong2020, Wu2021, Madsen2022}. NISQ devices are near-term quantum computers, with a limited number of qubits, and too few physical qubits to implement robust error correction schemes. Existing NISQ computers have already been shown to outperform conventional computers on a limited set of problems designed specifically to fit quantum computers' capabilities \cite{Arute2019, Zhong2020, Wu2021}. Algorithms running on these restricted devices may require only a small number of qubits, show some degree of noise resilience, and are often cast as hybrid algorithms, where some steps are performed on a quantum device and some on a conventional computer. In particular, the number of operations, or quantum gates, must remain moderate, as the longer it takes to implement them, the more errors are introduced into the quantum state, and the more likely it is to decohere. Due to these restrictions, there are severe limits on the scope of algorithms that can be considered. In particular, well-known quantum algorithms such as Shor's algorithm \cite{shorAlgorithmsQuantumComputation, Vandersypen2001, ChaoYang2007, Lanyon2007, Lucero2012, MartinLopez2012, Markov2013, Amico2019} for factoring prime numbers, or Grover's algorithm \cite{groverFastQuantumMechanical1996, Bennett1997, Cerf2000, Ambainis2004, Ambainis2007, Bernstein2010} for unstructured search problems, are not suitable.

The Variational Quantum Eigensolver (VQE) was originally developed by Peruzzo {\it et al.} \cite{Peruzzo2014}, and its theoretical framework was extended and formalized by McClean \textit{et al.} in Ref. \cite{mccleanTheoryVariationalHybrid2015}. The VQE is among the most promising examples of NISQ algorithms. In its most general description, it aims to compute an upper bound for the ground-state energy of a Hamiltonian, which is generally the first step in computing the energetic properties of molecules and materials. The study of electronic structures is a critical application for quantum chemistry (for instance: \cite{Deglmann2014, WilliamsNoonan2017, Heifetz2020}) and condensed matter physics (for instance: \cite{Continentino2021, VanderVen2020}). The scope of VQE is therefore very wide-ranging, being potentially relevant for drug discovery \cite{Cao2018_DD, Blunt2022}, material science \cite{Lordi2021}, chemical engineering \cite{Cao2019_QC}. Conventional computational chemistry, grounded in nearly a century of research, provides efficient methods to approximate such properties, but it becomes intractably expensive for very accurate calculations on increasingly large systems. This poses challenges in the practical application of such methods. One of the main reasons why computational chemistry methods can lack sufficient accuracy in molecular systems is an inadequate treatment of the correlations between constituent electrons. These interactions between electrons formally require computation that scales exponentially in the size of the system studied (i.e. the total time it takes to implement the computation is an exponential function of the system size), rendering exact quantum chemistry methods in general intractable with conventional computing. This limitation is well studied in the literature addressing simulation of quantum computers on conventional computers, Ref.~\cite{Zhou2020} provides an excellent example. 

This bottleneck is the motivation for investigating methods such as the VQE, with the anticipation that these could one day outperform the conventional computing paradigm for these problems \cite{Boixo2018, mccaskeyQuantumChemistryBenchmark2019}. In 1982, Richard Feynman theorized that simulating quantum systems would be most efficiently done by controlling and manipulating a different quantum system \cite{Feynman1982}. An array of qubits obey the laws of quantum mechanics, the same way an electronic wavefunction does. The superposition principle \cite{Silverman2008, Ballentine2008} of quantum mechanics means that it can be exponentially costly to encode the equivalent information on conventional devices, while it only requires a linearly growing number of qubits. In the context of electronic structure theory \cite{Helgaker2000, Kratzer2019, Li2020_EST} this is the appeal of quantum computing: it offers the possibility to model and manipulate quantum wavefunctions exactly, beyond what is possible with conventional computing. While largely dominated by electronic structure research, the VQE and its extensions have also been applied to several other quantum mechanical problems which face similar scaling issues. These notably include nuclear physics \cite{Miceli2019, DiMatteo2021} and nuclear structure problems \cite{Kiss2022, Romero2022}, high-energy physics \cite{Bauls2020, Bass2021, Bauer2022}, vibrational and vibronic spectroscopy \cite{McArdle2019_vibra, Sawaya2019, Ollitrault2020_vibrational, Jahangiri2020,Ltstedt2021, Sawaya2021}, photochemical reaction properties predictions \cite{Mitarai2020, Omiya2022}, periodic systems \cite{Liu2020, Yoshioka2022, Manrique2020}, resolution of non-linear Schr{\"{o}}dinger equations \cite{Lubasch2020}, and computation of quantum states of a Schwarzschild-de Sitter black hole or Kantowki-Sachs cosmology \cite{Joseph2022}.

The VQE starts with an initialized qubit register. A quantum circuit is then applied to this register to model the physics and entanglement of the electronic wavefunction. Quantum circuit refers to a pre-defined series of quantum operations that will be applied to the qubits \cite{nielsenQuantumComputationQuantum2010}. The number of consequential operations in a circuit is referred to as depth. This circuit is defined by two parts: (1) a structure, given by a set of ordered quantum gates, often referred to as an 'ansatz'; and (2) a set of parameters that dictates the behavior of some of these quantum gates. Once the quantum circuit has been applied to the register, the state of the qubits is designed to model a trial wavefunction. The Hamiltonian of the system studied can be measured with respect to this wavefunction to estimate the energy. The VQE then works by variationally optimizing the parameters of the ansatz in order to minimize this trial energy, constrained to always be higher than the exact ground state energy of the Hamiltonian by virtue of the variational principle \cite{Rayleigh1870, Ritz1908, Arfken1985}. For the VQE to be tractable, it is necessary that the number of quantum operations required to model the wavefunction is sufficiently low, imposing a relatively compact ansatz. The VQE admits wavefunction ans\"atze which cannot be efficiently simulated on conventional computers, indicating a possible advantage over conventional approaches if these quantum circuits are sufficiently accurate trial wavefunctions \cite{Peruzzo2014}. A first demonstration of the potential of these quantum ans\"atze was shown in \cite{Peruzzo2014}, where an ansatz with polynomially-scaling depth in the size of the qubit register was constructed using principles grounded in conventional quantum chemistry (namely, the Unitary Coupled Cluster, which is discussed in more detail in Sec. \ref{sec:UCCA}). Since then, many alternatives have been proposed, with ans{\"{a}}tze scaling as low a linearly \cite{Lee2019} in the size of the qubit register. It must be understood however that a shallower ansatz, i.e. with fewer necessary quantum operations, will in general cover an overall smaller span of the space of all possible wavefunctions, and could result in lower accuracy of the ground state energy.

While the design of the ansatz is core to the VQE and to the theoretical underpinnings of its potential advantage over conventional methods, there are many other parts of the algorithm that directly affect the cost and feasibility of the approach. We provide a discussion for each of these parts in our review, and briefly outline them here. To begin with, once one has chosen a quantum system to study, one must decide on how to construct its Hamiltonian. It is a requirement of the VQE that the Hamiltonian can be written as a sum of individual terms, where the number of terms increases with system size relatively slowly. This is a feature of electronic Hamiltonians, where the Coulomb interaction, a sum of two-body terms, scales polynomially in the system size \cite{Szabo1996}. Often there is considerable flexibility and freedom in the specific choice or how a Hamiltonian is represented mathematically. This step is critical, as it impacts all aspects of the VQE: number of qubits, depth of the ansatz, and the total number of measurements required. Once a Hamiltonian is constructed, one must translate it into operators that can be directly measured on a quantum computer (spin or Pauli operators), which can again impact the depth of the ansatz and number of measurements. The next step is to decide on an ansatz, which must be expressive enough to be able to approximately model the ground state wavefunction, but not so much that it results in circuits that become too deep, or parameterizations that are too complex to efficiently train. Once an ansatz is selected, one must choose an appropriate optimizer, which significantly impacts the convergence rate of the VQE optimization and the overall cost of the algorithm. Measurements on quantum devices are effectively stochastic \cite{nielsenQuantumComputationQuantum2010}, meaning that to obtain the expectation value of an operator, one must repeat measurements enough times to achieve a given level of precision. The number of measurements required is not only defined by the required precision, but also by the number of operators in the Hamiltonian. Adopting efficient measurement strategies is critical for controlling the implementation cost of VQE. Finally, to reduce the possible impact of quantum noise on the accuracy of the result, one can introduce error mitigation strategies into the individual measurements, which again need consideration of their associated computational cost. 

This review holds two main purposes: (1) provide an overview for each of the different components of the VQE algorithm outlined above, along with a suggestion for best practices found in the literature. Because of the large variety of quantum systems the VQE can handle, we propose sets of best practices for two different and fairly general types of systems: \textit{ab initio} molecular systems and spin-lattice models, though it is likely that some of the conclusions can be generalized to other Hamiltonians, including impurity models, condensed matter, vibrational spectroscopy or nuclear structure; (2) identify a list of open research questions regarding the future applicability of the VQE, and about individual aspects of the algorithm presented above. We identify four significant (and inter-related) hurdles, which impact the future viability of this method. These are presented here at a high level and discussed throughout the review.

Firstly, several studies have considered the scaling of the number of measurements as the system size increases \cite{Wecker2015, Elfving2020, Kuhn2019, Gonthier2020}. Despite scaling polynomially \cite{mccleanTheoryVariationalHybrid2015}, the number of measurements required can nonetheless rapidly become too large for the method to be viable \cite{Wecker2015, Elfving2020, Gonthier2020}. As a consequence, many studies have called for developments in efficient measurement schemes (for instance, Ref.~\cite{Gonthier2020}), with approaches to find compact representations of the Hamiltonian and judicious use of joint measurement of commuting observables an ongoing source of developments in this endeavor (see Sec.~\ref{sec:DecomposedInteractions}). 

Secondly, an alternative approach to the measurement problem could lie in the enormous potential for parallelization of VQE. This potential was clearly called for in the original VQE paper of Peruzzo {\it et al.} \cite{Peruzzo2014}, but has received relatively little consideration from the community, in particular with respect to efficient strategies for this parallelization and possible communication overheads (further discussion on these points is found in Sec.~\ref{sec:parallelization}). 

Thirdly, variational quantum algorithms suffer from a common pitfall in the challenge of parameter optimization: the so-called barren plateau problem \cite{Wecker2015, McClean2018, Cerezo2021_BP}. The barren plateau is a result of the phenomenon that gradients of the VQE parameters vanish exponentially with the number of qubits used in the model \cite{McClean2018}, in the overall expressibility of the ansatz \cite{Holmes2016}, degree of entanglement of the wavefunction modeled \cite{OrtizMarrero2020, Patti2021}, or non-locality of the cost function \cite{Cerezo2021_BP, Uvarov2020, Sharma2020}. This of course would prevent tractable optimization in larger systems. Many methods have been proposed to address this problem, but it is currently unclear whether these methods allow one to fully contain the barren plateau problem in the context of VQE (further discussions on this topic are found in Sec.~\ref{sec:barren_plateau}). A related point is that the complexity of the optimization landscape for the VQE remains poorly understood. Bittel and Kliesch~\cite{Bittel2021} characterized specific features of this landscape and demonstrated the risk of optimization being undermined by local minima and non-convex features. The underlying question here is the convergence rate of different optimizers, and the overall scaling of the number of iterations required to reach convergence as a function of the system size and entanglement in the wavefunction. 

Finally, the extent to which VQE is resilient to quantum noise and its ability to be mitigated in a tractable manner remains unclear. Variational algorithms have the advantage of exhibiting some ability to learn away certain types of systematic noise \cite{Enrico2021EvaluatingNoiseResilience,Enrico2020NoiseInducedBreakingOfSymmetries}. While it was also shown that the cost of several error mitigation methods may largely outweigh the benefits \cite{RyujiFundamentalLimits2021}, the question of the impact of quantum noise on VQE results and the predictability of these errors should be studied in much further depth than has been done so far. Overall, despite a number of impediments, the VQE has the potential to be among the first useful methods to be implemented on NISQ devices. However, as the hardware continues to progress, so must the theoretical underpinning of this approach, as well as the efficiency and robustness of the software and algorithms, to guarantee that the potential benefits of methods such as VQE can be tapped as early as possible. 

\paragraph{Structure of the review:} Sec. \ref{sec:overview} presents a more formal definition of the VQE, its potential advantages and limitations are discussed, and our assessment of the state-of-the-art VQE is presented. The following sections provide reviews of each of the components of the VQE pipeline (as also summarized in Fig. \ref{fig:VQE_pipeline}). The first step in the VQE process is to choose a representation for the molecular Hamiltonian (Sec. \ref{sec:Hamiltonian_representation}), this is followed by a description of the methods used to map fermionic operators to spin operators, therefore allowing the Hamiltonian operators to be directly measured on a quantum device (Sec. \ref{sec:Encoding}). The next section covers efficiencies that can be used to reduce the number of measurements required to estimate the expectation value of a Hamiltonian (Sec. \ref{sec:Grouping}). We then turn to detailing the various quantum circuit structures representing wave function ans\"atze that have been proposed as the core component of the VQE (Sec. \ref{sec:Ansatz}). This is followed by a presentation of the most important optimization methods and how they relate to the VQE (Sec. \ref{sec:Optimization}). Finally, we list and discuss the main error mitigation techniques that could improve the overall accuracy of the algorithm (Sec. \ref{sec:error-mit}). We close our review with a discussion of relevant extensions of the VQE for quantum chemistry applications (Sec. \ref{sec:Extensions_of_VQE}).

\paragraph{Additional resources and other reviews:} Nielsen and Chuang \cite{nielsenQuantumComputationQuantum2010} provide an introductory text detailing the fundamentals of quantum computing and quantum information which remains a seminal work on the topic. For an introductory overview of quantum computing and a pragmatic discussion of the current state of the technology, we recommend the review by Whitfield {\it et al.} \cite{Whitfield2022}. For a more compact version of introductory topics in quantum computing, we point readers to the review by McArdle {\it et al.} \cite{mcardleQuantumComputationalChemistry2018}. This latter reference also provides a comprehensive review of quantum computing methods for quantum chemistry and material sciences. Two reviews on the same topic are provided by Bauer {\it et al.} \cite{bauerQuantumAlgorithmsQuantum2020} and of Motta and Rice \cite{Motta2021_review}. For a review focused on general variational quantum algorithms we recommend the work of Cerezo {\it et al.} \cite{cerezoVariationalQuantumAlgorithms2020}, and for the wider domain of NISQ algorithms, we recommend the review of Bharti {\it et al.} \cite{Bharti2021_review}. There are also reviews dedicated to narrower topics relevant to the VQE, in particular, we recommend two reviews of ans{\"{a}}tze, which complement our analysis: that by Fedorov {\it et al.} \cite{Fedorov2021}, and the review of unitary coupled cluster based-ans{\"{a}}tze by Anand {\it et al.} \cite{Anand2021_review}. Finally, for an overview of the fundamentals of quantum chemistry, we recommend the book of Szabo and Ostlund \cite{Szabo1996}.

The VQE field is expanding rapidly, the current review only attempts to survey the literature to the end of May 2022.

%% file: 02_Overview_of_VQE.tex
\section{Overview of VQE} \label{sec:overview}

In this section, we aim to provide sufficient information for our reader to acquire a technical understanding of the VQE, and an appreciation of where the algorithm is positioned compared to both conventional electronic structure, and other quantum methods. We also provide an outline for the suggested best practices, collected from the remainder of the review, and a perspective on the overall resources that could be required for the VQE to achieve quantum advantage.

\subsection{A formal definition of the VQE}

The VQE was first presented in Ref. \cite{Peruzzo2014} and its theoretical framework was significantly extended in Ref. \cite{mccleanTheoryVariationalHybrid2015}. It is grounded in the variational principle (and more precisely in the Rayliegh-Ritz functional \cite{Rayleigh1870, Ritz1908, Arfken1985}), which optimizes an upper bound for the lowest possible expectation value of an observable with respect to a trial wavefunction. Namely, providing a Hamiltonian $\hat{H}$, and a trial wavefunction $\ket{\psi}$, the ground state energy associated with this Hamiltonian, $E_0$, is bounded by
\begin{equation}
    E_0 \leqslant \frac{\bra{\psi}\hat{H}\ket{\psi}}{\braket{\psi | \psi}}. \label{eq:ritzfunc}
\end{equation}
The objective of the VQE is therefore to find a parameterization of $\ket{\psi}$, such that the expectation value of the Hamiltonian is minimized. This expectation value forms an upper bound for the ground state energy, and in an ideal case should be indistinguishable from it to the level of precision desired. In mathematical terms, we aim to find an approximation to the eigenvector $\ket{\psi}$ of the Hermitian operator $\hat{H}$ corresponding to the lowest eigenvalue, $E_0$.

In order to translate this minimization task into a problem that can be executed on a quantum computer, one must start by defining a so-called ansatz wavefunction that can be implemented on a quantum device as a series of quantum gates. Given that we can only perform unitary operations or measurements on a quantum computer, we do this by using parametrized unitary operations (see Sec. \ref{sec:Ansatz}). We hence express $\ket{\psi}$ as the application of a generic parametrized unitary $U(\boldsymbol{\theta})$ to an initial state for $N$ qubits, with $\boldsymbol{\theta}$ denoting a set of parameters taking values in $(-\pi, \pi]$. The qubit register is generally initialized as $\ket{0}^{\otimes N}$, written as $\ket{\boldsymbol{0}}$ for simplicity, although low-depth operations can be performed for alternative initializations before the unitary is applied. Noting that $\ket{\psi}$ (as well as any $U(\boldsymbol{\theta}) \ket{\psi}$) is necessarily a normalized wavefunction, we can now write the VQE optimization problem as
\begin{equation} \label{eq:vqe_cost_function}
    E_{\mathrm{VQE}} = \min_{\boldsymbol{\theta}} \bra{\boldsymbol{0}} U^{\dagger}(\boldsymbol{\theta})\hat{H}  U(\boldsymbol{\theta}) \ket{\boldsymbol{0}}.
\end{equation}
Eq.~(\ref{eq:vqe_cost_function}) is also referred to as the cost function of the VQE optimization problem, which is a terminology inherited from the machine learning and optimization literature. 
We can continue this description by writing the Hamiltonian in a form that is directly measurable on a quantum computer, as a weighted sum of spin operators (Sec. \ref{sec:Hamiltonian_representation} shows how the Hamiltonian can be defined, and Sec. \ref{sec:Encoding} how it can be mapped to spin operators). Observables suitable for direct measurement on a quantum device are tensor products of spin operators (Pauli operators). We can define these as Pauli strings: $\hat{P_{a}} \in \{I, X, Y, Z\}^{\otimes N}$, with $N$ the number of qubits used to model the wavefunction. The Hamiltonian can be rewritten as
\begin{equation}
    \hat{H} = \sum_{a}^{\mathcal{P}} w_{a} \hat{P}_{a},
\end{equation}
with $w_a$ a set of weights, and $\mathcal{P}$ the number of Pauli strings in the Hamiltonian. Eq.~(\ref{eq:vqe_cost_function}) becomes
\begin{equation} \label{eq:vqe_hybrid_function}
    E_{\mathrm{VQE}} = \min_{\boldsymbol{\theta}} \sum_a^{\mathcal{P}} w_a \bra{\boldsymbol{0}} U^{\dagger}(\boldsymbol{\theta})\hat{P}_{a}  U(\boldsymbol{\theta}) \ket{\boldsymbol{0}},
\end{equation}
where the hybrid nature of the VQE becomes clearly apparent: each term $E_{P_a} = \bra{\boldsymbol{0}} U^{\dagger}(\boldsymbol{\theta})\hat{P}_{a}  U(\boldsymbol{\theta}) \ket{\boldsymbol{0}}$ corresponds to the expectation value of a Pauli string $\hat{P}_a$ and can be computed on a quantum device, while the summation and minimization $E_{VQE} = \min_{\boldsymbol{\theta}} \sum_a w_a E_{P_a}$ is computed on a conventional computer.

\subsection{The VQE pipeline}

\begin{figure*}
    \centering
    \includegraphics[width=\linewidth]{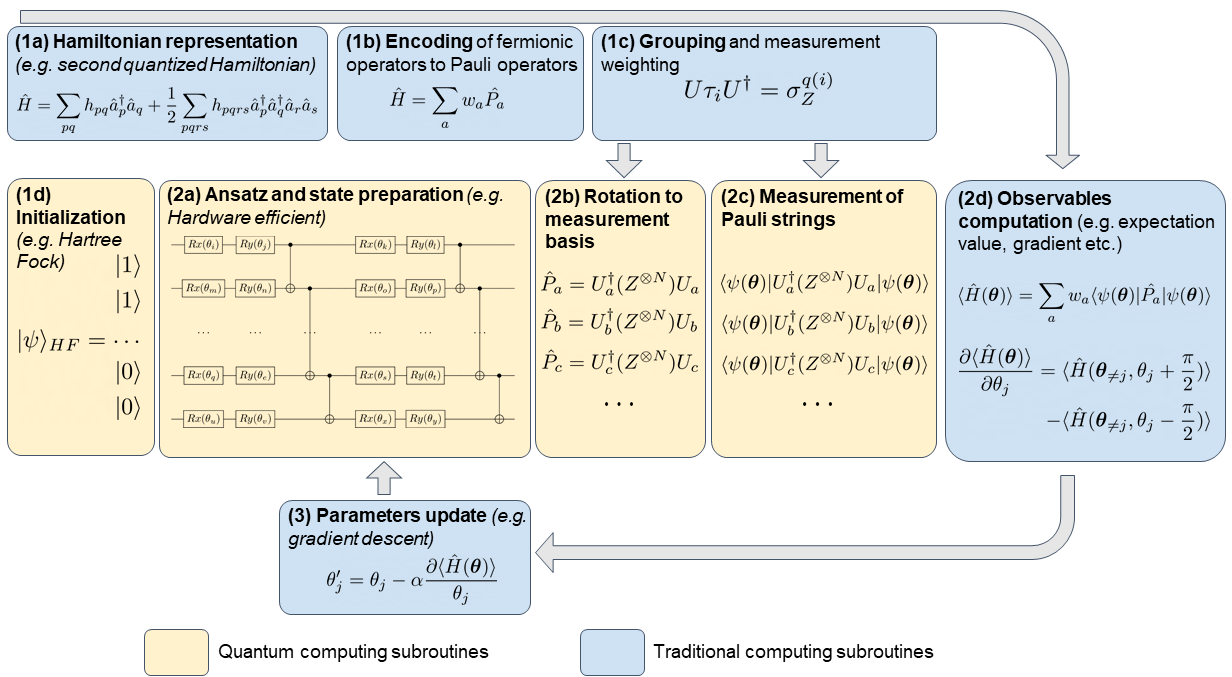}
    \caption{\textbf{The VQE Pipeline} - Formulas are illustrative and do not necessarily correspond to best practices. \textbf{(1) Pre-processing:}  \textbf{(a) Hamiltonian representation:} First pre-processing step of the VQE in which a set of basis functions is defined for the Hamiltonian to be expressed as a quantum observable of the electronic wave function (Sec. \ref{sec:Hamiltonian_representation}); \textbf{(b) Encoding:} Second pre-processing step of the VQE in which the Hamiltonian is encoded into a set of operators that can be measured on a Quantum Computer, using the qubit register wavefunction. To do this, fermionic operators in the Hamiltonian are mapped to spin operators using an encoding (Sec. \ref{sec:Encoding}); \textbf{(c) Grouping and measurement strategy:} Third step in the pre-processing, where operators defined in (b) are grouped in order to be measured simultaneously later on, usually requiring an add-on to the quantum circuit for each group in order to rotate the measurement basis in a basis in which all operators in the group are diagonalized. It is also the step in which we decide the measurement weighting strategy (Sec. \ref{sec:Grouping}); \textbf{(d) State initialization:} In this last pre-processing step, we decide how the state to which the ansatz is applied is initialized. In general, the Hartree-Fock wavefunction is used  \cite{Hartree1928, Slater1928, Gaunt1928, Hartree1935, Jensen2017}, but other options are briefly discussed in Sec. \ref{sec:ansatz_discussion} - \textbf{(2) The VQE loop:} \textbf{(a) Ansatz and trial state preparation:} The first step of the VQE loop is to apply the ansatz to the initialized qubit register, before the first iteration of the VQE all the parameters of the ansatz also need to be initialized (randomly or using a specific method, e.g. Ref. \cite{Grant2019, Patti2021}) (Sec. \ref{sec:Ansatz}); \textbf{(b) Basis rotation and measurement:} Once the trial wave function has been prepared, it must be rotated into the measurement basis of the operator of interest (quantum computers in general measure in the $Z$ basis), or a diagonal basis of a specific group of Pauli strings \textbf{(c) Observable computation:} The expectation value to be computed depend on the optimization strategy used, in any case however these are reconstituted by weighted summation on conventional hardware, or using a machine learning technique; \textbf{(d) Parameters update:} Based on the observables computed, and the optimization strategy, we can compute and apply updates to the ansatz parameters and begin a new iteration of the VQE (Sec. \ref{sec:Optimization}) - \textbf{(3) Post-processing, Error mitigation:} error mitigation is a layer of additional computation on measurement output (or directly on the quantum state prior to measurement) aimed at reducing the impact of quantum noise on the results (Sec. \ref{sec:error-mit})}
    \label{fig:VQE_pipeline}
\end{figure*}

The VQE, as presented using Eq.~(\ref{eq:vqe_hybrid_function}), can be decomposed into a number of components, which all entail significant choices that impact the design and overall cost of the algorithm. We refer to the layering of these different components as the VQE pipeline. Most choices made on specific elements of this pipeline have significant implications on the entire VQE process. We summarize the iterative process (and the main VQE loop) in Fig. \ref{fig:VQE_pipeline} to provide a schematic visual description of the algorithm and its main components. We list the key components below and provide a brief introduction to each of them and how they fit together:

\begin{itemize}
    \item \textbf{Hamiltonian construction and representation:} The first step in the VQE is to define the system for which we want to find the ground state. This can be an \textit{ab initio} molecular Hamiltonian for electronic structure \cite{Parr1990, Friesner2005, Szabo1996, Jensen2017}, a solid-state system \cite{nemoshkalenko1998computational, Martin2004, Marder2010, Continentino2021}, a spin lattice model \cite{Clark1997}, a nuclear structure problem Hamiltonian \cite{Romero2022}, or a Hamiltonian describing any other quantum system. For each of these, one starts with a specific geometry (or conformation) of the system, specifying for example the distance between each atom, or the geometry of the lattice. Constructing the Hamiltonian involves finding specific operators and their weights between basis functions spanning the physical problem, where the basis functions represent the individual single-particle degrees of freedom. Given the Hamiltonian defines the quantum observable for the total energy associated with a wavefunction, the choice of basis is critical to define the space its spans. It can have a significant impact on the accuracy and cost of the final result, as the type of basis and number of basis functions chosen both determine the size of the computation required and the accuracy of the representation. 
    In the case of electronic structure, these different representations could include, as examples, molecular orbitals from a prior mean-field calculation, plane-wave functions, or local atomic functions, all representing the spatial distributions (or `orbitals') for the single-particle Fock states, from which the many-body basis is formed \cite{Szabo1996}. 
    Additional complexity arises when specifically looking at electrons: following the Pauli exclusion principle \cite{Pauli1925, griffiths2005introduction} the wavefunction must be antisymmetric with respect to the exchange of two electrons. From a mathematical perspective, this means that we must decide whether we enforce this antisymmetry through the definition of the wavefunction or through the definition of the operators. These are referred to (for historical reasons) respectively as first and second quantization \cite{Szabo1996}.
    In second quantization the Hamiltonian is expressed in terms of fermionic operators, also known as creation ($\hat{a}^{\dagger}_j$) and annihilation ($\hat{a}_j$) operators. These correspond to the action of adding, or removing an electron from a given basis function with integer index $j$, respectively (e.g. an orbital or a lattice site), ensuring appropriate fermionic antisymmetry with respect to permutation of any two particles. We provide an introduction to Hamiltonian construction and discuss the wider implications of particular representation choices in Sec. \ref{sec:Hamiltonian_representation}.
    \item \textbf{Encoding of operators:} Qubit registers on quantum computers can only measure observables expressed in a Pauli basis, due to the two-level nature of spins: $\hat{P_{a}} \in \{I, X, Y, Z\}^{\otimes N}$, for $N$ qubits. In first quantization, the operators can be directly translated into spin operators that can be measured on quantum computers \cite{mcardleQuantumComputationalChemistry2018}, as they are not used to enforce antisymmetry of the wavefunction - we briefly outline how this is done in Sec. \ref{sec:first_quantization}. In second quantization the Hamiltonian is expressed as a linear combination of fermionic operators which are defined to obey this antisymmetry relationship, unlike Pauli operators. The role of a fermionic to spin encoding is therefore to construct observables, from Pauli operators, which maintain this relationship. A transformation of fermionic operators to spin operators that meets this criterion was demonstrated a long time ago \cite{Jordan1928}, and recent research has focused on improving on this initial work. The key factors determining the efficiency of an encoding are their Pauli weight (the maximum number of non-identity elements in a given spin operator), the number of qubits required, and the number of Pauli strings produced. We provide a list of the most relevant encodings for second quantized Hamiltonians in Sec. \ref{sec:Encoding}. It is worth noting that for certain ansatz choices, in particular those defined in terms of fermionic operators, the encoding can have significant implications on gate depth and trainability (Sec. \ref{sec:Ansatz}). Cases of encoding particles others than fermions (e.g. bosons), and which do not require antisymmetry to be enforced are far simpler and are presented directly with the relevant Hamiltonian representation in Sec. \ref{sec:Hamiltonian_representation}. 
    \item \textbf{Measurement strategy and grouping:} The next step in the VQE pipeline is to determine how measurements are distributed and organized to efficiently extract the required expectation values from the trial wave function. In general, to achieve a precision of $\epsilon$ on the expectation value of an operator, we are required to perform $\mathcal{O}(1/\epsilon^2)$ repetitions (usually denoted as shots) of the circuit execution, each completed with a measurement at the end \cite{mccleanTheoryVariationalHybrid2015}. The objective of the measurement strategy is to make the number of repetitions as low as possible. Several techniques are available to achieve this, in particular, the use of efficient weighting of the number of measurements across the operators \cite{Wecker2015, Rubin2018, Arrasmith2020}. This can be further optimized by using properties of the Lie algebra in which Pauli strings are defined. \textit{Via} processing of the encoded Pauli strings to be measured, it is possible to identify commuting groups of operators that can be measured jointly, and subsequently find the measurement bases in which all operators of a given group can be simultaneously measured \cite{Gokhale2019_long, Hamamura2020, Huggins2021} (see Sec.~\ref{sec:discussion_grouping}). In order to perform this joint measurement, a short quantum circuit must therefore be designed and applied for each group, to rotate the measurement basis and to perform this joint measurement. Alternatively, because of information overlap between different Pauli strings, one can also try to reduce the number of measurements required using inference methods from fewer shots \cite{Torlai2020, Huang2020}, with further details in Sec.~\ref{sec:Grouping}.
    \item \textbf{Ansatz and state preparation:} Once the Hamiltonian has been prepared such that its expectation value can be measured on a quantum device, we can turn to the preparation of the trial wavefunction. In order to do this, one must decide on a structure for the parametrized quantum circuit, denoted as ansatz. It is used to produce the trial state, with which the Hamiltonian can be measured. Upon successful optimization of the ansatz parameters, the trial state becomes a model for the ground state wavefunction of the system studied. A wide range of ans{\"{a}}tze are possible, and the appropriate choice depends on the problem being addressed. The key aspects of the ansatz are its expressibility and trainability. The expressibility defines the ability of the ansatz to span a large class of states in the Hilbert space \cite{Holmes2021, Nakaji2021}, defining the maximum accuracy its approximation of relevant low-energy states can achieve (assuming all parameters can be perfectly optimized). Its trainability describes the practical ability of the ansatz to be optimized using techniques tractable on quantum devices \cite{Cerezo2021_BP, Holmes2021} (related to the total number of parameters, their linear dependence, the structure of the optimization surface, and to the related concept of barren plateaus \cite{McClean2018}, which can arise when gradients almost vanish thereby preventing optimization). A good ansatz must be sufficiently expressive to guarantee that it can appropriately approximate the ground state wave function, however, it must not be so expressive that it renders the search for the target state intractable. Another important aspect of the ansatz choice is the scaling and complexity of its circuit depth with system size. This is particularly important for near-term application of the VQE, as it determines in great part the noise resilience of the method employed. Details about ansatz selection are presented in Sec. \ref{sec:Ansatz}. 
    \item \textbf{Parameter optimization:} The parameters of the ansatz used need to be updated iteratively until convergence. In general, this requires sampling the expectation value of the Hamiltonian several times for a given parameter set in the ansatz in order to define an update rule for the parameters (i.e. the updated value of the parameters is a function of the expectation value measured). The choice of optimization is critical for at least three main reasons: (1) it directly impacts the number of measurements required to complete an optimization step, as e.g. computing the numerical gradient of a quantum circuit can require value estimation of the Hamiltonian with respect to several slightly modified wave functions (this is also generally true for gradient-free methods) \cite{crooks_gradients_2019, izmaylov_analytic_2021}; (2) certain optimizers have been designed to alleviate specific optimization issues, such as the barren plateau problem \cite{haug_capacity_2021,haug_optimal_2021, nakanishi_sequential_2020,ostaszewskiStructureOptimizationParameterized2021,koczor_quantum_2020}; (3) it directly impacts the number of iterations required to reach convergence (if it allows for convergence to be reached at all) \cite{nakanishi_sequential_2020,ostaszewskiStructureOptimizationParameterized2021,koczor_quantum_2020}. We present the detailed description of the optimizers most relevant to the VQE in Sec. \ref{sec:Optimization}. 
    \item \textbf{Error mitigation:} Quantum noise is one of the main hurdles in the viability of the VQE, given that the method is to be used without error correction schemes on NISQ devices. Error mitigation aims to reduce the impact of quantum noise through post-processing of the measurement data (or occasionally through post-processing of the trial wave function ahead of measurements). Error mitigation techniques vary widely in terms of cost and benefits (see Sec. \ref{sec:error-mit}), and in general, a mix of these can be implemented jointly to achieve the best balance.
\end{itemize}

It is worth briefly outlining the distinction between the VQE pipeline and that of other variational quantum algorithms (VQAs) \cite{cerezoVariationalQuantumAlgorithms2020}. The key distinguishing feature of VQE is that it is restricted to finding the eigenstate of a quantum observable, which is not necessarily the case of other VQAs (such as Quantum Approximation Optimization Algorithms \cite{Farhi2014} or the Variational Quantum Linear Solver \cite{BravoPrieto2019, Pellow-Jarman2021}). As such, the process of encoding the Hamiltonian is peculiar to VQE. Similarly, while all VQAs would benefit from efficient measurements, the nature of the observable used in VQE (often scaling polynomially in the system size) mean that efficient grouping and measurements strategies will likely have a far greater impact on the overall scaling of the method. 

A similar distinction can be drawn with the field of machine learning using Quantum Neural Networks (QNN) \cite{Kwak2021, Garcia2022}. While such methods can be considered as variational algorithms, the reverse is not necessarily true. Most VQAs, and the VQE, aim at finding the solution to a given problem from initial inputs. Machine learning on QNN aims at abstracting, and generalizing \cite{Abbas2021}, a pattern from already solved problems used as initial input. As such the algorithmic pipeline and challenges of both methods are largely different. In the case of QNN, one will be less concerned about the representation of the observable and poorly scaling measurement requirements. Instead, deciding on the process for encoding conventional data into a quantum state (often referred to as quantum feature map) \cite{Schuld2019, Huang2021_powerofdata} will be critical in determining potential for quantum advantage and is completely absent from the VQE pipeline. 

\subsection{Advantage argument, assumptions, and limitations of the VQE} \label{sec:VQE_advantage}

Quantum supremacy is achieved when algorithms running on quantum computers can produce results that surpass those generated on conventional computing resources in accuracy and/or resources required \cite{Preskill2013}. It was demonstrated on a tailored sampling task by several research teams \cite{Arute2019, Zhong2020, Wu2021, Madsen2022}, although the magnitude of the advantage has also been contested \cite{Pednault2021, Liu2021}. Quantum advantage is in general used interchangeably with quantum supremacy, however we propose to use quantum advantage to refer to an instance of quantum supremacy where the advantage has relevant, tangible applications. This concept can rely on theoretical scaling arguments or practical demonstrations. A precise definition of the resources required and of the accuracy metrics is required for any specific demonstration of quantum advantage, which needs to include all overheads and initialization requirements. Computing resources can be defined in many different ways, including overall absolute runtime, time scaling, memory requirements, or indeed `indirect' metrics such as the financial cost of the computation or energy consumption (for a discussion on the energy consumption of Quantum Computers, we recommend Ref. \cite{Jaschke2022}).

The VQE allows for the probabilistic measurement of observables over certain classes of parameterized approximate wavefunctions, which can neither be sampled from nor have their properties computed efficiently (e.g. in polynomial time) on conventional devices as the system size gets large. Of course, this implies that the Hamiltonians studied can be written as a polynomially growing sum of independent observables \cite{Peruzzo2014}, as is commonly found in a number of fields such as quantum chemistry, condensed matter physics, or nuclear physics. We provide a more detailed description of these specific classes of wavefunctions (the ansatz of the VQE) in Sec. \ref{sec:Ansatz}, while a comparison to alternative wavefunction classes, which can admit polynomial complexity on conventional computing resources, are described in Sec. \ref{sec:vqe_vs_conventional}. If these wavefunction forms, accessible via the VQE yet practically inaccessible via conventional means, admit sufficient accuracy in their approximation to the ground state, quantum advantage can be considered within this paradigm. The argument outlined above defines a necessary condition for the VQE to become a practically useful method for computing properties of quantum systems. It is clear from the literature, and outlined in Sec. \ref{sec:sota_vqe}, that under certain assumptions this condition is theoretically achievable \cite{Peruzzo2014, Barkoutsos2018, OMalley2016, Wecker2015, Lee2019, Wiersema2020}.

There are however many restrictions of quantum computing that this approach does not take into account, and we therefore propose two more stringent conditions. The first one is that VQE must demonstrate similar or higher accuracy than any conventional method, but with lower computational time-to-solution. This condition takes into account possible limitations due to hardware runtime, potentially resulting in a large pre-factor for VQE computations. In this review, the pre-factor refers to the multiplier applied to a scaling rule to obtain the actual run time of the method. If the VQE has better asymptotic scaling than conventional methods, but a large pre-factor, this means an advantage could only be achieved in the asymptotic regime of very large systems (with runtime possibly too large for VQE to be realistically usable). This would make it difficult to demonstrate quantum advantage for practical moderately sized systems. The second condition, which is also the most stringent form of quantum advantage for the VQE, is to achieve at least as good accuracy, and with faster compute time, for a system of sufficient complexity to accurately model a real problem of physical and chemical relevance. This involves demonstrations on systems, where the approximation error in defining the specific Hamiltonian for the original problem is of smaller magnitude than its solution using the VQE. This could be as simple as ensuring that basis sets are sufficiently saturated \cite{Jensen2017}, or that the complexity of the interactions with a wider system were sufficiently resolved. For instance, consider computing the energy of a series of protein-ligand complexes (for which methods extending VQE have already been proposed \cite{Malone2021, Kirsopp2021}): even if the VQE achieves better accuracy in lower computation time, it is not guaranteed that these accuracy gains lead to a practical advantage. For example, the accuracy gains may still be insufficient to predict the most appropriate ligand in a physical experiment due to the approximations in the treatment of the environment in the Hamiltonian. Some researchers have attempted to estimate the tipping point for quantum computing-based quantum chemistry to overtake conventional methods. As one example, Elfving {\it et al.} \cite{Elfving2020} estimate the size of basis set (and hence the number of qubits) that would be required for a tangible quantum advantage of quantum computing based methods to lie somewhere between $19$ and $34$ molecular orbitals (or twice as many spin orbitals and hence twice as many qubits).

Despite sound theoretical arguments for the polynomial scaling of VQE \cite{Peruzzo2014, mccleanTheoryVariationalHybrid2015}, a number of potential limitations have been identified as well, which could prevent the VQE from achieving quantum advantage:

\begin{itemize}
    \item The VQE could be limited by a large pre-factor linked to the cost of accurate observable sampling. Several studies have analyzed the overall cost of VQE and whether it can reach a tipping point, at which it becomes advantageous compared to conventional methods \cite{Wecker2015, Elfving2020, Gonthier2020}. They have so far all concluded that given certain assumptions and the current state of research surrounding VQE functionality, the algorithm cannot outperform conventional methods within a remit of applications considered tractable. The main bottleneck identified in these studies (based on noiseless estimates) is the substantial number of measurements that are required to estimate the expectation value of the Hamiltonian using VQE (for further details see Sec. \ref{sec:resource_estimates}). The field of research is fast-moving however, and much research has been devoted to efficient operator sampling (summarized in Sec. \ref{sec:Grouping}). Using the parallelization potential of VQE (see Sec. \ref{sec:parallelization}) could also be a solution to this measurement problem but would require a paradigm shift in the way quantum hardware is conceived.
    \item The VQE involves solving an optimization problem. As such, to understand the true cost of implementing VQE, one needs to assess the complexity of the optimization process. The true cost and scaling are dependent on the optimizer and on the optimization landscape of the specific problem studied. While some optimizers have been shown to converge in polynomial time for convex cost functions, the VQE is far from having such a favorable landscape \cite{Bittel2021, Anschuetz2022}. In fact, the VQE optimization is shown to be NP-hard \cite{Bittel2021}, which means that in the worst possible case, finding the optimal solution to the problem is intractable. Of course, this is to be expected as all optimization problems can suffer from the same issue \cite{Krentel1988}. The key open question is to know whether VQE can be optimized heuristically in a polynomial number of iterations and converge to an approximate yet accurate enough solution. 
    \item Even if one could show VQE would theoretically converge in a tractable number of iterations, this would assume that expectation values and gradients are computed exactly. This assumption is not valid in the context of quantum computation, and it has been shown that the number of measurements required to accurately measure gradients could scale exponentially in certain parameterizations of systems, due to the barren plateau problem  \cite{McClean2018} (discussed in detail in Sec. \ref{sec:barren_plateau}). A number of mitigating methods have been proposed, such as the identity block initialization \cite{Grant2019} or the use of a local encoding for the Hamiltonian \cite{Cerezo2021_BP, Uvarov2020}. Nevertheless, the extent to which barren plateaus can indeed be managed for VQE remains an open question. 
    \item Related to both of the above, the extent to which VQE is resilient to quantum noise is also an open question, but actively mitigating errors will likely be unavoidable for relevant use of NISQ algorithms. Although error mitigation methods have shown great success in improving the accuracy of VQE on the current generations of quantum computers (for instance \cite{Barron2020, Sagastizabal2019, Nam2020, Arute2020, Tilly2021, Benfenati2021}), it can significantly increase the resource requirements of the algorithm (see Sec. \ref{sec:error-mit}). It is unclear whether this increase in resources is an acceptable cost or likely to be a critical limitation in larger-scale application of the VQE. A recent paper~\cite{RyujiFundamentalLimits2021} is rather pessimistic on this point, showing that the increase in cost is exponential when the ansatz circuit grows deeper. Conversely, it was suggested in the early days of VQE that variational algorithms possess inherent noise resilience since the optimization can effectively adapt to the noise~\cite{mccleanTheoryVariationalHybrid2015}. This resilience has helped VQE to be more successful than other algorithms on the current generation of quantum devices, and has been numerically demonstrated in small qubit simulations \cite{Enrico2021EvaluatingNoiseResilience}. However, it remains unclear whether this resilience from noise can be retained in larger quantum experiments, where one is confronted with a more complex ansatz, with more noise coming from the difficulty of controlling large numbers of qubits with precision.
\end{itemize}

An important additional point to stress here is that while in theory the exact state could in principle be spanned by a number of qubits that scales linearly with system size, this exactness is in general forgone in VQE via the imposed parameterization. At this point, a strictly exact limit within a defined Hamiltonian is only expected to be recovered with an exponential number of variational parameters (and hence circuit depth) \cite{Evangelista2019}. Therefore, to achieve advantage, the classes of states accessible within the VQE framework must admit superior approximations to quantum many-body systems of interest compared to accessible conventional descriptions of quantum states, as well as their scaling with system size. The key question regarding the realm of current applicability of the VQE is therefore whether it can achieve higher accuracy on at least some representative systems, with some appropriate resource metric, compared to conventional computational chemistry methods.

\subsection{VQE and conventional computational chemistry} \label{sec:vqe_vs_conventional}

The first step in any application of VQE to {\em ab initio} electronic structure is to define the basis functions determining the resolution and representation of the system. A common (but not required) approach to this would be to use the molecular orbitals obtained from a prior mean-field Hartree-Fock (for a description of this method, see  Ref.~\cite{Jensen2017}) or density functional theory (DFT) calculation \cite{Hohenberg1964, Levy1979, Vignale1987, Kohn1965} (for comprehensive reviews of DFT, see Refs.~\cite{Parr1995, Bagayoko2014}). These orbitals are used to define the representation of the Hamiltonian (see Sec. \ref{sec:Hamiltonian_representation}), and thus compute the operator weights of the resulting Pauli strings. In this way, the VQE already relies on the techniques of conventional quantum chemistry for its use. Furthermore, in order to clarify the challenge for quantum advantage, as well as the expected scope and applicability of the VQE in the context of computational chemistry, we provide a very brief review of existing methods in this domain.

Although exceptions exist, it should be noted that most conventional approaches for high accuracy {\em ab initio} ground-state energetic properties of molecular systems rely on wavefunction approximations, in keeping with the wavefunction approximation inherent in the VQE approach \cite{Eriksen2020}.
Other quantum variables (such as densities, density matrices, or Green's functions) can be used, but are in general unable to reach state-of-the-art accuracy for ground state energies \cite{Williams2020}. As such, methods like DFT which is widely used in material sciences, and offers a competitive cost-accuracy trade-off for large systems are not direct competitors to VQE, due to the lack of systematic improvability of their results. Despite some quantum algorithms for electronic structure presenting algorithms with scalings competitive or even lower than DFT (for example, Ref.~\cite{Babbush2019}), the most likely competitors for short to medium term applications of VQE are accurate wavefunction approaches, which can scale as high polynomial or even exponential, but which still are able to access comparatively large system sizes.

\subsubsection{Full Configuration Interaction} \label{sec:full_configuration_interaction}

Full configuration interaction (FCI) provides the benchmark for exact representation of a quantum state for a given Hamiltonian and basis set \cite{Szabo1996,DavidSherrill1999}. This results in the approach being in most cases intractable, with practical limits for {\em ab initio} systems currently being 18 orbitals \cite{Vogiatzis2017}. However, its numerically exact treatment of the correlated physics ensures that it occupies a unique and important position in quantum chemistry and electronic structure. FCI builds the variationally optimal wavefunction as a linear superposition of all possible configurations of electrons within the available degrees of freedom.
Whilst the inclusion of all possible configurations ensures that the final result is invariant to the precise single-particle representation of the orbitals considered, it is common to perform FCI in a basis of Hartree-Fock molecular orbitals to improve the convergence rate. The Hartree-Fock method provides the variationally optimized single Slater determinant, as appropriate for closed-shell systems \cite{Jensen2017}, approximating the ground state wavefunction at the mean-field level. In this basis, the orbitals have individual single-particle energies associated with them, since they diagonalize the single-particle Fock matrix. The structure of the FCI wavefunction then takes the following form, where the configurations can be classed by the number of particle-hole excitations they create in the reference Hartree-Fock configuration, as

\begin{equation} \label{eq:CI_wf}
    \ket{\psi}_{\text{FCI}} = c_0 \ket{\psi}_{\text{HF}} + \sum_{ia} c_{ia} \hat{a}_a^{\dagger} \hat{a}_i \ket{\psi}_{\text{HF}} + \sum_{ijab} c_{ij, ab} \hat{a}_a^{\dagger} \hat{a}_b^{\dagger} \hat{a}_j \hat{a}_i \ket{\psi}_{\text{HF}} + \dots ,
\end{equation}
where the first sum represents `singly-excited' configurations where an occupied spin-orbital, denoted by the indices $i, j, \dots$ is depopulated, and a virtual spin-orbital, denoted by the indices $a, b, \dots$ is populated (here we use terminology corresponding to electronic structure theory, however these considerations are valid for any quantum system expressed in a finite basis). This (de)population is achieved while preserving antisymmetry of the overall wavefunction, via the use of the fermionic second quantized operators, $\hat{a}^{(\dagger)}$, with more details of these found in Sec.~\ref{sec:second_quantization}. These number preserving excitations from the reference Hartree-Fock determinant can be extended to double excitations (second sum) all the way up to $m$-fold excitations, where $m$ is the number of electrons. This then spans the full space of configurations, and due to the linear parameterization, ensures that the minimization of the Ritz functional (Eq.~\ref{eq:ritzfunc}) can be written as a diagonalization of the full Hamiltonian in this basis \cite{Ross1952,Foresman1996}. Exact excited states (within the defined basis and resulting Hamiltonian) can then also be computed as successively higher-lying eigenvalues of the Hamiltonian matrix in this basis. 

While the FCI represents the `ground truth' solution for the defined combination of Hamiltonian and basis set, the core aim of much of electronic structure is to truncate the complexity of this FCI solution (ideally to polynomial scaling with system size), while minimizing the loss in accuracy resulting from this truncation \cite{Szabo1996}. It is also advantageous to have the ability to systematically relax this truncation of the approximate ansatz, allowing for improvable results when the situation demands (for instance, see the methods described in Ref.~\cite{Eriksen2020}). To this end, a large number of approximate parameterizations of the FCI wavefunction have been explored, which differ in their accuracy, scaling, functional form, and method of optimization. Many of these approaches have enabled chemical accuracy and beyond compared the FCI result to be routinely reached in systems far larger than those accessible by FCI \cite{White1992,Sharma2012,Tubman2016, Booth2010,Thomas2015,Li2018}. The technical definition of chemical accuracy is specifically the accuracy required to compute accurate enthalpies (heats) of reactions, which numerically corresponds to a method achieving an output within $1.6$ milli-Hartree (mE$_{\rm H}$) \cite{Peterson2012} of experimental results. It is widely used as a benchmark for numerical methods, although it is worth noting that other chemical properties need higher accuracy (sometimes by 2 to 3 orders of magnitude for instance in the case of spectroscopic properties). It is in general considered extremely difficult (or impossible) to reach for large systems due to the approximations which are made when constructing the model. Oftentimes computational methods aim at a correct qualitative description of the chemical properties instead. We refer to chemical precision as the benchmark for the precision at which the model is solved, irrelevant of the uncertainties and approximation made when constructing the model (see Ref.~\cite{Elfving2020} for a thorough discussion of chemical accuracy vs. precision in the context of computational chemistry). 

The considerations described in devising an effective parameterization also largely echo the developments of ans\"atze for the VQE, although the functional forms of ans\"atze which admit efficient evaluation on quantum devices are different. In the next section, we explore a few of these parameterizations which are used on conventional devices, and how these considerations have influenced and transferred over to the choice of ansatz developed in the context of the VQE.

\subsubsection{Efficient approximate wavefunction parameterizations for conventional computation}

While the complexity of the exact FCI ansatz (Eq.~\ref{eq:CI_wf}) is clearly combinatorial with the number of degrees of freedom, many accurate and more compact wavefunction forms have been established. It results in more efficient approaches than FCI which can access larger system sizes, with only small tradeoffs in accuracy. As an illustration of the capabilities of state-of-the-art methods, a recent study presents a blind test comparison of nine different methods applied to benzene on an active space of $30$ electrons and $108$ molecular orbitals \cite{Eriksen2020}. The root mean square deviation between the results produced by these methods was only $1.3 \text{mE}_{\text{h}}$, demonstrating a consistent and reliable level of accuracy between these methods, expected to be close to FCI accuracy. A similar (albeit not blind) study was conducted in Ref.~\cite{Williams2020}, with applications to transition metal systems, again showing excellent agreement between the most accurate wavefunction methodologies in these systems.

To rationalize some of these parameterizations, an obvious first approximation to Eq.~(\ref{eq:CI_wf}) can be made via truncation in the total number of excitations from the reference configuration, allowing retention of the efficient linear form. The most common of these is the configuration interaction with singles and doubles ansatz (CISD), where only up to double excitations are retained \cite{Szabo1996}. More recent adaptive, selective or stochastic inclusion of desired configurations exploit the sparsity in the optimized amplitudes, and can extend the ansatz further in accuracy, resulting in methods such as Adaptive Sampling CI (ASCI) \cite{Tubman2016, Tubman2020, Hait2019, Levine2020}, Semistochastic Heat-Bath CI (SHCI) \cite{Petruzielo2012, Holmes2016, Holmes2016_2, Sharma2017, Smith2017, Holmes2017, Li2018}, or Full Configuration Interaction Quantum Monte Carlo \cite{Booth2010, Cleland2012, Anderson2020, Booth2011, Blunt2019}. However, these truncated linear approximations can suffer from size-intensive total energies, where the energy does not scale appropriately with respect to the number of electrons, ensuring that the energy error per particle becomes increasingly large as systems grow in size \cite{Szabo1996}. Nevertheless, they can result in excellent variational approximations to FCI for small systems. An alternative approach is to construct a multi-linear approximation to the FCI wavefunction, which results in the matrix product state functional form. This form can be efficiently optimized within the density matrix renormalization group (DMRG), and can also yield accurate and systematically improvable approximations to FCI both in the case of \textit{ab initio} molecular Hamiltonian and for lattice models \cite{White1992, White1993, White1999, Mitrushenkov2001, Legeza2003, Chan2002, Chan2011, Sharma2012, OlivaresAmaya2015, Wouters2014, Yanai2014, Knecht2016}. More broadly, the development of tensor network theory \cite{Biamonte2017, Ors2019} and the use of matrix product states has resulted in significant improvements of methods for the resolution of lattice models (for instance \cite{Schneider2021, Kaneko2022}). Finally, it is worth mentioning that a larger class of approximate wavefunction ansatz are able to be optimized within the framework of `Variational Monte Carlo'. In these approaches, an approximate ansatz is chosen whose probability amplitude can be efficiently sampled at arbitrary electron configurations, but where a closed-form polynomial expression for the energy of the state is not accessible \cite{TOULOUSE2016285,Needs2020,Pfau2020}. Within this criteria, the parameters of the ansatz can be optimized via Monte Carlo integration of the energy functional in a very general framework, albeit with the necessity of controlling for stochastic errors and care in optimization of the parameters. Many of these considerations transfer through to the VQE.

Largely to correct for the size inconsistency in linear ans\"atze, the popular coupled-cluster ansatz truncates and then exponentiates the form of Eq.~(\ref{eq:CI_wf}). This results in an appropriately sized consistent method, with an excellent accuracy vs. cost balance \cite{Eriksen2016_I, Eriksen2016_II, Ying2019}. The coupled-cluster with single, double and perturbative triple excitations retained in the ansatz (known as CCSD(T)) is often referred to as the `gold standard' of quantum chemistry where the correlations are not too strong \cite{Bartlett2007}, while other approximate coupled-cluster forms suitable for stronger correlation effects have also been developed (for recent examples, see Ref.~\cite{Wang2020_BCCC, Lyakh2011}). The coupled cluster is also the motivation of the use of the {\em unitary} coupled cluster ansatz of VQE \cite{Evangelista2019}, where a similar structure of exponentiated excitations based around a reference configuration is constructed (see Sec.\ref{sec:UCCA}), with modifications to ensure efficient use on a unitary set of quantum circuits. Similar considerations of dynamic inclusion of additional excitations is also possible with the ADAPT-VQE ansatz \cite{Grimsley2019} (Sec.~\ref{sec:adapt-vqe}). Furthermore, the Efficient Symmetry Preserving ansatz \cite{Gard2020} looks to ensure the ability to systematically improve its span of the FCI description of Eq.~(\ref{eq:CI_wf}) ensuring the preservation of symmetries inherent in its form. However, ensuring this systematic coverage of the FCI ansatz means that this form remains exponential in the system size in a number of realistic cases, meaning that true FCI may also be out of reach for the VQE.

An important consideration in the application of these approximate conventional parameterizations is that the size of the errors is different for different systems. Over time and use, an understanding has emerged from both theoretical and numerical analysis for the domain of applicability of these approaches and physical properties of the state which enable their accuracy, e.g. low-rank excitations (coupled cluster), locality (DMRG) or sparsity in the state (selected CI, FCIQMC). This understanding has promoted their effective use in appropriate circumstances, and stimulated further developments to improve their accuracy and scope. This analysis of the errors in different systems from VQE ansatz for quantum simulation, as well as the reasons underpinning or limiting their accuracy, is only starting to be performed, with more work necessary to fully classify and numerically investigate the approximations made in their form \cite{Evangelista2019, Grimsley2019_UCC_Review}.

Overall, these established wavefunction methods based on conventional computing (some of which are briefly described in this section) constitute the state of the art in high-accuracy quantum chemistry, at least for the ground state energetics. It should be stressed again that these approaches, as opposed to FCI, constitute the ultimate benchmark on which the success of VQE should be measured, as they represent approaches to systematically achieve chemical accuracy but with greater efficiency than exact FCI. This constitutes a demanding target for VQE to meet, with many decades of research in this area. Furthermore, continuing research for other parameterizations suitable for conventional devices, such as the rapidly emerging field of machine-learning inspired ans\"atze \cite{Carleo2017, Nomura2017, Choo2018, Glielmo2020, Luo2019, Pfau2020, Choo2020, Hermann2020}, will continue to push the boundaries of accuracy on conventional devices to challenge the criteria for VQE superiority.

\subsection{VQE and Quantum Phase Estimation} \label{sec:vqe_vs_qpe}
 
The Quantum Phase Estimation algorithm (QPE) \cite{Kitaev1995, Abrams1997, abramsQuantumAlgorithmProviding1998, Cleve1998, AspuruGuzik2005} provides a method to find a given eigenvalue of a Hamiltonian from an approximated eigenstate (ground or excited). QPE can compute an eigenvalue to a desired level of precision with a probability proportional to how close the approximated eigenstate is to the true eigenstate \cite{McClean2014}. It does so however using quantum circuits of depths that are far beyond what is achievable in the NISQ era of quantum computing \cite{reiherElucidatingReactionMechanisms2017}. As part of our discussion on the VQE we briefly outline QPE and how the two compare.    

\subsubsection{Overview of the quantum phase estimation}

\begin{figure}[ht]
\centerline{
        \Qcircuit @C=2.0em @R=1.5em {
          &      &\gate{\Had} & \qw & \qw & \qw & \hdots &  & \ctrl{4} &  \multigate{3}{\QFT^{-1}} &\meter  \\
            \lstick{\raisebox{1.2em}{ $\ket{0}^{\otimes N_q}$}} &\vdots&   &    &&  & \vdots &   & & & \vdots\\
          &      &\gate{\Had} & \qw & \ctrl{2} & \qw & \hdots &  & \qw &  \ghost{\QFT^{-1}} &\meter  \\
          &      &\gate{\Had} & \ctrl{1} & \qw  & \qw & \hdots &  & \qw &   \ghost{\QFT^{-1}} &\meter  \\
          \lstick{\raisebox{1.2em}{ $\ket{\psi_{\sim \lambda_0}}$}} &   & \qw & \gate{U^{(0))}} & \gate{U^{(1)}} & \qw & \hdots &  & \gate{U^{(N_q-1)}} & \qw & \qw
}
}\caption{Quantum circuit for Quantum Phase Estimation}  \label{fig:qpe_circuit}
\end{figure}
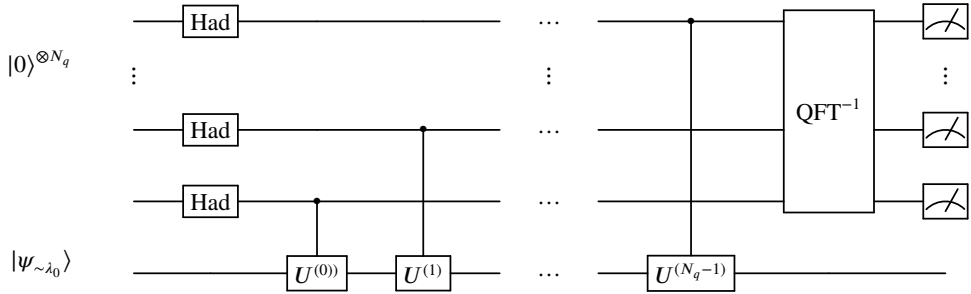

A representation of the quantum circuit used to implement QPE is presented in Fig. \ref{fig:qpe_circuit}, and the process can be described as follows (adapting the descriptions in Refs. \cite{nielsenQuantumComputationQuantum2010, mcardleQuantumComputationalChemistry2018}):
\begin{itemize}
    \item The objective of QPE, like the objective of VQE, is essentially to compute an eigenvalue of a Hamiltonian. In the case of QPE however, the problem presented in Eq. (\ref{eq:vqe_cost_function}) is slightly reformulated. For a given Hamiltonian $\hat{H}$, and a given eigenstate $\ket{\lambda_j}$ (usually the ground state:  $\ket{\lambda_0}$), one tries to find a eigenvalue $E_j$ such that:
    \begin{equation} \label{eq:qpe_problem}
       e^{i\hat{H}} \ket{\lambda_j}= e^{ i E_j} \ket{\lambda_j}.
    \end{equation}
    The Hamiltonian is exponentiated to obtain a unitary operator, and without loss of generality we can write $e^{iE_j} = e^{2 \pi i\theta_j}$, with $\theta_j$ the `phase' QPE aims at discovering.
    \item The only inputs available however are the Hamiltonian, and an approximation of the ground state $\ket{\psi_0} \sim \ket{\lambda_0}$, which can be generally expressed in the eigenbasis of the Hamiltonian as 
    \begin{equation}
        \ket{\psi_0} = \sum_{j=0}^{2^{N_q}-1} \alpha_j \ket{\lambda_j},
    \end{equation}
    where $N_q$ is the number of qubits used to represent the electronic wavefunction of the Hamiltonian (which therefore has a total of $2^{N_q}$ eigenstates.
    \item A register of ancilla qubits is used to map the eigenvalue sought-after, in general to a binary number. The number of ancillas required therefore depends on the type of implementation and desired precision (more ancillas mean a longer binary string, and therefore a higher precision \cite{nielsenQuantumComputationQuantum2010}). This ancilla register is initialized as an equally weighted superposition of all possible state in the computational basis (all possible binary strings). If we have a total of $N_a$ ancilla qubits, there are $2^{N_a}$ such basis elements. Starting from a register in state $\ket{0}^{\otimes N_a}$, a Hadamard gate ($\Had$) is applied to each qubit (for readers requiring a brief introduction to key quantum computing concepts and operations, we recommend Sec. II.A of Ref.~\cite{mcardleQuantumComputationalChemistry2018}). We recall that $\Had\ket{0} = (\ket{0} + \ket{1})/\sqrt{2}$, and $\Had\ket{1} = (\ket{0} - \ket{1})/\sqrt{2}$. After these operations, the state of the ancilla register is
    \begin{equation}
        \ket{\psi_{\mathrm{anc}}} = \frac{1}{\left(\sqrt{2}\right)^{N_a}}\sum_{x=0}^{2^{N_a} - 1} \ket{\operatorname{bin}(x)},
    \end{equation} 
    where $x$ represents integers from 0 to $2^N_a - 1$ and $\operatorname{bin}(x)$ the binary representation of of $x$. When both the ground state approximation and the ancilla register are considered together we get the total state of the qubit register $\ket{\psi_{\mathrm{tot}}} = \ket{\psi_{\mathrm{anc}}}\otimes \ket{\psi_0}$ such that
    \begin{equation}
        \ket{\psi_{\mathrm{tot}}} = \left(\frac{1}{\left(\sqrt{2}\right)^{N_{a}}} \sum _{x=0}^{2^{N_{a}} -1}\ket{\operatorname{bin} (x)}\right) \otimes \left(\sum _{j=0}^{2^{N_{q}} -1} \alpha _{j}\ket{\lambda _{j}}\right) = \frac{1}{\left(\sqrt{2}\right)^{N_a}} \sum_{j=0}^{2^{N_q}-1} \sum_{x=0}^{2^{N_a} - 1} \alpha_j \ket{\operatorname{bin}(x)} \otimes \ket{\lambda_j}.
    \end{equation}    

    \item In the superposition state above, there is no clear relation between an ancilla state $\ket{\operatorname{bin}(x)}$ and the $j$-th eigenstate $\ket{\lambda_j}$. That is, if one were to measure the ancillas resulting in a binary number $\operatorname{bin}(x)$, no information could be gained about the state of the wavefunction register which encodes $\ket{\lambda_j}$, and therefore no information can be gained about the associated eigenvalues. In the following, we will apply unitary gates to this superposition state such that there is a clear one to one correspondence between a measured binary number $\operatorname{bin}(x)$ and the eigenvector $\ket{\lambda_j}$. Consider the following unitary
    \begin{equation} \label{eq:qpe_unitary}
        U^{(k)} = \left(e^{i\hat{H}}\right)^{2^k},
    \end{equation}
    with $k$ an arbitrary number for the time being. Following Eq. (\ref{eq:qpe_problem}), if this unitary is applied to eigenstate $\ket{\lambda_j}$, it effectively results in a phase $e^{(2\pi i\theta_j 2^k)}$. Now suppose that $k$ is the index of the ancilla qubits, i.e. $k \in [0, N_a - 1]$ and that, for each $k$, $U^{(k)}$ is applied to the ground state approximation only if the ancilla qubit of index $k$ is in state $\ket{1}$. This operation can be performed by mean of a controlled unitary operation, which applies a unitary operation subject to the value of a control qubit \cite{nielsenQuantumComputationQuantum2010}. For a superposition instance $\ket{\operatorname{bin}(x)}$ of the ancilla register, this means that the unitary $U$ is applied $x$ times in total to the ground state (consider for example the ancilla superposition $\ket{\operatorname{bin}(5)} = \ket{101}$, here qubits are indexed from right to left to correspond to binary strings. The unitary is applied for $k = 0$, and for $k = 2$, hence following Eq. (\ref{eq:qpe_unitary}) it is applied $5 = 1 \cdot 2^2 + 1 \cdot 2^0$ times). We obtain the state
    \begin{equation} \label{eq:qpe_post_CU}
        \ket{\psi_{\mathrm{tot}}} = \frac{1}{\left(\sqrt{2}\right)^{N_a}} \sum_{j=0}^{2^{N_q}-1} \sum_{x=0}^{2^{N_a} - 1} e^{2\pi i \theta_j x} \alpha_j \ket{\operatorname{bin}(x)} \ket{\lambda_j}.
    \end{equation}
    \item The next step reduces the number of superposition instances by applying an inverse quantum Fourier transform (QFT) to the ancilla register. QFT is a transformation from the computation basis to the Fourier basis, mapping a single computational basis element $\ket{\operatorname{bin}(y)}$ to a superposition of all computational basis elements each with different relative phase (due to the periodicity of the phase, each relative phase is a different point on the $2\pi$ period, with a total of $2^{N_a}$ different points)
    \begin{equation}
        \QFT\ket{\operatorname{bin}(y)} = \sum_{y=0}^{2^{N_a}-1} e^{2\pi i (xy / 2^{N_a})} \ket{\operatorname{bin}(x)}.
    \end{equation}
    If we set $y = 2^{N_a}\theta_i$, we can observe that applying the inverse QFT to the ancilla register in Eq.(\ref{eq:qpe_post_CU}) results in
    \begin{equation}
        (\QFT^{-1} \otimes I^{\otimes N_q}) \ket{\psi_{\mathrm{tot}}} = \sum_{j=0}^{2^{N_q} - 1} \alpha_j \ket{\operatorname{bin}(2^{N_a}\theta_j)} \ket{\lambda_j},
    \end{equation}
    where for simplicity we have assumed $2^{N_a}\theta_i \in \mathbb{N}$.
    \item Measuring the ancilla register in the $Z$ basis returns the binary string $\operatorname{bin}(2^{N_a}\theta_i)$ with probability $|\alpha_j|^2$, from which $\theta_i$, and $E_i$ can be recovered easily. The complete qubit register then collapses to the state $\ket{\operatorname{bin}(2^{N_a}\theta_i} \ket{\lambda_j}$.
\end{itemize}

\subsubsection{Discussion and comparison}

Due to the ground state being measured directly in binary form in QPE, the number of ancilla qubits required is directly related to the precision $\epsilon$ targeted. Ancilla qubits provide one bit of information each, and as such, their number scales $\mathcal{O}(1/\log_2(\epsilon))$ in precision. The number of controlled unitaries is doubled for each ancilla qubit, and therefore scales $\mathcal{O}(1/\epsilon)$. These unitaries are effectively representing the action of the Hamiltonian on a state. The core component of QPE is therefore efficient Hamiltonian simulation (for an overview of relevant methods we recommend Ref.~\cite{Childs2018, Cao2019_QC}). Provided a non-restricted pool of qubits, Babbush \textit{et al.} \cite{Babbush2018} show that by engineering the correct Hamiltonian representation (namely, a plane wave basis in first quantization, see Sec. \ref{sec:Hamiltonian_representation}), one can achieve sub-linear scaling in the number of basis elements for Hamiltonian simulation.  

As mentioned above, QPE is only as likely to succeed as the fidelity of the input state to the unknown target eigenstate approaches one. In turn, this implies that using a randomized state as input is not an option as its expected fidelity to the target eigenstate approaches zero exponentially in the system size, resulting in QPE becoming exponentially costly with imperfect input state preparation \cite{McClean2014}. Numerous methods have indeed been proposed to prepare a good enough approximation of the target eigenstate in a tractable manner, often grounded in conventional quantum chemistry (for example\cite{tubman2018postponing, Sugisaki2018, Sugisaki2019, Motta2019, Murta2021}) or in adiabatic quantum computation (for example \cite{AspuruGuzik2005, Albash2018, Matsuura2020}). 

There have been a number of successful implementations of QPE on quantum devices \cite{Du2010, Lanyon2010, Li2011, Wang2015, OMalley2016, Paesani2017, Santagati2018}. These have only been on small systems, as large scale implementations require quantum resources which are not currently available. In particular, large scale controlled unitaries, required for QPE, cannot be reliably implemented on NISQ devices. This is also the case for the inverse QFT. Several numerical studies have been performed to assess the complete cost of implementing QPE on relevant systems, and estimate runtime on a fault tolerant quantum computer. The problem of nitrogen fixation has become a \textit{de facto} benchmark for this algorithm \cite{Beinert1997}. Reiher \textit{et al.} \cite{reiherElucidatingReactionMechanisms2017} estimates that the 54 electrons, 108 spin orbitals of $\mathrm{FeMo}$-co would require over $\mathcal{O}(10^{15})$ T gates, 200 millions qubits and would need to run for over a month to obtain quantitatively accurate results (assuming $100$ ns gate times and error threshold of $10^{-3}$). Berry \textit{et al.} rely on qubitization (a method which aims at transforming the evolution operator into a quantum walk) \cite{Low2019} to reduce the gate requirements to $\mathcal{O}(10^{11})$ Toffoli gates, despite an extended active space \cite{Li2019}. Lee \textit{et al.} \cite{Lee2021} further improve on these results and estimate they could perform this energy computation with four million physical qubits and under four days of runtime, with a similar $\mathcal{O}(10^{-3}$) error threshold. There have been many resource estimates performed for condensed matter models (for instance \cite{Babbush2018_Ham_Sim, Kivlichan2020}), with estimates as low as $\sim 500,000$ physical qubits running for a few hours to solve a $100$ site version of the Fermi-Hubbard model. Finally Elfving \textit{et al.} \cite{Elfving2020} estimate that with similar error rates, the chromium dimer ($\mathrm{Cr_2}$) with an active space of $52$ spin orbitals and $26$ electrons, would require $\mathcal{O}(10^{7})$ physical qubits running for about 110 hours. Research has progressed rapidly, and despite estimated runtimes and hardware requirements which remain daunting, offers a promising outlook for QPE, at least on targeted quantum chemistry tasks (examples of which are suggested in Ref.~\cite{Elfving2020}). 

The VQE trades off the depth and number of qubits required under QPE with a higher number of measurements and repetitions of the circuit, as well as the constraints of an approximate ansatz for the state. As presented in Ref.~\cite{Wang2019}, QPE requires $\mathcal{O}(1)$ repetitions with circuit depth scaling $\mathcal{O}(1/\epsilon)$ in precision $\epsilon$, VQE requires $\mathcal{O}(1/\epsilon^2)$ shots with circuit depth scaling $\mathcal{O}(1)$ in precision. While many other factors affect the overall time scaling of both methods, this point illustrates the asymptotic efficiency of QPE compared to VQE assuming access to fault tolerant quantum computers, but also the resource efficiency of VQE over QPE for NISQ-era devices. The frontier between NISQ and fault tolerant quantum computation is blurry, and as pointed out by Wang \textit{et al.} \cite{Wang2019} so is the frontier between VQE and QPE. They present an interpolation between the two algorithms, labeled Accelerated VQE (or $\alpha$-VQE), which uses smaller scale QPE calculations as sub-routines for the VQE. This method introduces a parameter $\alpha \in [0, 1]$ which allows tuning of the circuit depth $\mathcal{O}(1/\epsilon^{\alpha})$ and number of samples $\mathcal{O}(1/\epsilon^{2(1 - \alpha)})$ (one recovers the QPE scaling if $\alpha=1$, and the VQE scaling if $\alpha=0$). In general, rather than being mutually exclusive methods for solving an electronic structure problem, VQE and QPE are likely to provide the most benefit when combined as complementary approaches, offering algorithmic flexibility that can be adjusted depending on the progress of quantum hardware. 

\subsection{Our suggested best practices for VQE and their scaling assessment} \label{sec:sota_vqe}

In this section, we focus on combining compatible methods throughout the VQE pipeline, which offer the most promising scaling without compromising excessively on accuracy. The definition of a series of best practices for the VQE may suffer from many pitfalls since there remain many open research questions that affect the choice of optimal methods. It is also worth noting that it is likely that a method that is optimal for one system is not for another, and that this optimal compromise will change as quantum hardware improves.
With this in mind, we provide some suggestions for best practices on current devices for two broad families of systems. In particular, we can distinguish between lattice models \cite{Clark1997, nemoshkalenko1998computational,Marder2010, Continentino2021} and \textit{ab initio} molecular systems \cite{Parr1990, Friesner2005, Szabo1996, Jensen2017}. These two categories usually require different encodings, measurement strategies, and ans{\"{a}}tze. Table \ref{tab:sota_VQE} summarizes the most promising VQE methods that we have identified, together with their scaling. 
The key distinctive factor separating \textit{ab initio} molecular systems and lattice models is that the former makes no assumption on the range and type of interaction between the fermionic modes (beyond it being a two-body interaction), while the latter usually has a simplified and parameterized form which often only connects fermionic modes following a nearest-neighbor lattice structure and/or features a lower effective rank of interactions. 

As noted in the introduction, while the majority of the literature on VQE relates to electronic structure computation and lattice models, other applications have been proposed. Proposing best practices for these alternative applications is challenging as the research is sparse and therefore we avoid discussing these in this section. 

\begin{center}
\begin{longtable}{p{2.5cm} p{2cm} p{5cm} p{5cm}}
\caption{Summary of state of the art methods identified for the VQE for both \textit{ab initio} molecular systems and lattice models. These methods and scalings are indicative only, as there remain a number of uncertainties with respect to their behavior on large scale systems and in noisy environments.} \label{tab:sota_VQE} \\
\hline\hline
\textbf{Task} & \textbf{\phantom{placeholder}} & \textbf{\textit{ab initio} Molecular systems} & \textbf{Lattice models}
\\ \hline 
\endfirsthead
\hline
\multicolumn{4}{c}{\bfseries \tablename\ \thetable{} -- continued from the previous page} 
\\ \hline
\hline \textbf{Task} & \textbf{\phantom{placeholder}} & \textbf{\textit{ab initio} Molecular system} & \textbf{Lattice models}
\\ \hline 
\endhead
\hline \multicolumn{4}{r}{{Continued on next page}} \\ \hline
\endfoot
\hline \hline
\endlastfoot
\multirow{3}{=}{Hamiltonian construction \\(Sec. \ref{sec:Hamiltonian_representation})
} & Method & \textbf{Second quantization} & \textbf{Second quantization} 
\\
\phantom{} & Scaling & $\mathcal{O}(n^4)$ Hamiltonian terms and $N=\mathcal{O}(n)$ qubits, with $n$ number of basis functions & \textit{idem} \\
\phantom{} & Comments & First quantization could be advantageous on some systems, but further research is needed & \textit{idem} 
\\\hline 
\multirow{3}{=}{Fermion to spin encoding  \\(Sec. \ref{sec:Encoding})}
& Method & \textbf{Ternary tree Encoding} \cite{Jiang2020} & \textbf{Generalized Superfast Encoding} \cite{Setia2019} \\
\phantom{} & Scaling & Number of operators: $\mathcal{O}(N^4)$, Pauli weight $\mathcal{O}(\log_3(2N))$, & Qubit number: $N=\mathcal{O}(nd/2)$; for a regular lattice, number of operators scales $\mathcal{O}(ND)$, where $D$ is the lattice dimension; Pauli Weight: $\mathcal{O}(\log_2(d))$, with $d$ the fermionic-interaction graph maximum degree\\
\phantom{} & Comments & Low weight encodings could result in more resilience to barren plateau \cite{Cerezo2021_BP, Uvarov2020, Uvarov2020_frustrated}, and more compact ans\"atze, though there is at least some suggestions that it may increase impact of quantum noise \cite{Sawaya2016}  & \textit{idem}, the Compact encoding \cite{Derby2021, Derby2021_part2} results in a lower number of qubits for $D=2, 3$ but has not yet been generalized to higher dimensions. 
\\\hline 
\multirow{3}{=}{Grouping and measurement strategy \\ (Sec. \ref{sec:Grouping})}
& Method & \textbf{Decomposed interactions~} \cite{Huggins2021, Yen2021_Cartan} & \textbf{Qubit-wise commutation} \cite{mccleanTheoryVariationalHybrid2015, Kandala2017, Hempel2018, Rubin2018, Kokail2019, Izmaylov2019, Nam2020, Verteletskyi2020, Hamamura2020, Gokhale2019_long} \\
\phantom{} & Scaling & Operators to measure reduced to $\mathcal{O}(N)$; additional basis rotation circuit depth $\mathcal{O}(N/2)$ & Operators to measure reduced by a scalar, Additional basis rotation circuit depth $\mathcal{O}(1)$ \\
\phantom{} & Comments & Full rank optimization (in particular its extensions) \cite{Yen2021_Cartan} seem to achieve better overall measurement reduction for a given precision $\epsilon$ than the basis rotation method \cite{Huggins2021}, but cost scaling remains unclear. Classical shadows \cite{Huang2020} have been shown in at least one case \cite{OBrien2021} to outperform the scaling of decomposed interactions, though further numerical studies will be required to demonstrate dominance. & QWC grouping benefits from low Pauli weight encoding and comes at virtually no cost
\\\hline 
\multirow{3}{=}{Ansatz \\ (Sec. \ref{sec:Ansatz})} & Method & \textbf{k-UpCCGSD} \cite{Lee2019} & \textbf{Hamiltonian variational ansatz (HVA)~} \cite{Wecker2015, Wiersema2020}\\
\phantom{} & Scaling & Circuit depth of $\mathcal{O}(kN)$, number of parameters $\mathcal{O}(kN^2)$ & Circuit depth and number of parameters: $\mathcal{O}(k\tilde{C})$, with $\tilde{C}$ the number of commutative groups in the Hamiltonian (at most $\mathcal{O}(ND)$ for a regular lattice)\\
\phantom{} & Comments & Promising scaling, and good accuracy \cite{Lee2019, Grimsley2019_UCC_Review} but uncertainty remains for applications on large highly correlated systems. Uncertainty around $k$, the number of repetitions required. Adaptive ans\"atze \cite{Grimsley2019, Yordanov2020_IQEB, Tang2021} may perform better, but their scaling requires more investigation. & HVA has shown resilience to barren plateau and efficacy on lattice models \cite{Wiersema2020}. $k$ is the number of repetition of the anstaz required to reach the desired accuracy.
\\\hline 
\multirow{3}{=}{Optimizer \\ (Sec. \ref{sec:Optimization})} & Method & \textbf{Rotosolve} \cite{Vidal2018,nakanishi_sequential_2020,ostaszewskiStructureOptimizationParameterized2021} or \textbf{Fraxis} \cite{Watanabe2021WatanabeOptimizingSelection, Wada2021SimulatingCircuits} & \textbf{Rotosolve} \cite{Vidal2018,nakanishi_sequential_2020,ostaszewskiStructureOptimizationParameterized2021} or \textbf{Fraxis} \cite{Watanabe2021WatanabeOptimizingSelection, Wada2021SimulatingCircuits}\\
\phantom{} & Scaling & Requires sampling three values for each parameter at each step & \textit{idem} \\
\phantom{} & Comments & Some indication of faster convergence \cite{ostaszewskiStructureOptimizationParameterized2021}, but does not allow for full potential for parallelization of VQE, and requires more values to sample than most optimizers & \textit{idem}
\\\hline 
\multirow{3}{=}{Error mitigation strategy \\ (Sec. \ref{sec:error-mit})}
& Method & \textbf{Symmetry verification} \cite{mcardleErrorMitigatedDigitalQuantum2019,bonet-monroigLowcostErrorMitigation2018} and \textbf{extrapolation based methods} \cite{temmeErrorMitigationShortDepth2017,liEfficientVariationalQuantum2017} & \textit{idem} \\
\phantom{} & Scaling & Exponential with respect to the circuit depth & \textit{idem} \\
\phantom{} & Comments & 
The recommendation reflects the \textit{de facto} method of choice for experiments. A fair comparison of the performance and cost between different error mitigation methods requires further research.
& \textit{idem}
\\ \hline\hline
\end{longtable}
\end{center}

\subsubsection{Best practices for \textit{ab initio} electronic structure of molecular systems}

\paragraph{Hamiltonian construction:} In the case of an \textit{ab initio} molecular system, the Hamiltonian representing its electronic energy landscape is initially defined by a series of atoms and the spatial coordinates of their nuclei (see Sec. \ref{sec:Hamiltonian_representation}). The first choice to make is the basis in which the Hamiltonian is expressed. This directly impacts the number of qubits required for the implementation of VQE, which is proportional to the number of basis functions. Since the number of qubits is a limited resource in NISQ, we recommend using a molecular orbital basis, as it is in general more compact for a given target accuracy (compared to, for example, atomic or plane wave bases). Once a basis is decided upon, we must choose whether the Hamiltonian is prepared in first quantization (antisymmetry maintained by the wavefunction) or second quantization (antisymmetry maintained by the operators, see Sec. \ref{sec:Hamiltonian_representation}). The number of qubits in first quantization scales as $\mathcal{O}(m\log_2(n))$ \cite{Abrams1997, Berry2018}, with $m$ the number of electrons and $n$ the number of basis functions, against $\mathcal{O}(n)$ in second quantization \cite{Jordan1928}. The former also requires additional depth to enforce the antisymmetry of the wavefunction. There has not been, to the best of our knowledge, a rigorous study of the efficiency of using first quantization in VQE. While the scaling for first quantization could be advantageous in systems with few electrons and large basis sets (e.g. if a plane wave basis is used \cite{Babbush2018}), second quantization is generally preferred. 

\paragraph{Encoding:} The next decision to take is that of the mapping used to transform the fermionic, second quantized Hamiltonian into a weighted sum of Pauli operators (see Sec. \ref{sec:Encoding}). The most relevant encodings for \textit{ab initio} molecular system include Jordan-Wigner \cite{Jordan1928}, the Parity \cite{Seeley2012}, Bravyi-Kitaev \cite{Bravyi2002, Seeley2012, Tranter2015}, and ternary tree mappings \cite{Jiang2020} (all are explained in detail in Sec. \ref{sec:gen_encoding}). Out of these, the most promising encoding is the ternary tree mapping \cite{Jiang2020}, since asymptotically it has the lowest Pauli weight (maximum number of non-identity Pauli operators in the string), resulting in lower circuit depth and possibly higher resilience to the barren plateau problem \cite{Cerezo2021_BP, Uvarov2020}. It is however still unclear whether this lower circuit depth does indeed result in more noise resilience, as pointed out in Ref. \cite{Sawaya2016}. We also recommend that the resulting qubit Hamiltonian is further reduced using tapering off methods based on symmetries \cite{bravyi_tapering_2017, Setia2020, Kirby2021_CSVQE} (see Sec. \ref{sec:tappering_qubits}).

\paragraph{Measurement strategy:} The large number of measurements required to sample the numerous terms in the Hamiltonian is often cited as the most detrimental bottleneck of VQE \cite{Wecker2015, Elfving2020, Gonthier2020}. Deciding on an efficient strategy for grouping and measuring Hamiltonian terms can go a long way in reducing this bottleneck. The decomposed interactions methods \cite{Huggins2021, Yen2021_Cartan} provide on balance the most promising means to jointly measure the Hamiltonian. They allow measuring an entire molecular Hamiltonian with $\mathcal{O}(N)$ groups (for more details about this method, see Sec. \ref{sec:DecomposedInteractions}) and require additional circuit depth of $\mathcal{O}(N)$ to perform the necessary basis rotation, which is usually acceptable since this scaling is equivalent or better than for most ans{\"{a}}tze. While it was shown that methods using grouping based on general commutativity of Pauli strings (e.g. \cite{Gokhale2019_short, Hamamura2020}) require fewer shots to achieve a given accuracy \cite{Yen2021_Cartan} in some numerical studies (in particular when using the Sorted Insertion heuristic \cite{Crawford2021}), this reduction will likely not be worth the additional circuit depth scaling $\mathcal{O}(N^2)$ \cite{Gokhale2019_long} required to perform the joint measurements. It is also worth noting that among the decomposed interactions methods, the Variance-estimate Greedy Full Rank Optimization \cite{Yen2021_Cartan} appears to perform best, although it requires minimization search of decomposition parameters. While this cost could be tractable there has been, to the best of our knowledge, no thorough research on how it would scale on large systems. For this reason, the Basis Rotation Group methods \cite{Huggins2021}, which have a predictable cost, is a more cautious choice currently. For additional efficiencies and variance reduction, one can distribute shots according to the weights of each group in the Hamiltonian \cite{Wecker2015, Rubin2018,Arrasmith2020}. It is worth noting however that the $\mathcal{O}(N)$ scaling in number of groups is not an ideal proxy for the scaling in number of measurements required to achieve a given precision on observable estimation. This is due to possible covariances arising from the joint measurement of different operators \cite{mccleanTheoryVariationalHybrid2015}. Classical shadows \cite{Huang2020} is also a promising method for reducing measurement count in VQE and has been shown in one study to have a better asymptotic scaling than decomposed interactions \cite{OBrien2021}. Further numerical studies will be required to establish the true performance of classical shadows compared to grouping methods.  

\paragraph{Ansatz:} We now have to decide on an ansatz to model the electronic wavefunction on the qubit register. Deciding on an ansatz remains challenging because it is often unclear which is expressive and efficient enough to allow for a good approximation of the ground state. The ansatz with the best scaling, and some evidence for appropriate accuracy for the ground state representation \cite{Grimsley2019_UCC_Review}, is the Unitary paired Generalized Coupled Cluster Singles and Doubles (UpCCGSD) ansatz \cite{Lee2019} (see Sec. \ref{sec:Ansatz} for a detailed description). This ansatz scales linearly with the number of qubit $\mathcal{O}(kN)$, but may require to be repeated $k$ times to reach the desired accuracy. The scaling of required repetition of the ansatz $k$ has been partially studied \cite{Lee2019} but remains uncertain for large systems. This ansatz also has the advantage of only needing a fairly low number of parameters ($\mathcal{O}(kN^2)$). Adaptive ans\"atze (such as ADAPT-VQE \cite{Grimsley2019}, iterative Qubit Coupled Cluster \cite{Ryabinkin2020} and Cluster VQE \cite{Zhang2020}) are also promising, as they may provide resilience against barren plateaus. Their main drawback is that these adaptative methods come at the cost of selecting an operator to grow the ansatz (or Hamiltonian) and the need to fully re-optimize the ansatz at each iteration. Numerical studies have suggested that additional measurements may be required compared to fixed structure ans\"atze \cite{Claudino2020}, although further research is required to provide an exhaustive costs and benefits analysis.

\paragraph{Optimizer:} It is challenging to systematically compare different optimizers since no thorough large scale studies of their convergence rate have been conducted. For the time being, the Rotosolve optimizer \cite{Vidal2018,nakanishi_sequential_2020,ostaszewskiStructureOptimizationParameterized2021} (see Sec. \ref{sec:analytical_opt}) has been shown to converge significantly faster than several gradient based optimizers \cite{ostaszewskiStructureOptimizationParameterized2021}. It offers the advantage of not relying on any meta-parameters (such as a learning rate), which makes it a very easy optimizer to implement. However, Rotosolve presents two significant caveats: firstly, each iteration requires sampling three different values instead of two for most gradient based methods (one can avoid this overhead by finding one of the values from the optimization of the previous parameter, but this could result in correlated noise); secondly, parameters must be updated sequentially, thereby restricting the scope for parallelization of the VQE. The Fraxis method \cite{Watanabe2021WatanabeOptimizingSelection, Wada2021SimulatingCircuits} works in a similar manner and has been shown in some numerical studies to perform at least as good as Rotosolve or even outperform it. For the time being, given there are currently no optimizers that have been shown to have superior convergence rates, and given we do not expect that there will be a sufficient number of quantum computers to fully exploit the parallelization potential of the VQE in the NISQ era (see Sec. \ref{sec:parallelization}), we propose the Rotosolve / Fraxis optimizers over other alternatives (although it is worth noting that the Quantum Natural Gradient \cite{koczor_quantum_2020} has been shown to perform well and to benefit from resilience to barren plateaus \cite{stokes_quantum_2020,McArdle2019}, albeit at a significant cost \cite{koczor_quantum_2020}). 

\paragraph{Scaling:} Based on the discussion above, we can now construct a scaling estimate for a single iteration of the state-of-the-art VQE for \textit{ab initio} molecular systems. The overall scaling is expressed in terms of the number of quantum gate time steps that must be performed (i.e. several gates applied on disjoint sets of qubits can be implemented within the same time step). The computation of the expectation value of a single operator at a precision $\epsilon$ requires $\mathcal{O}(1/\epsilon^2)$ repetitions of the ansatz. In principle, $\epsilon$ should aim for chemical precision, generally accepted as 1.6 mE$_{\rm H} \approx 10^{-3} $E$_{\rm H}$. However, it is worth noting that, in practice gradients may become lower than chemical precision (due to the barren plateau problem for instance, described in Sec. \ref{sec:barren_plateau}). In this situation, estimating gradients may require a more precise $\epsilon$ and more measurements, but it also means that optimization may rapidly become impossible.  
If the k-UpCCGSD ansatz is chosen, this scales as $\mathcal{O}(kN)$, while choosing to use the decomposed interactions of Sec.~\ref{sec:DecomposedInteractions} requires $\mathcal{O}(N)$ different operators to be measured (and therefore a gate depth of $\mathcal{O}(N)$ for rotation to the joint measurement basis) resulting in a total scaling for a single estimation of the entire Hamiltonian of $\mathcal{O}(kN^2/\epsilon^2)$. 

There are $\mathcal{O}(kN^2)$ parameters in the k-UpCCGSD ansatz, hence this represents the cost scaling of updating each parameter using the Rotosolve optimizer. As this optimizer is not parallelizable, one may prefer to use a different method if sufficient sets of qubits are available. Overall, this gives us a total scaling for one iteration of the VQE of $\mathcal{O}(k^2N^4/\epsilon^2)$ without parallelization, and $\mathcal{O}(kN)$ with full parallelization (the circuit depth). This perfect parallelization would require $\mathcal{O}(kN^3/\epsilon^2)$ sets of $\mathcal{O}(N)$ qubits. Note that while qubits within one set need to be entangled for the course of a single measurement, there is no requirement for entanglement between qubits of different sets of parallel quantum compute nodes. The sets of qubits can therefore be either all within on one quantum computer, or else also distributed across different separated quantum computers (see Sec. \ref{sec:parallelization}). It should be noted that the precision $\epsilon$ is generally required to achieve chemical precision. However, if a barren plateau occurs, $\epsilon$ may need to be reduced by orders of magnitude to compute gradients accurately enough to achieve a satisfactory optimization.

So far we have only considered the scaling of one iteration. It is still an open research question how the number of iterations required to achieve convergence scales with system size for the VQE. This depends on numerous factors, including the ansatz, the optimizer used, and the system studied. One important point to note is that convergence tends to be rapid at the beginning of the optimization process, with large gradients that require only a low number of shots to be computed accurately enough to progress. It becomes more challenging close to the optimum, where gradients are smaller, requiring a larger number of shots to continue the optimization. As such the last few iterations of the VQE are likely orders of magnitude more expensive than the rest of the optimization, if the algorithm is implemented efficiently.  

There are other overheads that may be worth consideration in the initial setup of the system Hamiltonian. The computation of the Hamiltonian matrix elements generally has a polynomial scaling, while naive implementations of Hartree-Fock scale $\mathcal{O}(n^4)$ \cite{Koppl2016}, with $n$ the number of basis functions, and it can be reasonably assumed that $n=N$ for \textit{ab initio} molecular systems. Similarly, applying a decomposed interactions method to diagonalize operators and reduce measurements requires rewriting the Hamiltonian in a different basis \cite{Huggins2021, Yen2021_Cartan}. However, these costs only occur once at the beginning of the VQE process, and are unlikely to be a bottleneck. Despite possibly higher scaling than that of a VQE iteration, they are likely to have a significantly lower pre-factor (as implemented on conventional hardware), and as such are not likely to constrain  the application of the algorithm except far in the asymptotic realm. However, less investigated is that these joint measurement bases may result in covariance between measurements of different Hamiltonian terms, thereby requiring additional measurements \cite{mccleanTheoryVariationalHybrid2015, Crawford2021, Huggins2021, Yen2021_Cartan} which could significantly affect overall cost for the VQE.

\subsubsection{Best practices for lattice models}

Our suggestions for lattice models differ from their {\em ab initio} counterparts. Lattice models for the most part only include terms in the Hamiltonian between nearest-neighbors on their respective lattice, with interactions between more distant sites significantly truncated in range. 
In particular, this limited degree of connectivity of lattice models provides the option to construct mappings with much lower Pauli weight, enabling more compact ans{\"{a}}tze to be efficient, though this can come at the cost of additional qubits.

\paragraph{Hamiltonian construction:} In the case of a lattice model, the Hamiltonian is given as a small number of empirical parameters, requiring no prior computation of matrix elements. These models are generally most naturally and efficiently expressed in terms of fermionic operators in the site representation, where the locality of the interaction can be exploited to reduce the number of measurements. 

\paragraph{Encoding:} There is significant literature on the fermion-to-spin mapping for lattice models (see Sec. \ref{sec:lattice_encoding} for further details). These mappings are in general designed to minimize the required Pauli weight of the operators for a given lattice structure. The most important property of a lattice is the maximum degree of connectivity (coordination) of the sites, denoted $d$. For instance, a square lattice has $d=4$, and an equivalent hypercubic lattice of dimension $D$ has $d=2D$. 
If one is limited by the number of qubits available, the most appropriate mapping for a lattice is an adaptation of the Bravyi-Kitaev mapping (based on Fenwick trees) \cite{Havlek2017}, and which has a Pauli weight scaling as $\mathcal{O}(\log(v))$, where $v$ is the minimum number of sites in any one dimension for a $D=2$ lattice. It has the advantage of reducing the Pauli weight of the operators produced, compared to a naive implementation of Bravyi-Kitaev on a lattice, while maintaining the number of qubits required equal to the number of sites, $n$. If however the number of qubits is not a hard constraint, the Generalized Superfast Encoding method \cite{Setia2019} provides a lower Pauli weight scaling of $\mathcal{O}(\log_2(d))$ at the cost of requiring an increased number of qubits for each site, with an overall scaling of $\mathcal{O}(nd/2)$ qubits. The Compact encoding \cite{Derby2021, Derby2021_part2} requires a lower number of qubits ($\sim 1.5n$), but has not yet been generalized for regular lattices of more than three dimensions.  
Beyond their relationship to the resilience to barren plateaus in the optimization \cite{Cerezo2021_BP, Uvarov2020, Uvarov2020_frustrated}, the relevance of the Pauli weight in the context of VQE is also in how it affects the choice of ansatz, and in particular whether the ansatz is initially expressed in terms of fermionic operators. If the chosen ansatz is not dependent on fermionic terms, then Bravyi-Kiteav or Jordan-Wigner mappings are preferred. Furthermore, the number of qubits required to represent the Hamiltonian can be reduced using the tapering off methods based on symmetries as described in Sec.~\ref{sec:tappering_qubits} \cite{bravyi_tapering_2017, Setia2020}. 

\paragraph{Grouping and measurement strategy: } 
The number of operators in lattice models scales, in general, with the number of edges of the lattice graph, $E$. For example, for a hypercubic lattice of dimension $D$, the number of edges, and therefore, the number of operators will scale $\mathcal{O}(nD)$ (though it is worth noting that the pre-factor to this scaling may change significantly depending on the encoding used), as detailed in Sec. \ref{sec:lattice_encoding}. Because it is in general possible to reach low Pauli-weight encodings for lattice models (see Sec. \ref{sec:lattice_encoding}), qubit-wise commutativity (QWC) grouping \cite{mccleanTheoryVariationalHybrid2015, Kandala2017, Hempel2018, Rubin2018, Kokail2019, Izmaylov2019, Nam2020, Verteletskyi2020, Hamamura2020, Gokhale2019_long} could possibly offer significant potential for reductions in the number of terms. It is also worth considering the fact that ans\"atze for lattice models tend to be shallower, and as such, the cost of basis rotations in methods based on general commutativity of Pauli operators could be too significant to justify its use. As such we would therefore recommend QWC grouping until further research is conducted.

\paragraph{Ansatz:} Direct application of ans{\"{a}}tze suited for \textit{ab initio} molecular systems (such as Unitary Coupled Cluster, UCCSD and its extensions) have been shown to work in practice using generalized encodings such as Jordan-Wigner (see for instance, Ref.~\cite{Sokolov2020}). However, we note that the underlying physics motivating the ansatz is not ideally suited to strongly correlated lattice models, requiring care to ensure that they are efficient ansatz for these systems \cite{Evangelista2019}. Since these ans\"atze are formulated in a basis of Hartree-Fock or other mean-field orbitals, they do not allow using some of the low-weight encodings easily and do not enable exploitation of the low degree of connectivity of the model. Instead, one ansatz that has been shown to be very suitable for correlated lattice problems is the Hamiltonian variational ansatz (HVA) \cite{Wecker2015, Wiersema2020} (see Sec. \ref{sec:hva}). The ansatz leverages the more compact structure of the lattice model Hamiltonian and is built using fermionic operators, thereby making the most of low Pauli weight lattice encodings \cite{Setia2018, Setia2019, Derby2021, Derby2021_part2}. HVA was also shown in \cite{Wiersema2020} to be particularly resilient to the appearance of barren plateaus in the optimization problem. The ansatz has a depth and number of parameters scaling with the number of commutative Pauli groups in the Hamiltonian (though it may need to be repeated several times to account for lower expressibility compared to UCCSD). For a regular lattice, this can result in an overall scaling that is lower than $\mathcal{O}(nD)$. Extensions of HVA could also be considered for specific systems. For instance, Fourier Transform-HVA \cite{Babbush2018} could be very efficient on certain models (in particular jellium in Ref.~ \cite{Babbush2018}, which is a continuous model, but with the Hamiltonian defined by a single parameter). Symmetry breaking HVAs \cite{Vogt2020,Choquette2021} are also a promising avenue, and though numerical tests from Ref.~\cite{Choquette2021} show excellent fidelities of the state produced, results in Ref.~\cite{Vogt2020} show some instabilities of the ansatz. Overall, until further research is conducted, we consider that HVA is safest and most general option for lattice models. Finally, we have neglected consideration of adaptive ans{\"{a}}tze, as it is difficult to make a scaling argument for their efficiency in this domain, where the ability to justify the inclusion of some terms over others in the ansatz is likely to be diminished.

\paragraph{Optimizer: } Similar to our proposal for \textit{ab initio} molecular systems presented earlier, we find that Rotosolve \cite{Vidal2018, nakanishi_sequential_2020,ostaszewskiStructureOptimizationParameterized2021} and Fraxis \cite{Watanabe2021WatanabeOptimizingSelection, Wada2021SimulatingCircuits} currently have the best supporting evidence among the optimizers available.

\paragraph{Scaling:} Based on the above, we can make an argument for the scaling of VQE as implemented for lattice models. The number of shots required to achieve a precision of $\epsilon$ when computing an expectation value scales $\mathcal{O}(1/\epsilon^2)$, with $\epsilon$ the target precision. For lattice models however, the target is usually not chemical accuracy and instead depends on the aim of the calculation. Generally, the aim is to resolve some correlated features of the electronic structure (e.g. predicting the parameter regimes for different phases, or response properties Ref.~\cite{Fan1970, Marro1999, Stanislavchuk2015, Choi2019}). The HVA ansatz has a scaling capped by the number of terms in the lattice model Hamiltonian and the number of repetitions of the ansatz $k$, hence $\mathcal{O}(knD)$. However, there may be a few caveats to this. Since some encodings require additional qubits to reduce the Pauli weight, they would increase the depth and the number of terms in the ansatz, while at the same time reducing the depth per term of the ansatz due to the reduced Pauli weight. Overall these two effects are likely to only affect the pre-factor of the VQE, noting that we consider this an area of open research. Therefore, we maintain the previous scaling and estimate that a single energy evaluation would scale $\mathcal{O}(k(nD/\epsilon)^2)$. 

For a regular hypercubic lattice with only at most nearest-neighbor terms in the Hamiltonian, there are $\mathcal{O}(knD)$ parameters in the HVA, bringing the total scaling of computing gradients for a lattice model VQE to $\mathcal{O}((knD)^3/(\epsilon)^2)$. As noted previously, if the full parallelization potential is exploited, this would give a scaling of $\mathcal{O}(knD)$ for one iteration of the VQE. Discussions regarding the number of iteration are identical to the considerations raised in the previous section. It is worth noting that there has been so far no evidence that qubit-wise commutativity grouping can reduce the overall scaling of the number of operators to measure, and therefore we have not included it in this scaling analysis. Similarly, while the depth of HVA scales with the number of commutative groups in the Hamiltonian, we have not yet found an argument on whether this would reduce scaling below that of the number of terms $\mathcal{O}(nD)$.

\subsection{Resource estimate for VQE} \label{sec:resource_estimates}

\subsubsection{Cost and runtime estimates for VQE}

There have been several studies estimating the resources required to perform VQE on a system that is too large to be accurately treated using conventional methods. Wecker et al. \cite{Wecker2015} develops the Hamiltonian Variational Ansatz (HVA), and presents a numerical study of the accuracy of certain ans{\"{a}}tze (the HVA and various UCC based ans{\"{a}}tze) and the number of repetitions required. They also possibly unveiled the existence of the barren plateau problem ahead of it being characterized in Ref. \cite{McClean2018}, by numerically showing that more expressive forms of UCC cannot reach the same accuracy as less expressive forms on larger systems. They estimate that the total number of samples required to compute the ground state energy of $\mathrm{Fe_2S_2}$ to chemical precision (using the STO-3G basis with $n=112$ spin orbitals) is of the order $\mathcal{O}(10^{19})$, which is far beyond what could be considered tractable. Of course, conventional methods aiming at resolving exactly, and directly, the ground state of a $n=112$ spin orbitals Hamiltonian would also be intractable (this would equate to finding the lowest eigenvalue of a $2^{112}$ matrix). However this does not exclude the possibility that more refined conventional methods, accepting some approximation, could an acceptable level of accuracy on $\mathrm{Fe_2S_2}$ (as an example, please see Ref. \cite{Li2013}). The literature has progressed significantly since this 2015 study, and there are now more efficient ans\"atze (see Sec. \ref{sec:Ansatz}) and grouping methods (see Sec. \ref{sec:Grouping}) that may change this conclusion. 

K\"uhn \textit{et al.} \cite{Kuhn2019} numerically assess the number of qubits and circuit depths required for UCC based ans{\"{a}}tze. They show that to model a medium-sized organic molecule such as naphthalene ($\mathrm{C_{10}H_8}$, with $68$ electrons) would roughly require about $800$ to $1500$ qubits, and a number of two-qubit gates of about $\mathcal{O}(10^8)$ using UCCSD. This latter number may be significantly lowered if the k-UpCCGSD ansatz is used (assuming it can achieve the desired accuracy). They also claim that the run time for a VQE implementation would be impractical, even using full parallelization potential, without unfortunately providing more details about how this conclusion is reached. 

Gonthier {\it et al.} \cite{Gonthier2020} provide what probably constitutes the most comprehensive study of the VQE resource requirements to date by estimating the cost of combustion energy computation for nine organic molecules (including methane, ethanol, and propane). They provide a detailed estimate of the runtime for one energy estimation ranging from $1.9$ days for methane ($\mathrm{CH_4}$), which requires $104$ qubits for accurate treatment, to $71$ days for ethanol ($\mathrm{C_2H_6O}$), which requires 260 qubits (this estimate uses a frozen natural orbital basis, with 13 functions, i.e. 13 qubits for each electron). The analysis is rather exhaustive since it takes into consideration the joint measurements of Hamiltonian terms (see Sec. \ref{sec:Grouping}), and different optimization methods (see Sec. \ref{sec:measurements_strategies}). 

It is worth noting that the studies mentioned above do not take into account the three obstacles we listed at the beginning of this section (namely, the complexity of the optimization, the barren plateau problem, and the impact of quantum noise). At the same time, Refs.~\cite{Kuhn2019, Gonthier2020, Elfving2020} do not discuss the potential for parallelization (with the exception of \cite{Elfving2020} which touches upon it briefly). For instance, the runtime estimates of 1.9 to 71 days presented in \cite{Gonthier2020} can be parallelized efficiently, although this would require a significant quantity of qubits arranged in sets on which parallel computation can be performed, possibly resulting in a variety of new problems such as overhead communication cost and additional quantum noise (see Sec. \ref{sec:parallelization} for a discussion). 

We provide our estimated runtimes for the steps in the VQE for a representative example system in Table \ref{tab:runtime_noiseless_iteration}, including the general scaling estimate for such types of systems. The example system considered corresponds to the one proposed in Ref. \cite{Elfving2020}, and is the \textit{ab initio} computation of the chromium dimer ($\mathrm{Cr_2}$) with an active space of $26$ electrons in $26$ molecular orbitals ($52$ spin orbitals and $52$ qubits).

It is very difficult to estimate the pre-factor of the VQE, which would very much be dependent on the hardware, and a detailed numerical analysis is not within the scope of this review. To estimate the depth we compile a $52$ qubit version of k-UpCCGD assuming $k=1$ and full connectivity of the qubit register. We find a depth of $\sim 27,000$ timesteps, denoted $L$, and $170$ parameters, denoted $p$ (to illustrate the impact of the connectivity of the qubit register, we compiled the same ansatz assuming a linear qubit register, and find that the depth required is increased by more than tenfold). Note that choosing $k=1$ is likely insufficient \cite{Lee2019, Grimsley2019_UCC_Review}. Gate time is assumed to be $T = 100$ ns (similar to what is presented in \cite{Gonthier2020}, which itself refers to Table 1 in \cite{Kjaergaard2020}), which is better than what can  currently be achieved for superconducting qubits ($\sim 500$ ns), but is probably achievable over the next few years. We assume readout and reset times are negligible compared to the circuit runtime. The pre-factor for the number of operators ($\mathcal{P}$) to measure can easily be assumed to be $16$ as each fermionic operator result in two Pauli strings under generalized mappings, and there are four fermionic operators in each two-body term in the Hamiltonian. Using a form of decomposed interactions we estimate that the number of operators is $\sim 16N = 832$ (we do not consider the impact of covariances may have on the noise of the estimates, though point out that $16$ is a conservative pre-factor), and assume that the required circuit depth for basis rotation is negligible. Finally, we set the target precision $\epsilon$ to $10^{-3}$ mE$_{\rm H}$, which is close to chemical precision ($1.6$ mE$_{\rm H}$) \cite{Peterson2012} and roughly assume that $S = 1/\epsilon^2=1,000,000$ shots are used for the estimation. A much lower number of shots would be sufficient to progress the initial part of the optimization, and this high number of shots is only required in the last iterations of a VQE close to convergence to reach chemical precision (note however that this number of shots may need to be much higher in case of barren plateaus). It is therefore likely that the last few iterations before convergence are the most costly and time-consuming, largely dominating the cost. However, despite some optimistic assumptions listed above, it is clear that the time cost of VQE implemented on a single set of qubits remains orders of magnitude too large to be realistically viable, pointing to the dependence of the method on parallelization. 

\begin{table}
\caption{Indicative estimates of the run time of one iteration of the VQE making the following assumptions: gate time: $100$ ns. This assumes an active space of $26$ molecular orbitals for $\mathrm{Cr_2}$ spanned over $N=52$ qubits, and a gate runtime of $T=100$ ns.} \label{tab:runtime_noiseless_iteration}
\begin{tabularx}{\linewidth}{Xccc}
\toprule
\\
Operation & Scaling & Formula & Runtime \\
\\ \midrule \\
Single shot (using k-UpCCG SD, k = 1) & $\mathcal{O}(kN)$ &  $L \times T$ & $3$ ms \\
\\ \hline \\
One expectation at $\epsilon = 10^{-3}$ mE$_{\rm H}$ (using decomposed interactions methods) & $\mathcal{O}( \frac{kN^2}{\epsilon^2})$ &  $L \times T \times \mathcal{P} \times S$ & $25$ days \\
\\ \hline \\
Full iteration using Rotosolve & $\mathcal{O}(\frac{k^2N^4}{\epsilon^2})$ &  $L \times T \times \mathcal{P} \times S \times 3p$ & $35$ years \\
\\ \hline \\
Full iteration using gradient based method & $\mathcal{O}(\frac{k^2N^4}{\epsilon^2})$ &  $L \times T \times \mathcal{P} \times S \times 2p$  & $24$ years \\
\bottomrule
\end{tabularx}
\end{table}

\subsubsection{Parallelization potential of the VQE} \label{sec:parallelization}

The potential for parallelization of the VQE was already identified in the initial paper by Peruzzo {\it et al.} \cite{Peruzzo2014} and subsequently mentioned in many VQE papers, although an in-depth study is lacking. Parallelism is however critical for the viability of the method. Parallelism of the VQE offers a direct way to convert runtime cost into hardware cost by splitting the shots required onto different sets of qubits (which can be arranged in different threads on a single quantum computer, or multiple, disconnected quantum computers). To illustrate this point, we adapt the estimates presented in Table \ref{tab:runtime_noiseless_iteration} assuming that perfect parallelization is possible, and present the results in Table \ref{tab:parallel_computation}. 

\begin{table}
\caption{Indicative estimates of the run time and number of quantum computers used for one iteration of the VQE assuming perfect parallelization of the method can be achieved and neglecting any communication overheads - using the same assumptions stated in Table \ref{tab:runtime_noiseless_iteration}} \label{tab:parallel_computation}
\begin{tabularx}{\linewidth}{Xcccc}
\toprule
\\
Operation & Time scaling & Runtime & Scaling sets of qubits & Sets of qubits (here, $N=52$)\\
\\ \midrule \\
Single shot (using k-UpCCGSD, k = 1) & $\mathcal{O}(kN)$ & $3$ ms & $\mathcal{O}(1)$ & $1$\\
\\ \hline \\
One expectation at $\epsilon = 10^{-3}$ mE$_{\rm H}$ (using decomposed interactions methods) & $\mathcal{O}(kN)$ &  $3$ ms & $\mathcal{O}(\frac{N}{\epsilon^2})$ & $\sim 800 \times 10^6$ \\
\\ \hline \\
Full iteration using Rotosolve & $\mathcal{O}(k^2N^3)$ & $0.5$ s & $\mathcal{O}(\frac{N}{\epsilon^2})$ &$\sim 2,500 \times 10^6$ \\
\\ \hline \\
Full iteration using gradient based method & $\mathcal{O}(kN)$ & $3$ ms & $\mathcal{O}(\frac{kN^3}{\epsilon^2})$ & $\sim 280 \times 10^9$  \\
\bottomrule
\end{tabularx}

\end{table}

It is clear that parallelization will be a critical part of any future success of the VQE method. Broad availability of quantum computers with increasing number of qubits could therefore significantly speed-up the VQE process, however there are significant caveats to that. One key observation is that full parallelization would require a number of quantum computers cores (or threads) that scales $O(p N^4 / \epsilon^2)$, with $p$ the number of parameters in the ansatz. This could clearly be a prohibitive number for large computation given the current state of hardware technology, and it is possible that fault-tolerant technology could arrive before we are able to produce such large quantities of devices. 

Even if it was possible to build large quantities of quantum computers, there are many caveats to the potential of parallelization for the VQE. First, as it is the case for conventional parallel computing, parallel quantum computing will suffer from communication overheads. These overheads are the computational cost of coordinating the parallel tasks, which can include the likes of synchronization cost, data aggregation and communication (possibly latency if the different sets of qubits are connected through the cloud). Second, parallelization could result in higher noise levels. We note two possible sources of additional noise: (1) if parallelization is done on multi-threaded quantum computers, there is higher chance of cross-talk between qubits; (2) variational algorithms are considered to be somewhat noise resilient as they can learn the systematic biases of a given hardware ~\cite{mccleanTheoryVariationalHybrid2015, Enrico2021EvaluatingNoiseResilience} - if the algorithm is run on multiple quantum computers these noise resilient effects may disappear, as systematic biases on one set of qubits, which differs on another, may no longer be learned through the variational process. 

\subsubsection{Distribution of resources between quantum and conventional computation} \label{sec:TBD}

A final remark on the overall computational cost of VQE is worth raising. The vast majority of the research in the field is focused on the simple divide between sampling observables on the quantum computer, and performing the parameter updates on a conventional device. This approach however ignores the excellent scaling of the simulation of certain families of quantum states on conventional computers (for example, see Ref.~\cite{Zhou2020}). 

It is very likely that a future relevant application of VQE will require a more insightful split between quantum and conventional resources. Several avenues have already been proposed in this respect. For instance, Stenger \textit{et al.} \cite{Stenger2022} propose partial solve a many-body Hamiltonian on a conventional computer before performing a quantum based VQE. Okada \textit{et al.} \cite{Okada2022} alternatively show how one can perform a classical optimization of the VQE for local-interaction states, leaving quantum sampling only for measurement of global quantities. Methods of Hamiltonian dressing such as ClusterVQE \cite{Zhang2021} (presented in Sec. \ref{sec:Hamiltonian_dressing}) work differently but aim at the same objective by absorbing in the Hamiltonian parts of the ansatz that are prone to quantum noise and worse gate scaling.

%% file: 03_hamiltonian_main.tex
\section{Hamiltonian Representation}\label{sec:Hamiltonian_representation}

Before studying any methods relating directly to implementing the VQE, we must turn towards the definition of the problem itself. As written before, our aim is to find the lowest eigenvalue of a Hermitian matrix, or stated in quantum chemistry terms, we are trying to find an approximation of the ground state energy of an interacting Hamiltonian. 
In this section, we provide details about how Hamiltonian can be constructed, and how choices in certain freedoms in expressing this Hamiltonian may impact the remainder of the VQE pipeline. Many different forms of Hamiltonians exist in physics and chemistry; we begin with a presentation of the electronic structure Hamiltonian and construction of the eigenvalue problem. This is followed by examples of other applications, namely lattice models, vibrational spectroscopy, and periodic adaption of electronic structure problems. Unlike the other models, the electronic structure Hamiltonian is in general known and the choices related to its construction are aiming at reducing the complexity of the eigenvalue problem. In the case of lattice models, and vibrational spectroscopy Hamiltonian assumptions are made to define the Hamiltonian. 

\subsection{The electronic structure Hamiltonian}

\subsubsection{The \textit{ab initio} molecular Hamiltonian}

This represents the operator for the total energy of an arbitrary molecular system defined in terms of its atomic composition, and the relative positions of the nuclei. From this geometrical definition, also referred to as a conformation, one must determine the correlated probability amplitudes of the electrons in the space surrounding the nuclei, i.e. the electronic wavefunction, which has the lowest energy: the ground state energy. 
 
In a non-relativistic settings and following the Born–Oppenheimer approximation (which assumes that the motion of the nuclei can be neglected, as they are much heavier than the electrons) the electronic Hamiltonian depends parametrically
on the nuclear positions $\mathbf{R}_{k}$.
Up to a constant (given by the nuclear-nuclear repulsion energy) the electronic Hamiltonian can be written as
\begin{equation}
\label{eq:molecularhamiltonian3}
\hat{H} = \hat{T}_e + \hat{V}_{ne} + \hat{V}_{ee},
\end{equation}
where
\begin{align} \label{eq:molecularhamiltonian2}
&\hat{T}_e = -\sum_{i} \frac{\hbar^{2}}{2 M_{i}} \nabla_{i}^{2}, \\
&\hat{V}_{ne} = -\sum_{i, k} \frac{e^{2}}{4 \pi \epsilon_{0}} \frac{Z_{k}}{\left|\mathbf{r}_{i}-\mathbf{R}_{k}\right|}, \\
&\hat{V}_{ee} = \frac{1}{2} \sum_{i \neq j} \frac{e^{2}}{4 \pi \epsilon_{0}} \frac{1}{\left|\mathbf{r}_{i}-\mathbf{r}_{j}\right|},
\end{align} 
with, $\mathbf{r}_{i}$ is the position of electron $i$, $M_{i}$ its mass, $Z_k$ is the atomic number of nucleus $k$, $e$ is the elementary charge, $\hbar$ is the reduced Plank constant, and $\nabla_{i}^2$ is the Laplace operator for electron $i$. 

\subsubsection{Construction of the wavefunction}

One needs to define a basis in which to represent the electronic wavefunction. A number of possible types of basis functions exist. Given the number of qubits required scales as a function of the number of basis functions (see Sec.~ \ref{sec:Encoding}), choosing a basis that is compact, but yet provides an accurate description of the system studied, is critical for an efficient implementation of VQE. 
Basis elements, functions or orbitals describe the probability distribution of a single electron. 
For {\em ab initio} systems, these have primarily been built from parameterized atom-centered Gaussian (`atomic') orbitals, with the majority of research to date in VQE using minimal sized Stater Type Orbitals (STO) (for example \cite{Peruzzo2014, Kandala2017, Lee2019}). These basis functions are defined as a weighted sum of Gaussian functions to provide approximately the right radial distribution, long-range behaviour, and nuclear cusp conditions for each atom \cite{Hehre1969,Stewart1970}. 
As an illustration, the radial component of the minimal STO-3G basis for each atom is constructed from three Gaussians such that an atomic orbital is given by 
\begin{equation}
    \chi(r) = c_1 \gamma_1(r) + c_2 \gamma_2(r) + c_3 \gamma_3(r)
\end{equation}
where $\gamma_i$ are Gaussian functions, $r$ the distance of the electron from the nucleus, and $c_i$ are fitted weight parameters. 
These basis sets are often called minimal basis sets as they only include orbitals necessary to represent the valence shell of an atom. For an appropriate treatment of the correlation, it is essential in real systems to enlarge the basis set to include higher energy atomic orbitals, allowing for additional polarization and diffuse functions, and higher angular momentum functions which are required to build in flexibility to describe the correlated positions of the electrons \cite{Helgaker2000}.
An example of these larger basis sets commonly used for correlated calculations are the correlation-consistent polarized Valence n-Zeta (cc-pVnZ) basis sets \cite{Dunning1989}, which allow for a systematic (and extrapolatable) expansion in terms of the cardinal number of the basis set \cite{Varandas2000}. These basis sets have been used in some VQE research (for example \cite{Kuhn2019, Kottmann2021_1, Tilly2021}), but their additional size can limit the size of the systems treated. Alternative basis sets have been considered in the context of VQE. For example, plane wave basis have been used as a mean to construct a compact ansatz for certain models which naturally exploit the translational symmetry of certain models \cite{Babbush2018}. Alternatively, a grid of points in real space make a natural representation to enforce locality and enable a sparse representation of the Hamiltonian. However grid representations of these basis functions generally require a significantly larger qubit count, and therefore their use is limited in NISQ applications and for the VQE \cite{mcardleQuantumComputationalChemistry2018,Wiesner1996, Zalka1998, Lidar1999, Kassal2008, Ward2009, Jones2012, Kivlichan2017}.

These non-orthogonal atomic orbitals are generally linearly combined into `molecular' orbitals before use, which constitutes an orthonormal set of delocalized basis functions which can no longer be assigned to a particular atomic site. The Hamiltonian is then expressed within this molecular basis by way of a transformation of the matrix elements. Overwhelmingly, this transformation is obtained via a mean-field (generally Hartree-Fock) calculation, which produces this rotation to the molecular orbital basis, and additionally provide an energy measure for each single-particle molecular orbital (for a description of this, see Ref.~\cite{Jensen2017}). However, given the constraints on qubit numbers and therefore the size of the single particle basis, there is also research in the use of further contractions of molecular orbitals to a more suitable and compact basis for subsequent correlated calculations. These are often based on approximate correlated treatments in order to truncate to frozen natural orbitals, which has been favorably suggested by Verma \textit{et al.} \cite{Verma2021}, Mochizuki \textit{et al.} \cite{Mochizuki2019} and showed to potentially resulting in computation cost reductions for VQE by Gonthier \textit{et al.} \cite{Gonthier2020}. Furthermore, self-consistent active space approaches also optimize the set of molecular orbitals within the correlated treatment, to optimally span this correlated physics, and are considered further in Sec.~\ref{sec:multiscale}.

Once the single-particle basis functions have been selected, the many-body basis for the electronic wavefunction is constructed from products of these functions. For a non-interacting Hamiltonian, the solution is given as a single many-body basis function with optimized orbitals, which is the principle behind the Hartree-Fock and other mean-field methods. 
In addition, following the Pauli exclusion principle, the electronic wavefunction must be antisymmetric, meaning that the exchange of any two electrons changes the sign of the wavefunction. 
To account for this, these many-body basis functions can be formally written as Slater determinants, which for a wavefunction of $n$ occupied orbitals can be formally written as
\begin{equation}
 \psi(\mathbf{x}_1, \mathbf{x}_2, \ldots, \mathbf{x}_n) =
  \frac{1}{\sqrt{n!}}
  \begin{vmatrix} \phi_1(\mathbf{x}_1) & \phi_2(\mathbf{x}_1) & \cdots & \phi_n(\mathbf{x}_1) \\
                      \phi_1(\mathbf{x}_2) & \phi_2(\mathbf{x}_2) & \cdots & \phi_n(\mathbf{x}_2) \\
                      \vdots & \vdots & \ddots & \vdots \\
                      \phi_1(\mathbf{x}_n) & \phi_2(\mathbf{x}_n) & \cdots & \phi_n(\mathbf{x}_n)
  \end{vmatrix} ,
\label{eq:slaterdeterminant}
\end{equation}
where $\phi_j(\mathbf{x}_j)$ denotes a spin-orbital of the chosen basis, with the variable $\mathbf{x}_j = (\mathbf{r}_j, \sigma_j)$ subsuming both the associated spatial and spin indices. As a shorthand for this, we can write an $n$-electron many-body basis function as
\begin{equation} \label{eq:slaterdeterminant_wf}
    \ket{\mathbf{\psi}} = \ket{\phi_{1} \phi_{2} \ldots \phi_{n}},
\end{equation}
where $\phi_j = 1$ means that the $j$-th basis function $\phi_j(\mathbf{x})$ is occupied, and $\phi_j = 0$ is unoccupied. This representation can be simply encoded in the qubit register, with an occupied spin-orbital denoted by an up-spin, and an unoccupied orbital by a down spin. This defines a many-body Hilbert space in which the correlated wavefunction can be expanded, since the correlations will ensure that the state can no longer be written as a single Slater determinant, but rather a linear combination over the space. This Hilbert space of electron configurations has a well-defined inner product between these many-body basis states, as $\braket{\mathbf{\psi}_a | \mathbf{\psi}_b} = \delta_{a,b}$.

There are commonly two ways that antisymmetry condition can be met in practice: one can do so by enforcing it in the construction of the wavefunction, where the Schrodinger equation is typically written in real space and is generally referred to as first quantization; alternatively, one can enforce antisymmetry through enforcing antisymmetry within the commutation relations of the operators for expectation values, and is traditionally referred to as second quantization. The choice of representation has significant implications for the resources required to implement VQE, and is the topic we turn to in Sec. \ref{sec:Encoding}. 

\subsubsection{Hamiltonian quantization}

\paragraph{First quantization: } \label{sec:first_quantization}

First quantization is most commonly used in quantum mechanics and is a direct adaption of classical mechanics to quantum theory by quantizing variables such as energy and momentum. It was the method of used in early quantum computing research of Hamiltonian simulation  \cite{Lloyd1996, Abrams1997, Zalka1998, Boghosian1998, Lidar1999, Wiesner1996, Kassal2008}, and has seen a resurgence in recent years due to promising scaling \cite{Babbush2019, Su2021}.
Following the definition of the wavefunction in terms of Slater determinant Eq. (\ref{eq:slaterdeterminant_wf}), one must ensure that the antisymmetry of the wavefunction is verified:

\begin{equation}
    \ket{\mathbf{\sigma}(\phi_{1} \phi_{2} \ldots \phi_{n})} = (-1)^{\pi(\mathbf{\sigma})}\ket{\phi_{1} \phi_{2} \ldots \phi_{n}}, 
\end{equation}
where $\pi(\mathbf{\sigma})$ is the parity of a given permutation $\sigma$ on a set of basis functions.

As the antisymmetry is addressed in the wavefunction, the Hamiltonian can be constructed by simple projection onto the single particle basis function. If we project the Hamiltonian over the space spanned by  $\{\phi_i(\mathbf{x}_i)\}$ which we assume to be orthonormal, we obtain (using the Slater–Condon rules \citep{Slater1929, Condon1930}) the one and two body integrals. For the one-electron terms matrix elements $h_{p q}$:
\begin{equation}
\label{eq:HFintegral1}
\begin{aligned}
h_{p q} &= \bra{\phi_{p}} \hat{T}_e + \hat{V}_{ne} \ket{\phi_{q}}\\
&=\int \mathrm{d} \mathbf{x~} \phi_{p}^{*}(\mathbf{x})\left(-\frac{\hbar^{2}}{2 m_{e}} \nabla^{2}-\sum_{k} \frac{e^{2}}{4 \pi \epsilon_{0}} \frac{Z_{k}}{\left|\mathbf{r}-\mathbf{R}_{k}\right|}\right) \phi_{q}(\mathbf{x})
\end{aligned}
\end{equation}
And for the two-electron interaction terms we obtain the matrix elements $h_{p q r s}$:
\begin{equation}
\label{eq:HFintegral2}
\begin{aligned}
h_{p q r s} &= \bra{\phi_{p} \phi_{q}}\hat{V}_{ee}\ket{\phi_{r} \phi_{s}}\\
&= \frac{e^{2}}{4 \pi \epsilon_{0}} \int \mathrm{d} \mathbf{x}_{1} \mathrm{d} \mathbf{x}_{2} \frac{\phi_{p}^{*}\left(\mathbf{x}_{1}\right) \phi_{q}^{*}\left(\mathbf{x}_{2}\right) \phi_{r}\left(\mathbf{x}_{2}\right) \phi_{s}\left(\mathbf{x}_{1}\right)}{\left|\mathbf{r}_{1}-\mathbf{r}_{2}\right|}.
\end{aligned}
\end{equation}

The first quantized form of the Hamiltonian can therefore be written directly in the single particle basis:

\begin{equation} \label{eq:first_quantized_hamiltonian}
    \hat{H} = \sum_{i=1}^m \sum_{p, q = 1}^n h_{p q} \ket{\phi_p^{(i)}} \bra{\phi_q^{(i)}} + \frac{1}{2} \sum_{i \neq j}^m \sum_{p, q, r, s = 1}^n  h_{p q r s} \ket{\phi_p^{(i)}\phi_q^{(j)}} \bra{\phi_r^{(i)}\phi_s^{(j)}}.
\end{equation}

Following the method described in \cite{Abrams1997}, one can map these $n$ single-particle basis functions binary numbers and to $\log_2(n)$ qubits. For instance, qubit state $\ket{\phi_1} = \ket{00...00}$ would represent function $\phi_1(\mathbf{x}_1)$, $\ket{\phi_2} = \ket{10...00}$ would represent function $\phi_2(\mathbf{x}_2)$, and so on. Given $m$ electrons in the wavefunction we are trying to model, and given each electron can occupy at most one basis function (represented by $\log_2(n)$ qubits), one can model any product state with $N=m\log_2(n)$ qubits. 
As discussed before however, in first quantization the antisymmetry is enforced directly through the wavefunction. Procedures have been developed to maintain this requirement using an ancilla qubit register of $\mathcal{O}(m\log_2(n))$ qubits, and a circuit depth of $\mathcal{O}(\log^c_2(m)\log_2(\log_2(n)))$, where $c \geq 1$ depends on a choice of sorting network \cite{Berry2018}. This additional depth would be considered acceptable in contrast to the scaling of most VQE ans\"atze (see Sec. \ref{sec:Ansatz}).  

Translation of the Hamiltonian operators (e.g. $\ket{\phi_p^{(i)}} \bra{\phi_q^{(i)}}$) into operators that can be measured directly on quantum computers is fairly straightforward (also clearly explained in Ref.~\cite{mcardleQuantumComputationalChemistry2018}). Each operator can be re-written into a tensor product of four types of single qubit operators: $|0 \rangle \langle 0 |$, $|0 \rangle \langle 1 |$, $|1 \rangle \langle 0 |$, and $|1 \rangle \langle 1|$ which can be mapped to Pauli operators as follows: 
\begin{equation} \label{eq:first_quantized_paulis}
    \begin{aligned}
        &|0 \rangle \langle 0 | = \frac{1}{2} (I + Z) \\
        &|0 \rangle \langle 1 | = \frac{1}{2} (X + iY) \\
        &|1 \rangle \langle 0 | = \frac{1}{2} (X - iY) \\
        &|1 \rangle \langle 1 | = \frac{1}{2} (I - Z).
    \end{aligned}
\end{equation}
From Eq. (\ref{eq:first_quantized_hamiltonian}) one can see that the number of operators in the Hamiltonian will scale with the number of two body terms $\mathcal{O}(n^4m^2)$, as there is one term for each combination of $4$ spin-orbitals and $2$ electrons. Each spin-orbital is represented by $\log_2(n)$ qubits, hence for the two body terms we have $2\log_2(n)$ tensored qubit outer products (e.g. $|0 \rangle \langle 1 |$), each of them composed of two Pauli operators. This results in a sum of $2^{2\log_2(n)}= n^2$ Pauli strings each acting on up to $2\log_2(n)$ qubits (hence the Pauli weight in first quantization), for every operator in the Hamiltonian. This implies that the scaling of Pauli strings of the Hamiltonian in first quantization is $\mathcal{O}(n^6m^2)$.

A few points are worth noting with respect to the use of first quantization for quantum computing. Firstly, to the best of our knowledge, there are no publications studying the usage of this method within the context of VQE or any other NISQ algorithm. Secondly, it is very clear that while the ancilla qubits do not change the overall scaling of the method, it bears a significant cost for NISQ devices, in addition to having to compute all spatial integrals on the quantum computer \cite{Moll2018}. Finally it is very important to note that in general first quantization can be advantageous on systems which require a very large number of basis functions compared to the number of electrons (due to the logarithmic scaling of the number of qubits in $n$). This is the case in particular when a plane wave basis is selected, as it usually requires several orders of magnitude more functions than the molecular basis to achieve equivalent accuracy \cite{Babbush2018} - plane wave basis in first quantization has been shown to bring significant scaling advantages for in fault tolerant quantum simulation of chemical systems \cite{Babbush2019}. Overall, despite offering clear promise for fault tolerant quantum computing, it appears for the moment that first quantization is on balance too costly for the NISQ era.

\paragraph{Second quantization: } \label{sec:second_quantization}

Second quantization distinguishes itself from first quantization in that it enforces antisymmetry through the construction of its operators, rather than through the wavefunction. As such, the operators used to construct the Hamiltonian must abide by certain properties. 
The action of the operators must also allow moving a particle from one basis function to another (e.g. from an occupied orbital to a virtual orbital). In particular, while in first quantization this action (e.g. $\ket{\phi_p^{(i)}} \bra{\phi_q^{(i)}}$) straightforwardly acts as moving an electron, operators in second quantization additionally need to verify antisymmetric properties. These operators are often referred to as fermionic creation ($\hat{a}^{\dagger}_p$) and annhiliation ($\hat{a}_p$) operators.  In their 1928 paper \cite{Jordan1928}, Jordan and Wigner introduced the canonical fermionic anti-commutation relation:
\begin{equation} \label{eq:anticommutation}
\begin{aligned}
&\left\{\hat{a}_p, \hat{a}_q^{\dagger}\right\}= \delta_{pq}, \\
&\left\{\hat{a}_p^{\dagger}, \hat{a}_q^{\dagger}\right\} = \left\{\hat{a}_p, \hat{a}_q\right\} = 0, 
\end{aligned}
\end{equation}

from where one can also derive the commutation relations:

\begin{equation}
\begin{aligned}
&\left[\hat{a}_p, \hat{a}_q \right] = -2\hat{a}_q\hat{a}_p, \\
&\left[\hat{a}_p, \hat{a}_q^{\dagger} \right] = \delta_{pq} - 2\hat{a}_q\hat{a}_p.
\end{aligned}
\end{equation}

These allow to re-write the Slater determinants as
\begin{equation}
    \ket{\psi} = \prod_i (\hat{a}_i^\dagger)^{\phi_i} \ket{\mathrm{vac}} 
    = (\hat{a}_1^\dagger)^{\phi_1} (\hat{a}_2^\dagger)^{\phi_2} \cdots (\hat{a}_n^\dagger)^{\phi_n} \ket{\mathrm{vac}},
\end{equation}
where $\ket{\mathrm{vac}}$ is the special vacuum state which disappears after any operation by the annihilation operator, 
\begin{equation}
    \hat{a}_j\ket{\mathrm{vac}}=0.
\end{equation}
With such definition, $\hat{a}_{p}^{\dagger}$ acts on the unoccupied $p$-th orbital to make it occupied, and $\hat{a}_{p}$ acts on the occupied $p$-th orbital to make it unoccupied. Specifically, it is straightforward to show that
\begin{equation} \label{eq:JWladder}
\begin{split}
    \hat{a}_{j}^{\dagger }\ket{\phi_{1} \phi_{2} \cdots } & =\begin{cases}
    0 , &\phi_{j} =1\\
    s_p \ket{\phi_1\phi_2\cdots 1_j \cdots} , &\phi_{j} =0
    \end{cases}\\
    \hat{a}_{j}\ket{\phi_{1} \phi_{2} \cdots } & =\begin{cases}
    0 , &\phi_{j} =0\\
    s_p \ket{\phi_1 \phi_2 \cdots 0_j \cdots}, &\phi_{j} =1
    \end{cases}
\end{split}
\end{equation}
where $s_p$ is the parity of $p$-th orbital, i.e. $s_p$ is $1$ or $-1$ when the number of occupied orbitals up to and not including the $p$-th orbital is even or odd:
\begin{equation}
    p = ( -1)^{\sum _{i=1,2,\cdots j-1} \phi_{i}}.
\end{equation}

Fermionic operators allow re-writing the Hamiltonian presented in Eq. (\ref{eq:first_quantized_hamiltonian}), using the one and two body integrals given in Eq.~(\ref{eq:HFintegral1}) and Eq.~(\ref{eq:HFintegral2}):
\begin{equation}
\label{eq:molecularhamiltonianladder}
\hat{H} =\sum_{p q} h_{p q} \hat{a}_{p}^{\dagger} \hat{a}_{q}+\frac{1}{2} \sum_{p q r s} h_{p q r s} \hat{a}_{p}^{\dagger} \hat{a}_{q}^{\dagger} \hat{a}_{r} \hat{a}_{s},
\end{equation}
providing a second quantized form of the molecular Hamiltonian. Lattice Hamiltonians can also be written using these operators. We can  project the electronic coordinates into a basis set in order to define the two-body reduced density matrix (RDM):
\begin{equation}
    \Gamma_{pqrs}
    \equiv
    \bra{\psi} \hat{a}_p^{\dagger}\hat{a}_q^{\dagger} \hat{a}_r \hat{a}_s \ket{\psi},
    \label{eq:two_body_rdm}
\end{equation}
with other rank RDMs defined equivalently, and where the indices $p, q, \dots$ label spin-orbital degrees of freedom. In this example, the partial trace down to the one-body RDM can then be written as
\begin{equation} \label{eq:one_body_rdm}
    \gamma_{pr} = \frac{1}{n-1}\sum_q\Gamma_{pq,rq}.
\end{equation}
Despite tracing out large numbers of degrees of freedom, these two-body RDMs still contain all the information about a quantum system required for physical observables of interest which depend on (up to) pairwise operators, including the total energy.

Because, unlike first quantization operators, the fermionic operators are not defined explicitly, there could be a number of different ways to define them in terms of explicit Pauli operators (the reference to Jordan and Wigner's 1928 research is one of them). We have dedicated Sec. \ref{sec:Encoding} to detailing numerous methods used to explicitly defined these operators. 

\subsection{Other Hamiltonian models}

\subsubsection{Lattice Hamiltonians} 

Rather than defining a Hamiltonian based on atomic configurations, lattice models assumes a number of sites organized along a lattice, which could be a one-dimensional chain, a two-dimensional lattice of various geometries, or any higher dimensional graph. Here we consider electrons as the “particles moving in this discretized space. Note that if one considers bosonic instead of fermionic particles, this would result in much simpler representation and encoding, since the fermionic antisymmetry relationships described in the following section are not required. 
Lattice models are widely used in condensed matter physics to model phenomenological properties of certain materials, such as electronic band structures \cite{nemoshkalenko1998computational, Harrison2004, Marder2010} or phase transitions \cite{Vojta2003, Sachdev2009}. There exist a number of lattice models, here we only describe a few examples briefly: 
\begin{itemize}
    \item Hubbard / Fermi-Hubbard~\cite{Simon2013}:
    \begin{equation}
    \label{eq:Hubbardmodel_maintext}
    \hat{H} = -t \sum_{\sigma} \sum_{\langle p, q \rangle} ( \hat{a}_{p,\sigma}^{\dagger} \hat{a}_{q,\sigma} + \hat{a}_{q,\sigma}^{\dagger} \hat{a}_{p,\sigma}) + U \sum_{p} \hat{a}_{p,\uparrow}^{\dagger}
    \hat{a}_{p,\uparrow} \hat{a}_{p,\downarrow}^{\dagger} \hat{a}_{p,\downarrow},
    \end{equation}
    where the sum $\langle p, q \rangle$ is only taken for neighboring lattice sites, with $\hat{a}_{p,\sigma}$ denoting second quantized electron annihilation operators on site $p$ with $\sigma \in \{\alpha, \beta\}$ an index for the spin of the electron, as discussed in more detail in Sec.~\ref{sec:second_quantization}.
    
    \item The Anderson impurity model used within the dynamical mean field theory~\cite{Bauer2016} (see Sec. \ref{sec:Extensions_of_VQE}):
    \begin{equation}
    \begin{aligned}
    \hat{H} & =\hat{H}_{\mathrm{imp}} +\hat{H}_{\mathrm{bath}} +\hat{H}_{\mathrm{mix}}\\
    \hat{H}_{\mathrm{imp}} & =\sum _{\alpha \beta } t_{\alpha \beta } a_{\alpha }^{\dagger } a_{\beta } +\sum _{\alpha \beta \gamma \delta } U_{\alpha \beta \gamma \delta } a_{\alpha }^{\dagger } a_{\beta }^{\dagger } a_{\gamma } a_{\delta }\\
    \hat{H}_{\mathrm{mix}} & =\sum _{\alpha i}\left( V_{\alpha i} a_{\alpha }^{\dagger } a_{i} +\overline{V}_{\alpha i} a_{i}^{\dagger } a_{\alpha }\right)\\
    \hat{H}_{\mathrm{bath}} & =\sum _{i} t'_{i} a_{i}^{\dagger } a_{i}
    \end{aligned}
    \end{equation}
    where indices $\{\alpha ,\beta ,\gamma ,\delta \}$ refer to the impurity sites, index $i$ refers to bath sites, and$\{t_{\alpha \beta } ,U_{\alpha \beta \gamma \delta } V_{\alpha i} ,t'_{i}\}$ parameterize the impurity model. $\hat{H}_{\mathrm{imp}}$ is a general Hamiltonian describing the local correlation of the impurity, though this is also often approximated to have Hubbard-like interactions. $\displaystyle \hat{H}_{\mathrm{mix}}$ describes the hopping between the impurity and the bath, which is taken in this example in a particular `star' geometry, and $\hat{H}_{\mathrm{bath}}$ is the non-interacting Hamiltonian of the bath sites. 
    
    \item Spin Hamiltonians, such as the Heisenberg model~\cite{bosse2021probing,kattemolle2021variational}:
    \begin{equation}
        \hat{H} = J \sum_{\langle p,q \rangle} \hat{\boldsymbol{S}}_p \cdot \hat{\boldsymbol{S}}_q
    \end{equation}
    where again $ \langle p,q \rangle$ denotes a sum over neighboring pairs of sites on the lattice, $J$ is the positive exchange constant, $\hat{\boldsymbol{S}}_p = (\hat{S}^x_p, \hat{S}^y_p, \hat{S}^z_p)$ is the three spin-$1/2$ angular momentum operators on site $p$. Note that the spin-$1/2$ matrices are related to Pauli matrices by $(\hat{S}^x, \hat{S}^y, \hat{S}^z)=\frac{\hbar}{2}(X, Y, Z)$.

\end{itemize}

\subsubsection{Vibrational Hamiltonian model} 

The vibrational Hamiltonian is the nuclear-motion counterpart to the electronic Hamiltonian within the Born-Oppenheimer approximation. The Hamiltonian describes the nuclear motions resulting from bond vibration or rotation of the whole molecule. Example of vibrations will include bond stretching, or bending. The vibrationals modes is the number of possibility for a molecule to undergo vibrational tensions. For a molecule with $N$ atoms, each atom can move along three dimensions resulting in a total of $3N$ degrees of freedom. However, three degrees of freedom correspond to translations of the molecule in 3D space, and three degrees of freedom correspond to rotations of the molecule (of course, if the molecule is linear, there are only two rotational modes). This results in a $3N - 6$ for molecules ($3N - 5$ for linear molecule). 

Vibrational Hamiltonian have far more options in their constructions than molecular Hamiltonians \cite{03PeHa} resulting in a large variety of Hamiltonian representations. The simplest example, presented here is that of the harmonic oscillator, suppose a molecule with $V = 3N - 6$ vibrational modes. We can provide an example of an effective nuclear Hamiltonian following McArdle \textit{et al.} \cite{McArdle2019_vibra}, once the electronic Hamiltonian has been solved (or estimated), as: 
\begin{equation}
\label{eq:vibrational_Hamiltonian}
\hat{H} = \frac{\boldsymbol{p}^2}{2} + V_s(\boldsymbol{q}),
\end{equation}
where $\boldsymbol{q}$ represent the nuclear positions, $\boldsymbol{p}$ the nuclear momenta, and $V_s(.)$ the nuclear potential which is dependent on the electronic potential energy surface. Noting $\omega_i$ the harmonic frequency of vibrational mode $i$, the Hamiltonian in Eq. \ref{eq:vibrational_Hamiltonian} can be approximated as a sum of independent harmonic oscillators: 

\begin{equation}
\label{eq:vibrational_Hamiltonian_harmonic}
\hat{H} = \sum_i \omega_i \hat{a}^{\dagger}_i \hat{a}_i.
\end{equation}

The accuracy of results can be improved by adding anharmonic terms, however raising the complexity of the computation. In this case, the annhiliation and creation act differently than in the case of fermionic operators, instead they represent transitions between different eigenstate of a single mode harmonic oscillator Hamiltonian. Suppose we consider $d$ eigenstates of $\hat{h}_i = \omega_i \hat{a}^{\dagger}_i \hat{a}_i$. In Ref. \cite{McArdle2019_vibra}, two means of encoding such Hamiltonians in qubits are proposed: in the first one (called the direct mapping) each eigenstate $\ket{s}$, $s \in \{0, d - 1\}$ requires $d$ qubits, and only qubit $s$ is equal to $1$ for state $\ket{s}$. The resulting creation operator is: 
\begin{equation}
\label{eq:vibrational_direct_creation}
\hat{a} = \sum_0^{d - 2} \sqrt{s + 1} \ket{0}\bra{1}_s \otimes \ket{1}\bra{0}_{s + 1}.
\end{equation}
In the second case (called the compact mapping), each eigenstate is encoded in binary form, with each requiring $\log_2(d)$ qubits. Which results in the following creation operator: 
\begin{equation}
\label{eq:vibrational_compact_creation}
\hat{a} = \sum_0^{d - 2} \sqrt{s + 1} \ket{s + 1}\bra{s}.
\end{equation}
From this representation, the Hamiltonian of Eq.~(\ref{eq:vibrational_Hamiltonian_harmonic}) (or its extension to anharmonic terms) can be mapped to Pauli operators in the same way as the electronic Hamiltonian in first quantization (see Sec. \ref{sec:first_quantization}). Given $V$ vibrational modes, the direct mapping requires $Vd$ qubits, and if the anharmonic terms up to order $k$ are included the Hamiltonian will have $\mathcal{O}(V^kd^k)$ Pauli terms. The compact mapping requires $V\log_2(d)$ qubits, and $\mathcal{O}(V^kd^{2k})$ terms in the Hamiltonian \cite{McArdle2019_vibra}. 

\subsubsection{Periodic systems} 

Periodic system Hamiltonians aim at representing the energetic behavior of solid state systems, and periodic materials. As such, they are very similar to the definition of the molecular electronic structure Hamiltonian, but with the addition of boundary conditions defining the periodicity of the system. In second-quantized form (as described in Ref. \cite{Yoshioka2022, Manrique2020}), the crystal Hamiltonian is given by: 

\begin{equation}
\label{eq:molecularhamiltonianladder_periodic}
\hat{H} =\sum_{p q} \sum_{\boldsymbol{k}} h_{p q}^{\boldsymbol{k}} \hat{a}_{p\boldsymbol{k}}^{\dagger} \hat{a}_{q\boldsymbol{k}}+\frac{1}{2} \sum_{p q r s} \sum_{\boldsymbol{k}_p\boldsymbol{k}_q\boldsymbol{k}_r\boldsymbol{k}_s} h_{p q r s}^{\boldsymbol{k}_p\boldsymbol{k}_q\boldsymbol{k}_r\boldsymbol{k}_s} \hat{a}_{p\boldsymbol{k}_p}^{\dagger} \hat{a}_{q\boldsymbol{k}_q}^{\dagger} \hat{a}_{r\boldsymbol{k}_r} \hat{a}_{s\boldsymbol{k}_s},
\end{equation}

where $h_{p q}^{\boldsymbol{k}}$, and $h_{p q r s}^{\boldsymbol{k}_p\boldsymbol{k}_q\boldsymbol{k}_r\boldsymbol{k}_s}$ are the one and two body integrals of the periodic system. In this representation, the one electron integrals is diagonal in $\boldsymbol{k}$ and the two-electron integrals follows $\boldsymbol{k}_p + \boldsymbol{k}_q - \boldsymbol{k}_r - \boldsymbol{k}_s = \boldsymbol{G}$, where $\boldsymbol{G}$ is a reciprocal lattice vector \cite{Manrique2020, McClain2017}.

%% file: 04_encoding_main.tex
\section{Fermionic space to spin space transformations}\label{sec:Encoding}

The core part of the VQE algorithm is the measurement of the expectation value of the Hamiltonian operator (Eq. \ref{eq:molecularhamiltonianladder}) with respect to a parameterized ansatz wavefunction.  As mentioned in the previous section, in the NISQ era, with a limited number of qubits available, second quantization formalism has been favored over first quantization for quantum simulation. The fermionic creation and annihilation operators in second quantization formalism obey the anti-commutation algebra (Eq. \ref{eq:anticommutation}).

Qubits in comparison are spin-$1/2$ objects, and as such, the only operators that can, in general, be directly measured on QPUs are spin-operators (or the Pauli operators, $X$, $Y$, and $Z$) which obey a different algebra specified by their Lie bracket. This means that before starting the VQE loop, one must transform the Hamiltonian in second quantization into a linear combination of Pauli strings (tensor product of Pauli operators on multiple qubits). 

As discussed in Sec. \ref{sec:second_quantization} however, operators in second quantization must enforce the antisymmetry of the wavefunction and as such must obey anti-commutation rules as detailed in Eq.~(\ref{eq:anticommutation}). Pauli operators do not naturally obey these relationships and therefore specific cares must be brought to the mapping of fermionic operators to spin operators in the case of second quantization. 

All transformations can be formalized as 
\begin{equation}
    \mathcal{T}: \mathcal{F}_n \rightarrow (\mathbb{C}^2)^{\otimes N},
\end{equation}
where a transformation $\mathcal{T}$ maps the space of operators acting on Fock states of $n$ spin orbitals, $\mathcal{F}_n$, to the Hilbert space $(\mathbb{C}^2)^{\otimes N}$ of operators acting on spin states of $N$ qubits. The most important feature of the transformation is that it must maintain the anti-commutation of fermionic operators. Conveniently, Jordan and Wigner \cite{Jordan1928} showed long before Quantum Computers were even conceptualized that an isomorphism exists between fermionic space and qubit space, maintaining the algebraic structure. 

It is important to note that the mappings described in this section not only affect the operators measured when computing the expectation value of the Hamiltonian, but also the construction of an ansatz that is initially defined in fermionic terms (for instance the Unitary Coupled Cluster ans{\"{a}}tze, see Sec. \ref{sec:UCCA}). These ans{\"{a}}tze must be transformed into a series of Pauli operators exponentials which can be implemented as quantum gates.

In this section, we present a number of such transformations. There are three main characteristics relevant to deciding on a specific encoding, although it is worth noting that these are not necessarily independent from each other:
\begin{itemize}
    \item \textbf{Number of qubits:} The number of qubits required to represent the electronic wavefunction. In general, the number of qubits is directly proportional to the number of spin-orbitals or sites considered. However, several techniques have been developed to concentrate the information held in the wavefunction to as few qubits as possible \cite{Moll2016, bravyi_tapering_2017, Steudtner2018, Setia2020, Kirby2021_CSVQE}. These methods generally rely on symmetries of the Hamiltonian and are presented in the last part of this section. 
    \item \textbf{Pauli weight:} The maximum number of qubits on which each Pauli string produced act, i.e. the maximum number of non-identity operators in any Pauli string produced by the mapping. This is, in general, referred to as Pauli weight. Many features throughout the VQE pipeline are affected by this number. Firstly, low-weight encodings result in lower depth ans{\"{a}}tze and lower circuit construction costs \cite{Havlek2017, Clinton2020, Cade2020}. This is in part because a few operators that act locally on different qubit subsets can be implemented in parallel, in other parts because operators acting on local qubit subsets require fewer entangling gates than non-local qubit subsets. Secondly, it has been shown that using low-weight operators as observable in the VQE cost function provides some resilience against the barren plateau problem \cite{Cerezo2021_BP} (see Sec. \ref{sec:barren_plateau}). Finally, the lower the Pauli weight, the lower is the overall probability of readout error. This is obvious, as identity operators do not need to be measured, and as such the probability of measurement error increases with the Pauli weight \cite{Huggins2021}.
    \item \textbf{Number of Paul-strings:} The number of different Pauli strings resulting from the mapping. The number of Pauli strings directly impacts the cost of implementing a VQE. As such, one should always prefer to have the lowest number of Pauli strings to measure. In general, we see that this number of strings scales $\mathcal{O}(n^4)$ for molecular Hamiltonian, and with the number of edges for lattice models.  
\end{itemize}

We can distinguish two main families of fermion-to-spin encoding. The first one concerns itself with being as general as possible and directly encodes the entire Fock space. The mappings it includes tend to be more relevant for \textit{ab initio} molecular systems, with their main drawback being the non-locality of the operators produced (see Sec. \ref{sec:gen_encoding}). The second one uses the specific geometry of the system studied to try to minimize the Pauli weight of the operators produced, at the cost of using additional qubits. These mappings tend to be more relevant for low degree lattice models (for instance the 2-dimensional Hubbard model) as they allow capping the Pauli weight (see Sec. \ref{sec:lattice_encoding}).

\subsection{Generalized encodings} \label{sec:gen_encoding}

In this section, we are concerned with encodings that map the entire Fock space in which the Hamiltonian is expressed. As such, the number of qubits $N$ required from these encodings is in principle equal to the number of spin orbitals $n$ in the Hamiltonian considered (without taking into consideration possible use of symmetries to reduce the number of qubits required, as presented in Sec. \ref{sec:tappering_qubits}). These encodings are the most general, in that they are not tailored for specific Hamiltonian structures. They are agnostic to the degree of connectivity of the Hamiltonian graph (the maximum number of fermionic operators linking one spin orbital to another) and are in general better suited for \textit{ab initio} molecular systems than to lattice models. One recurrent issue related to these encodings is the fact that they transform one and two-body fermionic operators which are local (act only on up to four spin orbitals at the time) into Pauli strings which are in general non-local and therefore have either high Pauli weight or Pauli weight that scales with the number of spin orbitals in the system. It also means that high connectivity on the qubit lattice may be required to efficiently implement fermionic operator based ans{\"{a}}tze without relying on a large number of entangling gates. A summarized comparison of these mappings is presented in Table \ref{tab:generalized_encodings}.

\begin{table} [ht]
\caption{Overview and comparison of the generalized encodings. $n$ represents the number of fermionic mode in the Hamiltonian considered. All mappings in the table use by default $n$ qubits, and produce a number of operators scaling $\mathcal{O}(n^4)$. These operators are counted assuming an \textit{ab initio} molecular Hamiltonian in second quantized form.}
\begin{tabularx}{\textwidth}{p{0.20\linewidth}|cp{0.6\linewidth}}
\toprule
\\  Method &  Pauli Weight & Comments \\\\
\midrule\\
    Jordan-Wigner \cite{Jordan1928} & $\mathcal{O}(n)$ & Most commonly used mapping. Encodes orbital occupation directly and locally onto qubits.  \\\\
\hline\\
    Parity \cite{Seeley2012} & $\mathcal{O}(n)$ &  Encodes orbital parity directly and locally onto qubits.  \\\\
\hline\\
    Bravyi-Kitaev \cite{Bravyi2002, Seeley2012} & $\mathcal{O}(\log_2(n))$ & Focuses on minimizing the Pauli weight by mixing occupation and parity encoding. Usually results in lower gate depth \cite{Tranter2015, Tranter2018, Setia2018}, but not necessarily higher noise resilience \cite{Sawaya2016}. \\\\
\hline\\
    Optimal general encoding on ternary trees \cite{Jiang2020} & $\mathcal{O}(\log_3(2n))$ & Achieves optimal Pauli weight asymptotically. Little to no benchmarking of actual applications compared to other mapping invites to further investigation. \\\\
\bottomrule
\end{tabularx}
 \label{tab:generalized_encodings}
\end{table}

\subsubsection{The Jordan-Wigner encoding} \label{sec:jordanwigner}

The Jordan-Wigner mapping encodes the electronic wavefunction in an array of qubits by mapping the occupation number of spin orbitals in qubits. The occupation number of $n$ spin orbitals is stored in $N$ qubits. $|0\rangle_{j}$ corresponds to the $j$-th spin orbital being unoccupied and $|1\rangle_{j}$ corresponds to the $j$-th spin orbital being occupied (where once again, $j$ merges the indices of spatial and spin orbitals. The theoretical foundations of this mapping lie in the 1928 Jordan-Wigner transformation \cite{Jordan1928} using spin-$1/2$ operators to explicitly describe fermionic ladder operators. Its inverse can therefore be used as a means to simulate fermions on a Quantum Computer \cite{Ortiz2001, Somma2002, Somma2003, AspuruGuzik2005}. 

As mentioned in Sec. \ref{sec:second_quantization}, Jordan and Wigner \cite{Jordan1928} introduced the canonical fermionic anti-commutation relation given in Eq.~(\ref{eq:anticommutation}) and showed how one can go about making the identification between fermionic operators and spin operators.

Considering the case of a single orbital, we can write the fermionic operators actions presented in Eq. \ref{eq:JWladder} on the $j$-th qubit as spin operators:
\begin{equation}
\label{eq:JW1}
\begin{aligned}
\hat{a}_{j}^{\dagger} \xrightarrow{?} |1\rangle \langle 0|_j
&=\left[\begin{array}{ll}
0 & 0 \\
1 & 0
\end{array}\right]=\frac{X_{j}-i Y_{j}}{2} \\
\hat{a}_{j} \xrightarrow{?} |0\rangle \langle 1|_j
&=\left[\begin{array}{ll}
0 & 1 \\
0 & 0
\end{array}\right]=\frac{X_{j}+i Y_{j}}{2}, \\
\end{aligned}
\end{equation}
where $X_j$ and $Y_j$ are Pauli gates acting on the $j$-th qubit.
These Pauli operators do not enforce the fermionic sign prescription in Eq.~(\ref{eq:JWladder}) (or equivalently the anticommutation relation of Eq. (\ref{eq:anticommutation})) which is critical to preserve the structure of the algebra through the transformation.

We can fix this by upgrading the definition in Eq. (\ref{eq:JW1}) to include a string of $Z$ operators $Z_{0} \otimes \dots \otimes Z_{j-1}$ acting on each of the other qubits up to the $j$-th.
Indeed we have that $Z_{0} \otimes \dots \otimes Z_{j-1}$ has eigenvalue $+1$ for states with an even number of occupied orbitals up to the $j$-th, and eigenvalue $-1$ for states with an odd number of occupied orbitals up to the $j$-th, restoring the fermionic sign prescription and providing the Jordan-Wigner transformation:
\begin{equation}
\label{eq:JW2}
\begin{aligned}
\hat{a}_{j}^{\dagger} &\to \frac{X_{j}-i Y_{j}}{2} \otimes Z_{0} \otimes \dots \otimes Z_{j-1}\\
\hat{a}_{j} &\to \frac{X_{j}+i Y_{j}}{2} \otimes Z_{0} \otimes \dots \otimes Z_{j-1}.
\end{aligned}
\end{equation}

One can easily verify that this also enforces the anticommutation relation of Eq. (\ref{eq:anticommutation}). Indeed for $i \neq j$ we have that $\hat{a}_i$ and $\hat{a}_j^\dagger$ anticommute:
\begin{equation}
\hat{a}_i \hat{a}_j^{\dagger} = -\hat{a}_j^{\dagger} \hat{a}_i,
\end{equation}
and the definition in Eq. (\ref{eq:JW2}) gives us the correct anticommutation relation as the $Z_j$ operator anticommutes with both $X_j$ and $Y_j$ and hence anticommutes with $|0\rangle \langle 1|_j$ and $|1\rangle \langle 0|_j$).

The direct consequence of adding the $Z$ strings is that the Pauli weight of Jordan-Wigner mapping scales $\mathcal{O}(N)$. This makes the mapping relatively costly with regards to the number of entangling gates required to simulate fermionic based ans{\"{a}}tze. However, it is worth noting that some ans{\"{a}}tze include spin operator level parameters \cite{Yordanov2021}. This allows the VQE to variationally learn the anti-commuting relationship and therefore allows forgoing the $Z$ strings at the cost of additional circuit parameters. This approach caps the Pauli weights of the Jordan-Wigner mapping to four (one for each fermionic operator in two-body terms) and significantly reduces the number of entangling gates required.  

Another feature of the Jordan-Wigner mapping that is worth discussing is the number of Pauli strings produced. From Eq.~(\ref{eq:JW2}), it is clear that each fermionic operator results in two Pauli strings. As such, the number of Pauli strings scales at the same rate as the number of fermionic operators in the system considered. For a second quantized Hamiltonian, the number of fermionic operators scales with the number of two-body fermionic terms, or $\mathcal{O}(n^4)$, and therefore, the number of strings in the Jordan-Wigner mapping would scale  $\mathcal{O}(N^4)$. The Jordan-Wigner mapping is by design applicable only to qubits, encoded from SU(2) fermions. However a generalized form of Jordan-Wigner, applicable to SU(N) fermions was proposed by Consiglio \textit{et al.} \cite{Consiglio2021}.

\subsubsection{The parity encoding}

If instead of using $j$-th qubit to encode directly the information of whether the $j$-th orbital is occupied, we use it to encode information about the parity of the orbitals up to the $j$-th, we obtain the parity encoding. The parity mapping was explicitly defined in \cite{Seeley2012}. We have $|0\rangle_{j}$ if the number of orbitals up to and including the $j$-th that are occupied is even and $|1\rangle_{j}$ if it is odd. Given this definition, if we are given the fermionic state $\left|\nu_{0} \nu_{1} \ldots \nu_{n}\right\rangle$ we can translate it to the qubit state $\left|p_{0} p_{1} \ldots p_{n}\right\rangle$ in the parity encoding by
\begin{equation}
\label{eq:parity1}
\begin{aligned}
p_{i} &= \sum_{j \leq i} \nu_{j} &\pmod 2\\
&= \sum_{j}\left[\pi_{n}\right]_{i j} \nu_{j} &\pmod 2,
\end{aligned}
\end{equation}
or:
\begin{equation}
\label{eq:parity1.1}
\ket{\boldsymbol{p}} =  \ket{\pi_{n}(\boldsymbol{\nu})} \quad \pmod 2,
\end{equation}
where addition is taken modulo 2, and $\pi_{n}$ is the $n \times n$ matrix defined as
\begin{equation}
\label{eq:parity2}
\pi_{n}=\left[\begin{array}{cccc}
1 & 0 & \cdots & 0 \\
1 & 1 & \cdots & 0 \\
\vdots & \vdots & \ddots & \vdots \\
1 & 1 & \cdots & 1
\end{array}\right].
\end{equation}
If the parity changes at the $j$-th position (that is if we have either $|0\rangle_{j - 1}$ and $|1\rangle_{j}$ or $|1\rangle_{j - 1}$ and $|0\rangle_{j}$) then we know that the $j$-th orbital is occupied. Conversely if the parity does not change (that is if we have either $|0\rangle_{j - 1}$ and $|0\rangle_{j}$ or $|1\rangle_{j - 1}$ and $|1\rangle_{j}$) the $j$ is unoccupied. 

Moreover, in the parity representation, we already have the information on the parity of the qubits up to the $j$-th encoded in the ($j - 1$)-th qubit. Hence we know that following Eq. (\ref{eq:JWladder}) the operators $\hat{a}^\dagger_j$ and $\hat{a}_j$ give a minus sign when we have $|1\rangle_{j - 1}$, and a plus sign when we have $|0\rangle_{j - 1}$.  
We can thereby define mappings for isolated fermionic operators as
\begin{equation}
\label{eq:parityladder1}
\begin{aligned}
\hat{a}_{j}^{\dagger} &\xrightarrow{?} |01\rangle \langle 00|_{j-1, j} - |10\rangle \langle 11|_{j-1, j} \\
&=\frac{Z_{j-1} \otimes X_{j} -i Y_{j}}{2} \\
\hat{a}_{j} &\xrightarrow{?} |00\rangle \langle 01|_{j-1, j} - |11\rangle \langle 10|_{j-1, j} \\
&=\frac{Z_{j-1}\otimes X_{j} + i Y_{j}}{2}. 
\end{aligned}
\end{equation}

However because we are adding or removing an electron in the $j$-th orbital and hence changing the parity encoded in the qubits that follow the $j$-th we also need to flip those qubits. We can update the parity by means of a string of $X$ operators acting on the qubits that follow the $j$-th qubit, $X_{j + 1} \otimes \dots \otimes X_{n - 1}$, so that the parity transformation is given by
\begin{equation}
\label{eq:parityladder2}
\begin{aligned}
\hat{a}_{j}^{\dagger} &\to \frac{Z_{j-1}\otimes X_{j} -i Y_{j}}{2} \otimes X_{j + 1} \otimes \dots \otimes X_{n-1} \\
\hat{a}_{j} &\to \frac{ Z_{j-1}\otimes X_{j} + i Y_{j}}{2} \otimes X_{j+1} \otimes \dots \otimes X_{n-1} \\
\end{aligned}
\end{equation}

The Pauli weight of the parity mapping scales similarly to that of the Jordan-Wigner mapping: $\mathcal{O}(N)$. It also results in the same operator scaling: $\mathcal{O}(N^4)$.  

\subsubsection{The Bravyi-Kitaev encoding} \label{sec:bravyi-kitaev}

The theory of the Bravyi-Kitaev encoding was first given in Ref. \cite{Bravyi2002} and then the encoding was explicitly constructed in Ref.~\cite{Seeley2012} (and revisited later in Ref.~ \cite{Tranter2015}). The motivation for this encoding comes from the desire to lower the Pauli weight of the qubit operators. The Bravyi-Kitaev encoding achieves lower Pauli weight by storing a combination of parity and occupation number in qubits. This produces $\mathcal{O}(N^4)$ spin operator terms, with a maximum Pauli weight of $\mathcal{O}(\log_2(N))$. 

We can define an encoding matrix for the Bravyi-Kitaev mapping, similar to the one proposed for the parity mapping in Eq.~(\ref{eq:parity2}). It is produced recursively by defining blocks of increasing sizes and combining them. If we set $x$ as the recursive index, the block $\beta$ of index $x$ has size $2^x$. Therefore, as the final matrix must be $N\times N$, the recursion stops whenever $x \geq \log_2(N)$ - the matrix rows and columns greater than $N$ can be discarded. 

Starting from  $\beta_1 = \left[\begin{array}{c} 1 \end{array}\right]$, the recursion rule can be defined as 
\begin{equation}
\label{eq:bravyikitaev2}
\beta_{2^x} =\left[\begin{array}{c|c}
\beta_{2^{x-1}} & 0 \\
\hline \begin{array}{c} 0 \\ \leftarrow 1 \rightarrow \end{array} & \beta_{2^{x-1}}
\end{array}\right],
\end{equation}
where the bottom-left block is a $2^{x-1} \times 2^{x-1}$ matrix of all zeros except for the bottom row.

For example, the Bravyi-Kitaev block for $x = 3$ matrix acting on $8$ qubits $\beta_8$ is given by
\begin{equation}
\beta_8 = \left[\begin{array}{llllllll}
1 & 0 & 0 & 0 & 0 & 0 & 0 & 0 \\
1 & 1 & 0 & 0 & 0 & 0 & 0 & 0 \\
0 & 0 & 1 & 0 & 0 & 0 & 0 & 0 \\
1 & 1 & 1 & 1 & 0 & 0 & 0 & 0 \\
0 & 0 & 0 & 0 & 1 & 0 & 0 & 0 \\
0 & 0 & 0 & 0 & 1 & 1 & 0 & 0 \\
0 & 0 & 0 & 0 & 0 & 0 & 1 & 0 \\
1 & 1 & 1 & 1 & 1 & 1 & 1 & 1
\end{array}\right].
\end{equation}
From here, it is easy to notice that within each block, the $j$-th qubit encodes the occupancy of the $j$-th orbital when $j$ is even, and it encodes information on the parity of a set of orbitals and occupancy of the $j$-th orbital combined when $j$ is odd. Regarding the parity information, we can distinguish two cases: if $j$ is odd but different from a power of $2$, minus 1 (since we index from $0$) it encodes the parity of orbitals up to the $j$-th orbital within a block; if it is equal to a power of $2$ minus 1, it encodes the parity of the entire set of orbitals up to and including $j$. 

To be able to write down expressions for the representations of the $\hat{a}^\dagger_j$ and $\hat{a}_j$ operators we first need to consider four qubit sets (for a thorough and formal definition of these sets and how to obtain them recursively, we recommend \cite{Seeley2012, Tranter2015}). These sets are defined based on index $j$ and determine the behavior of qubits of other indices whenever an operator acts on qubit $j$ :
\begin{itemize}
  \item \textbf{The update set $U(j)$} includes qubits that are dependent on the occupation of orbital $j$. Because even-index qubits encode occupancy, they are never part of this set. Odd qubits are included if they encode the parity of orbital $j$. Intuitively, it corresponds to the nonzero elements in the $j$-th column of the Bravyi-Kitaev transformation matrix, excluding the $j$-th qubit itself (for example, Eq. (\ref{eq:bravyikitaev2}). 
  \item \textbf{The parity set $P(j)$} are the qubits that determine the parity of the set of orbitals up to, but excluding, $j$ (and therefore the fermionic sign, Eq. \ref{eq:JWladder}), when an operator acts on qubit $j$). 
  \item \textbf{The flip set $F(j)$} is a subset of the parity set which determines whether qubit $j$ is equal or opposite to orbital $j$ of the fermionic basis. Of course, this set is always empty for even qubits which encode occupancy. 
  \item \textbf{The remainder set $R(j)$} are the qubits that are part of the parity set, but not the flip set.
  \begin{equation}
      R(j) = P(j) \setminus F(j)
  \end{equation}
\end{itemize}
For a visual interpretation of how these sets can be constructed, one can refer to the Fenwick tree construction of the Bravyi-Kiteav mapping presented in Ref.~\cite{Havlek2017} (we also provided a brief description in Appendix \ref{sec:Fenwick_trees}).

With this in mind, we now look at how to express fermionic operators in the Bravyi-Kitaev mapping. These differ depending on whether they act on qubits that encode occupancy, or qubits that encode some parity information. When considering the definition of the qubit sets above, one can observe that due to the recursive definition of the Bravyi-Kiteav transformation matrix, the maximum number of qubits included in any set scales $\mathcal{O}(\log_2(N))$. The Pauli weight of this encoding follows from this observation and the construction of the operators presented below.

\paragraph{Operators for occupancy qubits:} as explained above, these operators apply to even-index qubits. For individual qubits, the creation and annihilation operators are identical to those initially formulated for the Jordan-Wigner mapping (Eq. \ref{eq:JW1}). We now need to incorporate the relevant actions on the reminder of the wavefunction, namely: we need to flip all the qubits above index $j$ that are affected by an action on qubit $j$ (i.e. the qubits from the update set $U(j)$, and we need to apply the correct fermionic sign, depending on the qubits that define the parity of the wavefunction up to orbital $j$ (the qubits from the parity set $j$). The former requires $X$ gates, the latter requires $Z$ gates. As such, we can write the operators acting on even qubits as
\begin{equation}
\label{eq:BKladder1}
\begin{aligned}
\hat{a}_{j}^{\dagger} &\rightarrow Z_{P(j)} \otimes |1\rangle \langle 0|_j \otimes X_{U(j)} \\
&= \frac{1}{2}  Z_{P(j)} \otimes \left(  X_{j} - iY_{j} \right) \otimes X_{U(j)}  \\
\hat{a}_{j} &\rightarrow Z_{P(j)} \otimes |0\rangle \langle 1|_j \otimes X_{U(j)}\\
&= \frac{1}{2} Z_{P(j)} \otimes \left(X_{j} + i Y_{j}\right) \otimes X_{U(j)} .
\end{aligned}
\end{equation}

\paragraph{Operators for parity qubits:} When $j$ is odd, a creation operator, which in the case of occupancy only excites qubits from $\ket{0}$ to $\ket{1}$, could flip a qubit from $\ket{0}$ to $\ket{1}$ instead depending on the parity of the wavefunction up until $j$ (if the parity is $1$ until $j$, and if qubit $j$ is in state $\ket{0}$ it means orbital $j$ is occupied as the parity is flipped). Recall that the flip set determines whether qubit $j$ is equal, or opposite to orbital $j$. Two cases are possible, either the number of qubits equal to $\ket{1}$ in the flip set is even, in which case qubit $j$ is equal to the orbital occupancy (the parity flips if qubit $j = \ket{1}$, and vice-versa); or the number of $\ket{1}$ in the flip set is odd, and qubit $j$ is opposite to the orbital occupancy (the parity flips if qubit $j = \ket{0}$, and vice-versa). Following this, one can define two projectors onto the even and odd states of the flip set:

\begin{equation}
\begin{aligned}
E_{F(j)} &= \frac{1}{2}\left(I^{\otimes N} + Z_{F(j)} \right)\\
O_{F(j)} &= \frac{1}{2}\left(I^{\otimes N} - Z_{F(j)} \right).
\end{aligned}
\label{eq:BKprojection}
\end{equation}
Therefore, action on an isolated qubit can be described as follows:

\begin{equation}
\label{eq:BKladder3}
\begin{aligned}
\hat{a}_{j}^{\dagger} &\xrightarrow{?}  E_{F(j)} \otimes |1\rangle \langle 0|_{j}  +  O_{F(j)} \otimes |0\rangle \langle 1|_{j}   \\
&= \frac{X_{j} - i Z_{F(j)} \otimes Y_{j}}{2} \\
\hat{a}_{j} &\xrightarrow{?}  E_{F(j)} \otimes |0\rangle \langle 1|_{j}  + O_{F(j)} \otimes |1\rangle \langle 0|_{j}   \\
&= \frac{X_{j} +   i Z_{F(j)} \otimes Y_{j}}{2}.
\end{aligned}
\end{equation}
Just like in the case of even $j$, our creation and annihilation operators also need to flip all the qubits in the update set $U(j)$ by applying $X$ Pauli gates on them and to enforce the correct fermionic sign by applying $Z$ operators to the qubits given by the parity set $P(j)$. And because $Z^2 = I$ we have $Z_{F(j)} \otimes Z_{P(j)} = Z_{P(j) \setminus F(j)} = Z_{R(j)}$, so that
\begin{equation}
\label{eq:BKladder4}
\begin{aligned}
\hat{a}_{j}^{\dagger} &\rightarrow \frac{Z_{P(j)} \otimes X_{j} \otimes X_{U(j)} -i  Z_{R(j)} \otimes Y_{j} \otimes X_{U(j)} }{2}\\
\hat{a}_{j} &\rightarrow \frac{ Z_{P(j)} \otimes X_{j} \otimes X_{U(j)}+ i Z_{R(j)} \otimes Y_{j} \otimes  X_{U(j)}}{2}
\end{aligned}
\end{equation}

\subsubsection{Optimal general encoding based on ternary trees} \label{sec:ternary_tree_encoding}

The mapping proposed by Jiang \textit{et al.} \cite{Jiang2020} offers an optimal scaling of Pauli weights of $\mathcal{O}(\log_3(2n + 1))$, with $n$ the number of fermionic modes, outperforming Bravyi-Kitaev in this respect. It is optimal in the sense that for a Hamiltonian for which fermionic modes are fully connected, it achieves the minimum average Pauli weight possible. It organizes qubits along with ternary trees \cite{Vlasov2019} and relies on the definition of the second quantized Hamiltonian in terms of Majorana fermions.

Majorana fermions are theorized particles, which act as their own antiparticle  \cite{Majorana1937, Majorana2006}. Most of the Standard  Model fermions are known not to behave like Majorana particles, at least at low energy, except the neutrino for which the question is still open \cite{Kayser2009, Baha2019, Hirsch2018, Bilenky2020}. Bound Majorana fermions have been shown to appear composed of several particles in condensed matter physics \cite{Wilczek2009}. Formally, this means that creation and annihilation operators for Majorana fermions are identical $\hat{\gamma}_i^{\dagger} = \hat{\gamma}_i$. These can also be expressed in terms of ordinary fermionic operators as
\begin{equation} \label{eq:majorana}
\begin{aligned}
\hat{\gamma}_{2j} &= \hat{a}_j + \hat{a}^\dagger_j\\
\hat{\gamma}_{2j + 1} &= -i(\hat{a}_j - \hat{a}^\dagger_j).
\end{aligned}
\end{equation}
There are two Majorana operators for each fermionic operator. These must anticommute if they are of different indices and commute otherwise.  

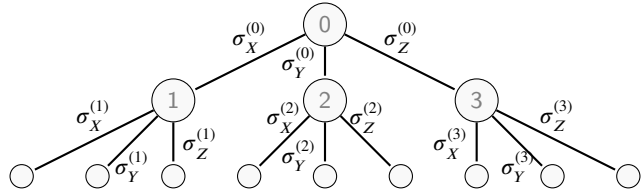
\begin{figure} [h]
\centering
\begin{tikzpicture}[
    shorten > = 1pt,
node distance = 3cm and 4cm,
    el/.style = {inner sep=2pt, align=left, sloped},
every label/.append style = {font=\tiny}
                    ]
\node[shape=circle,draw=black,fill=gray!5] (0) at (4,2) {0}; 
\node[shape=circle,draw=black,fill=gray!5] (1) at (2,1) {1};
\node[shape=circle,draw=black,fill=gray!5] (2) at (4,1) {2};
\node[shape=circle,draw=black,fill=gray!5] (3) at (6,1) {3};
\node[shape=circle,draw=black,fill=gray!5] (4) at (0,0) {};
\node[shape=circle,draw=black,fill=gray!5] (5) at (1,0) {};
\node[shape=circle,draw=black,fill=gray!5] (6) at (2,0) {};
\node[shape=circle,draw=black,fill=gray!5] (7) at (3,0) {};
\node[shape=circle,draw=black,fill=gray!5] (8) at (4,0) {}; 
\node[shape=circle,draw=black,fill=gray!5] (9) at (5,0) {}; 
\node[shape=circle,draw=black,fill=gray!5] (10) at (6,0) {}; 
\node[shape=circle,draw=black,fill=gray!5] (11) at (7,0) {}; 
\node[shape=circle,draw=black,fill=gray!5] (12) at (8,0) {};

\path [-] (0) edge[color=black,thick] node[pos=0.5, above] {$\sigma^{(0)}_X$} (1);
\path [-] (0) edge[color=black,thick] node[pos=0.5, left]  {$\sigma^{(0)}_Y$} (2);
\path [-] (0) edge[color=black,thick] node[pos=0.5, above]  {$\sigma^{(0)}_Z$} (3);
\path [-] (1) edge[color=black,thick] node[pos=0.5, above] {$\sigma^{(1)}_X$} (4);
\path [-] (1) edge[color=black,thick] node[pos=0.5, below] {$\sigma^{(1)}_Y $} (5);
\path [-] (1) edge[color=black,thick] node[pos=0.5, right] {$\sigma^{(1)}_Z $} (6);
\path [-] (2) edge[color=black,thick] node[pos=0.5, above] {$\sigma^{(2)}_X$} (7);
\path [-] (2) edge[color=black,thick] node[pos=0.8, left] {$\sigma^{(2)}_Y $} (8);
\path [-] (2) edge[color=black,thick] node[pos=0.5, above] {$\sigma^{(2)}_Z $} (9);
\path [-] (3) edge[color=black,thick] node[pos=0.5, left] {$\sigma^{(3)}_X$} (10);
\path [-] (3) edge[color=black,thick] node[pos=0.5, below] {$\sigma^{(3)}_Y$} (11);
\path [-] (3) edge[color=black,thick] node[pos=0.5, above] {$\sigma^{(3)}_Z$}(12);

\end{tikzpicture}
\caption{Ternary tree structure in the case of a $4$ fermionic modes system. Labeled nodes represent qubits, while unlabeled nodes are empty and added only to complete the edges from the previous level. We note $h$ the height of the tree (here we have $h=2$). The operators are placed on edges but apply to the qubit at the origin of each edge (i.e. the qubit with index corresponding to the node from which the edge start). By convention, $X$ applies to left edges, $Y$ to central edges, and $Z$ to right edges.}
\label{fig:ternary_tree}
\end{figure}

To build this encoding, one must first map the qubits to the vertices of a ternary tree (a tree that splits into three edges after each vertex) as presented in Fig. \ref{fig:ternary_tree}. For any path $\boldsymbol{p}$ in the tree, one can define the following operators:
\begin{equation}
    A_{\boldsymbol{p}} = \bigotimes_{l=0}^{h-1} \sigma^{(\nu)}_{\alpha},
\end{equation}
where $h$ is the height of the tree, $\nu$ is the qubit index on path $\boldsymbol{p}$ and $\alpha$ corresponding to $X$, $Y$ or $Z$ depending on whether the path follows the left, central or right edge respectively after qubit $\nu$.

These operators clearly obey the same anti-commutator relationships as Majorana operators: $\{ A_{\boldsymbol{p}}, A_{\boldsymbol{q}}\}= 0$ if $\boldsymbol{p} \neq \boldsymbol{q}$, and $ A_{\boldsymbol{p}}^2 = \unit$. Given there are $2n + 1$ distinct path in the ternary tree, and we need $2n$ Majorana operators, Jiang et al. \cite{Jiang2020} propose mapping each $A_{\boldsymbol{p}}$ to a single Majorana operator, and these operators have a maximum Pauli weight equal to the height of the tree ($h$), hence $\log_3(2n + 1)$. 

To construct the Hamiltonian, one needs to first transform the fermionic operators into Majorana operators and re-write the Hamiltonian, and then decide on an allocation of the Pauli operators defined above to the Majorana operators \cite{Jiang2020}. Similar to the encodings defined previously defined, the number of qubits $N$ is equal to the number of fermionic modes $n$, and the scaling of the number of Pauli operators is the same as that of the two body-terms in the second quantized Hamiltonian $\mathcal{O}(N^4)$.

\subsubsection{Discussion on generalized encondings}

Before drawing a comparison of the different generalized encodings mentioned above, we would like to raise two relevant points regarding this type of encoding: 
\begin{itemize}
    \item All the encoding mentioned above can be constructed in terms of Fenwick trees \cite{Havlek2017}, providing an opportunity to define them differently and possibly optimize the encoding to specific fermionic models. We included a description of Fenwick trees construction (similar to the presentation in Ref.~\cite{Havlek2017}) and their relation to encodings in Appendix \ref{sec:Fenwick_trees}. Fenwick trees are also relevant in the context of optimizing the Bravyi-Kitaev mapping for application to 2-dimensional lattice models \cite{Havlek2017}. 
    \item Steudtner et al. \cite{Steudtner2018} propose three additional generalized encodings which allow the number of qubits to be reduced using symmetries at the cost of additional entangling gates. Because the method presented in Ref.~\cite{bravyi_tapering_2017} and Ref.~\cite{Setia2020} addresses these symmetries without significant additional gate cost (see Sec. \ref{sec:tappering_qubits}), it is in general preferred. We nonetheless encourage readers interested in going deeper in the knowledge of these general encodings to go through Ref.~\cite{Steudtner2018} as it discusses the relevant theory in depth. 
\end{itemize}

The key metric for comparing the four mappings presented above is their Pauli weight: while Jordan-Wigner and Parity scale $\mathcal{O}(N)$, Bravyi-Kitaev scales $\mathcal{O}(\log_2(N))$ and the optimal ternary tree encoding scales $\mathcal{O}(\log_3(2N))$ (which is asymptotically the better one). The number of qubits is the same across all these mappings and equates to the number of fermionic modes ($N=n$). These mappings can also benefit from the same amount of qubit reduction using symmetries (see Sec. \ref{sec:tappering_qubits}). While the number of operators may vary, it has also been shown numerically that in general applying grouping strategies (presented in Sec. \ref{sec:Grouping}) to these operators results in a very similar number of total operators to measure \cite{Hamamura2020}. 

Because the Bravyi-Kitaev and optimal ternary tree mappings have lower Pauli weight, they are in theory less subject to read-out errors. However, this becomes invalid once the grouping of the Pauli operators for joint measurements is introduced. This is because, for most groups, the entire register of qubits needs to be measured in the $Z$ basis. The main advantage of the $\log_2(N)$ or $\log_3(Nn)$ Pauli weight is in using the mapping to construct an ansatz based on fermionic operators (such as the Unitary Coupled Cluster ansatz, UCC, see Sec. \ref{sec:UCCA}). Numerical studies looking at the use of Bravyi-Kitaev mapping in the context of Quantum Phase Estimation have consistently shown that it results in a significant reduction in the number of entangling gates required \cite{Tranter2015, Tranter2018, Setia2018}, without impacting the overall accuracy of the result in noiseless simulations. It is clear these results also translate to applications of the UCC ansatz in the context of the VQE and are likely to extend to the optimal ternary tree mapping. It is important to note that the relative impact of low Pauli weight is very much dependent on the degree of connectivity of the qubit register, and limited connectivity (e.g. qubits placed in a line), could result in a significant increase in the number of entangling gates required in a low Pauli weight encoding.

A final point to note is that a recent study \cite{Sawaya2016} numerically tested that UCC ansatz constructed for VQE with the Bravyi-Kitaev mapping could result in slightly higher sensitivity to quantum noise channels (see Appendix \ref{sec:common_noise_models}) than the Jordan-Wigner mapping, despite lower gate depth. Sawaya et al.\cite{Sawaya2016} conjecture that this could be because occupation numbers are stored locally in the Jordan-Wigner mapping: a single qubit error only impacts the result of one orbital, while for the Bravyi-Kitaev mapping, it could affect several. This also implies that occupation number errors create greater errors in the energy than parity errors \cite{Sawaya2016}. One could expect the parity and ternary tree mappings to be affected by the same phenomenon, although we are aware of any further research on this topic.

\subsection{Lattice model tailored encoding} \label{sec:lattice_encoding}

The mappings defined above are oftentimes inefficient for lattice models. This is because they assume full connectivity between the different fermionic modes, and translating this into spin-operators results in non-local operations. Instead, the mappings presented in this section are concerned about being as efficient as possible for a given lattice model, where fermionic modes are attached to a specific geometry. The earliest explicit definition of fermion to qubit mapping was indeed an application to the Hubbard model \cite{Abrams1997}. The literature is divided into two mains approaches for doing so: auxiliary fermion schemes \cite{Verstraete2005, Ball2005} and Loop-Stabilized Bravyi-Kitaev (LSBK), also known as Superfast Encoding (SFE) \cite{Bravyi2002}. Maintaining low Pauli weights comes at a cost of additional qubits for all the methods mentioned in this section. However, it has been shown that all these methods also allow for at least some degree of error correction. A summary of the key metrics of each encoding is presented in Table~\ref{tab:lattice_encoding}.

\begin{table} [ht]
\caption{Comparative summary of lattice tailored encodings. $d$ represents the degree of the Hamiltonian graph, $v$ and $h$ respectively the vertical and horizontal dimensions of a 2-dimensional lattice (we set $v \leq h$), $n$ is the number of fermionic modes / sites on the lattice. The number of operators scales linearly with $E$, the number of edges. We have $E = h(v - 1) + v (h - 1)$ for a 2-dimensional regular lattice, and $E = \sum_i^D \left[ (n_i - 1) \prod_{j \neq i}^D n_j \right]$ for an regular lattice of dimension $D$ and $n_i$ the number of sites along the $i^{th}$ dimension.} \label{tab:lattice_encoding}
\begin{tabularx}{\linewidth}{X|m{0.15\linewidth}m{0.15\linewidth}X}
\toprule
Method &  Pauli Weight & Qubits & Comments \\
\midrule
    Jordan-Wigner (snake pattern) \cite{Verstraete2005} & $\mathcal{O}(2v)$ & $n$ & Optimal direct application of the Jordan-Wigner mapping to a 2-dimensional lattice.  \\\\
\hline\\
    Bravyi-Kitaev (Fenwick tree lattice mapping) \cite{Havlek2017} & $\mathcal{O}(\log(v))$ & $n$ & Optimal direct application of the Bravyi-Kitaev mapping to a 2-dimensional lattice.  \\\\
\hline\\
    Auxiliary Fermion Scheme \cite{Verstraete2005} & $4$ (2-dim.) & $2n$, ($n = vh$) & Uses auxiliary fermion/qubit registers to create operators that cancel-out $Z$ strings in the Jordan-Wigner mapping  \\\\
\hline\\
    Superfast Bravyi-Kitaev \cite{Bravyi2002, Havlek2017, Setia2018} & $\mathcal{O}(2d)$& $\mathcal{O}(nd/2)$ & Relies on stabilizers formalism to define an efficient encoding with cost dependent on the degree of the Hamiltonian graph  \\\\
\hline\\
    Generalized Superfast Encoding \cite{Setia2019} & $\mathcal{O}(\log_2(d))$ & $\mathcal{O}(nd/2)$ &Extension of Superfast BK, optimizing Pauli weight and offering better opportunities for error corrections  \\\\
\hline\\
    Compact encoding \cite{Derby2021, Derby2021_part2} & $3$ (2-dim.), $4$ (3-dim.) & $\mathcal{O}(1.5n)$ &Modifies the stabilizer formalism used in Superfast BK to optimize the number of qubits required. So far limited to 2 and 3-dimensional lattices \\\\
\bottomrule
\end{tabularx}
\end{table}

All the mappings presented in this section will exhibit a similar scaling in the number of operators produced, which will be linearly proportional to the number of edges $E$ in the lattice. For example, consider a regular lattice of dimension $D$, with $n_i$ the number of sites along the $i^{th}$ dimension we easily obtain
\begin{equation}
    E = \sum_i^D \left[ (n_i - 1) \prod_{j \neq i}^D n_j \right].
\end{equation}
We also have the total number of sites $n = \prod_{j}^D n_j$ and therefore a number of edges is capped below $\mathcal{O}(nD)$. An additional pre-factor should be included to account for the fact that some encoding require auxiliary sites (and hence auxiliary operators), however this does not change the overall scaling.

\subsubsection{Auxiliary fermion schemes}

Suppose a 1-dimensional spin-chain, with only nearest neighbor interactions. All operators of interest are of the form $\hat{a}_i^{\dagger}\hat{a}_j$, with $i = j \pm 1$. Therefore, under the Jordan-Wigner mapping all operators have the form
\begin{equation} \label{eq:hopping}
\hat{a}_{i \pm 1}^{\dagger} \hat{a}_{i}=\frac{X_{i \pm 1}-i Y_{i \pm 1}}{2} \otimes \frac{X_{i}+i Y_{i}}{2}.
\end{equation}
One can note that the $Z$ strings required to maintain the anticommuting relationship have canceled out, and are no longer necessary: the Jordan-Wigner mapping preserves operator locality in one-dimensional systems. Auxiliary Fermion Schemes answer the question of how to maintain operator locality in mappings such as Jordan-Wigner when the lattice model considered has higher dimensions.  

A first example of encoding tailored for lattice model was independently proposed in \cite{Verstraete2005} and \cite{Ball2005}. As described above (both grounded in \cite{Levin2003, Wen2003}), it can be seen as a Jordan-Wigner encoding optimized for rectangular lattice problems. It avoids the need for strings of $Z$ operators in interaction terms built out of creation and annihilation operators thereby optimizing the Pauli weight, but at the cost of increasing the number of qubits. This is referred to in the literature alternatively as the Ball-Verstraete-Cirac (BVC) encoding (for example \cite{Whitfield2016}.

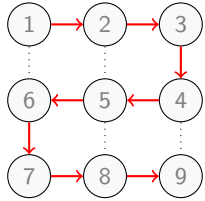
\begin{figure} [h]
\centering
\begin{tikzpicture}
\node[shape=circle,draw=black,fill=gray!5] (1) at (0,2) {1}; 
\node[shape=circle,draw=black,fill=gray!5] (2) at (1,2) {2};
\node[shape=circle,draw=black,fill=gray!5] (3) at (2,2) {3};
\node[shape=circle,draw=black,fill=gray!5] (6) at (0,1) {6};
\node[shape=circle,draw=black,fill=gray!5] (5) at (1,1) {5};
\node[shape=circle,draw=black,fill=gray!5] (4) at (2,1) {4};
\node[shape=circle,draw=black,fill=gray!5] (7) at (0,0) {7};
\node[shape=circle,draw=black,fill=gray!5] (8) at (1,0) {8};
\node[shape=circle,draw=black,fill=gray!5] (9) at (2,0) {9}; 

\path [->] (1) edge[color=red,thick] (2);
\path [->] (2) edge[color=red,thick] (3);
\path [->] (3) edge[color=red,thick] (4);
\path [->] (4) edge[color=red,thick] (5);
\path [->] (5) edge[color=red,thick] (6);
\path [->] (6) edge[color=red,thick] (7);
\path [->] (7) edge[color=red,thick] (8);
\path [->] (8) edge[color=red,thick] (9);
\path [-] (1) edge[color=black,dotted] (6);
\path [-] (2) edge[color=black,dotted] (5);
\path [-] (5) edge[color=black,dotted] (8);
\path [-] (4) edge[color=black,dotted] (9);

\end{tikzpicture}
\caption{Example of a 2-dimensional spin lattice with nine sites. Edges represent connections between the different fermionic modes. Red arrows are connections that provide an example of a 'snake ordering' of the fermionic modes in a 1-dimensional pattern, while dotted lines show the connections missed by the ordering.}
\label{fig:nine_site_lattice}
\end{figure}

Suppose a rectangular spin lattice, such as the 2-dimensional Hubbard model. One could nearly map it to a one-dimensional system by ordering the operators (for example, as the 'snake pattern' presented in Fig. \ref{fig:nine_site_lattice}). Along with this ordering (i.e. considering only the operator connections it covers), one can easily implement a local version of the Jordan-Wigner mapping. The issue however is that some connections are not covered and cannot be directly expressed with local operators using the Jordan-Wigner mapping (consider for example the dotted line between site 1 and 6, representing operators $\hat{a}^{\dagger}_1 \hat{a}_6$ and transpose in the figure, which would require $Z$ strings in this ordering).

The BVC mapping provides a means to encode operators along the edges not covered in the ordering locally. In order to do so, it defines an auxiliary Hamiltonian composed of a weighted sum or interaction operators which are constructed using an alternative set of fermionic operators $\hat{b}_i$ such that
\begin{equation}
    \hat{\mathcal{H}}_{\mathrm{aux}} = \sum_{\{i, j\}} \hat{h}_{ij}^{(\mathrm{aux})} = \sum_{\{i, j\}} - (\hat{b}_i + \hat{b}_i^{\dagger})(\hat{b}_j - \hat{b}_j^{\dagger}), 
\end{equation}
which we note can be re-written in terms of Majorana operators (see Eq. \ref{eq:majorana}): 
\begin{equation}\label{eq:auxiliary_couplings}
    \hat{\mathcal{H}}_{\mathrm{aux}} = \sum_{\{i, j\}} i \hat{\gamma}_i \hat{\gamma}_{j+1}. 
\end{equation}

The fermions are configured to be in the ground state of this Hamiltonian, denoted $\ket{\chi}$. The auxiliary Hamiltonian sites match those of the physical Hamiltonian. The overall system is now composed of $2n$ fermions which we can order as $1, 1^{\prime}, 2, 2^{\prime}, \dots n, n^{\prime}$.   

For notational simplicity, indices of the auxiliary fermionic system are primed. From there, the edges of the physical lattice Hamiltonian which are not covered by the initial ordering of fermions into a 1-dimensional system (indexed here by $\{p, q \}$), can be modified as follows:

\begin{equation}
    \hat{a}^{\dagger}_p\hat{a}_q \rightarrow \hat{a}^{\dagger}_p\hat{a}_q \hat{h}_{pq}^{(\mathrm{aux})} = \hat{a}^{\dagger}_p\hat{a}_q \left(i \hat{\gamma}_{p^{\prime}} \hat{\gamma}_{q^{\prime} + 1} \right).
\end{equation}

It is important to note, that because any of the vertical hopping terms in the physical Hamiltonian commute with the auxiliary Hamiltonian, the operation described above does not modify the ground state of the system. The consequence is that when now applying the Jordan-Wigner transformation to the new, joint Hamiltonian, the $Z$ strings of the original, non-covered edges are canceled out by those of the auxiliary terms by which they have been modified. Note that using Jordan-Wigner, the auxiliary Hamiltonian presented in Eq. (\ref{eq:auxiliary_couplings}) can be non-local, though this can be addressed easily by substitution of the non-local operators with local ones (see Ref.~\cite{Verstraete2005} for details).
The main advantage of this method is that, as long as only nearest-neighbor interactions are allowed, it caps the Pauli weights to 4 for hopping terms (basically following Eq.~\ref{eq:hopping}), and to $2$ for Coulomb terms. Suppose a rectangular lattice of $h \times v$ sites, (we set $v \leq h$), a naive implementation of a Jordan-Wigner mapping would result in a maximum Pauli weight of $\mathcal{O}(2v)$ (consider for example the mapping of the hopping term between site 1 and 6 in Fig. \ref{fig:nine_site_lattice}). This benefit comes at the cost of doubling the number of qubits required due to the introduction of the auxiliary Hamiltonian, hence a total of $2vhL$ qubits as initially one is required for each fermionic mode.   

Whitfield {\it et al.} \cite{Whitfield2016} extend the theory of the BVC mapping presented in \cite{Verstraete2005, Ball2005}, by showing that there is a range of possible choices of the auxiliary coupling operators, which are not required to be Majorana operators. They also show that the number of auxiliary modes required for each fermionic mode in a lattice grows as $D - 1$, with $D$ the dimension of the lattice. Further improvements were also proposed in \cite{Steudtner2019}.

An adaptation of the auxiliary fermion scheme to the Bravyi-Kitaev encoding is also proposed by Havl{\'{\i}}{\v{c}}ek {\it et al.} ~\cite{Havlek2017}, thereby optimizing it for rectangular lattice models. Havl{\'{\i}}{\v{c}}ek {\it et al.} propose to first build the Bravyi-Kitaev mapping into a data structure using a Fenwick tree and then map the connections of this Fenwick tree to a fermion lattice. Because sites that are connected through the tree (either by mean of the parity set or the update set) only require local operations, non-local operations are restricted to the mapping of connections on the lattice which are not included on the tree. By doing so, they show that with optimal mapping of the Fenwick tree to the lattice, the Pauli weight of their adapted Bravyi-Kitaev encoding scales as $\mathcal{O}(\log(v))$, with $v$ the smallest side of the lattice.

\subsubsection{Superfast Encoding / Loop Stabilizer Encodings} \label{sec:superfast_encoding}
The encoding concept developed in this section was initially presented in \cite{Bravyi2002} and is based on a graph representation of the Hamiltonian. Of course, any lattice models (or \textit{ab initio} molecular system) can be represented as a graph, where each fermionic mode is a vertex, and each interaction with another fermionic mode is a weighted edge in the graph. Unlike the mappings presented so far, in this mapping, qubits are used to encode interactions between fermionic modes rather than the state of the fermionic modes themselves. It is referred to in the literature as the Superfast Bravyi-Kitaev (SFBK) or alternatively, Loop-Stabilized Bravyi-Kitaev (LSBK) \cite{Havlek2017} (we use SFBK as it is more generally used in the literature).

The overall advantage of this method is that, similar to the auxiliary qubit scheme, the Pauli weight of the qubit operators corresponding to the fermionic operators does not depend on the number of fermionic modes in the Hamiltonian but instead depends on the degree of the interaction graph. The number of qubits required depends on the total number of edges which is a function of the total number of fermionic modes and the degree of the interaction graph. This makes Superfast Encoding more suitable for lattice based models. We first present the SFBK mapping, as theorized in \cite{Bravyi2002}, and further developed in \cite{Whitfield2016, Havlek2017, Setia2018}, and we then turn to extensions developed based on this encoding \cite{Setia2019, Jiang2019, Derby2021, Derby2021_part2}.

\begin{figure} [h]
\centering
\begin{tikzpicture}
\node[shape=circle,draw=black,fill=gray!5] (D) at (0,0) {D};  \node[shape=circle,draw=black,fill=gray!5] (A) at (0, 2) {A};
\node[shape=circle,draw=black,fill=gray!5] (B) at (2, 2) {B};
\node[shape=circle,draw=black,fill=gray!5] (C) at (2, 0) {C};

\path [->] (A) edge[color=red,thick] (B);
\path [->] (B) edge[color=red,thick] (C);
\path [->] (C) edge[color=red,thick] (D);
\path [->] (D) edge[color=red,thick] (A);

\node[shape=rectangle, draw=black,fill=black, text=white] (4) at (1, 0) {4};
\node[shape=rectangle,draw=black,fill=black, text=white] (2) at (0, 1) {2};
\node[shape=rectangle,draw=black,fill=black, text=white] (1) at (1, 2) {1};
\node[shape=rectangle,draw=black,fill=black, text=white] (3) at (2, 1) {3};
\end{tikzpicture}
\caption{A graph corresponding to the SFBK encoding defined from the $H_2$ Hamiltonian in a minimal basis set: fermions in the physical system of interest correspond to the vertices (grey circles) of the graph; qubits (the black rectangles) in the quantum computer are associated to the edges (red arrows) on the graph; the vertices and edges the graph are used to define the edge and vertex fermionic operators $\hat{A}_{jk}$ and $\hat{B}_j$, which in turn correspond to edge and vertex qubit operators $\tilde{A}_{jk}$ and $\tilde{B}_j$}
\label{fig:h2moleculegraph1}
\end{figure}
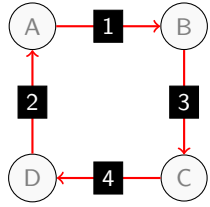

\paragraph{Background and definition of SFBK:}
Given the fermionic Hamiltonian an interaction graph can be constructed. The set of edges, $E$ for the interaction graph correspond to the excitation, number excitation and double excitation operators (e.g. Fig. \ref{fig:h2moleculegraph1}). In SFBK, the qubits are placed on the edges of the interaction graph, and the Fock space of the fermionic modes is associated with a codespace of a stabilizer code defined on the qubits on the edges. The qubit indices are defined based on two fermionic indices where a connection exists (e.g. $(jk)$). To define the codespace and the fermionic operators in terms of the qubit operators acting on qubits on the edges of the interaction graph, it is helpful to work with the Majorana modes defined in terms of Fermionic modes as in Eq. (\ref{eq:majorana}). These operators are Hermitian and satisfy
\begin{align}
\hat{\gamma}_{j}\hat{\gamma}_{k}+\hat{\gamma}_{k}\hat{\gamma}{j}=2\delta_{jk} \label{eq:Maj_modes}
\end{align}
Using Majorana modes, the following vertex operator and edge operator can be defined as
\begin{equation}
\begin{aligned}
\hat{B}_j &= -i \hat{\gamma}_j\hat{\gamma}_{j+1}, \quad
\hat{A}_{jk} &= -i\hat{\gamma}_j\hat{\gamma}_k.
\end{aligned}
\end{equation}
The vertex operator $\hat{B}_j$ only acts on fermionic mode $j$, while the edge operator $\hat{A}_{jk}$ connects fermionic mode $j$ to mode $k$. These operators satisfy the following algebraic relations:
\begin{align} \label{eq:algebra_sfbk_bb}
&\left[ \hat{B}_k, \hat{B}_l \right] = 0,\\
&\hat{A}_{(jk)}\hat{B}_l = (-1)^{\delta_{(jl)} + \delta_{(kl)}} \hat{B}_l \hat{A}_{(jk)}, \label{eq:algebra_sfbk_ab} \\
&\hat{A}_{(jk)}\hat{A}_{(ls)} =  (-1)^{\delta_{(jl)} + \delta_{(js)} + \delta_{(kl)} + \delta_{(ks)}}\hat{A}_{(ls)}\hat{A}_{(jk)}, \label{eq:algebra_sfbk_aa}
\end{align}
and the loop condition:
\begin{align} \label{eq:sfbk_loop}
(i)^{p} \hat{A}_{(j_0 j_1)}\hat{A}_{(j_1 j_2)} \dots \hat{A}_{(j_{p-1} j_p)}\hat{A}_{(j_p j_0)} = \unit,
\end{align}
with $j_0 \dots j_p$ any closed loop over the vertex indices.

It can be shown that the vertex and the edge operators satisfying the above relations generate the algebra of the physical operators. This means that any physical Hamiltonian with even fermionic modes can be represented in terms of these edge and vertex operators. The fermionic creation and annihilation operators can be converted into Majorana modes which can then be converted to edge and vertex operators. Table \ref{tab:SFBKoperatortable} gives the edge and vertex operators expressions for various Hamiltonian terms (with computation initially presented in \cite{Setia2018}). 
Using the algebraic relations for the edge and vertex operators we can define the qubit operators for the edge and vertex operators. As we will see through the construction, it is also possible to define the codespace for the simulation.

Starting from the vertex operator, we note that $\hat{B}_k$ obeys an additional constraint on the excitation parity when considering an even number of fermions in the system, for a total of $\mathrm{V}$ vertices, the fermion parity operator is

\begin{equation} \label{eq:vertex_sfbk}
\begin{aligned}
\prod_{k \in \mathrm{V}} \hat{B}_k = \unit
\end{aligned}
\end{equation}

If the number of particles is instead odd, one can always split the Hamiltonian in even and odd sectors, the odd sector can be simulated by changing the sign of the mapping as described below in Eq.~(\ref{eq:vertex_sfbk_qubit}) \cite{Jiang2019}. From Eq. \ref{eq:vertex_sfbk} and the rule defined in  Eq.~(\ref{eq:algebra_sfbk_bb}), one can easily create a corresponding definition for the qubit version vertex operator $\tilde{B}_k$:
\begin{equation}\label{eq:vertex_sfbk_qubit}
\tilde{B}_{k} =\bigotimes_{j \in n(k)} Z_{(k j)},
\end{equation}
where $n(j)$ is the set of vertices connected to mode $j$ by an edge, and the pair $(kj)$ indexes a qubit. This definition meets the constraint of Eq. \ref{eq:vertex_sfbk} as each edge is connected to exactly two vertices. We note that the Pauli weight for these operators is capped at the degree of the Hamiltonian graph. 

For the edge operator $\hat{A}_{(jk)}$, we start by defining an ordering of the vertices and related direction of edges, such that $\epsilon_{jk} = 1$ when $j > k$, and $\epsilon_{kj} = - 1$. Looking first at the rule of Eq.~(\ref{eq:algebra_sfbk_ab}), in the case where $l \neq j$ and $l \neq k$ operators $\hat{A}_{(jk)}$ and $\hat{B}_{l}$ must commute, and hence given the definition of $\tilde{B}_{l}$ from Eq.~(\ref{eq:vertex_sfbk_qubit}) all qubit operators representing edges adjacent to $j$ or $k$ (except for $(jk)$ itself) must be either $\unit$ or $Z$, as $l$ could be one of the connected vertices. 
If $l = j$ or $k$, then $\hat{A}_{(jk)}$ and $\hat{B}_{l}$  must anticommute. Following the same reasoning as above, the qubit operator on the edge connecting $l$ to the other vertex ($j$ or $k$) must anticommute with $Z$: it must be either $X$ or $Y$. 

The last commutation rule that must be met is Eq.~(\ref{eq:algebra_sfbk_aa}), which states that two edge operators anticommute if they share a single vertex (they commute otherwise). All these conditions, together can be met by the following qubit operator as defined in \cite{Bravyi2002, Havlek2017, Setia2018}:
\begin{equation}
\tilde{A}_{(j k)} =\epsilon_{j k} X_{(j k)} \bigotimes_{l < k}^{n(j)} Z_{(l j)} \bigotimes_{s < j}^{n(k)} Z_{(s k)},
\end{equation}
where the $Z$ operators on vertices either connected and directed into $j$ or from $k$, and the $X$ operator on vertex $(jk)$ enforce the rule of Eq.~(\ref{eq:algebra_sfbk_ab}), and the phase factor $\epsilon_{j k}$ enforce the rule of Eq.~(\ref{eq:algebra_sfbk_aa}).

The qubit operator representation of vertex and edge operators $\tilde{B}_{j}$ and $\tilde{A}_{(j k)}$ do not satisfy the loop condition defined in Eq.~(\ref{eq:sfbk_loop}). And this condition lets us define the codespace. Using the qubit operator representation of the edge operator, $\tilde{A}_{jk}$, we can define a set of stabilizers corresponding to the loops in the graph and define the codespace to be the subspace where the loop condition defined in Eq.~(\ref{eq:sfbk_loop}) hold. The loop operator is defined as follows:
\begin{equation} \label{eq:loop_operator}
    \tilde{A}(\eta) \equiv  (i)^{p} \hat{A}_{(\eta_0 \eta_1)}\hat{A}_{(\eta_1 \eta_2)} \dots \hat{A}_{(\eta_{p-1} \eta_p)}\hat{A}_{(\eta_p \eta_0)}, 
\end{equation}
with $\eta$ defining a loop of length $p$ on the Hamiltonian graph. 
This definition of the loop operator generates an Abelian group of all the stabilizers (all necessary properties of this group are explored and demonstrated in \cite{Setia2019}). The number of independent loops $s$ in the graph is equal to the number of edges, minus the number of vertices plus one, or recalling $N$ the number of qubits, and $n$ the number of fermionic mode: $s = N - n + 1$. Therefore the code space defined by the stabilizer group encodes $N - s = n - 1$ logical qubits, into $N$ physical qubits. This code space effectively restricts operators to the even parity subspace of the fermionic Fock space \cite{Bravyi2002, Setia2019}. Once this restriction is applied to $\hat{B}_{j}$ and $\hat{A}_{(j k)}$, one can defined an encoded qubit Hamiltonian.

The total number of qubits required for this encoding is proportional to the number of edges in the Hamiltonian graph. For regular lattice models, the number of edges itself is a direct multiple of the graph degree and of the number of sites. Since the Pauli weight of vertex and edge operators scales as $O(d)$, where $d$ is the degree of the interaction graph, the Pauli weight of each term in the transformed Hamiltonian also scales as $O(d)$.

\begin{table*}
\caption{Molecular Hamiltonian operators in second quantized form and in the corresponding vertex and edge operator form used in the Superfast Bravyi-Kitaev encoding.}
\begin{tabular}{ l | cc}
  \toprule			
Operator & Second quantized form & Vertex and edge operator form \\    \midrule
Number operator & $h_{pp} \hat{a}_p^\dagger \hat{a}_p$ & $\frac{1 - \hat{B}_p}{2}$ \\ 
Coulomb/exchange operators & $h_{pqqp} \hat{a}_p^\dagger \hat{a}_q^\dagger \hat{a}_q \hat{a}_p$ & $h_{pqqp} \frac{(1 - \hat{B}_p)(1 - \hat{B}_q)}{4}$ \\ 
Excitation operator & $h_{pq} (\hat{a}_p^\dagger \hat{a}_q + \hat{a}_q^\dagger \hat{a}_p$) & $-h_{pq} \frac{i}{2} (\hat{A}_{pq} \hat{B}_q + \hat{B}_p \hat{A}_{pq}) $ \\  
Number-excitation operator & $h_{pqqr} (\hat{a}_p^\dagger \hat{a}_q^\dagger \hat{a}_q \hat{a}_r + \hat{a}_r^\dagger \hat{a}_q^\dagger \hat{a}_q \hat{a}_p)$ & $- h_{pqqr} \frac{i}{4} (\hat{A}_{pr} \hat{B}_r + \hat{B}_p \hat{A}_{pr}) (1 - B_q) $ \\  
Double excitation operator & $h_{pqrs} (\hat{a}_p^\dagger \hat{a}_q^\dagger \hat{a}_r \hat{a}_s + \hat{a}_s^\dagger \hat{a}_r^\dagger \hat{a}_q \hat{a}_p)$ & $\frac{h_{pqrs}}{8} \hat{A}_{pq} \hat{A}_{rs} (-1 - \hat{B}_p \hat{B}_q + \hat{B}_p \hat{B}_r +$ \\ && $\hat{B}_p \hat{B}_s + \hat{B}_q \hat{B}_s - \hat{B}_r \hat{B}_s - \hat{B}_q + \hat{B}_p \hat{B}_r \hat{B}_s)$\\
  \bottomrule	
\end{tabular}
\label{tab:SFBKoperatortable}
\end{table*}

SFBK was applied to the 2-dimensional Hubbard model in Ref.~\cite{Havlek2017} and was shown to be applicable to \textit{ab initio} molecular systems in \cite{Setia2018}. In both cases, it is clear that for a Hamiltonian graph of degree $d$, and a number of fermionic modes $n$, the Pauli weight of SFBK scales $\mathcal{O}(2d)$ and the number of qubits required scales $\mathcal{O}(nd/2)$.

\paragraph{Generalized Superfast Encoding:}
The Abelian stabilizer group defined using the loop operators given in Eq.~(\ref{eq:loop_operator}) defines the codespace for the simulation. Since the edge and vertex operators commute with the stabilizer group, a state initialized within the codespace remains in codespace through the action of edge or vertex operators. Any operation that moves the state out of the codespace is not valid and can be considered as an error. 
The motivation for the Generalized Superfast Encoding was to come up with a modified version of SFBK that can detect all the errors. The number of qubits required for GSE is the same as SFBK, but the Pauli weight of qubit operator representation of edge and vertex operator is lower and scales as $\mathcal{O}(\log(d))$.

In contrast to SFBK, where qubits are placed on the edges, GSE places them on vertices which consequently modifies the operators used for the encoding. For each fermionic mode $j$ (which are assumed to be even in GSE), one must use $d^{(j)}/2$ qubits, where $d^{(j)}$ is the degree of the vertex corresponding to the mode $j$. Hence, it has the same scaling of the number of qubits as for SFBK: $\mathcal{O}(nd/2)$.

In the GSE, a vertex $j$ with $d/2$ qubits encodes $d$ Majorana modes $\hat{\gamma}{j, 1}, \hat{\gamma}{j, 2},...\hat{\gamma}{j, d}$. A procedure to construct these Majorana modes can be found in \cite{Setia2018}. The vertex and edge operator for qubits can be reformulated for the GSE as follows: 
\begin{equation}\label{eq:vertex_gse_qubit}
\tilde{B}_{j} = (-i)^{d^{(j)}/2}\hat{\gamma}{j, 1}\hat{\gamma}{j, 2} \dots\hat{\gamma}{j, d^{(j)}},
\end{equation}
\begin{equation} \label{eq:edge_gse_qubit}
\tilde{A}_{(j k)} =\epsilon_{j k} \hat{\gamma}{j, p} \hat{\gamma}{k, q},
\end{equation}
where $p$ is the $p$-th local Majorana mode on the site $j$, and $q$ is the $q$-th local Majorana mode on the site $k$.  These operators can also be restricted to the even-partiy subspace, using the same stabilizers  definition based on the loop operator of Eq.~(\ref{eq:loop_operator}) \cite{Setia2019}. It can be shown using Fenwick tree encoding \cite{Havlek2017} (see appendix \ref{sec:Fenwick_trees}) that a Majorana mode $\hat{\gamma}{j, p}$ can be encoded into Pauli operators of weight $\log_2 d^{(j)}$. This means that operators $\tilde{A}_{(j k)}$ of Eq.~(\ref{eq:edge_gse_qubit}) have a maximum Pauli weight of $2\log_2 d$, while operators $\tilde{B}_{j}$ (Eq. \ref{eq:vertex_gse_qubit}) have a Pauli weight of $1$ as it requires a single $Z$ string \cite{Havlek2017}. This shows a key advantage of GSE over SFBK. 

As mentioned above, it is shown in \cite{Setia2019} that GSE additionally allows use of the code space defined as part of the encoding as a means of performing error correction. This property is applicable from $ d^{(j)} \geq 6$ and catches single-qubit errors. So far, the error threshold tolerance of this error mitigation method has not been estimated.

Setia {\it et al.} \cite{Setia2019} also showed that SFBK cannot provide single-qubit error correction for Hamiltonian graph of degree $ d^{(j)} \leq 6$.  The Majorana Loop Stabilizer Code (MLSC) was introduced in \cite{Jiang2019} as a means to allow addressing single-qubit errors on 2D square lattices (hence with vertex degree equal to $4$) while preserving a local encoding. This encoding is still dependent on further research to be extended to higher-dimensional systems. These ideas have been further generalized in \cite{Chien2020}.    

\paragraph{Compact mappings:}

The so-called compact mapping \cite{Derby2021} is another extension of SFBK which does not concern itself error correcting properties, but focuses on minimizing the Pauli weight while reducing the number of qubits required in the methods previously mentioned. Derby and Klassen \cite{Derby2021} present application of the method to square and hexagonal lattices, and extend it in  Ref.~\cite{Derby2021_part2} to uniform lattices of degree less than $4$ and cubic lattices. 

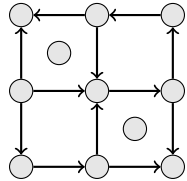
\begin{figure} [h]
\centering
\begin{tikzpicture}
\node[shape=circle,draw=black,fill=black!10] (1) at (0,2) {}; 
\node[shape=circle,draw=black,fill=black!10] (2) at (1,2) {};
\node[shape=circle,draw=black,fill=black!10] (3) at (2,2) {};
\node[shape=circle,draw=black,fill=black!10] (4) at (0,1) {};
\node[shape=circle,draw=black,fill=black!10] (5) at (1,1) {};
\node[shape=circle,draw=black,fill=black!10] (6) at (2,1) {};
\node[shape=circle,draw=black,fill=black!10] (7) at (0,0) {};
\node[shape=circle,draw=black,fill=black!10] (8) at (1,0) {};
\node[shape=circle,draw=black,fill=black!10] (9) at (2,0) {}; 

\node[shape=circle,draw=black,fill=black!10] (10) at (0.5,1.5) {}; 
\node[shape=circle,draw=black,fill=black!10] (11) at (1.5,0.5) {}; 

\path [->] (2) edge[color=black,thick] (1);
\path [->] (4) edge[color=black,thick] (1);
\path [->] (4) edge[color=black,thick] (5);
\path [->] (2) edge[color=black,thick] (5);
\path [->] (3) edge[color=black,thick] (2);
\path [->] (5) edge[color=black,thick] (6);
\path [->] (6) edge[color=black,thick] (3);
\path [->] (4) edge[color=black,thick] (7);
\path [->] (7) edge[color=black,thick] (8);
\path [->] (8) edge[color=black,thick] (5);
\path [->] (8) edge[color=black,thick] (9);
\path [->] (6) edge[color=black,thick] (9);

\end{tikzpicture}
\caption{Example of qubit placement and edge orientation in the compact mapping. Each gray dot represents a qubit, only those connected by edges correspond to sites, the others correspond to odd lattice faces.}
\label{fig:compact_encoding}
\end{figure}

In the compact encoding, one must first assign a qubit to each vertex $j$. In a 2D lattice, each square of four vertices (referred to as a face), is defined as even or odd in a checkerboard pattern. Edges also need to be given an orientation. The orientation recommended in Ref.~\cite{Derby2021_part2} is to set the orientation anticlockwise for even faces of the lattice, orientation on odd faces follows from completion. A schematic is presented in Fig. \ref{fig:compact_encoding}.

From this allocation of qubits, we can already see that the number of qubits for this encoding is capped to $1.5$ times the number of sites (or number of fermionic mode, $n$). We first define the edge operators: for a given edge $(i, j)$, oriented $i$ to $j$, at most one of the adjacent faces is odd, in which case we index the related qubit with $(i, j)$. We have for a downward edge  \begin{equation}
    \tilde{A}_{(ij)} = X_i Y_j X_{(i,j)},
\end{equation}
for an upward edge: 
\begin{equation}
    \tilde{A}_{(ij)} = - X_i Y_j X_{(i,j)},
\end{equation}
and for a horizontal edge:
\begin{equation}
    \tilde{A}_{(ij)} = X_i Y_j Y_{(i,j)}.
\end{equation}
When the edge is on a boundary of the lattice with no odd face adjacent, there is no qubit $(i,j)$, and therefore the relevant Pauli operator can be omitted. 

Vertex operators are defined as
\begin{equation}
    \tilde{B}_{j} = Z_j.
\end{equation}
It is easy to verify that these operators meet the conditions set for SFBK in Eq.~(\ref{eq:algebra_sfbk_bb}), \ref{eq:algebra_sfbk_ab}, and \ref{eq:algebra_sfbk_aa}. As for SFBK, the loop condition in Eq.~(\ref{eq:sfbk_loop}) must also be met. The key feature of the compact mapping is that it does so in a way that allow to avoid the parity condition set by Eq.~(\ref{eq:vertex_sfbk}), and with it the need to have a Pauli weight equal to the degree of the graph. 

Considering first the stabilizers for odd faces, defining $a,b,c, d$ as the four vertices (for instance consider the first face in Fig. \ref{fig:compact_encoding}), we have 
\begin{align}
    \tilde{A}_{(ab)} & = (X_bY_aY_{(a,b)}) \\ \nonumber
    \tilde{A}_{(bc)} & = (X_bY_cX_{(b,c)}) \\ \nonumber
    \tilde{A}_{(cd)} &=  (X_dY_cY_{(c,d)}) \\ \nonumber
    \tilde{A}_{(dc)} &= (-X_dY_aX_{(d,a)})   \\ \nonumber
    \tilde{A}_{(ab)}\tilde{A}_{(bc)}\tilde{A}_{(cd)}\tilde{A}_{(dc)} &= - (Y_{(a,b)}X_{(a,b)})(Y_{(a,b)}X_{(a,b)}) \\ \nonumber
    & = - (-iZ_{(a,b)})(-iZ_{(a,b)}) \\ \nonumber
    &= \unit
\end{align}
as of course, the qubit with index $(a, b)$, is the same as with index $(b, c)$, $(c, d)$, and $(d, a)$.

For even faces however, the stabilizers are non-trivial and therefore impose a constraint on the operators. Construction of these constraints are detailed in \cite{Derby2021, Derby2021_part2}. Extensions of this method to other uniform lattices of degree up to 4 and cubic lattices can be found in \cite{Derby2021_part2}. In addition, it was shown in \cite{Bausch2020} that the compact encoding can also be used to detect a large proportion of single-qubit errors. 

\subsubsection{Discussion on lattice tailored encodings}

There are three main metrics that are worth discussing when comparing the different lattice tailored mapping: the number of qubits they require, their Pauli weights, and their capacity to mitigate or even correct errors resulting from quantum noise. 

The number of qubits required appears to be a trade-off for the two other features mentioned above. All mappings from this section considered, the compact encoding from Ref.~\cite{Derby2021} offer the most advantageous combination of low number of qubits and low Pauli weights for systems of dimensions up to $3$. Extension of the compact encoding to higher dimensions requires further research. In the meantime, the GSE \cite{Setia2019} provides the best performance in terms of Pauli weight but does require that the number of qubits increases linearly with the degree $d$ of the graph (recalling $d = 2D$) and the number of sites $n$.

While we found no comprehensive studies on the matter, these would very likely result in the most compact ansatz. However, as noted in \cite{Sawaya2016} when comparing the Jordan-Wigner and Bravyi-Kitaev mappings, lower gate depth does not necessarily translate to an overall higher resilience to quantum noise.  

An overall question that remains pending regarding all the lattice mappings considered is their efficacy at realizing error mitigation or error correction. For example, while the GSE demonstrates some error correcting properties, it is still unclear what quantum noise threshold is tolerable for these properties to be useful. In particular, such assessment is required to fully compare these encoding like-for-like. For example, the compact encoding could require additional quantum resources from error mitigation techniques (see Sec. \ref{sec:error-mit}) to achieve an accuracy comparable to what can be achieved when relying on the error correcting properties of other mappings such as SFBK or GSE. 

This point also extends to applications of methods to \textit{ab initio} molecular systems. Ref.~\cite{Setia2018} shows that SFBK results in lower gate depth than Jordan-Wigner, but higher than Bravyi-Kitaev on a small molecular system. Chien \textit{et al.} \cite{Chien2019} also show that Jordan-Wigner can outperform SFBK in terms of gate count under certain conditions on the Hamiltonian studied. One can expect GSE to perform better due to its lower Pauli weight, however for a relevant comparison, one would need to incorporate equivalent error mitigation techniques into Jordan-Wigner or Bravyi-Kitaev to balance quantum resources required to achieve a given accuracy. Overall, the question of the real impact of the error mitigating and correcting properties of lattice tailored encodings could be critical in determining their future applicability. 

\subsection{Reducing qubit requirements} \label{sec:tappering_qubits}

Several methods have focused on reducing the overall number of qubits required to encode a specific Hamiltonian \cite{Moll2016, bravyi_tapering_2017, Steudtner2018, Setia2020, Kirby2021_CSVQE}. These methods usually come at limited additional costs and can therefore be used to significantly improve the efficiency of VQE. Bravyi {\it et al.} \cite{bravyi_tapering_2017}] formulate several proposals designed to reduce (`taper off') the number of qubits required to simulate fermionic systems for variational quantum algorithms. In particular, they find that it is possible to reduce qubit requirements in encodings where $\mathbb{Z}_2$ symmetries are present. This results in halving the dimension of the Hilbert space of the system considered for each qubit removed. The starting point of the proposal in Ref.~\cite{bravyi_tapering_2017} is the observation that in some encodings, some qubits are representing a conserved quantity of the molecular system, and should therefore never change. One example is the last qubit in the parity mapping which encodes the parity of the entire wavefunction and is directly dependent on the electron number.

By definition, a symmetry is an operator which leaves the Hamiltonian invariant when acting on it (i.e. it must commute with the Hamiltonian) \cite{messiah2014quantum}. Hence one can always find a common eigenbasis for the operator corresponding to the symmetry and the Hamiltonian. As the qubit Hamiltonian is written as a weighted sum of Pauli strings, if all Pauli strings commute with an operator, it is a symmetry operator. As such, given any initial encoding, one can find a certain number of operators, the symmetry operators $\tau_i$, that commute with each of the Pauli strings in the Hamiltonian and between themselves. This can be done by finding the kernel of the check matrix that corresponds to the Pauli terms \cite{bravyi_tapering_2017}. Although only symmetries that can be expressed as tensor products of Pauli operators can be found by this procedure. We can then define a change of basis, represented by a unitary operator $U$ such that for each symmetry $\tau_i$ we have
\begin{equation}
    U \tau_i U^\dagger = Z_{q(i)}.
\end{equation}
In this basis the symmetry operator acts as a $Z$ operator on the $q(i)$-th qubit. The matrix corresponding to $U$ can be found as the product $U = \prod_i U_i$, where each unitary $U_i$ is defined as
\begin{equation}
    U_i = \frac{Z_{q(i)} + \tau_i}{\sqrt{2}}
\end{equation}
In a similar fashion to what we have done before, we can remove each $q(i)$-th qubit and replace it in each Pauli term in the Hamiltonian by its eigenvalue. Setia et al \cite{Setia2020} show that symmetry operators can be identified using molecular point group symmetries \cite{cotton1990chemical} for the case of quantum chemistry simulations. One can remove the qubits corresponding to each member of an Abelian (commutative) subgroup of the molecular point group describing the symmetries of the molecule under consideration. It is shown in Ref.~ \cite{Setia2020} that the number of qubits that can be removed is either higher or the same as the method proposed in Ref.~\cite{bravyi_tapering_2017}. For instance, $\mathrm{CO_2}$ on $30$ qubits, $\mathrm{C_2H_2}$ on 24 qubits, $\mathrm{BeH_2}$ on 14 qubits, can all be reduced by $5$ qubits using the method presented in Ref.~\cite{Setia2020}. It is worth noting that a similar approach is briefly outlined in \cite{Zhang2021_shallow}, along with a proposed method to identify point-group symmetries. 

The Contextual Subspace VQE (CSVQE) \cite{Kirby2021_CSVQE}, proposes to separate the expectation value of the Hamiltonian into two contributions: a contextual part, computed using VQE, and a non-contextual part, computed using a conventional computer. A set of Pauli strings observables is considered non-contextual if it is possible to measure and assign value to them simultaneously without contradiction \cite{Kirby2021_CSVQE, Raussendorf2013, Howard2014, Cabello2014, Cabello2015, Ramanathan2014, Kirby2019, Kirby2020_classical_sim}. While this technique can introduce an approximation in the energy computed, it does so with a significant reduction in the number of qubits, and can also be applied after using the qubit reduction technique presented above for further efficiency gains\cite{Kirby2021_CSVQE}.

%% file: 05_group_main.tex
\section{Efficient grouping and measuring strategies} \label{sec:Grouping}

One of the key challenges possibly holding back the VQE is the very large amount of samples that are required to accurately compute the relevant values of the algorithm. There are two main aspects to manage for efficiently sampling these expectation values: the number of terms in the Hamiltonian cost functions (computed using the mappings presented in Sec. \ref{sec:Encoding}), and the number of shots required to sample an expectation value at a certain level of accuracy. It is worth noting that the level of accuracy required changes throughout the optimization process. This could be because of the accuracy needed for computation of the final output of the VQE, but also most importantly because optimizer gradients (or optimizer steps in gradient-free methods) must be estimated precisely enough to be distinguished from one another when the optimization landscape flattens. An extreme case of the latter is the barren plateau problem (\cite{McClean2018}, see Sec. \ref{sec:barren_plateau}). We first discuss strategies for optimal sampling error reduction (Sec. \ref{sec:measurements_strategies}), and then we discuss methods used to reduce the number of Pauli strings that are necessary to separately measure (Sec. \ref{sec:pauli_grouping}).

\subsection{Scaling of shot numbers in VQE}  \label{sec:measurements_strategies}

In this section, we discuss methods that have been proposed to reduce the impact of the denominator in Eq.~(\ref{eq:measurement_scaling}) on the total number of measurements required for each energy evaluation in the VQE, and in particular for each estimation within a gradient estimation. Namely how to minimize the number of measurements required to achieve a given level of precision, established by a target standard error on the measurement: $\epsilon$. 

\subsubsection{Overall scaling of measurements:}

To get an idea of the scaling in sampling requirements, let us first consider the scaling of the output from the mapping methods presented in the previous section (Sec. \ref{sec:Encoding}). As seen previously, generalized mappings for molecular Hamiltonians result in $\mathcal{P} \sim \mathcal{O}(n^4)$ distinct Pauli strings to estimate. For mappings tailored to lattice models, this scales with the number of edges in the lattice (as an illustration, for a regular square lattice Hubbard model of dimension $D$, the number of edges scales $\mathcal{O}(nD)$).  

With this in mind, let us consider the number of shots required to achieve a given precision. In any sampling experiment, the standard error is equal to $\epsilon = \sigma/\sqrt{S}$, where $\sigma$ is the population standard deviation, and $S$ is the experimental sample size, in our case, the number of shots also noted $S$ (for a general introduction to statistical theory, we recommend \cite{Dekking2005}). This means that the number of times an experiment needs to be repeated to achieve a given expected error $\epsilon$ goes as $O(1/\epsilon^2)$. 
More specifically, when measurements are distributed optimally among the different Pauli strings, such that the variance is minimized with respect to a given precision $\epsilon$, the number of measurements required is upper-bounded by
\begin{equation} \label{eq:measurement_upper_bound}
    S \leqslant \left( \frac{\sum_a^{\mathcal{P}} w_a}{\epsilon} \right)^2,
\end{equation}
where $w_a$ are the weights of the Pauli strings in the Hamiltonian \cite{Wecker2015, Rubin2018}. 

As a result, for a given level of accuracy for each Pauli string measured independently, the overall scaling of the number of shots required for an energy estimation is:
\begin{equation} \label{eq:measurement_scaling}
    \mathcal{O}\left(\frac{N^4}{\epsilon^2}\right).
\end{equation}

In the context of quantum chemistry, successful computing methods are expected to produce results within a precision of $\epsilon = 1.6$ mE$_{\rm H}$ \cite{Peterson2012} to the target. When results obtained numerically are within this level of precision to experimental results, the simulations is deemed to reach chemical accuracy. This metric can be used as a bound for target precision in the VQE context (for an excellent discussion of precision vs. accuracy in this context, we direct readers to Eflving {\it et al.}~\cite{Elfving2020}). One should be cautious however not to assume too much of a relationship between this number and the number of shots required to perform VQE. That is because the key bottleneck of VQE optimization is not the estimation of the wavefunction itself but the estimation of gradients and in particular the difference between these gradients (which allow for the optimization step to be performed reliably). This difference may be orders of magnitude smaller than the chemical precision threshold in a barren plateau, requiring that many more measurements (\cite{McClean2018}, see Sec. \ref{sec:barren_plateau}). 
While polynomial in scaling, it has been pointed out on several occasions that the number of shots required to accurately compute a VQE optimization process rapidly becomes unmanageable \cite{Elfving2020, Wecker2015, Gonthier2020}, suggesting the method might be unable to compete with its conventional computing counterparts \cite{Elfving2020}. Ref. \cite{Wecker2015} estimates (though conservatively) that simulating the energy of $\mathrm{Fe_2S_2}$ ferredoxin (in STO-3G basis with $N=112$ spin-orbitals) using a VQE would require a total of $\mathrm{O}(10^{19})$ shots based on the upper bound defined in Eq.~(\ref{eq:measurement_upper_bound}). As such, a significant amount of effort has been devoted to finding solutions that reduce the pre-factor for the number of shots required. 

\subsubsection{Measurement weighting}

\paragraph{Uniform distribution of measurements:} The variance of  Pauli strings can be computed easily as they are self-inverse. In particular: 
\begin{align}
    \operatorname{Var}[\hat{P}_a] &= \bra{\psi}\hat{P}_a^2\ket{\psi} - \bra{\psi}\hat{P}_a\ket{\psi}^2 \nonumber \\
    &= 1 - \bra{\psi}\hat{P}_a\ket{\psi}^2 \leqslant 1.  
\end{align}
Measurements of different Pauli operators (when not grouped) are independent and therefore uncorrelated, resulting in mean squared error for the energy estimate given a total of $\mathcal{P}$ Pauli strings of \cite{Rubin2018}:
\begin{align} \label{eq:standard_error}
    \epsilon  &= \sqrt{\sum_a^{\mathcal{P}} \frac{w_a^2 \operatorname{Var}[\hat{P}_a]}{S_a}} \nonumber \\
    &= \sqrt{\sum_a^{\mathcal{P}} w_a^2 \left(\frac{1 -  \bra{\psi}\hat{P}_a\ket{\psi}^2}{S_a}\right)},
\end{align}
where $S_a$ is the number of shots used to measure the expectation value of each Pauli string $\hat{P}_a$, which has weight $w_a$, and such that $S = \sum_a S_a$. Assuming uniformly distributed shots across all Pauli strings, we can rearrange this result to showcase the number of measurements required to achieve a target standard error \cite{Arrasmith2020}: 

\begin{align} \label{eq:num_measurements_for_precision}
    S  = \mathcal{P}\sum_a^{\mathcal{P}} \frac{w_a^2 \operatorname{Var}[\hat{P}_a]}{\epsilon^2}.
\end{align}
This particular distribution of shots could be considered optimal in the special case where $\sqrt{\operatorname{Var}[\hat{P}_a]} \propto  1/|w_a|$ \cite{Rubin2018, Arrasmith2020}. This is in general not the case and therefore further methods have been developed to distribute measurements to optimally reduce estimation variance \cite{Arrasmith2020}. The methods outlined below aim to reduce $S$, the total number of shots, while maintaining a given precision $\epsilon$.

\paragraph{Weighted distribution of measurements:} An alternative to uniform distribution of shots is to focus on measuring more precisely the operators which contribute most to the total variance of the expectation value estimated. With a given shot budget, one can improve the overall precision of measurement by distributing these shots towards specific operators. 

A straightforward manner to distribute these shots is to simply weight them with respect to the Pauli strings weights ($|w_a|$) in the Hamiltonian \cite{Wecker2015}. When looking at Eq.~(\ref{eq:num_measurements_for_precision}), we can easily see that reducing the number of shots on strings contributing less to the total energy estimate (with lower $|w_a|$ value) and adding these to strings that contribute the most reduces total variance, as long as $\operatorname{Var}[P_a]$ are similar for all $a$ \cite{Rubin2018}. Rubin \textit{et al.} \cite{Rubin2018} indeed show that $S_a \propto |w_a|\sqrt{\operatorname{Var}[\hat{P}_a]}$, is optimal (although $\operatorname{Var}[\hat{P}_a]$ might not be easily accessible), while Arrasmith {\it et al.}  \cite{Arrasmith2020} show numerically that when considering random states, variations in $|w_a|$ tend to be higher than variation in $\operatorname{Var}[\hat{P}_a]$, resulting in the weight pro-rata distribution (where $S_a \propto |w_a|$) of measurements outperforming the uniform distribution in most cases. 
To address cases in which the number of shots is limited (i.e. there are so few shots, that each one could create a bias in the energy estimate), it is proposed in Ref. \cite{Arrasmith2020} to perform measurements on Pauli strings randomly, with probabilities proportional to $|w_a|\sqrt{\operatorname{Var}[\hat{P}_a]}$, thereby allowing unbiased estimates even with a low total shot number. 

Rubin {\it et al.}~\cite{Rubin2018} further use fermionic marginals, and $N$-representability constraints to determine optimized measurements distributions for Hamiltonian estimates. The idea of optimizing shots distributions was also merged to optimization strategy by designing specific optimizers aiming to balance their optimization-per-shot cost~\cite{Arrasmith2020,Kubler2020adaptiveoptimizer}. 

\paragraph{Term truncation:} Another approach to consider is to remove from measurement scope terms that have contributions significantly below the error tolerance threshold $\epsilon$ \cite{mccleanTheoryVariationalHybrid2015}. This method has been shown to significantly reduce the cost of quantum chemistry calculations with negligible impact on accuracy \cite{McClean2014}.

To implement this, one must observe that the contribution of any Pauli observable to the final energy estimate is bounded by the absolute value of its associated weight: $|\bra{\psi}w_a \hat{P}_a \ket{\psi}| \leqslant |w_a|$.  By ordering these contributions in ascending order, one can construct a partial sum of the $k \leq \mathcal{P}$ smallest contributor: 
\begin{equation}
    e_k = \sum_a^k |w_a|.
\end{equation}
From there, one can choose a constant $C \in [0, 1 [$, and include in the partial sum the terms up to an index $k$ that verify: $e_k < C \epsilon$. This method  \cite{mccleanTheoryVariationalHybrid2015} introduces a bias in total energy estimation, and as such the key to implement successfully is to pick a constant $C$ such that the truncation bias is lower than the mean square error reduction from measurements added to the remaining terms. McClean {\it et al.} \cite{mccleanTheoryVariationalHybrid2015} present an adjusted estimate for the number of shots required to achieve $\epsilon$, to be contrasted with Eq.~(\ref{eq:num_measurements_for_precision}):
\begin{align} \label{eq:num_measurements_for_precision_2}
    S  = (\mathcal{P} - k)\sum_{a=k+1}^{\mathcal{P}} \frac{w_a^2 \operatorname{Var}[\hat{P}_a]}{(1 - C^2)\epsilon^2}.
\end{align}
If the expected number of shots is lower than without term truncation (Eq. \ref{eq:num_measurements_for_precision}), then the method provides an improvement regarding the precision to measurement cost ratio.

\subsection{Pauli string groupings, and other joint measurement strategies} \label{sec:pauli_grouping}

The methods described below works with the idea that measuring a given Pauli string $a$: 
\begin{align}
    &\hat{P}_a = \sigma_{p^{(a)}_1} \otimes \sigma_{p^{(a)}_{2}} \otimes ... \otimes \sigma_{p^{(a)}_{N-1}} \otimes \sigma_{p^{(a)}_N} \\
    &p^{(k)}_i \in \{I, X, Y, Z\}, \nonumber
\end{align}
provides information about other Pauli strings which have overlapping Pauli elements (i.e. $p^{(a)}_j = p^{(b)}_j$) (for instance \cite{mccleanTheoryVariationalHybrid2015}). In essence, all of these methods target the same information gathering optimization. As such, there are a number of incompatibilities between them. In particular, General Commutativity \cite{Yen2020, Hamamura2020, Gokhale2019_short}, Unitary \cite{Izmaylov2020a, Zhang2020} and decomposed interactions \cite{Huggins2021} based grouping are aimed at diagonalizing a set of Pauli strings (i.e. rotate them so that $\forall i$, $p^{(a)}_j \in \{I, Z \}$), they exhaust all of the information that can be gained from inference methods (e.g. \cite{Huang2020, Torlai2020}. It is not the case however for Qubit-Wise Commutation based grouping \cite{mccleanTheoryVariationalHybrid2015, Kandala2017, Hempel2018, Rubin2018, Kokail2019, Izmaylov2019, Nam2020, Verteletskyi2020, Hamamura2020, Gokhale2019_long} which can also be used with inference methods.

\subsubsection{Inference methods}
\label{sec:pauli_grouping-inference-methods}

Inference methods all seek to recover the expectation values of an observable from a restricted set of operators, rather than the complete basis in which it can be measured. They are extensions built upon the theory of Quantum State Tomography (for some literature on the topic, we recommend: \cite{MauroDAriano2003, Cramer2010, Christandl2012, Bisio2009, ODonnell2016, ODonnell2017, Haah2017}) aiming at achieving a target precision $\epsilon$ with a minimum number of shots $S$ using the structure of the Lie algebra in which Pauli strings are defined \cite{Hall2015}. 

\paragraph{Methods for low Pauli weight Hamiltonians:} Restricting the problem of tomography to the estimation of a Hamiltonian can however significantly reduce the cost of measurements. For instance, taking into account the Pauli weight of the Hamiltonian can help reduce the sampling requirements. It is worth noting that both methods described in this paragraph are initially designed for Hamiltonian characterization rather than ground state estimation, and as such may not be directly optimized for use within the context of the VQE. 

One such method is Quantum Overlapping Tomography (QOT) \cite{Cotler2020} which aims at efficiently estimating all $\mathrm{k}$-body operators for a given quantum state (a $\mathrm{k}$-body qubit operator is computed in a manner that is similar to the one- and two-body RDM presented in Sec. \ref{sec:second_quantization}, but replacing the fermionic operators with computational basis elements), which in turn allows computing the expectation of an observable having up to $\mathrm{k}$ Pauli weight. They observe that complete tomography on a $\mathrm{k}$-qubit state grows exponentially in $\mathrm{k}$ \cite{ODonnell2016, Haah2017} (namely $3^{\mathrm{k}}$), and that there are $\binom{N}{\mathrm{k}}$  $\mathrm{k}$-body reduced density operators to measure for an $N$ qubit state, bringing the cost of a naive measurement of all $\mathrm{w}$-body operators to $\sim e^{\mathcal{O}(\mathrm{w})}\binom{N}{\mathrm{w}}$. However, one can also use the fact that many of these operators can be overlapped and measured in parallel on the same qubit register. Cotler et al. \cite{Cotler2020} show that this can result in a significant reduction of the number of shots required to estimate all $\mathrm{k}$-qubit reduced density matrices within a precision of $\epsilon$, to:
\begin{equation}
    S \sim \mathrm{k}e^{\mathcal{O}(\mathrm{k})} \left( \frac{\log(N)}{\epsilon} \right)^2,
\end{equation}
with the detail process on how to optimally allocate this measurements being described in Ref.~\cite{Cotler2020}. An interesting research question would be to study whether this method lends itself well to \textit{ab initio} molecular systems, and lattice models with high connectivity - this is because QOT does not necessarily take into account pre-existing structures of the Hamiltonian, for which in principle $\mathcal{O}(n^4)$ and $\mathcal{O}(nD)$  operators are normally needed to be estimated respectively, instead of the $3^{\mathrm{k}}$ for each $\mathrm{k}$-body reduced density matrix.

Another such method that is worth mentioning briefly is Bayesian Hamiltonian Learning \cite{Evans2019} which can infer the value of $\mathcal{P}$ Pauli strings with high success probability using $\mathcal{O}(3^{\mathrm{k}} \log(\mathcal{P}) / \epsilon^2) $ parallel shots. 
\paragraph{Shadow tomography, classical shadow, and locally-biased classical shadow:} The concept of shadow tomography was initially presented in \cite{Aaronson2019, Aaronson2019_prooceed}. It describes the task of predicting certain properties of a quantum state without conducting full tomography on the state. In particular, it was shown that an exponential number of target functions (for instance, computing the expectation value of an operator) can be predicted from a polynomial number of shots. It was highlighted however in Ref.~\cite{Huang2020} that this method requires exponential depth in the quantum circuit and access to quantum memory, thereby rendering it on balance too costly for NISQ algorithms such as VQE. 

The concept of classical shadow \cite{Huang2020} is an attempt to extend the idea of shadow tomography and address some of these caveats. It aims at efficiently learning a classical sketch $\tilde{\rho}$ of an unknown quantum state $\rho$, that is then used to predict arbitrary linear functions of that state (for instance, the expectation value of an operator: $\langle \hat{O} \rangle = \mathrm{Tr}[\hat{O} \rho]$) using a median-of-means protocol \cite{Jerrum1986, Nemirovsky1983}. The classical shadow is beneficial if it can be constructed with a tractable number of measurements and such that $\mathbb{E}[f(\tilde{\rho})] = f(\rho)$, with $f$ any of the aforementioned functions. Applied to the case of the expectation value of an observable, measurement of the classical shadow produces a random variable whose expectation value must match the expectation value of the observable with respect to the quantum state:
\begin{equation}
    \mathbb{E}[\langle \tilde{O} \rangle] = \mathbb{E}[\mathrm{Tr}(O \tilde{\rho})] = \operatorname{Tr}[O\rho].
\end{equation}

To construct a classical shadow, one must first produce an instance of the state $\rho$ and apply a random unitary $U$ taken from an ensemble $\mathcal{U}$ (defined by its elements and a probability rule for picking each element) to rotate it before performing a measurement. The measurement returns a N-bit measurement vector $\ket{\tilde{b}} \in \{0, 1 \}^N $. The conjugate of the unitary can then be applied back to measurement vector, producing $U^{\dagger}\ket{\tilde{b}}\bra{\tilde{b}}U$. 

From this point, we can define the linear map $\mathcal{M}$ (or measurement channel), which transforms $\rho$ to the expectation value of $U^{\dagger}\ket{\tilde{b}}\bra{\tilde{b}}U$ over both any unitary in $\mathcal{U}$ and all measurement outcome possible $\ket{\tilde{b}}$. We have: 
\begin{align}
    \mathcal{M}(\rho) &= \mathbb{E}_{U\in\mathcal{U}} \sum_{b \in \{0, 1 \}^N} \bra{b}U\rho U^{\dagger}\ket{b} U^{\dagger}\ket{b}\bra{b} U \nonumber \\
    & = \mathbb{E}\left[U^{\dagger}\ket{\tilde{b}}\bra{\tilde{b}}U  \right],
\end{align}
where Born's rule was used: $\mathrm{Pr}[\tilde{b} = b] = \bra{b} U \rho U^{\dagger} \ket{b}$. The linear map admits a unique inverse (this requires Tomogaphic completeness, please refer to the Supplementary materials of \cite{Huang2020} for details) which can be applied through conventional post-processing, producing a classical shadow, such that:
\begin{equation}
    \tilde{\rho}_i = \mathcal{M}^{-1} \left( U_i^{\dagger}\ket{\tilde{b}_i}\bra{\tilde{b}_i}U_i \right).
\end{equation}
This procedure is repeated $T$ times, to produce an $T$-sized classical shadow array: $S(\rho, T) = \{ \tilde{\rho_1}, ..., \tilde{\rho_T} \}$. Huang \textit{et al.} \cite{Huang2020} report that the use of classical shadow allows scaling measurements required for the desired precision logarithmically in the number of operators to measure. This can be combined with measurements weighting methods \cite{Hadfield2020} to avoid exponential scaling of required classical shadow in case of non-local observables. Low weight derandomization strategy, presented in Ref.~\cite{Huang2021} (and independently Ref.~ \cite{acharya2021informationally}) expand on the ideas of shadow tomography and classical shadows. It removes the need to draw unitaries (rotating the measurement basis) randomly from a pool, and the need for median-of-means predictions, by progressively and deterministically selecting Pauli strings that have the highest impact on narrowing a given confidence bound. An adaptive process to reduce computational cost of the low-weight derandomization strategy was also proposed by Hadflied \cite{Hadfield2021}.

Two questions central to implementing classical shadows are the choice of the unitary ensemble $\mathcal{U}$ to draw from, and the construction of the inverse linear map $\mathcal{M}^{-1}$. Clifford ensembles were initially proposed \cite{Huang2020}, but alternative ensembles have also been discussed, such as unitaries corresponding to time evolution of a random Hamiltonian \cite{Hu2022}, unitary ensembles defined through locally scrambled quantum dynamics \cite{Hu2021} achieving a lower tomography complexity compared to Clifford based methods, and in the case of fermionic states, a discrete group of fermionic Gaussian unitaries \cite{Zhao2021, OBrien2021}. Bu \textit{et al.} \cite{Bu2022} propose to use Pauli-invariant unitary ensembles (unitary ensembles that are invariant under multiplication by a Pauli operator), a class which includes both Clifford ensembles and locally scrambled unitary ensembles. They also provide an explicit formula for the inverse linear map corresponding to these ensembles.

Classical shadows can be applied to study several aspects of quantum states, in particular for energy estimation as part of a VQE \cite{Hadfield2020, Hadfield2021, OBrien2021, Lukens2021}, and though it remains unclear how it can perform over other efficient measurement schemes numerical studies have already suggested superior empirical scaling over methods such as Basis Rotation Grouping \cite{Huggins2021} (described below).
These methods have also been studied in the presence of noise and modifications to the scheme have been shown to render the method resilient to quantum noises \cite{Chen2021_shadow, Koh2020}. Finally, classical shadows have been shown to be useful as a mean to mitigate barren plateaus \cite{Sack2022}. 

\paragraph{Neural Network tomography:} Another type of method proposed to reduce the number of shots required to achieve a given precision level on the measurement of a Hamiltonian is to use Machine Learning on a series of shot outputs to decrease the variance of the expectation value.     
A first example is presented in Ref. \cite{Torlai2020}, where Torlai et al. present a method to learn a mock of the state produced by an ansatz using an unsupervised restricted Boltzmann machine (RBM) \cite{Ackley1985} then used to compute expectation values of quantum observables. RBMs have been shown as successful models to represent quantum states in the field of condensed matter physics \cite{Melko2019, Torlai2020_MLQS}. 

RBMs are in general composed of two layers of binary-valued neurons, a visible layer composed of $i$ units equal to the input size, noted $\boldsymbol{v}$ (in this case, it would be $N$) and a hidden layer composed of $j$ units, noted $\boldsymbol{h}$. The two layers are connected by a set of weights (a matrix noted $\boldsymbol{W}$, where entry $W_{ij}$ connects unit $v_i$ to unit $h_j$), and units within each layer are not connected (unlike in general Boltzmann machines). Each unit also has a bias weight $b_i^{(v)}$ and $b_j^{(h)}$ ($\boldsymbol{b}^{(v)}$ and $\boldsymbol{b}^{(h)}$ in vector notation), such that the energy of the RBM can be expressed as: 

\begin{align} \label{eq:RBM_energy}
    E(\boldsymbol{v}, \boldsymbol{h}) = - \boldsymbol{b}^{(v)T}\boldsymbol{v} - \boldsymbol{b}^{(h)T}\boldsymbol{h} - \boldsymbol{v}^{T}\boldsymbol{W}\boldsymbol{h}.
\end{align}

In the case of representing a many-body wavefunction, the network parameters $\boldsymbol{\lambda} = \{\boldsymbol{b}^{(v)}, \boldsymbol{b}^{(h)}, \boldsymbol{W}\}$ are complex valued \cite{Carleo2017}. In addition, following \cite{Torlai2020}, the energy of the RBM in Eq.~(\ref{eq:RBM_energy}) is modified to represent a wavefunction dependent on the network parameters and a binary vector $\boldsymbol{v}$ of size $N$:

\begin{align} \label{eq:RBM_qmb_energy}
    \psi_{\boldsymbol{\lambda}}(\boldsymbol{v}) = e^{\boldsymbol{b}^{(v)T}\boldsymbol{v}}e^{\sum_j \log \cosh{\sum_i W_{ij} v_i + b^{(h)}_j} }
\end{align}

The aim of the RBM is to train $\boldsymbol{\lambda}$ such that given an element of the computational basis $\ket{\boldsymbol{v}}$:

\begin{align}
    \psi_{\boldsymbol{\lambda}}(\boldsymbol{v}) = \langle \boldsymbol{v} | \psi(\boldsymbol{\theta}) \rangle, 
\end{align}

where $\ket{\psi(\boldsymbol{\theta})}$ is the output state of the ansatz so that RBM approximates the probability distribution of measuring the output state. Otherwise said, the RBM is trained so that the output of the RBM energy function (Eq. \ref{eq:RBM_qmb_energy}) is equal to the amplitude of the quantum circuit output state with respect to the basis element $\ket{\boldsymbol{v}}$.

To train this RBM, suppose you have a pool of single-qubit measurement basis operators $\boldsymbol{p} = \bigotimes_i^N  p_i$, with $p_i = \{ X, Y, Z \}$. For each $\boldsymbol{p}$, there exist a set of binary vectors $\{ \boldsymbol{v}^{\boldsymbol{p}} \}$ corresponding to the possible measurement outcomes of $\boldsymbol{p}$ by state $\ket{\psi(\boldsymbol{\theta})}$. As such, following Born's rule we can define the probability of measuring each of $\boldsymbol{v}^{\boldsymbol{p}}$ as $P(\boldsymbol{v}^{\boldsymbol{p}}) = | \langle \boldsymbol{v}^{\boldsymbol{p}} | \psi(\boldsymbol{\theta}) \rangle |^2 $. We note $P_{\boldsymbol{\lambda}}(\boldsymbol{v}^{\boldsymbol{p}}) = |\psi_{\boldsymbol{\lambda}}(\boldsymbol{v}^{\boldsymbol{p}})|^2$, the amplitude square of the output of the RBM given $\boldsymbol{v}^{\boldsymbol{p}}$ as input and $\boldsymbol{\lambda}$ as parameters. Torlai et al. \cite{Torlai2020} suggest to use the extended Kullback-Leibler (KL) divergence as cost function for the RBM: 
\begin{align}
    \mathcal{C}_{\boldsymbol{\lambda}} &= \sum_{\boldsymbol{p}} \sum_{\boldsymbol{v}^{\boldsymbol{p}}} P(\boldsymbol{v}^{\boldsymbol{p}}) \log \frac{P(\boldsymbol{v}^{\boldsymbol{p}})}{P_{\boldsymbol{\lambda}}(\boldsymbol{v}^{\boldsymbol{p}})}  \nonumber \\
    & \approx - \sum_{\boldsymbol{p}} \sum_{\boldsymbol{v}^{\boldsymbol{p}}} P(\boldsymbol{v}^{\boldsymbol{p}}) \log P_{\boldsymbol{\lambda}}(\boldsymbol{v}^{\boldsymbol{p}}),
\end{align}
where the numerator of the log has been discarded in the second equality as it does not depend on $\boldsymbol{\lambda}$ and therefore does not impact the optimization process. This formulation is of course intractable as the set $\{ \boldsymbol{p} \}$ scales $3^N$ and each set $\{ \boldsymbol{v}^{\boldsymbol{p}} \}$ scales $2^N$. Ref. \cite{Torlai2020} bypass these exponential sums by restricting  $\{ \boldsymbol{p} \}$  to the set of Pauli strings already included in the Hamiltonian decomposition, and restricting the measurement outputs $\{ \boldsymbol{v}^{\boldsymbol{p}} \}$  to a set of finite size measurements $\mathcal{D}$ such that: 

\begin{equation}
    \mathcal{C}_{\boldsymbol{\lambda}} = -  \sum_{\boldsymbol{v}^{\boldsymbol{p}} \in \mathcal{D}} P(\boldsymbol{v}^{\boldsymbol{p}}) \log P_{\boldsymbol{\lambda}}(\boldsymbol{v}^{\boldsymbol{p}})
\end{equation}

\subsubsection{Hamiltonian partitioning based on commutativity}

The premise of this section, and the methods that follow, is that an Abelian group of Pauli strings can be simultaneously diagonalized through a unitary rotation of the measurement basis \cite{griffiths2005introduction}, thereby reducing the number of terms that need to be measured to accurately compute the expectation value of a Hamiltonian. This implies that single measurement values for all the terms in a given Abelian group can be inferred from a single \textit{joint measurement} (a measurement that simultaneously assesses multiple Pauli operators) of the complete qubit register. This principle derives from the stabilizer theory (for a review of stabilizer theory, we recommend: Refs. \cite{Gottesman1997, nielsenQuantumComputationQuantum2010}). First, we need to identify a set of generators of the Abelian group, $\{ \tau_i\}$. From there, we know that there exist a unitary $U$, such that we can write: 

\begin{equation}
    U \tau_i U^{\dagger} = \sigma_Z^{q(i)},
\end{equation}

where $q(i)$ maps generator index $i$ to a unique address in the qubit register. This means that after applying $U$ to a given quantum circuit, obtaining the expectation value of generator $\tau_i$ can be done by measuring the expectation value of $ \sigma_Z^{q(i)}$. Finding the unitary $U$ for an Abelian group therefore allows to measure all of its generators by simply measuring $\sigma_Z$ on all qubits (the expectation values of all the Pauli terms within the group, which are not generators, can be also be recovered, from the data gathered from the generators measurements). \\

\paragraph{Qubit-wise commutativity (QWC) or Tensor Product Basis (TPB) groups:} Two of Pauli strings are said to be QWC if each Pauli operator in the first commute with the Pauli operator of the second one that has the same index. Generally speaking, that would be any group where, any given Pauli operator in any Pauli string has an index such that, all the operators of the same index across all the other Pauli strings in the group are either the same Pauli operator, or the identity (for example, $XI$, $IZ$, and $XZ$ are altogether QWC). The term TPB refers to the same idea, all the Pauli strings in the group can be diagonalized simultaneously in a joint tensor product basis with no entanglement. 

This basis for grouping terms has been widely used and studied \cite{mccleanTheoryVariationalHybrid2015, Kandala2017, Hempel2018, Rubin2018, Kokail2019, Izmaylov2019, Nam2020, Verteletskyi2020, Hamamura2020, Gokhale2019_long}. It allows performing joint measurements more efficiently. In particular, Gokhale et al. \cite{Gokhale2019_long} found that this method reduces the pre-factor for the number of Pauli terms to be measured by about three, without however changing its asymptotic scaling (see also Ref. \cite{Yen2020}). It is worth noting that the process for partitioning for QWC is significantly cheaper computationally \cite{Gokhale2019_long} than for the General Communitativity (GC) rule which we present later on.

Another key advantage is that the basis rotation used to conduct the joint measurements only requires a circuit of depth 1. To achieve this, we need to find the unitary $U_{QWC}$ which rotates all the Pauli strings in a given QWC group into a basis in which they are all diagonalized. This is a straightforward process as any individual Pauli operator can be rotated in the $Z$ basis with one single-qubit operation as follows: 

\begin{align}
    &Z = \mathrm{Ry}\left(-\frac{\pi}{2}\right) X \mathrm{Ry}\left(-\frac{\pi}{2}\right)^{\dagger} \nonumber \\
    &Z = \mathrm{Rx}\left(\frac{\pi}{2}\right) Y \mathrm{Rx}\left(\frac{\pi}{2}\right)^{\dagger}
\end{align}
This method of grouping and joint measurement is therefore relatively cheap to implement and allows for significant savings in the number of shots required to complete a VQE, although without changing the overall scaling. 

\paragraph{General Commutativity (GC), or entangled measurements:} This method uses the same logic as the QWC groupings, however rather than allowing for qubit-wise commutation groupings, it allows for grouping of any Pauli strings that generally commute. There has been a number of independent research producing similar results \cite{Yen2020, Hamamura2020, Gokhale2019_short}. From here it is obvious that the set of QWC relationships is included in the set of GC relationships (if two Pauli strings Qubit-Wise commute, they necessarily generally commute, the reverse is not true). 

Consider two Pauli strings $\boldsymbol{\nu}$ and $\boldsymbol{\eta}$, each composed of $N$ Pauli / identity tensored operators (noted $\nu_i$ and $\eta_i$ for $i \in [1, N]$. If two Pauli operators do not commute we have: $\nu_i\eta_i = - \eta_i\nu_i$, and  $\nu_i\eta_i = \eta_i\nu_i$ if they commute. This implies that:
\begin{equation}
    \boldsymbol{\nu}\boldsymbol{\eta} = (-1)^k \boldsymbol{\eta}\boldsymbol{\nu},
\end{equation}
where k is the number of index-wise Pauli operators that do not commute. As such, two Pauli strings generally commute if they comprise an even number $k$ of index-wise operators that do not commute (for example, $XYZ$, $XZY$, and $ZIX$ altogether are GC). The interesting aspect of this approach is that it reduces the scaling of the number of separate terms to be measured from $\mathrm{O}(N^4)$ to $\mathrm{O}(N^3)$ \cite{Gokhale2019_short, Yen2020, Jena2019} in the case of \textit{ab initio} molecular Hamiltonian, providing a significant advantage to VQE optimization. \\

The problem of building appropriate unitaries for joint measurements of the groups identified is more complicated than in the QWC case. The objective is the same: finding a unitary rotation that simultaneously diagonalizes all the elements of a given group. The measurement basis however is more complex than a TPB, and requires entangled measurements \cite{Hamamura2020}. As for QWC, the qubit register is entirely measured in the $Z$ basis and therefore the unitary performing the basis rotation itself must include non-local, entangling operations. This requires careful circuit design, which has been thoroughly explained in Ref.~\cite{Gokhale2019_long}, and we strongly recommend readers that are interested in building their implementation to refer back to this article, or to the "CZ" construction (based on Ref.~\cite{VandenNest2004}) and "CNOT" construction (based on Ref.~\cite{Aaronson2004, Patel2008}) proposed in Ref.~\cite{Crawford2021}. These two latter methods have the advantage of explicitly treating the case where the number of independent operators in a group, $k$, is strictly less than $N$ (as we have in all cases  $k \leqslant N$). It is worth noting that while not explicitly covered in Ref.~\cite{Gokhale2019_long}, this case can also be addressed using the former method.

Gokhale et al. \cite{Gokhale2019_long} also point out that the number of gates in the circuits scales $\mathcal{O}(N^2)$ (with some gate parallelization possible, making this a worst-case for depth), while Ref. \cite{Yen2020} shows a gate scaling of $\mathcal{O}(N^2/\log (N))$. The methods presented in \cite{Crawford2021} have a number of two-qubit gates that in the worst-case scale:
 \begin{align}
     &u_{\mathrm{CZ}}(k, N) = kN - k(k+1)/2 \nonumber \\
     &u_{\mathrm{CNOT}}(k, N) = \mathcal{O}(kN/\log(k)).
 \end{align}
In all cases, it can be considered negligible compared to the scaling of most ans{\"{a}}tze, but not all, as linearly scaling ans\"tze such as UpCCGSD \cite{Lee2019} could see their cost become negligible compared to the required basis rotation (see Sec. \ref{sec:Ansatz}). As such, the decision on whether this method should be used depends on the type of ansatz used and the quantum cost (circuit depth) one is willing to reduce the computing time (number of repeated measurements). 

\paragraph{Simple example:} As an example, consider the arbitrary set of Pauli strings covering 4 qubits presented in Table  \ref{tab:eg_pauli_set}

\begin{table} [ht]
\caption{Example set of Pauli strings}\label{tab:eg_pauli_set}
\begin{tabularx}{\textwidth}{p{\linewidth}}
\toprule
$IIXX, YXII, XYII, YXYY, YXXX, XYYY, XYXX, IIYY, XXYX, IIYX, YYYX, IIXY, XXII, YYII,$ \\ 
$XXXY, YYXY$\\
\bottomrule
\end{tabularx}
\end{table}

From this set, one can apply one of the heuristics presented in Sec. \ref{sec:grouping_heuristics} to decompose the set into commutative groups. As an example, one can identify the commutative group presented in Table  \ref{tab:eg_commutative_group}.

\begin{table} [ht]
\caption{Example set of a commutative group extracted from Table \ref{tab:eg_pauli_set}}
\begin{tabularx}{\textwidth}{p{\linewidth}}
\toprule
$XXYX, IIYX, YYYX, IIXY, XXII, YYII, XXXY, YYXY$
\\
\bottomrule
\end{tabularx}
 \label{tab:eg_commutative_group}
\end{table}

The next step is to identify a minimum set of generators of the multiplicative group defined - for instance, we can observe that $(XXYX) \times (IIYX) = XXII$. Ref. \cite{Gokhale2019_long} suggests using Gaussian elimination to achieve this. There are of course more than one possibility. The set presented in Table  \ref{tab:eg_commutative_generators} is an example of generators identified.

\begin{table} [ht]
\caption{Example set of generators for the commutative from Table  \ref{tab:eg_commutative_group}}
\begin{tabularx}{\textwidth}{c}
\toprule
$XXYX, IIYX, YYYX, IIXY$
\\
\bottomrule
\end{tabularx}
\label{tab:eg_commutative_generators}
\end{table}

The final step is to construct the unitary $U$ that maps each of the Pauli strings above to single-qubit measurements, such that:

\begin{align} \label{eq:grouping_GC_example}
    U^{\dagger} \left( XXYX \right) U  = ZIII \nonumber \\
    U^{\dagger} \left( IIYX \right) U = IZII \nonumber \\
    U^{\dagger}\left( YYYX \right) U = IIZI \nonumber \\
    U^{\dagger}\left( IIXY \right) U = IIIZ
\end{align}

\begin{figure}[ht]
\centerline{
\Qcircuit @C=1em @!R {
   \lstick{\ket{q_0}}   &   \qw   &   \gate{H}   &    \qswap    &  \gate{S} & \control \qw  & \qw    & \qw & \qw  &   \gate{H}   & \meter          \\
   \lstick{\ket{q_1}}   &   \qw   &   \qw   &    \qswap \qwx    &  \qswap & \ctrl{-1} & \gate{S} & \control \qw & \qw  &  \gate{H}    & \meter            \\
   \lstick{\ket{q_2}}   &   \qw   &   \gate{H}   &    \qw    &  \qswap \qwx & \qw & \qw & \ctrl{-1} & \gate{S} &    \gate{H}    & \meter                \\
   \lstick{\ket{q_3}}   &   \qw   &   \qw   &    \qw   &  \qw &  \gate{S} & \qw  & \qw  &  \qw &  \gate{H}  & \meter                 \\
} 
}\caption{Example quantum circuit for the realization of the relationships defined in Eq.~(\ref{eq:grouping_GC_example}), following the method presented in \cite{Gokhale2019_long}} \label{fig:grouping_GC_circuit}
\end{figure}
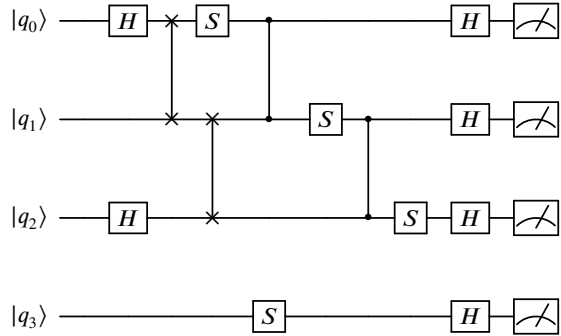

The quantum circuit realizing this unitary can be produced for example following the method presented in Ref. \cite{Gokhale2019_long}. For the example above, an example of realization for this unitary is given in Fig. \ref{fig:grouping_GC_circuit}.

From this, we can see that half of the Pauli strings set presented in Table  \ref{tab:eg_pauli_set} can be jointly measured through the operator $ZZZZ$.

\subsubsection{Unitary partitioning of the Hamiltonian:} 

An alternative to the GC rule as a connection system for grouping Hamiltonian terms is to connect anti-commuting terms. In essence, this is the complement to the GC connections (if the GC relationships for Hamiltonian terms are defined in terms of a graph, the anti-commuting relationships are the complementary graph). It offers however a convenient property: it allows partitioning the Hamiltonian as a weighted sum of a minimum (or close to minimum) number of unitary operators \cite{Izmaylov2020a}.

To see this, consider that while Pauli strings are hermitian unitary, a weighted sum of Pauli strings generally is not. The conditions for a weighted sum to be unitary however can be recovered through anti-commuting grouping. In particular, for any operator expressed as a Pauli sum:
\begin{equation} \label{eq:weighted_sum_operator}
    \hat{O} = \sum_{a}^{\mathcal{P}} w_{a} \hat{P_{a}},
\end{equation}
to be unitary, we must have: 
\begin{align}
    \mathrm{Im}(w_{a}^{\dagger}w_{b}) = 0, \label{eq:real_condition}\\
    \sum_a |w_{a}|^2 = 1,\label{eq:normalized_condition}
\end{align}
and: 
\begin{equation}
    \{\hat{P_{a}}, \hat{P_{b}} \} = 2 \delta_{ab}. \label{eq:anticommute_condition}
\end{equation}

The first two conditions are easily met. Given $w_a$ weights are always real, Eq.~(\ref{eq:real_condition}) is always verified in the VQE context. The normalization condition in Eq.~(\ref{eq:normalized_condition}) can easily be engineered by extracting a normalization factor $\boldsymbol{w}$ such that 
\begin{equation}
    \mathrm{w} = \left( \sum_a |w_a|^2 \right)^{\frac{1}{2}}, 
\end{equation}
and
\begin{equation}
    \hat{O} = \mathrm{w} \sum_{a} \frac{w_{a}}{\mathrm{w}} \hat{P_{a}}.
\end{equation}
By selecting the Pauli strings in $\hat{O}$ such that they anti-commute, we can re-write the above equation as 
\begin{equation}
    \hat{O} = \mathrm{w} \hat{U},
\end{equation}
with $\hat{U}$ a unitary operator. Therefore, by partitioning the Hamiltonian following anti-commuting operator it can be re-written as a weighted sum of $\mathcal{U}$ unitary operators as 
\begin{equation} \label{eq:unitary_hamiltonian}
    \hat{H} = \sum_a^{\mathcal{U}} \mathrm{w}_a \hat{U}_a.
\end{equation}
This grouping premise has been shown to allow a linear reduction in the number of terms in the Hamiltonian, similar to what can be achieved with GC grouping \cite{Izmaylov2020a, Zhao2020, Ralli2021}.

The next step is to implement the joint measurement of all the elements of each unitary group identified. Unlike the groupings based on QWC and GC, the groups produced based on unitary partitioning can be directly implemented as quantum circuits. To see how this is done, one can follow the method in Ref. \cite{Izmaylov2020a} starting by incorporating the partitioned Hamiltonian from Eq.~(\ref{eq:unitary_hamiltonian}) into the VQE optimization problem (Eq. \ref{eq:vqe_hybrid_function}):
\begin{equation} \label{eq:vqe_pb_unitary}
    E_{\mathrm{VQE}} = \min_{\boldsymbol{\theta}} \sum_a^{\mathcal{U}} \mathrm{w}_a \bra{\psi(\boldsymbol{\theta})} \hat{U}_{a} \ket{\psi(\boldsymbol{\theta})}.
\end{equation}
The unitary group cannot be directly measured jointly. However, as the observable is a unitary, one can transform Eq.~(\ref{eq:vqe_pb_unitary}) into an overlap measurement problem (rather than an observable measurement problem). Using the fact that a unitary observable has the same expectation value as its complex conjugate, we have
\begin{align} \label{eq:unitary_and_conj}
    E_{\mathrm{VQE}} &= \frac{1}{2} \sum_a^{\mathcal{U}} \mathrm{w}_a \bra{\psi(\boldsymbol{\theta})} \hat{U}_{a} \ket{\psi(\boldsymbol{\theta})} + \bra{\psi(\boldsymbol{\theta})} \hat{U}_{a}^{\dagger} \ket{\psi(\boldsymbol{\theta})} \nonumber \\
    & = \frac{1}{2} \sum_a^{\mathcal{U}} \mathrm{w}_a \left( \langle \boldsymbol{0} | \Psi_a \rangle + \langle \Psi_a | \boldsymbol{0} \rangle \right) \nonumber \\
    & = \sum_a^{\mathcal{U}} \mathrm{w}_a Re\left( \langle \boldsymbol{0} | \Psi_a \rangle \right), 
\end{align}
where $\ket{\boldsymbol{0}}$ is the initial qubit register (which can be replaced with the Hartree-Fock wavefunction) and we have set $\ket{\Psi_a} = U_{\boldsymbol{\theta}}^{\dagger} U_a   U_{\boldsymbol{\theta}} \ket{\boldsymbol{0}}$ (and of course $\ket{\psi(\boldsymbol{\theta})} = U_{\boldsymbol{\theta}} \ket{\boldsymbol{0}}$). The real part of the overlap in Eq.~(\ref{eq:unitary_and_conj}) can be computed using a Hadamard test (as presented in Appendix [\ref{sec:hadamard-test}]). 

In terms of the cost of the circuit add-on, one should first note that the ansatz used to prepare the trial state is used twice therefore at least doubling circuit depth. In addition, one must take into account the depth required to implement $U_a$. As a sum of Pauli operators, it can be implemented using $2(\tilde{q} - 1)$ CNOT gates per operator, where $q$ is the Pauli weight of the operator. Hence, $U_a$ takes a total of $2L(\tilde{q} - 1)$, where L is the number of term is $U_a$ (and as such, on average $L \approx N$ given there are $\mathcal{O}(N^3)$ groups). This gives a total CNOT scaling of $\mathcal{O}(\tilde{q}N^2)$ \cite{Izmaylov2020a}. 
 
Ref. \cite{Zhao2020} show that unitary groups can also be jointly measured in a manner similar to commutative groups by appending the unitary at the end of the ansatz which rotates the unitary group into a diagonal basis. Their method requires entangling gate scaling of $\mathcal{O}(qL)$, with $q$ the maximum Pauli weight in the Hamiltonian, and $L$ the number of terms in each unitary group. As discussed before, this implies a scaling of $\mathcal{O}(N^2)$ using Jordan-Wigner, and $\mathcal{O}(N \log(N))$ using Bravyi-Kitaev. Zhang \textit{et al.} \cite{Zhao2020} also show that the circuit depth can be reduced further using ancilla qubits. These methods of partitioning the Hamiltonian into unitary groups have been tested in an implementation on a quantum computer by Ralli \textit{et al.} \cite{Ralli2021}.

\subsubsection{Decomposed interactions for efficient joint measurements:} \label{sec:DecomposedInteractions}

\paragraph{Basis Rotation Grouping:}
The method proposed in Ref.~\cite{Huggins2021}, referred to as "Basis Rotation Grouping" is based on a tensor decomposition of the two-body operator. It details how to significantly reduce the overall number of (joint) terms to measure in the Hamiltonian, down to a linear number with system size. This same decomposition has also been used to reduce the total gate depth of the full UCCSD Ansatz as well as Trotter steps in Ref.~\cite{Motta2021} and is based on a two-stage decomposition of the interaction tensor originally proposed in Ref.~\cite{Peng2017}. It also provides a large improvement in the noise resilience of mappings with high Pauli weight such as Jordan-Wigner (See Sec.~\ref{sec:Encoding}). The price to pay for this is that the measurement has to take place in a different basis for each term, necessitating an additional $\mathcal{O}(N)$ gate depth before measurement to implement this orbital rotation for each grouped term of this decomposed Hamiltonian (which remains much less than $\mathcal{O}(N^2)$ required for general commutative grouping).

The first step in this method is to re-write the second-quantized Hamiltonian into a factorized form through decomposition of the two-electron integral tensor \cite{Berry2019, Motta2021, Huggins2021}. Starting from Eq.~(\ref{eq:molecularhamiltonianladder}) in the basis of spin-orbitals, we can rewrite the two-electron part as
\begin{align} \label{eq:two_body_decomp}
    V &= \frac{1}{2} \sum_{pqrs=1}^{n} h_{pqrs} \hat{a}_{p}^{\dagger} \hat{a}_{q}^{\dagger} \hat{a}_{r} \hat{a}_{s} \\
    &= \frac{1}{2} \sum_{pqrs=1}^{n} h_{ps,qr} (\hat{a}_{p}^{\dagger} \hat{a}_s \hat{a}_{q}^{\dagger} \hat{a}_r - \hat{a}_p^{\dagger} \hat{a}_r \delta_{qs}) = V' + S ,
\end{align}
where $S$ is an additional one-body operator, and $h_{ps,qr}=(ps|qr)$ is a representation of the two-body integrals in the conventional `chemists' notation. The positive-definite super-operator $h_{ps,qr}$ can be decomposed in a low-rank spectral decomposition:

\begin{align} \label{eq:spectral_decomposition_two_RDM}
    V' = \sum_{pqrs} \sum_{l=1}^L v_{ps}^{(l)} v_{qr}^{(l)} \hat{a}_p^{\dagger} \hat{a}_s \hat{a}_q^{\dagger} \hat{a}_r.
\end{align}
This decomposition has a long history in quantum chemistry, and this low-rank factorized form can be directly constructed, avoiding an explicit diagonalization (which would scale as $\mathcal{O}(N^6)$), through techniques such as density fitting or Cholesky decomposition \cite{Pedersen2009, Beebe1977, Koch2003, Purwanto2011, Mardirossian2018, Peng2017} (which is generally accepted to scale $\mathcal{O}(N^3)$). It has long been known that the decomposed form is not of full rank, with a number of terms $L = \mathcal{O}(N)$, which is a sufficient description of the system in the case of arbitrary basis quantum chemistry \cite{Pedersen2009}. It is worth noting that in special representations where the Coulomb operator is diagonal, as demonstrated for the plane wave basis and dual basis in Ref.~\cite{Babbush2018}, this can be rigorously $L=1$.

Further eigendecomposition of the resulting matrix for each value of $l$ is possible, with
\begin{equation}
    v_{ps}^{(l)} a_p^{\dagger} a_s = \sum_{i=1}^{\rho_l} U_{pi}^{(l)} \lambda_i^{(l)} U_{si}^{(l)} \hat{a}_p^{\dagger} \hat{a}_s ,
\end{equation}
where $U$ denote single-particle unitary operators.
These can be combined with a decomposition of the one-body part into a final doubly factorized Hamiltonian form of
\begin{equation}
    \hat{H} = U^{(0)} \left( \sum_i g_i \hat{n}_i \right) (U^{(0)})^{\dagger} + \sum_{l=1}^L U^{(l)} \left( \sum_{ij}^{\rho_l} g_{ij}^{(l)}\hat{n}_i \hat{n}_j\right)(U^{(l)})^{\dagger},
\end{equation}
where $\hat{n}_i = \hat{a}_i^{\dagger} \hat{a}_i$ and $g_{ij}^{(l)}=\frac{\lambda_i^{(l)} \lambda_j^{(l)}}{2}$ are scalars constructed from absorbing all relevant weights. The single-particle unitaries $U^{(0)}$ and  $U^{(l)}$ implement the orbital basis change and can be applied to the prepared state before measurement. 
The key is that all $\hat{n}_i$ and $\hat{n}_i \hat{n}_j$ for a given $l$ commute, and thus can be measured simultaneously, resulting in $L \sim \mathcal{O}(N)$ separate terms to estimate, but additionally requiring the change in measurement basis for each term increasing the gate depth by $\mathcal{O}(N)$. This method was tested in Ref.~\cite{Gonthier2020} and showed clear superiority of basis rotation method compared to QWC, achieving a significant scaling reduction in the number of measurements required.

\paragraph{Full rank optimization (FRO):}
Yen and Izmaylov \cite{Yen2021_Cartan} propose an a more general method for decomposition of the two body operator. Starting from $V$, as defined in Eq. (\ref{eq:two_body_decomp}), one can write
\begin{align} \label{eq:full_rank_opt}
    V &= \frac{1}{2}\sum_{\alpha = 1}^L U_{\alpha}^{\dagger} \left[ \sum_{ij}^n \lambda_{ij}^{(2, \alpha)} \hat{a}^{\dagger}_i\hat{a}_i \hat{a}^{\dagger}_j\hat{a}_j  \right]  U_{\alpha} \nonumber \\
    &= \frac{1}{2}\sum_{\alpha = 1}^L \sum_{ij}^n \lambda_{ij}^{(2, \alpha)} [ U_{\alpha}^{\dagger}\hat{a}^{\dagger}_i\hat{a}_i U_{\alpha}][U_{\alpha}^{\dagger}\hat{a}^{\dagger}_j\hat{a}_j U_{\alpha}],
\end{align}
with $\boldsymbol{\lambda}$ a tensor which must be discovered. The transformation unitary can be written as: 
\begin{align}
    U_{\alpha} &= \exp\left[ \sum_{i>j}^n -i \zeta_{ij}^{(\alpha)} ( \hat{a}_i^{\dagger}\hat{a}_j + \hat{a}_j^{\dagger}\hat{a}_i) +  \eta_{ij}^{(\alpha)} ( \hat{a}_i^{\dagger}\hat{a}_j - \hat{a}_j^{\dagger}\hat{a}_i) \right],
\end{align}
where $\boldsymbol{\zeta}$ and $\boldsymbol{\eta}$ are tensors which must be discovered. The process for estimating tensors  $\boldsymbol{\lambda}$,  $\boldsymbol{\zeta}$, and  $\boldsymbol{\eta}$ involves re-writing Eq.(\ref{eq:full_rank_opt}) to express the two-body coefficients $h_{pqrs}$ as function of these tensors. The system of equation can then be resolved using a difference minimization (for which the Broyden-Fletcher-Goldfarb-Shanno, BFGS, optimizer is suggested \cite{broyden_convergence_1970,fletcher_new_1970,goldfarb_family_1970,shanno_conditioning_1970}), and is described in further details in Ref.~\cite{Yen2021_Cartan}. 

Two variants of this method are also proposed in Ref.~\cite{Yen2021_Cartan}. The Greedy FRO (GFRO) discovers the tensors mentioned above iteratively by starting from $\alpha=1$. The variance-estimate GFRO (VGFRO) also takes into account the variance of the operators for the discovery of the tensors. Overall, Yen and Izmaylov find that FRO achieves a much lower number of partitions than the Basis rotation group method \cite{Huggins2021} (less than half, on six different systems ranging from $4$ to $20$ qubits), though FRO and VGFRO result in significantly more partitions (between $5$ and $7$ times more). The comparative analysis is however pushed further to take into account the risk for co-variances resulting from joint measurement of operators \cite{mccleanTheoryVariationalHybrid2015} (more details below), and it is found \cite{Yen2021_Cartan} that General Commutativity (GC) with Sorted Insertion \cite{Crawford2021} (presented below) performs the best in most systems. VGFRO, GFRO, and the basis rotation methods perform similarly, and far better than FRO. Yen and Izmaylov \cite{Yen2021_Cartan} conjecture that the relative performance of VGFRO, GFRO and basis rotation methods will improve compared to GC with Sorted Insertion as the size of the system studied increases. 

\subsubsection{Grouping heuristics} \label{sec:grouping_heuristics}

We have seen that defining whether two Pauli strings are QWC, GC, or anti-commuting is straightforward. However, given neither QWC, GC or anti-commutation properties are transitive (if two Pauli strings, $A$ and $B$ commute, and $A$ commutes with a third one, $C$, $B$ does not necessarily commute $C$), finding optimal groups of Pauli strings that together all meet one of these criteria out of the $\mathrm{O}(N^4)$ terms in the Hamiltonian can be a challenging process. 

Before detailing grouping methods directly dedicated to Pauli strings, we briefly outline conventional heuristics that have been used in the context of grouping Pauli strings and the studies that have been conducted to compare these. 

\paragraph{Conventional grouping heuristics used in VQE context:}

The problem of grouping Pauli terms that have been connected following a specific rule (QWC, GC, or AC), can be straightforwardly mapped to a graph problem. Namely, it can be mapped to the Minimum Clique Cover (MCC) problem \cite{Hamamura2020, Jena2019, Crawford2021, Verteletskyi2020, Izmaylov2020a, Zhao2020, Yen2020} (or equivalently the graph coloring problem \cite{Garey1979}), which aims at finding the minimum number of fully connected subgraphs in an initial input graph. In this case, we can define a graph $G( V, E)$, where $V$, the vertices are representing the Pauli strings, and $E$, the edges, are representing the connections established using one of the rules defined above. It is in general NP-hard \cite{Karp1972} and therefore should be solved using heuristics. The problem of grouping Pauli strings can therefore be addressed using the same heuristics as for MCC like problems: 

\begin{itemize}
    \item Largest Degree First Colouring (LDFC) Algorithm: In this algorithm, edges represent anti-commuting relationships. It works by first assigning a color to the vertex $V$ with the highest degree (colors are represented by integers, starting with $1$). Following this step, the vertex among those remaining with the highest degree is assigned the lowest color that is not already attributed to one of its neighbors. The process is repeated iteratively until all colors have been assigned \cite{Welsh1967} (Used for Pauli grouping in \cite{Hamamura2020}).
    \item Smallest first: Identical to the LDFC except for the ordering of colors allocation, which starts from the vertex with the smallest degree.  \cite{Matula1972}
    \item DSatur: The degree of saturation of a vertex is defined as the number of different colors it is adjacent to. With that in mind, the DSatur algorithm functions broadly like the LDFC algorithm, albeit by attributing colors along an ordering of the degree of saturation of the remaining uncolored vertices. The first color is attributed based on the largest degree. \cite{Brelaz1979}
    \item Independent-operator sorting algorithm: Vallury et al. \cite{Vallury2020} propose to group Pauli strings, using a QWC relationship, by ranking them according to the number of identity operators in each string. The method starts with the Pauli string having the lowest number of identity operators. Following strings are iteratively sorted: if a string QWC with all elements of an existing TBP group, it is added to this group, otherwise a new group is created. 
    \item Others of note include: Dutton and Brigham, \cite{Dutton1981}, COSINE \cite{Hertz1990}, Ramsey \cite{Boppana1992}, \cite{Tomita2006}, and Connected Sequential d.f.s \cite{Kubale2004}, Recursive largest first \cite{Leighton1979}.
\end{itemize}

For a thorough review of graph coloring methods, we recommend \cite{Kubale2004}.

All the heuristics mentioned above, except the BKT algorithm, are polynomial in scaling with respect to the number of graph vertices \cite{Izmaylov2020a}. The number of vertices being equal to the number of Pauli strings in the Hamiltonian these heuristics, in general, scale $\mathcal{O}(N^{k4})$, with $k \geqslant 2$ an integer corresponding to the respective scaling of each method. For instance, LDFC scales quadratically in the number of vertices in the graph \cite{Kubale2004}, therefore the method's time complexity is $\mathcal{O}(N^8)$ for a graph built from the second quantized Hamiltonian.

An alternative has been proposed in \cite{Gokhale2019_short}, where the problem is treated as finding a Maximal Flow in a Network Flow Graph. Gokhale et al. propose to identify the commuting relationships at the fermionic operator level, basing themselves on spin orbital indices of the fermionic terms. This is only treated under the Jordan-Wigner mapping They show that they can create partitions of size $N$, thereby reducing the total number of terms to measure to $N^3$, using the Baranyai construction approach which has a computational cost of $\mathcal{O}(N^5\log N) \leqslant \mathcal{O}(N^{k4})$. They also note that the groupings can be re-used across multiple Hamiltonian of the same sizes, which does not necessarily occur when grouping at the spin-operator level.

Several papers have drawn comparisons in the ability of these different heuristics to approach the MCC in the context of Pauli strings grouping. In particular, \cite{Izmaylov2020a} shows that RLF and DB tend to result in the lowest number of cliques for Unitary based grouping, on systems up to 14 qubits. \cite{Verteletskyi2020} show similar results for GC based grouping, with Largest First and SL also performing equivalently (on systems up to 14 qubits for all methods, and up to 36 for Largest first). \cite{Crawford2021} compares a lower number of methods (and also performs analysis on the co-variance implications, which is discussed in the next paragraph) and also finds that LDFC finds a lower number of cliques against Connected d.f.s, DSATUR, and Sorted insertion. 

\paragraph{Sorted Insertion, heuristic dedicated to Pauli strings and VQE applications:} 

The method presented in Ref.~\cite{Crawford2021} stems from the observation that grouping Pauli strings can result in measurement co-variances, thereby increasing the number of measurements required to achieve the desired precision \cite{mccleanTheoryVariationalHybrid2015}. It targets grouping based on optimization of the number of measurements required rather than on the number of terms to measure. Sorted Insertion works by allocating Pauli strings to commuting groups in descending order of their absolute weights in the Hamiltonian. The complexity of implementing this grouping heuristic is capped to $\mathcal{O}(N\mathcal{P}^2)$, with $\mathcal{P}$ the number of terms in the Hamiltonian \cite{Crawford2021} - hence in the general second quantized Hamiltonian case  $\mathcal{O}(N^9)$.

In addition, Crawford et al. \cite{Crawford2021} show that breaking commuting groups into smaller groups cannot reduce the variance under an optimal measurement strategy. They also define a useful figure of merit for a grouping strategy: the ratio of the minimum number of measurements required to achieve a desired precision (see Eq. \ref{eq:num_measurements_for_precision}) in the cases where Hamiltonian terms are not grouped, over the cases in which they are. They propose an approximate version of this metric which can be computed analytically from the Hamiltonian and the grouping. Given a Hamiltonian $\hat{H}$, which can be decomposed into $k$ operators $\hat{h}_i = \sum_j w_{ij}P_{ij}$ where all $P_{ij}$ for a given $i$ commute, the figure of merit is given by
\begin{equation}
    \tilde{R}:= \left(\frac{\sum_i^k \sum_j |w_{ij}|}{\sum_i^k \sqrt{\sum_j |w_{ij}|^2}}\right)^2.
\end{equation}
It is shown that Sorted Insertion achieves a significantly higher $\tilde{R}$ score than LDFC, DSatur, Connected Sequential d.f.s. and Independent Set on a number of molecular systems. Yen and Izmaylov \cite{Yen2021_Cartan} also showed that GC grouping using Sorted Insertion achieves the highest reduction in the number of shots required for a given precision for several systems up to $16$ qubits (one system of $20$ qubits is tested in Ref. \cite{Yen2021_Cartan}, and for which decomposed interactions methods \cite{Huggins2021, Yen2021_Cartan} perform better, see \ref{sec:DecomposedInteractions}).

\subsection{Discussion on measurement strategies and grouping methods} \label{sec:discussion_grouping}

The definition of the 'best possible grouping method' is not straightforward. While it is clear that aiming for the lowest number of groups possible is advantageous, it is not the only metric to take into consideration. In particular, it was shown in \cite{mccleanTheoryVariationalHybrid2015} that grouping terms, both under GC and QWC based grouping, suffers from co-variances arising from the joint measurement (thereby changing the formula presented in Eq. \ref{eq:standard_error}). This covariance effects increase the sampling noise and as such the total number of measurements required at a given level of precision. ~Joint measurements under the Unitary grouping or the decomposed interaction methods suffer from the same issue \cite{Yen2021_Cartan}. Therefore, the total number of measurements required to achieve a given precision should be taken into consideration as figure of merit for a grouping strategy \cite{Crawford2021}. Another cost to consider is the additional quantum noise resulting from the circuit used to rotate the measurement basis which could be significant in both the case of Unitary grouping and GC grouping. Further resources may be required to mitigate these additional errors (Sec. \ref{sec:error-mit}). Finally, an important point to note is that the scaling of most grouping heuristics could end up being somewhat prohibitive (for instance, we recall that LDFC scales $\mathcal{O}(N^8)$) for large systems. A possible way to dampen this issue is to use grouping methods which can be re-used across different Hamiltonians of the same active space sizes, by applying heuristics at the fermionic operator level \cite{Gokhale2019_long}, or by relying on two-body reduced density matrices \cite{Tilly2021}. 

Inference methods could perform better for each of the costs listed above but could also face their own pitfalls, in particular when an additional machine learning model requires training. A thorough numerical analysis of the multiple methods that have been proposed would be an interesting avenue for future research. In the meantime, we consider that the decomposed interactions \cite{Huggins2021, Yen2021_Cartan} methods have the most supporting arguments for the treatment of molecular Hamiltonians. While it is shown numerically in Ref.~\cite{Yen2021_Cartan} that GC with Sorted Insertion \cite{Crawford2021} tends to be most efficient with respect to the number of measurements required to achieve a given level of precision, decomposed interactions methods perform almost at the same level (and it appears that the gap in performance narrows as the system increases in size \cite{Yen2021_Cartan}), and requires significantly less depth to perform the required unitary transformations ($\mathcal{O}(N)$ against $\mathcal{O}(N^2)$ for GC). It is worth noting that the basis rotation group method \cite{Huggins2021} currently has a much more predictable implementation cost than FRO and its extensions. The latter requires solving a minimization problem to perform the decomposition, which could come at a significant computational cost or loss in accuracy, though further research will be required to investigate this point. 

%% file: 06_ansatz_main.tex
\section{Ansatz selection and construction}\label{sec:Ansatz}

Ansatz selection is a central part of the VQE pipeline. The right choice of ansatz is critical to obtain a final solution that is close to the true state of interest. To achieve this, it is essential to maximize the span of the ansatz in parts of the Hilbert space that contain the solution (i.e., a state that is sufficiently close to the desired state which globally minimizes the expectation value of the Hamiltonian). The span of possible states an ansatz can reach is referred to as its expressibility. However, performing a variational optimization on an ansatz with high expressibility could easily become intractable due to the number of parameters required to allow reaching many these different states, the number of iterations required for convergence, or the number of shots required to achieve sufficient gradient accuracy to continue the optimization. Whether an ansatz can be optimized in a tractable manner is referred to as its trainability. In practice, it is better to choose an ansatz spanning a smaller subspace, but remaining trainable. Designing an efficient ansatz for a given number of qubits hence involves finding an optimal trade-off between expressibility and trainability.

\paragraph{Expressibility:} The expressibility of an ansatz describes its span across the unitary space of accessible states \cite{Sim2019, Holmes2021, Nakaji2021}.
One can quantify the expressibility of an ansatz by assessing the distance between the distributions of the unitaries that can be generated by the ansatz, and the uniform distribution of unitaries in the corresponding Hilbert space \cite{Sim2019}, also known as the Haar measure. A given ansatz is called a t-design if it is indistinguishable from the Haar measure up to the $t^{th}$ moment. A 2-design ansatz can produce any possible state in the Hilbert space considered, from any input state: it is maximally expressive. As a side note, Hubregtsen et al. \cite{Hubregtsen2021} also study the relationship between the expressibility of quantum neural networks and their accuracy in a classification task. 

More formally, one can define as $\mathbb{U}$ the set of unitaries accessible by an ansatz, and $\mathcal{U}(N)$ the complete unitary group in which the ansatz is expressed (with $N$ the number of qubits it spans), such that $\mathbb{U} \subseteq \mathcal{U}(N)$ \cite{Sim2019, Cerezo2021_BP, Holmes2021, Nakaji2021}.
The following super operator, representing the second order difference between the Haar measure on $\mathcal{U}(N)$ and the uniformity distribution of $\mathbb{U}$ can be constructed (we follow the formalism in \cite{Holmes2021}):
\begin{align} \label{eq:expressibility_integral}
    \mathcal{A_{\mathbb{U}}}(\cdot) := \int_{\mathcal{U}(n)} d_{\mu}(V)& V^{\otimes 2}(\cdot)(V^{\dagger})^{\otimes 2} \nonumber \\
    & - \int_{\mathbb{U}} dU U^{\otimes 2}(\cdot)(U^{\dagger})^{\otimes 2},
\end{align}
with $d_{\mu}(V)$ the volume element of the Haar measure, and $dU$ the uniform distribution over $\mathbb{U}$, $V \in \mathcal{U}_N$ and $U \in \mathbb{U}$. If $\mathcal{A}_\mathbb{U}(\hat{O}) \rightarrow 0$, then the ansatz producing $\mathbb{U}$ approaches a 2-design and therefore offers maximal expressibility. From this super-operator, one can compute a metric for expressibility of an ansatz as
\begin{align}
    &\varepsilon_{\mathbb{U}}^{\rho} := || \mathcal{A}_\mathbb{U}(\rho^{\otimes 2})||_2 \label{eq:expressibility_density} \\
    &\varepsilon_{\mathbb{U}}^{\hat{P}} := || \mathcal{A}_\mathbb{U}(\hat{P}^{\otimes 2})||_2. \label{eq:expressibility_operator} 
\end{align}
Consequently, the expressibility of an ansatz can be expressed with respect to an initial input state ($\rho$), or with respect to a measurement operator ($\hat{P}$). Following the equations above, one can interpret that if $\varepsilon = 0$ the ansatz is maximally expressive, while expressibility decreases as $\varepsilon$ increases.  
Expressibility has also been shown to be a convenient metric for assessment of parametrized quantum circuits more generally \cite{Nakaji2021}. In addition, several methods have been proposed to remove redundant parameters from quantum circuits without decreasing expressibility \cite{Rasmussen2020} or reducing the set of states that can be generated through the circuit \cite{Funcke2021}.

\paragraph{Trainability:} The trainability of an ansatz refers to the ability to find the best set of parameters of the ansatz by (iteratively) optimizing the ansatz with respect to expectation values of the Hamiltonian in a tractable time \cite{Cerezo2021_BP, Holmes2021}.
More specifically, an ansatz is considered trainable if its expected gradient vanishes at most polynomially as a function of the various metrics of the problem (e.g. system size, circuit depth). On the other hand, if the gradient vanishes exponentially, it is said to suffer from the barren plateau problem. 
In this section, we first provide a review of the barren plateau problem \cite{McClean2018}, which is the main known obstacle to ansatz trainability, and discuss its implication for the VQE in Sec. \ref{sec:barren_plateau}. We then describe of the most relevant fixed-structure ansatz for the VQE in Sec. \ref{sec:fixed_struct}, followed by a description of the most relevant adaptive structure ansatz in Sec. \ref{sec:adaptive_ansatz}. Finally, we provide a discussion regarding ansatz selection for the VQE in Sec. \ref{sec:ansatz_discussion}. For alternative recent reviews of ansatz selection for the VQE, we recommend Refs.~\cite{Fedorov2021, Anand2021_review}.

\subsection{The barren plateau problem} \label{sec:barren_plateau}

A key issue that is inherent to all types of variational quantum algorithms is the risk of vanishing gradients, either during training or as a result of a random initialization \cite{McClean2018}. This refers to the risk of the cost function gradients vanishing exponentially as a function of specific properties of the optimization for a problem. McClean \textit {et al.} \cite{McClean2018} provide the first formal characterization of this barren plateau problem (some early numerical evidence of this problem are outlined in Ref.~\cite{Wecker2015}, without a characterization being provided), and show that cost function gradients are vanishing exponentially in the number of qubits in the quantum register when provided with random initialization of the circuit parameters. Even though this problem is akin to the vanishing gradient problem in machine learning, it has two striking differences that make it significantly more impactful on the prospects of variational quantum algorithms \cite{McClean2018}:
\begin{itemize}
    \item The estimation of the gradients on a quantum device is essentially stochastic. Any observable can only be measured to a certain precision, increasing as the inverse square root of the number of shots (see Sec. \ref{eq:measurement_scaling}). If gradients are exponentially approaching zero, it means that distinguishing between a positive and a negative gradient becomes increasingly difficult. Failing to establish the sign of the gradient reliably transforms the optimization into a random walk, overall requiring an exponential number of shots to continue optimization.
    \item The barren plateau problem is dependent on the number of qubits (while the problem is dependent on the number of layers for the vanishing gradient problem). Additional research also shows that it can be linked to other factors specific to quantum circuits, including expressibility of the ansatz \cite{Holmes2021}, degree of entanglement of the wavefunction \cite{OrtizMarrero2020, Patti2021}, non-locality of the wavefunction \cite{Cerezo2021_BP, Uvarov2020, Sharma2020}, or quantum noise \cite{Wang2020}.
\end{itemize}

Before describing key drivers of the barren plateau problem in more detail, and potential methods to address it, it is worth briefly discussing the typical cost function landscape for single parameters in the variational quantum eigensolver. Another problem that affects this landscape is that of 'narrow gorges' (initially characterized in \cite{Cerezo2021_BP}). It refers to the fact that the local minimum (well defined by the region starting from the end of a barren plateau and going towards a local minimum) contracts exponentially in the number of qubits. Interestingly, it was shown that these two problems are equivalent \cite{Arrasmith2021}. An alternative way to present the barren plateau problem is that it implies the expectation value of an observable with respect to a random state concentrates exponentially around the mean value of that observable \cite{McClean2018}, rendering intractable optimization away from the mean. 

In the context of the VQE, the barren plateau problem can be formally characterized as follows. Consider a VQE optimization problem with cost function:
\begin{equation} \label{eq:vqe_cost}
E(\boldsymbol{\theta}) = \bra{\psi(\boldsymbol{\theta)}} \hat{H} \ket{\psi(\boldsymbol{\theta})},
\end{equation}
with $\hat{H}$ the molecular Hamiltonian operator, and $\ket{\psi(\boldsymbol{\theta})}$ the parametrized wave function with a vector $\boldsymbol{\theta}$ of parameters.
This cost function exhibits a barren plateau if, for any $\theta_i \in \boldsymbol{\theta}$ and for any $\epsilon > 0 $ there is $b > 1$ such that:
\begin{equation}
    Pr( |\partial_{\theta_i} E(\boldsymbol{\theta})| \geqslant \epsilon ) \quad \leqslant \quad \mathcal{O}(\frac{1}{b^N}),
\end{equation}
which is an immediate consequence of Chebyshev's inequality and the result from above (for the expectation value and variance) \cite{Cerezo2021_BP}. This means that the probability of a gradient being above a certain threshold (which could be arbitrarily small), can always be upper-bound by a number that decreases exponentially in the system size $N$. It is however important to note that while defined with respect to a cost gradient, the barren plateau problem also affects gradient-free optimizers \cite{Arrasmith2020_BP_GF, OrtizMarrero2020} (e.g. COBYLA, Powel, Nelder-Mead, RotoSolver, see Sec. \ref{sec:Optimization}). It is easy to understand, as gradient-free optimizers usually rely on sampling the cost landscape of specific parameters. If the variance across the landscape is too small, then it becomes impossible to accurately progress through the optimization step.

\subsubsection{Drivers of the barren plateau problem}

\paragraph{System size and random initialization \cite{McClean2018}:} The barren plateau problem refers to the fact that the gradient of a cost function incorporating a layered ansatz has an exponentially vanishing variance, and values approaching zero in the number of qubits, provided ansatz parameters are initialized randomly. A layered ansatz for a random parametrized quantum circuit can be described as \cite{McClean2018}
\begin{equation} \label{eq:layered_ansatz}
    U(\boldsymbol{\theta}) = \prod_{l = 1}^L U_l(\theta_l) \mathcal{W}_l,
\end{equation}
where $U(\theta)_l = e^{-i\theta_l \hat{V}_l}$, with $\hat{V}_l$ a hermitian operator, and $\mathcal{W}_l$ a generic non-parametrized unitary. The cost function is as described in Eq.~(\ref{eq:vqe_cost}), taking $\ket{\psi(\boldsymbol{\theta})} = U(\boldsymbol{\theta}) \ket{0}$. The gradient of this cost function with respect to any given parameter $\theta_i$ can be conveniently computed as
\begin{equation} \label{eq:layered_ansatz_gradient}
    \partial_{\theta_i} E = i \bra{0} U^{\dagger}_{1 \rightarrow (i - 1)} \left[\hat{V}_i, U^{\dagger}_{i \rightarrow L} \hat{H} U_{i \rightarrow L}  \right] U_{1 \rightarrow (i - 1)} \ket{0}.
\end{equation}
where $1 \rightarrow (i - 1)$ represent the ansatz layers from layer index $1$ to layer index $(i - 1)$, and $i \rightarrow L$ represent the ansatz layers from layer index $i$ to layer index $L$. From the computation of the gradient, McClean et al. \cite{McClean2018} show that if both $U_{1 \rightarrow (i - 1)}$ and $U_{i \rightarrow L}$ are 2-designs, the variance of the gradient is clearly vanishing exponentially in the system size:
\begin{equation}
    \mathrm{Var}[\partial_{\theta_i} E] \approx \frac{1}{2^{(3N - 1)}} Tr\left[\hat{H}^2\right] Tr\left[\rho^2\right] Tr\left[\hat{V}^2\right] 
\end{equation}
Cases in which either of $U_{1 \rightarrow (i - 1)}$ or $U_{i \rightarrow L}$ is not a 2-design are also addressed in Ref.~\cite{McClean2018}, with similar outcomes (we direct readers to this reference for a full demonstration, as well as detailing of the rules needed to compute the expected value of a variance over an ansatz). Further analysis conducted by Napp in Ref. \cite{Napp2022} shows additional analytical bounds for unstructured variational ans{\"{a}}tze, moving away from the layered ansatz described above. 

\paragraph{Expressibility \cite{Holmes2021}:} Holmes et al. show that trainability and expressibility of the ansatz are inversely related. In other words, the more expressive an ansatz is, the more prone it is to barren plateaus. This does not mean that low-expressivity ans\"atze are not affected by barren plateaus, as other drivers can otherwise trigger the problem (for instance, system size and random initialization, as above, or a very non-local cost function \cite{Cerezo2021_BP, Uvarov2020, Sharma2020}). This observation implies that one cannot lower-bound gradients as a function of expressibility, but it can be upper-bounded.

This is shown in Ref.~\cite{Holmes2021} by extending the expression of the barren plateau problem as explained in Ref. \cite{McClean2018}, and setting an upper bound for the variance of the cost gradient as a function of the ansatz' distance to a 2-design. As such, they use the same layered ansatz template (Eq. \ref{eq:layered_ansatz}) and resulting gradients (Eq. \ref{eq:layered_ansatz_gradient}). As an illustration, Ref.~\cite{Holmes2021} find a generalized bound for gradient variance as a function of expressibility (we encourage the reader to read the finer details directly from the source material), as
\begin{equation}
    \mathrm{Var}[\partial_{\theta_i}, E] \leqslant \frac{g (\rho, \hat{P}, U)}{2^{2N}- 1} + f(\varepsilon^{\hat{P}}_L, \varepsilon^{\rho}_R),
\end{equation}
where the first half of the bound corresponds to the maximally expressive ansatz (and $g(.)$ is a function defining a pre-factor on this expressibility, defined in detail in Appendix E of Ref.~\cite{Holmes2021}). The function $f(.)$ is the extra bound resulting from the expressibility (or lack thereof) of the ansatz, defined as
\begin{equation}
    f(\varepsilon_x, \varepsilon_y) := 4\varepsilon_x \varepsilon_y + \frac{2^{N + 2} ( \varepsilon_x ||\hat{P}||^2_2 + \varepsilon_y||\rho||^2_2) }{2^{2N} - 1},
\end{equation}
and in which $\varepsilon^{\hat{P}}_L$ (the expressibility metric for the part of the ansatz to the left of the parameter $i$, where the gradient is taken with respect to the measurement operator), and  $\varepsilon^{\rho}_R$ (the expressibility metric for the part of the ansatz to the right of the parameter $i$, where the gradient is taken with respect to the state density matrix), have been used as arguments. From this equation, one can see that the gradient variance admits an upper-bound approaching $\mathcal{O}(\varepsilon^{\hat{P}}_L \varepsilon^{\rho}_R)$ as $N \rightarrow \infty$. As a result of the definitions in Eqns.~(\ref{eq:expressibility_integral}-\ref{eq:expressibility_operator}), it shows that high expressibility (low $\varepsilon$) lowers the gradient variance bound and therefore limits the trainability of the ansatz. For further information on expressibility of ans\"atze, Nakaji et al. \cite{Nakaji2021} provide a study of the expressibility of the shallow alternating layer ansatz. 

\paragraph{Cost function non-locality: \cite{Cerezo2021_BP, Uvarov2020, Sharma2020}:} In Ref.~\cite{Cerezo2021_BP}, Cerezo et al. show that an ansatz trained on local cost functions are more resilient to the barren plateau problem than those trained on global cost functions. They illustrate this point by comparing a cost function constructed around the expectation value of a global observable: $\hat{O}_G = \unit - \ket{0}\bra{0}^{\otimes N}$ to a cost function constructed around the expectation value of a local observable $\hat{O}_L = \unit - \frac{1}{N} \sum_i^N \ket{0}\bra{0}_i \otimes \unit_{\neq i}$, the latter being local as each component of the observable only applies to a single qubit. It is shown in particular that while alternative layered ans{\"{a}}tze trained on the global cost function are never trainable (this is not necessarily true on other types of  ans{\"{a}}tze), ans{\"{a}}tze trained on the local cost function are trainable if their depth scales logarithmically with the circuit width (i.e. $\mathcal{O}(\log(N))$ or below). Cerezo et al. \cite{Cerezo2021_BP} also show that ans{\"{a}}tze with a scaling  $\mathcal{O}(\mathrm{poly}(\log(N)))$ could also be either trainable or not. Ref. \cite{Uvarov2020} extend these findings to a wider range of cost functions, and Ref.~\cite{Sharma2020} demonstrated the occurrence of this phenomenon in the case of Dissipative Perceptron-Based Quantum Neural Networks. 

An important consequence for the VQE, as pointed out in \cite{Cerezo2021_BP} is that local encoding such as Bravyi-Kitaev \cite{Bravyi2002, Seeley2012, Tranter2015} (Sec. \ref{sec:bravyi-kitaev}), ternary tree encoding \cite{Jiang2020} (Sec. \ref{sec:ternary_tree_encoding}) or Generalized Superfast Encoding \cite{Setia2019} (Sec. \ref{sec:superfast_encoding}) with lower Pauli weight, would offer more resilience than encodings such as Jordan-Wigner \cite{Jordan1928} (Sec. \ref{sec:jordanwigner}) which has Pauli weights scaling of $\mathcal{O}(N))$, therefore resulting in a very non-local VQE cost function. It is also worth noting that the best known scaling for a VQE ansatz is linear (e.g. k-UpCCGSD \cite{Lee2019}, Sec. \ref{sec:UCCA}, Fourier Transform-HVA \cite{Babbush2018} on some systems, Sec. \ref{sec:hva}). Uvarov \textit{et al.} \cite{Uvarov2020_frustrated} compares numerically the impact of using Jordan-Wigner compared to Bravyi-Kitaev on a Hubbard-like model and find that in this case the latter results in gradient variance nearly one order of magnitude larger than the former. This should be caveated by the fact that the numerical results in Ref.~\cite{Uvarov2020_frustrated} also show that the number of layers used in the ansatz (in this case, a symmetry preserving ansatz is used, similar to the one presented in Ref.~ \cite{Barkoutsos2018}) ultimately dominates and reduces gradient variance to a negligible number in either case.

\paragraph{Noise induced barren plateau (NIBP):} Wang \textit{et al.} \cite{Wang2020} show that incorporating quantum noise in a variational optimization can accelerate the occurrence of barren plateaus, and additionally result in vanishing of the amplitude of the expectation value. Their main result is a bound on the value of a parameters' gradient (with notation slightly changed from Ref.~\cite{Wang2020}), as
\begin{equation}
    \left| \frac{\partial \langle \hat{H}(\boldsymbol{\theta}) \rangle}{\partial \theta_i}  \right| \leq G(N)q^{L+1},
\end{equation}
where $q < 1$ is a parameter representing the strength of the noise model (the lower it is, the more noise there is), $G(N) \sim \mathcal{O}(2^{-\alpha N})$ with $\alpha$ an arbitrary, positive constant, and $L$ represents the number of layers in the ansatz. A few important points can be raised as a consequence of this bound \cite{Wang2020}. First, the noise-induced barren plateau is independent of parameter initialization, or locality of the cost function, meaning that some of the strategies listed in Sec. \ref{sec:bp_mitigation} will not work in a noisy setting. Second, this bound is conceptually different from the previously described drivers of barren plateaus as it is a bound on the gradient, rather than on the variance of the gradient. Rather than the flattening optimization landscape described previously, NIBP results in the vanishing of the amplitude of the expectation value function and a bias away from the minimum. 

\paragraph{Large degrees of entanglement \cite{OrtizMarrero2020, Patti2021}:} The degree of entanglement of the trial wavefunction has also been shown to be associated with the barren plateau problem. In particular, Patti \textit{et al.} show that one can link vanishing gradients to the entanglement entropy \cite{nielsenQuantumComputationQuantum2010} of the trial state wavefunction even at low circuit depth. Ortiz Marrero \textit{et al.} reach a similar conclusion by first showing that entanglement between visible and hidden units in a Quantum Neural Network reduces trainability. The result is then extended to unitary networks (very much similar to UCC based ans\"atze) and quantum Boltzmann machines. 
A final point to note is that it was also shown that higher-order derivatives of the cost function are also affected by the barren plateau problem, and therefore cannot be used as a means to circumvent it \cite{Cerezo2021_BP_higher_order}.

\subsubsection{Methods to address barren plateau problem} \label{sec:bp_mitigation}

It follows from Refs.~\cite{McClean2018} and \cite{Holmes2021} that addressing the barren plateau problem can be done through modification of the ansatz. In particular, techniques focusing on selectively reducing the expressibility of the ansatz, or in other words, avoiding a 2-design (which would be the maximally expressive unitary on a given Hilbert space) are expected to be more resilient to barren plateaus. In the context of the VQE, this can be done by restricting the span of the ansatz to a section of the Hilbert space of interest. In particular, it was shown that adaptive ans{\"{a}}tze (Sec. \ref{sec:adaptive_ansatz}) exhibit some resilience to the barren plateau problems. In addition, there are optimization methods which aim at tempering this problem (for example Ref.~\cite{haug_optimal_2021}) that are considered further in Sec.~\ref{sec:Optimization}. Another means available to contain the barren plateau problem is to select a local encoding (see Sec. \ref{sec:Encoding}) with low Pauli weights, as discussed above and explained in Ref.~\cite{Cerezo2021_BP}.

Some methods have also been developed specifically to address barren plateaus. A first example consists of initializing ansatz parameters such that subsections of the ansatz (as split when computing the gradient) do not form a 2-design, at the very least avoiding to start the optimization process in a barren plateau \cite{Grant2019}. Starting from the layered ansatz in Eq.~(\ref{eq:layered_ansatz}), we can divide the ansatz into $K$ blocks of depth $D$ (such that the total depth $L = KD$). The depth $D$ of each block considered in isolation needs to be shallow enough to ensure that the block does not approach a 2-design. Each block $U_k(\boldsymbol{\theta}_k)$, parametrized by a vector $\boldsymbol{\theta}_k$ can then be split into two parts of equal depth, such that
\begin{equation}
    U_k(\boldsymbol{\theta_k}) = \prod_{d=1}^{D/2 - 1} U_d(\theta^k_{d, 1}) W_d \prod_{d=D/2}^{D} U_d(\theta^k_{d, 2}) W_d,
\end{equation}
where $\theta^k_{d, 1}$ can be initialized at random, but where $\theta^k_{d, 2}$ are initialized such that $U_d(\theta^k_{d, 2}) W_d  = (U_d(\theta^k_{d, 1}) W_d)^{\dagger}$. The result is that, for all $k$, before any optimization, we have
\begin{equation}
    U_k(\boldsymbol{\theta}_k) = I_k, \quad  \mathrm{and} \quad U(\boldsymbol{\theta}_{init}) = I.
\end{equation}

Grant et al. \cite{Grant2019} show that this parameter initialization strategy could slow down the optimization process of the VQE as the initial state produced by the circuit would have no entanglement. They propose to initialize the qubit register with a random entangled state, using a shallow random unitary which remains constant throughout the optimization process. While showcasing promising results on small systems, the method is however quite challenging to implement in practice. Identifying a block initialization is not directly possible for all ansatz structures: for example it is not straightforward for unitary coupled cluster (see Sec. \ref{sec:UCCA}) based ansatz without repeating some operators, and it is in general not possible exactly with hardware efficient ans{\"{a}}tze. 

Nonetheless, the idea of adjusting parameter initialization to improve resilience to the barren plateau problem has been extended to alternative mitigation techniques. In Ref.~\cite{Sauvage2021}, Sauvage \textit{et al.} propose to select optimal initial parameters with the help of a machine learning model (FLexible Initializer for arbitrarily-sized Parametrized quantum circuits, or FLIP). The model is trained to identify structures of parameters that best suit specific families of quantum circuit optimization problems, and is numerically shown to provide significant improvements. Similarly, Kulshrestha and Safro show that initializing ansatz parameters by picking them from a beta distribution reduces the impact of the barren plateau problem compared to picking them from a uniform distribution \cite{Kulshrestha2022}. They also show that adding perturbations to the parameters between each optimization step also helps in mitigating vanishing gradients.

Several additional methods have been developed for the barren plateau in the general case of parametrized quantum circuits. In Ref.~\cite{Volkoff2021}, Volkoff et al. show that one can reduce the dimensionality of the parameter space by using spatially and temporally correlated parameterized quantum gates, resulting in higher resilience to barren plateaus. The ansatz can also be trained layer by layer to the same effect \cite{Skolik2021}, though limitations of this method were shown in Ref.~\cite{Campos2021}. This is somewhat akin to adaptive ans{\"{a}}tze (see Sec. \ref{sec:adaptive_ansatz}), but generalized to any quantum neural network optimization. Patti \textit{et al.} \cite{Patti2021} also propose several additional mitigating methods including an alternative initialization strategy in which two qubit registers are initially not entangled, regularization on the entanglement, the addition of Langevin noise, or rotation into preferential cost function eigenbases. Sack \textit{et al.} \cite{Sack2022} showed that barren plateaus can be partially mitigated as part of a classical shadow measurement scheme. Wu \textit{et al.} \cite{Wu2021_mit} propose to mitigate the impact of NIBP by defining an alternative cost function with the same optimal state but without sensitivity to vanishing gradients, by identifying and eliminating the dominant term in the Pauli representation of the observable measured.
Finally, though not directly relevant to VQE, Pesah et al. \cite{Pesah2020} show numerically that Quantum Convolutional Neural Networks exhibit natural resilience to the barren plateau problem. Similarly, Zhang \textit{et al.} \cite{Zhang2020} show that Quantum Neural Networks with tree tensor structure and step-controlled architectures have gradients that vanish at most polynomially in the system size. 

\subsubsection{Comments on barren plateau in the context of the VQE} \label{sec:bp_for_vqe}

In Ref.~\cite{Holmes2021}, Holmes et al. point out that the problem structure of VQE can be used to limit the impact of barren plateaus. The features in question include symmetries of the problem, which can be used to reduce the portion of the Hilbert space, or physically-motivated ans\"atze targeting a restricted part of the Hilbert space in which good approximations of the ground state are expected to be. Such relevant structures are in general more difficult to find when considering the wider field of quantum neural networks. 
Adopting an initialization strategy such as the ones presented in Refs.~\cite{Grant2019, Sauvage2021, Patti2021} would allow an ansatz to begin optimization away from a barren plateau, and as such, away the target operator mean. One could argue that given the optimization problem aims at finding a minimum, a reliable optimizer should always move a state away from the mean expectation value and therefore away from the barren plateaus regions. This point however could be invalidated by several aspects of the optimization process. These include the existence of local minima, the increase in entanglement of the trial wavefunction as the optimization progresses, or the presence of noise which results in vanishing of the value function amplitude. 

Focusing on local encodings (see Sec.\ref{sec:Encoding}) has been shown to provide some resilience to barren plateaus \cite{Cerezo2021_BP, Uvarov2020, Uvarov2020_frustrated}, suggesting that using VQE on lattice models with limited dimensions could be performed relatively better than on a molecular Hamiltonian in that respect. As pointed out in Ref.~\cite{Cerezo2021_BP}, ans\"tze scaling logarithmically in the system size, and measured on local observable are resilient to barren plateaus. At this stage, however, there is no known VQE ansatz scaling as such that also guarantees an accurate description of the ground state. 

In the remainder of this section, we discuss options for ans{\"{a}}tze in the context of VQE. We look at two different types of ansatz: fixed structure ansatz (Sec. \ref{sec:fixed_struct}) which are set at the beginning of the optimization process and remain unchanged afterward, and adaptive structure ansatz (Sec. \ref{sec:adaptive_ansatz}) which are constructed iteratively as part of the optimization process. We can present three key metrics to compare ans\"tze: 
\begin{itemize}
    \item The depth of the ansatz (number of sequential operations required for the implementation), which impacts the overall runtime of the method, its resilience to noise and to barren plateaus.
    \item The number of parameters, which significantly influence the overall runtime of the implementation (although this cost can in theory, and in most cases, be entirely parallelized, see Sec. \ref{sec:parallelization}), and the complexity of the optimization process.
    \item The number of entangling gates, which is, in general, the main source of noise resulting from execution of a quantum circuit.
\end{itemize}
Of course, while these metrics can be objectively defined and provide a useful tool to compare different ans\"atze, they are only relevant if the ans\"atze studied can indeed provide an accurate description of the ground state wavefunction. This latter point is far more difficult to assess, beyond the formal considerations at the beginning of this chapter.

\subsection{Fixed structure ans{\"{a}}tze} \label{sec:fixed_struct}

Fixed structure ans{\"{a}}tze are initialized at the beginning of the VQE process and remain unchanged (aside from the value of their parameters) throughout the optimization. Table  \ref{tab:ansatz_scaling} provides a comparative summary of these ans{\"{a}}tze. Some fixed structure ans{\"{a}}tze (such as the Qubit Coupled Cluster \cite{Ryabinkin2018}, and Qubit Coupled Cluster Singles and Doubles \cite{Xia2020}) are only discussed in the adaptive ans{\"{a}}tze section as their adaptive version tend to be more relevant for discussion.

\begin{table} [ht]
\caption{Summary of circuit depth, parameters and entangling gates scaling across most fixed structure ans{\"{a}}tze reviewed. The scaling in the number of entangling gates assumes full connectivity of the qubit lattice.}
\label{tab:ansatz_scaling}
\begin{tabularx}{\linewidth}{p{0.15\linewidth}|p{0.14\linewidth}p{0.14\linewidth}p{0.14\linewidth}X}
\toprule
\\  Method & Depth & Parameters & Entangling gates & Comments \\
\midrule
    Hardware Efficient Ansatz (HEA) \cite{Kandala2017} & $\mathcal{O}\left(L \right)$ & $\mathcal{O}\left(NL \right)$ & $\mathcal{O}\left((N - 1)L \right)$ & $L$ is an arbitrary number of layers, its scaling for exact ground state is unknown, exponential in the worst case (i.e. if the entire Hilbert space needs to be spanned to find the ground state) \\
\hline
    UCCSD \cite{Peruzzo2014, Romero2019}& $\mathcal{O}\left((N - m)^2m\tau \right)$ &$\mathcal{O}\left((N-m)^2m^2 \tau \right)$& $\mathcal{O}\left(2(\tilde{q} -1)N^4 \tau\right)$ & $\tilde{q}$ is the average Pauli weight across the operators used to build the ansatz. As an indication, maximum Pauli weight under Jordan-Wigner is $N$, and $log(N)$ under Bravyi-Kitaev. $\tau$ is the number of Trotter steps used\\
\hline
    UCCGSD \cite{Peruzzo2014, Wecker2015, Lee2019}& $\mathcal{O}\left(N^3 \tau \right)$ & $\mathcal{O}\left(N^4 \tau \right)$& $\mathcal{O}\left(2(\tilde{q} -1)N^4 \tau\right)$ & As above\\
\hline
    k-UpCCGSD \cite{Lee2019}&$\mathcal{O}\left(kN \tau \right)$& $\mathcal{O}\left(k\tau N^2/4 \right)$& $\mathcal{O}\left(k\tau(\tilde{q} -1) N^2/2 \right)$& $k$ is an arbitrary constant which determines the accuracy of the result. Scaling is unknown. Rest is as above. \\
\hline
    OO-UCCD \cite{Mizukami2020} & $\mathcal{O}\left((N - m)^2m\tau \right)$ &$\mathcal{O}\left((N-m)^2m^2 \tau \right)$& $\mathcal{O}\left(2(\tilde{q} -1)N^4 \tau\right)$ & Same as UCCSD. It is worth noting that OO-UCCD is a nested loop between orbital (one-body terms) optimization, done on a conventional machine, and two-body terms optimization done on the quantum computer.
\\
\hline
    Symmetry Preserving \cite{Barkoutsos2018} & $\mathcal{O}\left( (N-1)L \right)$&$\mathcal{O}\left(2(N-1)L \right)$&$\mathcal{O}\left(3(N-1)L \right)$& This ansatz spans a wider range of the Hilbert space than the EPS. It is therefore likely it requires more circuit resources and as such we suspect that $L$ grows exponentially in $N$ for exact resolution of the ground state. This logic also applies to HEA. \\
\hline
     Efficient Symmetry Preserving (EPS) ansatz \cite{Gard2020} & $\mathcal{O}\left(2\binom{N-1}{m} \right)$ & $\mathcal{O}\left(2\binom{N}{m} - 2 \right)$ & $\mathcal{O}\left(3\binom{N}{m} \right)$& Scaling range from linear when $m=1$, or $m=N-1$, to exponential if $m\sim N/2$\\
\hline
    Hamiltonian Variational Ansatz \cite{Wecker2015, Wiersema2020} & $\mathcal{O}\left(\tilde{C} L \right)$ & $\mathcal{O}\left(\tilde{C} L \right)$& $\mathcal{O}\left(2(\tilde{q} -1)C L\right)$ & $L$ represents the number of repetition of the ansatz required to achieve the desired accuracy. $C$ is the number of terms in the Hamiltonian, and $\tilde{C}$ the number of commutative groups among these terms.
\\
\bottomrule
\end{tabularx}
\end{table}

\subsubsection{Hardware-efficient ansatz (HEA)} \label{sec:HEA}

\begin{figure}[ht]
\centerline{
\Qcircuit @C=1em @!R {
   \lstick{\ket{q_1}}   &   \qw   &   \gate{R_x(\theta_i)}   &    \gate{R_y(\theta_j)}    &  \ctrl{1} &  \qw  & \qw                    \\
   \lstick{\ket{q_2}}   &   \qw   &   \gate{R_x(\theta_p)}   &    \gate{R_y(\theta_q)}    &  \targ & \ctrl{1} & \qw                     \\
   \lstick{\ket{q_3}}   &   \qw   &   \gate{R_x(\theta_m)}   &    \gate{R_y(\theta_n)}    &  \qw & \targ & \qw                     \\
}
} \caption{Example of one block of a HEA, using $R_x$ and $R_y$ as rotation gates and $CNOT$ as entangling gates, on three qubits} \label{fig:HEA}
\end{figure}
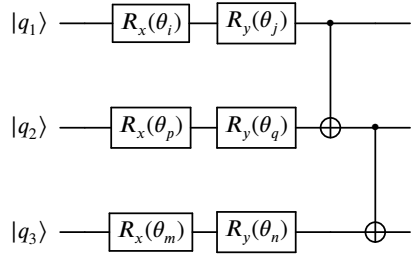

The initial motivation for the Hardware-Efficient ansatz (HEA) was for the trial state of VQE to be parametrized by quantum gates directly tailored to the quantum device on which the experiment is run \cite{Kandala2017}. Several flavors of the HEA have been proposed, however, they all follow the same logic: the ansatz is constructed by repeating blocks of interweaved single qubit, parametrized, rotation gates, and ladders of entangling gates (see Fig.~\ref{fig:HEA} for an example). A generic representation of the HEA can be presented as follows:

\begin{equation} \label{eq:HEA_unitaries}
\ket{\psi (\theta)} = \left[ \prod_{i=1}^d U_{rota}(\theta_i) \times U_{ent} \right] \times U_{rota}(\theta_{d + 1}) \ket{\psi_{init}},
\end{equation}
where $U_{rota}(\theta_i) = \prod_{q, p}^{N, P} R_p^q(\theta_i^{pq})$, with $q$ the qubit addresses and $p \in \{X, Y, Z\}$, the selected set of parametrized Pauli rotations.

Both the set of rotation gates and entangling gates can vary depending on the native gate set of the device, and the complexity of the target state sought \cite{Mitarai2019}. The ansatz' entangling section was originally presented as a quantum analogue task \cite{Kandala2017}. The main promise of HEA is that it can be flexibly tailored to the specific native gate set of the device used, while at the same time being highly expressive. Its versatility and ease of construction resulted in it being widely used for numerous small-scale quantum experiments~\cite{Kandala2017, Mitarai2019, higgottVariationalQuantumComputation2019, Tilly2020, Tilly2021, Chen2021}.

There are however many known limitations of HEA. A first obvious issue is the fact that it must span a very large portion of the Hilbert space to guarantee that an accurate enough representation of the ground state wavefunction can be produced. HEA can therefore be quite inefficient with respect to its required Hilbert space coverage, requiring in worst cases an exponential depth (though further research would be required to verify how the accuracy of ground state representations would respond to arbitrary depth reduction and shrinkage of the Hilbert space coverage). An immediate consequence is that HEA is significantly limited by barren plateaus \cite{Holmes2021} (see Sec. \ref{sec:barren_plateau}, \cite{McClean2018}). It also implies that a large number of parameters need to be optimized and managed (for a numerical example of the number of parameters that HEA requires compared to other ans{\"{a}}tze, one can refer to Table II in Ref.~\cite{Choquette2021}, where $198$ parameters are required for a $12$ qubit version of $\mathrm{H_2O}$, against $8$ for UCCSD detailed below). Regarding this last point, Wu \textit{et al.} have shown that certain architectures of HEA provide higher expressibility and can be overall more efficient in terms of quantum resources \cite{Wu2021_hea}.
With some caveats, Bravo-Prieto \textit{et al.} \cite{BravoPrieto2020} demonstrate that HEA type structures can provide reasonable accuracies on spin chain models such as Ising or XXZ chains using a number of layers at least linear in the system size. 
Overall, the HEA is a convenient tool for a proof of principle for a given algorithm or optimization strategy. It is, however, unlikely to be suitable for general larger-scale chemical problems. One therefore would rather turn toward more problem-tailored ans{\"{a}}tze.

\subsubsection{The Unitary Coupled Cluster (UCC) Ansatz and extensions}  \label{sec:UCCA}

The Unitary Coupled Cluster (UCC) ansatz is arguably the most studied ansatz for VQE. It featured in the initial VQE work of Peruzzo et al. \cite{Peruzzo2014} and has occupied an important position in the literature since then. As such, we have split this section into three parts: a presentation of the general UCC framework, a description of the extensions that have been proposed, and a discussion on the benefits and issues of the ansatz. An extensive review dedicated to this family of ans{\"{a}}tze was recently published by \cite{Anand2021_review}, we direct reader in search for further information towards it.\\

It is worth noting that while initially designed in the context of electronic structure computation, variants of UCC have also been applied in other contexts. For example, it is been applied to variational quantum computation of ground and excited states for periodic systems \cite{Yoshioka2022, Manrique2020, Liu2020}, and for vibrational spectroscopy (with the Unitary Vibrational Coupled Cluster ansatz proposed in Ref. \cite{McArdle2019_vibra}, and applied in Ref. \cite{Ollitrault2020_vibrational}).

\textbf{General framework} \\

The UCC theory \cite{Bartlett1989, Taube2006, Filip2020} stems from adapting the Coupled Cluster (CC) theory. CC is a post Hartree-Fock method that aims at recovering a portion of electron correlation energy by evolving an initial wave function (usually the Hartree-Fock wave function) under the action of parametrized excitation operators \cite{Coester1960, Cizek1966, Paldus1972, Paldus1977, Cizek1980, Bartlett2007}.

In general, these are single-electron excitations and double electron excitations, resulting in CC Single and Double (CCSD),  however, these can be extended systematically to higher-order (e.g. CCSDT). Only excitation operators allowing transitions from an occupied orbital to an unoccupied orbital are traditionally included in CC (note that we use $i$, $j$, $k$, and $l$ for occupied orbital indices, $a$, $b$, $c$, and $d$ for virtual orbital indices and $p$, $q$, $r$, and $s$ for either).
The action of these operators on the initial state is performed through exponentiation of part of the cluster operator $T$. For $\nu$ denoting the maximum allowed excitation, we have:
\begin{equation}
    \hat{T} = \hat{T}_1 + \hat{T}_2 + ... \hat{T}_{\nu},
\end{equation}
with for example the single and double excitation operators:
\begin{align}
\hat{T}_1 &= \sum_{ia} t_{i}^a\hat{a}^{\dagger}_a \hat{a}_i \label{eq:T_one_body} \\
\hat{T}_2 &= \sum_{ijab} t_{ij}^{ab}\hat{a}^{\dagger}_a \hat{a}^{\dagger}_b \hat{a}_j \hat{a}_i. \label{eq:T_two_body}
\end{align}
Using the Hartree-Fock state as reference state, the CC ansatz wave function is given by
\begin{equation}
\ket{\psi} = e^{\hat{T}} \ket{\psi_{HF}}.
\end{equation}

The conventional resolution method for CCSD scales $\mathcal{O}(m^2(N - m)^4)$, with $m$ the number of electrons, and $N$ the number of spin orbitals. CCSD is however in general not variational and has been reported to fail in numerous cases, in particular in systems with strong correlation, with possible solutions to avoid these failures usually scaling exponentially in the system size (for further reading on CC theory, failures and how they can be addressed, we direct the reader toward Refs.~\cite{Paldus1984, Piecuch1995, Small2012, Small2014, Lee2017, Lee2018}). Another issue in the context of the VQE is that the operator $e^{\hat{T}}$ is not unitary, and therefore the CC ansatz cannot be implemented as a series of quantum gates.

The UCC method was developed as a way to address these limitations \cite{Bartlett1989, Taube2006, Filip2020}. It is based on the fact that for any linear operator $\hat{T}$, the expression $(\hat{T} -\hat{T}^{\dagger})$ is an anti-Hermitian operator. The exponential of an anti-Hermitian operator is a unitary operator, and the difference between the cluster operator and its complex conjugate can be used as an evolution operator to form a unitary version of CC. Elements of the truncated cluster operator for UCC are identical to that of CC, as presented in Eq.~(\ref{eq:T_one_body}) and Eq.~(\ref{eq:T_two_body}).
\begin{equation}
    \ket{\psi} = e^{\hat{T} - \hat{T}^\dagger} \ket{\psi_{HF}}.
\end{equation}
The energy can then be evaluated using a variational approach based on the Ritz functional (see Sec.~\ref{sec:overview}). Exact evaluation of this functional for UCC on classical resources is known to scale exponentially in the system size \cite{Taube2006}. It is therefore natural to bring this ansatz to quantum computation.

By grouping the individual excitation operators in $\hat{T}$ with their corresponding conjugate in $\hat{T}^{\dagger}$, and denoting them as $\tau$, we can obtain the parametrized version of the UCC ansatz:
\begin{equation} \label{eq:ucca}
U(\vec{t}) = e^{\sum_j t_j(\tau_j - \tau^\dagger _j)},
\end{equation}
where $t_j$ correspond to the amplitude weights presented in Eq.~(\ref{eq:T_one_body}) and Eq.~(\ref{eq:T_two_body}), and with $j$ spanning all the excitation operators included. Akin to our description of CC, UCC can accept several level of excitations (UCCSD, UCCSDT).

Using any of the mappings described in Sec. \ref{sec:Encoding} we can re-write each of the fermionic operators as

\begin{equation} \label{eq:ucc_operators}
\tau_j - \tau^\dagger _j = i \sum_k \hat{P}_{k, j},
\end{equation}
where the subterm $\hat{P}_{k, j}$ is a tensor product of Pauli operators (technically, this term can actually be a sum of mutually commuting Pauli operators, as commuting groups do not need to be Trotterized). Resulting in the following expression of the ansatz:

\begin{equation} \label{eq:ucca_pauli}
U(\vec{t}) = e^{\sum_j \sum_k i t_j  \hat{P}_{k, j}},
\end{equation}

The next step is to convert the UCC ansatz into a series of parametrized quantum gates which can directly be implemented on a quantum computer. A usual step in the process of building the UCC ansatz is to use a Trotter-Suzuki decomposition (a process also referred to as Trotterization) to separate the sum in the exponent of Eq.~(\ref{eq:ucca_pauli}) into a product of exponentials, such that
\begin{equation}
U(\vec{t}) \sim  U_{\text{Trotter}}(\vec{t}) = \left( \prod_{j, k} e^{\frac{i t_j}{\rho}\hat{P}_{k, j}} \right)^\rho.
\end{equation}
The Trotter number $\rho$ defines the precision of the approximation, but also enters as a pre-factor impacting the overall depth of the quantum circuit required for its implementation (technically, $\hat{P}_{k, j}$ can actually be a sum of mutually commuting Pauli operators, as the exponential of a sum of commuting groups does not need to be Trotterized and can be directly converted to a product of exponentials). Assuming $\rho=1$, this reduces Eq.~(\ref{eq:ucca}) to
\begin{equation} \label{eq:pauli_exponential}
U_1(\vec{t}) = \prod_{j, k} e^{\frac{i t_j}{\rho}\hat{P}_{k, j}} .
\end{equation}

This illustrates the basic framework surrounding UCC and its application to VQE \cite{Peruzzo2014, Wecker2015, Romero2019}. The required circuit depth for this version of UCC has been shown to scale polynomially in the system size. More specifically, with $\eta$ the number of electrons and $N$ the number of spin-orbitals, the circuit depth is shown to scale $\mathcal{O}((N - m)^2m)$ for each Trotter step \cite{Lee2019}. It was also numerically shown that a single Trotter step is sufficient for an accurate description of the ground state in simple molecular systems \cite{Barkoutsos2018}, because variational optimization can absorb some of the Trotterization error \cite{OMalley2016, Romero2019} (though it does imply that operator ordering can impact accuracy \cite{Grimsley2019_UCC_Review}, see discussion below for more details). 
The UCC ansatz can also be used in the context of restricted active space methods \cite{Roos1980}, considered in Ref.~\cite{Romero2019}. Finally, it has been shown that operators in the UCC ansatz contribute to the expressibility of the ansatz to different extents, allowing for approaches where some are discarded to increase efficiency \cite{Zhang2021_shallow} (a method similar to what is proposed in ADAPT-VQE \cite{Grimsley2019}, see Sec \ref{sec:adaptive_ansatz}). The remainder of the section covers several extensions which aim at improving both the precision and efficiency of UCC based ansatz.\\

\textbf{Extensions of UCC} \\

\paragraph{Generalized UCC:} The first extension of the UCC theory applied in the context of the VQE is the Generalized UCC (UCCG) \cite{Lee2019}. Unlike conventional CC and UCC which only include excitation operators that correspond to transitions from occupied to virtual orbitals, UCCG is agnostic to the orbital character in the (generally Hartree-Fock) reference state it is applied to. Therefore the single and double excitations in Eq.~(\ref{eq:T_one_body}) and Eq.~(\ref{eq:T_two_body}) are re-written in the context of the UCCGSD \cite{Lee2019} as

\begin{align}
\hat{T}_1 &= \sum_{pq} t_{p}^q\hat{a}^{\dagger}_q \hat{a}_p \label{eq:T_one_body_G} \\
\hat{T}_2 &= \sum_{pqrs} t_{pq}^{rs}\hat{a}^{\dagger}_r \hat{a}^{\dagger}_s \hat{a}_q \hat{a}_p, \label{eq:T_two_body_G}
\end{align}
where we recall that $p$, $q$, $r$, and $s$ are used as indices spanning both occupied and virtual orbitals. Generalized CC, was actively discussed in the conventional computing literature, though a unitary version was only mentioned in \cite{Nooijen2000} and never extensively studied. Lee et al. \cite{Lee2019} provide an in-depth study of the method and showed that the associated UCCGSD ansatz depth scales $\mathcal{O}(N^3)$ in the number of qubits $N$ while at the time providing example showing that this method is more robust to noise and accurate than UCCSD.

\paragraph{Paired UCC (k-UpCCGSD):} A second extension of UCC is the Unitary pair CC with Generalized Singles and Doubles (UpCCGSD) \cite{Lee2019}. This version is based on the paired coupled-cluster method \cite{Stein2014, Limacher2013}. The key difference with CCSD is that it only includes two body terms that move pairs of opposite-spin electrons together from doubly-occupied to fully unoccupied spatial orbitals, ensuring no single occupancy in the described states. Recalling that we use $i$ for an occupied orbital and $a$ for virtual orbitals, the $\hat{T}_2$ terms included in UpCCSD can be described as
\begin{equation} \label{eq:T2_pair_CC}
    \hat{T}_2 = \sum_{ia} t_{i_{\alpha} i_{\beta}}^{a_{\alpha} a_{\beta}}\hat{a}^{\dagger}_{a_{\alpha}} \hat{a}^{\dagger}_{a_{\beta}} \hat{a}_{i_{\beta}} \hat{a}_{i_{\alpha}},
\end{equation}
with $\alpha$ and $\beta$ the spin up and spin down indices. The one body terms remain identical to Eq.~(\ref{eq:T_one_body}). Lee et al. \cite{Lee2019} however showed that using a unitary version of the method failed to achieve chemical accuracy. They therefore adapted it by using generalized one- and two- body excitation (see Eq. \ref{eq:T_one_body_G}) (where the indices are not constrained by whether they are occupied or not) while maintaining the spatial orbital pairing), hence the 'G' in UpCCGSD. They further improved the method by allowing repetitions of the ansatz (the number of repetitions referred to by the $k$ parameter) allowing more flexibility for the wave function. This forms the $k$-UpCCGSD ansatz, which can be described as
\begin{equation} \label{eq:kUpCCGSD}
    \ket{\psi}_{k-UpCCGSD} = \prod_{l=1}^k (e^{\hat{T}^{(l)}} - e^{\hat{T}^{(l)\dagger}}) \ket{\psi_{HF}}.
\end{equation}

The key advantage of this method is to allow a linear scaling ansatz \cite{Lee2019}, namely that the ansatz depth scales as $\mathcal{O}(kN)$ with $N$ the number of spin-orbitals, and a reasonable quadratic scaling in the number of parameters $\mathcal{O}(kN^2)$ \cite{Huggins2020}. In this method, the parameters for each $l$ block of the ansatz are treated independently. A few drawbacks of this method have been raised in Ref. \cite{Grimsley2019_UCC_Review}. In particular, it is shown that while the ansatz error can be arbitrarily reduced by increasing $k$, it also involves a nonlinear increase in the optimization burden due to the energy landscape becoming more complex from adding new parameters. The result found through this ansatz also seems to be dependent on the parameter initialization \cite{Grimsley2019_UCC_Review}.
However, Lee et al. ~\cite{Lee2019} show numerically that k-UpCCGSD finds better energies than UCCSD (also for excited-state calculations, see Sec.~\ref{sec:excited_states}). These results were corroborated in Ref.~\cite{GreeneDiniz2020}, which studies the accuracy of UCC-based ans{\"{a}}tze. 

\paragraph{Pair-natural orbital-UCC (PNO-UpCCGSD):}The idea behind k-UpCCGSD was further extended in Refs.~\cite{Kottmann2021_1, Kottmann2021_2} by restricting excitations to those between pair-natural orbitals (PNO) and the occupied orbitals of the reference state. PNOs are an efficient basis for modeling of the most relevant virtual states for a given occupied orbital pair \cite{Edmiston1968, Ahlrichs1975, Meyer1973, Meyer2009} initially developed for Coupled Electron-Pair Approximations, but subsequently used in CC \cite{Neese2009_1, Neese2009_2}, and found to significantly reduce the size of of the virtual orbital space in CC calculations (for example \cite{Nagy2019}). PNOs are found as the eigenvectors of a pair density matrix, often found from a lower level of theory.

One can define a set of orthonormalized PNOs $\mathcal{S}_{ij}$ from each possible pair $\{i, j\}$ of canonical HF orbitals. The PNO-UpCC(G)SD ansatz is constructed similarly to any UCC ansatz, but only with operators allowing excitations from occupied indices (i.e  $\mathcal{S}_{ii}$) to unoccupied orbitals of the PNO set. It is also possible to extend beyond excitations from a reference orbital to create a generalized version of the ansatz. The single excitation term then allows excitations from any PNO to any PNO. Ref.~\cite{Kottmann2021_1} shows numerically that the PNO method significantly reduces the number of parameters and of CNOT gates compared to k-UpCCGSD (by a factor of $4$ to $8$ in both cases). This method was used to develop an ansatz tractable on conventional computers \cite{Kottmann2021_2}.

\paragraph{OO-UCC:} A third example of an extension of UCC is the Orbital Optimized UCC (OO-UCC) (independently proposed in both Ref.~\cite{Mizukami2020} and Ref.~\cite{Sokolov2020} with minor differences). The logic behind OO-UCC is to optimize the one-body part of the ansatz using conventional computation, as it is a tractable optimization of orbital rotations. In particular, the one-body terms act as orbital rotation on the initial input state (usually the Hartree-Fock wave function), therefore if we start from the UCCSD ansatz and apply and use a single Trotter step approximation ($\rho = 1$) we can write:
\begin{align}
&\ket{\psi}_{UCCSD} = e^{\hat{T}_2+ \hat{T}_1}\ket{\psi_{HF}} \nonumber \\
& \approx e^{\hat{T}_2} e^{\hat{T}_1}\ket{\psi_{HF}}.
\end{align}
With the one body operator only changing the Hartree Fock determinant to another initial wave function determinant, such that $\ket{\psi_{\mathrm{init}}} = e^{\hat{T}_1}\ket{\psi_{HF}} $, we have the OO-UCCD ansatz:
\begin{align}
\ket{\psi}_{OO-UCCD} = e^{\hat{T}_2} \ket{\psi_{\mathrm{init}}}
\end{align}
The one body terms $\hat{T_1}$ are used to optimize single-orbital rotations on a conventional computer, effectively resulting in a change to the orbital definitions on which the VQE is subsequently applied at each step \cite{Mizukami2020, Sokolov2020}. In this way, OO-UCC is bears similarities to multi-configurational self-consistent field (MCSCF) methods, as it requires an iterative optimization of the orbitals on a conventional computer coupled to the VQE on the quantum computer (for MCSCF proposals on a quantum computer see, for instance, Refs.~\cite{Yalouz2021, Tilly2021}). 

Sokolov \textit{et al.} \cite{Sokolov2020} propose two variants of OO-UCC. The first one (OOpUCC) only allows paired excitations in the two-body terms (similar to UpCCGSD \cite{Lee2019}), grounded in the success of the analogous CC method \cite{Bozkaya2011,Bozkaya2013, Stein2014} (and labeled OO-pUCC). The second one adapts the singlet CC method (CCD0) \cite{Bulik2015}, where the two-body excitation operators are split into singlet and triplet terms, and singlet operators are kept in the ansatz. For this variant, Sokolov \textit{et al.} \cite{Sokolov2020} propose to either reduce operators by leveraging index symmetries in singlet terms (OO-UCCD0) or to use the full set of singlet operators (OO-UCCD0-full), akin to the methods proposed in Ref.~\cite{Bulik2015}. Mizukami et al. \cite{Mizukami2020} point out that, while OO-UCC requires more computation than UCC, it trades it off for a simpler quantum circuit, allowing all wave function parameters to be treated fully variationally, and allowing incorporation of correlations outside of the active space with a reduced the number of qubits (as discussed for MCSCF methods in Ref.~\cite{Yalouz2021, Tilly2021}). \\

\paragraph{Downfolded Hamiltonian and Double Unitary Coupled Cluster (DUCC):} Downfolded Hamiltonian methods were initially developed in the context of CC theory and shortly thereafter extended to UCC \cite{Kowalski2018, Bauman2019, Bauman2019_2, Kowalski2020, Kowalski2021}, to reduce the space in which the ansatz is constructed without impacting the quality of the representation (and usually referred to as DUCC formalism). The idea was adapted for implementation as part of VQE using a downfolded Hamiltonian and a Generalized UCC ansatz \cite{Bauman2021} or UCCSD ansatz \cite{Metcalf2020}. Here we report on the construction of the effective Hamiltonian under the DUCC formalism, as any other UCC ansatz one might want to use). To begin with, the clusters usually used in CC theory are split into $\hat{T}_{int}$ and $\hat{T}_{ext}$ which represents excitations within, and outside the target active space respectively \cite{Bauman2019, Bauman2019_2}. Unitary operators are then constructed using the UCC formalism:
\begin{equation}
    \begin{aligned}
        &\hat{\sigma}_{int} = \hat{T}_{int} - \hat{T}^{\dagger}_{int} \\
        &\hat{\sigma}_{ext} = \hat{T}_{ext} - \hat{T}^{\dagger}_{ext}. \\
    \end{aligned}
\end{equation}
The DUCC wavefuncion can be constructed by exponentiation these operators and applying them to a reference state:
\begin{equation}
    \ket{\psi_{DUCC}} = e^{\hat{\sigma}_{ext}}e^{\hat{\sigma}_{int}}\ket{\psi_0}.
\end{equation}
The part of the ansatz external to the  target active space (or the part that one aims to fold into the effective active space Hamiltonian) is then absorbed into the Hamiltonian giving the following operator $\hat{H}_{ext}^{DUCC} = (e^{\hat{\sigma_{ext}}})^{\dagger} \hat{H}e^{\hat{\sigma_{ext}}}$. Because the external operators represent higher energy excitations, they can be approximated without signficant loss in accuracy \cite{Metcalf2020, Bauman2019}. The downfolded Hamiltonian can be constructed in the active space by defining a projector on the reference state $\hat{P}$ and on the orthogonal determinants of the target active space $\hat{Q}_{int}$ such that: $\hat{H}_{ext}^{eff-DUCC} = (P + Q_{int})\hat{H}_{ext}^{DUCC}(P + Q_{int})$. The DUCC wavefunction can then be written as
\begin{equation}
    \ket{\psi_{DUCC}} = Ee^{\hat{\sigma}_{int}}\ket{\psi_0},
\end{equation}
with the remaining internal excitations being implemented using a UCC ansatz and trained with VQE \cite{Metcalf2020, Bauman2021}. 

The method was shown to recover a large proportion of correlation energy on several small molecules while saving significant quantum resources \cite{Metcalf2020}. In Ref.~\cite{Bauman2021}, Bauman et al. also show that using UCCGSD provides a material improvement over using UCCSD in this context, bringing results close to FCI on $\mathrm{H_2O}$, of course at the cost of additional parameters from the Generalized UCC.  

\paragraph{Other UCC-based methods:} Several other extensions of UCC are also worth mentioning briefly. The Unitary Cluster-Jastrow \cite{Matsuzawa2020} ansatz uses Jastrow factors, traditionally used in Quantum Monte Carlo \cite{Foulkes2001} to include the impact of cusps from the Coulomb potential in the wavefunction and adapted for CC theory in \cite{Neuscamman2013}. The method shows similar scaling to k-UpCCGSD and excellent results in numerical analysis \cite{Matsuzawa2020}. It also does not require trotterization and thereby is less prone to errors due to operator ordering \cite{Grimsley2019_UCC_Review}). The multicomponent UCC \cite{Pavosevic2018}  extends the UCC ansatz to systems that include coupling to bosonic statistic modes. It was also applied to photon number states to cover polaritonic chemistry \cite{Pavosevic2021}. Finally, the Low Depth Circuit Ansatz (LDCA) \cite{Dallaire2019} is grounded in the Bogoliubov (or quasi-particle) CC \cite{Stolarczyk2010, Rolik2014, Signoracci2015}, for which a unitary version was developed in Ref.~\cite{Dallaire2019}. While the Bogoliubov UCC can be implemented as an ansatz for VQE, Dallaire et al. \cite{Dallaire2019} propose some additional transformations to exactly parametrize the ansatz in a linear depth circuit (but with quadratic parameter scaling). The parameter-efficient circuit training (PECT) optimizer proposer in Ref.~\cite{Sim2021}, appears to provide an effective method to optimize LDCA on larger systems. \\
\\

\paragraph{Unitary Selective Coupled-Cluster Method:} The USCC method, proposed by Fedorov \textit{et al.} \cite{Fedorov2022} could be categorized both as a UCC extension and as an adaptive ansatz. To implement this method, one starts with the reference Hartree-Fock state. From there, one and two body excitations (from occupied to unoccupied orbitals) which have corresponding Hamiltonian matrix elements above a given threshold $\tau_1$ are included in the ansatz. A VQE is then performed on this ansatz until convergence - the first iteration of the method is complete. For the following iterations, the threshold is update such that: $\tau_i = \tau_{i-1} / 2$, and new, higher order excitations are generated by applying one and two body excitations to those already included in the ansatz. The metric to compare to the threshold for inclusion of these excitations is obtained by multiplying the parameters obtained after optimization at the previous step with the Hamiltonian matrix element corresponding to the single or double excitation added. As per previous iterations, coefficients that are above the threshold are included in the ansatz and then optimized through VQE.
The optimization continues until a convergence criteria is met or sufficient excitations have been included. Fedorov \textit{et al.} \cite{Fedorov2022} note that the method does not provide a wavefunction form that is as compact as can be achieved through ADAPT-VQE, but is likely to require significantly less measurements to perform and converge. It is worth noting that as an adaptive method USCC would likely also present some resilience to the barren plateau problem. 

\textbf{UCC discussion}\\

Overall, it is clear that there are several advantages of using UCC-type ansatz compared to HEA. First and foremost, the number of parameters in the former, ranging from $\mathcal{O}(N^3)$ to $\mathcal{O}(kN^2)$ ($k$ an arbitrary constant) is significantly more advantageous than in the HEA which spans a much wider part of the Hilbert space. It was also shown numerically that the UCC ansatz can provide exact parametrization for an arbitrary electronic wavefunction \cite{Evangelista2019}, subject to certain conditions on optimization. The accuracy of the family of ansatz and its ability to approach FCI level results on a variety of small systems was numerically established in Ref.~\cite{GreeneDiniz2020} (namely $\mathrm{H_2}$, $\mathrm{H_3}$, $\mathrm{H_4}$, $\mathrm{NH}$, $\mathrm{OH^{+}}$, $\mathrm{CH_2}$, $\mathrm{NH_2^+}$). Similarly, Ref.~\cite{Sokolov2020} establishes that UCCSD outperforms conventional CCSD methods in accuracy on $\mathrm{H_4}$, $\mathrm{H_2O}$, $\mathrm{N_4}$ and on the one-dimensional Fermi-Hubbard model. It is further shown in the same study that orbital optimized methods (namely p-OO-UCCSD, OO-UCCD0 and OO-UCCD0-full) achieve results within $10$ to $20$ mH, showing a minor loss in accuracy compared to UCCSD, but a significant reduction in quantum resources required.

A clear drawback of UCC-type ans{\"{a}}tze is specifically that these are not hardware efficient. All UCC ans\"atze are designed in a manner that is agnostic to the connectivity of the device. The number of CNOT gates for each of the exponentiated Pauli strings required in the ansatz scales with its Pauli weight $q$ (number of non-identity operators in the string) when full connectivity is allowed. Namely, each Pauli strings exponential requires $2(q - 1)$ CNOT gates \cite{nielsenQuantumComputationQuantum2010}. Assuming double excitations are used, this implies a scaling of $\mathcal{O}(2N^4(\tilde{q} -1))$ CNOT gates (with $\tilde{q}$ the average Pauli weight across all strings) if no further action is taken to make the ansatz more efficient (some of these gates can be canceled out at compilation). However, a significant overhead in the number of CNOT gates is required if connectivity is limited to bridge entanglement between qubits that are not directly connected on the device. Connectivity limitation of the qubit register can be addressed by introducing SWAP networks to effectively relocate qubits together when $q$-qubit operators need to be implemented on a non-local set of qubits \cite{OGorman2019}. This allows maintaining the depth scaling of UCC ans\"atze by parallelizing on a single qubit register several operators that would otherwise not commute, irrelevant of the degree of connectivity of the device. It does however come at the cost of additional SWAP (or CNOT) gates required to build the SWAP network compared to a fully connected device. This is also related to a similar drawback of UCC: while the scaling is advantageous, the ansatz generally requires a very large pre-factor in the number of gates \cite{Grimsley2019_UCC_Review, Romero2019, Choquette2021}. A similar method to construct a more efficient gate fabric, but with the added benefit of preserving quantum number symmetry is found in Ref.~\cite{Anselmetti2021}. 

Overall, the UCC ans\"atze has a large pre-factor for the depth and number of entangling gates. As an example, Ref.~\cite{Choquette2021} show that to achieve similar accuracy on a $12$-qubit version of $\mathrm{H_2O}$, UCCSD requires $528$ entangling gates, hence at the very best a depth of about $100$ as one can fit at most $6$ entangling gates in one time step, against $88$ entangling gates for HEA, or a circuit depth of about $40$ (In Ref.~\cite{Choquette2021}, Choquette et al. also present an ansatz which extends the Hamiltonian Variational Ansatz presented in the following section, and which requires $108$ CNOT gates for the same molecular model). It is fair to assume that k-UpCCGSD would perform significantly better than UCCSD in terms of depth. The Qubit-excitations \cite{Xia2020, Yordanov2020_QE, Yordanov2021}, and qubit-operators \cite{Ryabinkin2018, Ryabinkin2020, Tang2021} Coupled Cluster based ansatz (see Sec. \ref{sec:adaptive_ansatz}), are a potential answer to address the scaling of CNOT gates while preserving some of the benefits of UCC ans{\"{a}}tze. 

Another point worth mentioning about UCC-based ans\"atz is their dependence on operator ordering \cite{Grimsley2019_UCC_Review, Izmaylov2020b}. Ordering of fermionic operators within the ansatz significantly impacts its accuracy, in particular for highly correlated electron wave functions (which is, as Grimsley et al. \cite{Grimsley2019_UCC_Review} and Wecker et al. \cite{Wecker2015} point out, the primary target of interest for quantum computers, as low correlation wave functions can easily be treated with conventional methods - see Sec. \ref{sec:vqe_vs_conventional}). This is particularly noticeable when dissociation limits are approached \cite{Grimsley2019_UCC_Review, Matsuzawa2020}, but can be in great part addressed by repeating the ansatz (i.e. setting $k$ to a higher number for k-UpCCGSD). As pointed out in Ref.~\cite{Lee2019} and corroborated by results from Ref.~\cite{Grimsley2019_UCC_Review}, k-UpCCGSD does appear to be largely resilient to problems of operator order as long as one allows for $k$ to be sufficiently high to provide the variational flexibility needed. 

Finally, another issue that can happen in both HEA and UCC ansatz is the appearance of 'kinks' in the Potential Energy Surface (PES). While already apparent in Ref.~\cite{Kandala2017}, these are very well characterized in Ref.~\cite{Ryabinkin2019}. These kinks are due to the energy ordering of the electronic states and lack of preservation of certain of symmetries (e.g. electron number, total spin) when building the ansatz, and can be addressed using symmetry preserving methods, including as an additional constraint placed in the VQE cost function \cite{Ryabinkin2019} (see Sec. \ref{sec:cost_function}).

\subsubsection{Symmetry-preserving methods} \label{sec:symmetry_preserving_methods}

A proposed solution to a common pitfall of HEA and UCC, is to build the ansatz in a manner that preserves certain symmetries of the system studied (though it is worth noting that UCC ans\"tze in general preserve particle number symmetry). There are two main benefits related to this:
\begin{itemize}
    \item It restricts the size of the Hilbert space in which the ground state can be searched while remaining largely agnostic to the system studied, potentially speeding up the optimization and reducing the risk of barren plateaus.
    \item It avoids risks of localized inaccuracies that have been observed as 'kinks' in the dissociation curves of molecules studied \cite{Kandala2017}, and that are related to local minima states with a different number of electrons \cite{Ryabinkin2019}
\end{itemize}

It is worth pointing out that symmetry preservation can also be achieved using an adjustment to the cost function \cite{mccleanTheoryVariationalHybrid2015, Ryabinkin2019} (further details are presented in Sec. \ref{sec:cost_function}), and can be used as error mitigation method (see Sec. \ref{sec:mit-symmetry-verification}).

The core objective of symmetry-preserving ans{\"{a}}tze in general, is the preservation of the electron number in the modeled wave function. A common theme in these methods is the use of exchange-type gates (or iSWAP gates)  (for a description and hardware implementation of these gates, we refer readers to Refs.~\cite{Roth2018, Egger2019}). These two-qubit gates aim at allowing relative phase changes between superpositions of $\ket{01}$ and $\ket{10}$ states, while leaving $\ket{00}$ and $\ket{11}$ unchanged. In the context of a VQE ansatz, their use can therefore never alter the number of occupied orbitals in the modelled wave-function.

There exist two published versions of this type of gate. The most commonly used one is a two-parameter version, originating from Ref.~\cite{Egger2019}, which allows further flexibility in the output wavefunction:

	\begin{equation} \label{eq:two_param_xchange_gate}
     A(\theta, \phi) =
        \begin{pmatrix}
        1& 0 & 0 & 0 \\
        0 & \cos\theta & e^{i\phi}\sin\theta & 0\\
        0 & e^{-i\phi}\sin\theta & -\cos\theta  & 0 \\
        0 & 0 & 0 & 1
        \end{pmatrix}.
	\end{equation}

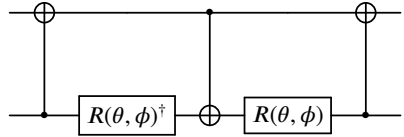
\begin{figure}[ht]
\centerline{
    $$
    \Qcircuit @C=1em @!R {
    & \targ     & \qw                               & \ctrl{1}& \qw                    & \targ     & \qw \\
    & \ctrl{-1} & \gate{R(\theta, \phi)^{\dagger}}  & \targ   & \gate{R(\theta, \phi)} & \ctrl{-1} & \qw   \\
    }
    $$
    } \caption{Quantum circuit for realization of gate $A(\theta, \phi)$, Eq.~(\ref{eq:two_param_xchange_gate}). With $R(\theta, \phi) = R_z(\phi + \pi)R_y(\theta + \pi/2) $.} \label{fig:A_gate}
\end{figure}

It can be obtained with a quantum circuit composed of three CNOT gates and two rotation gates, as presented in Fig. \ref{fig:A_gate},

A special, single parameter version of this gate has also been presented (see Refs.~\cite{McKay2016, Sagastizabal2019})

\begin{equation}\label{eq:one_param_xchange_gate}
    A(\theta) =
    \begin{pmatrix}
    1& 0 & 0 & 0 \\
    0 & \cos2\theta & -i\sin2\theta & 0\\
    0 & -i\sin2\theta & \cos2\theta  & 0 \\
    0 & 0 & 0 & 1
    \end{pmatrix}
\end{equation}

It is worth noting that this approach to particle conservation in the wave function is only valid when considering certain types of fermion-to-spin mappings (see Sec.~\ref{sec:Encoding}). In particular, while these gates necessarily preserve the number of excited qubits under Jordan-Wigner \cite{Gard2020} and the Parity mappings, it is easy to see that it does not under Bravyi-Kitaev, as swapping an excited qubit with a non-excited qubit across two blocks could require changing the state of all the subsequent `parity' encoding qubits within a block (see Sec.~\ref{sec:bravyi-kitaev}).

A first relevant use case is presented in Ref. \cite{Barkoutsos2018}, where the HEA (Sec.~\ref{sec:HEA}) is adapted to preserve the particle number. The method transforms the HEA into the particle/hole representation of the Hamiltonian (see Sec. \ref{sec:Hamiltonian_representation}). The ansatz in Ref.~\cite{Barkoutsos2018} is structured like the HEA, with a series of single-qubit rotations and entangling blocks, similar to that presented in Eq.~(\ref{eq:HEA_unitaries}). Each constituent block however can be composed of interlaced exchange-conserving gates, and as these are parametrized, the rotation unitaries can be absorbed in the entangling unitaries.

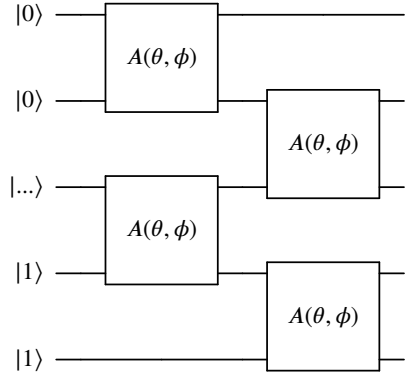
\begin{figure}[ht]
\centerline{
\Qcircuit @C=1em @!R {
   \lstick{\ket{0}}   &   \qw   &   \multigate{1}{~A(\theta, \phi)~}   & \qw & \qw                              & \qw          \\
   \lstick{\ket{0}}   &   \qw   &   \ghost{~A(\theta, \phi)~}          & \qw & \multigate{1}{~A(\theta, \phi)~} & \qw                     \\
   \lstick{\ket{...}} &   \qw   &   \multigate{1}{~A(\theta, \phi)~}   & \qw & \ghost{~A(\theta, \phi)~}        & \qw \\
   \lstick{\ket{1}}   &   \qw   &   \ghost{~A(\theta, \phi)~}          & \qw & \multigate{1}{~A(\theta, \phi)~} & \qw                     \\
   \lstick{\ket{1}}   &   \qw   &   \qw                                & \qw & \ghost{~A(\theta, \phi)~}        & \qw \\
}} \caption{Single layer of symmetry preserving ansatz presented in Ref. \cite{Barkoutsos2018}, based in two-parameters exchange conserving gates $A(\theta, \phi)$, Eq.~(\ref{eq:two_param_xchange_gate})} \label{fig:circuit_for_two_param_sym}

\end{figure}

\begin{figure}[ht]
\centerline{
\Qcircuit @C=1em @!R {
   \lstick{\ket{0}}   &   \qw   &   \gate{Rz(\phi_N)}      &   \qw   &   \multigate{1}{~A(\theta)~}   & \qw & \qw                        & \qw \\
   \lstick{\ket{0}}   &   \qw   &   \gate{Rz(\phi_{N-1})}  &   \qw   &   \ghost{~A(\theta)~}          & \qw & \multigate{1}{~A(\theta)~} & \qw \\
   \lstick{\ket{...}} &   \qw   &   \gate{Rz(\phi_{...})} &   \qw   &   \multigate{1}{~A(\theta)~}   & \qw & \ghost{~A(\theta)~}        & \qw \\
   \lstick{\ket{1}}   &   \qw   &   \gate{Rz(\phi_2)}      &   \qw   &   \ghost{~A(\theta)~}          & \qw & \multigate{1}{~A(\theta)~} & \qw \\
   \lstick{\ket{1}}   &   \qw   &   \gate{Rz(\phi_1)}      &   \qw   &   \qw                          & \qw & \ghost{~A(\theta)~}        & \qw \\
}}\caption{Single layer of symmetry preserving ansatz presented in Ref. \cite{Barkoutsos2018}, based in one-parameters exchange conserving gates $A(\theta)$, Eq. \ref{eq:one_param_xchange_gate}.} \label{fig:circuit_for_one_param_sym}

\end{figure}
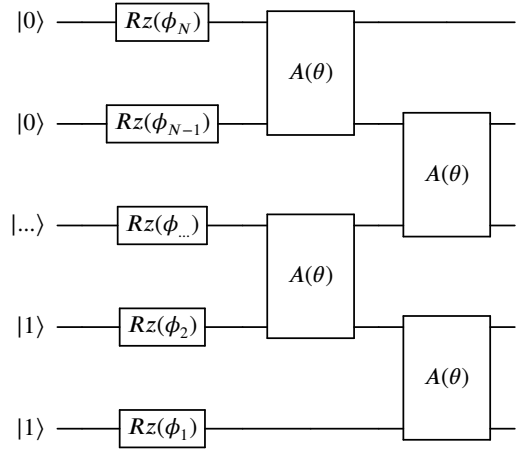

Two versions are proposed in Ref.~\cite{Barkoutsos2018}. The first one makes use of the two-parameter exchange-conserving gate (Eq. \ref{eq:two_param_xchange_gate}), as presented in Fig. \ref{fig:circuit_for_two_param_sym}, the second one relies on the one-parameter exchange-conserving gate (Eq. \ref{eq:one_param_xchange_gate}), as presented in Fig.\ref{fig:circuit_for_one_param_sym} (note these require additional rotation gates in each block). Barkoutsos et al. \cite{Barkoutsos2018} also provide a comparison of these two approaches against a more typical HEA method (where the entangling block is composed of a ladder of CNOT gates), and show numerically that particle conserving blocks achieve better accuracy when computing the dissociation curve of $\mathrm{H_2O}$ (with a slight advantage for the two-parameters gates). This approach is further studied and tested successfully on a superconducting quantum device in Ref. \cite{Ganzhorn2019}.

In Ref. \cite{Gard2020}, Gard et al. propose to address the wider set of symmetries (particle number, time reversal, total spin, spin magnetization) by creating an Efficient Symmetry-Preserving (ESP) ansatz for VQE. It uses the same two-parameters exchange-conserving building blocks as in Ref.~\cite{Barkoutsos2018} and a very similar structure to Fig. \ref{fig:circuit_for_two_param_sym} with some optimization. The idea behind this ansatz is to build a structure that spans the largest part of the Hilbert space while constraining symmetries as much as possible. As such, it is more efficient than HEA, but is less restrictive than UCC, guaranteeing that the ground state is indeed in the spanned Hilbert space \cite{Gard2020}. In that respect, we consider that this is the VQE ansatz that is closest to providing the span of FCI (see Sec. \ref{sec:full_configuration_interaction}).

\begin{figure*}[ht]
\centerline{
\Qcircuit @C=1em @!R {
   \lstick{\ket{0}} & \qw      & \multigate{1}{~A(\theta_1, \phi_1)~} & \qw                                  & \multigate{1}{~A(\theta_4, \phi_4)~} & \qw                                  & \meter  \\
   \lstick{\ket{0}} & \gate{X} & \ghost{~A(\theta_1, \phi_1)~}        & \multigate{1}{~A(\theta_3, \phi_1)~} & \ghost{~A(\theta_4, \phi_4)~}        & \multigate{1}{~A(\theta_6, \phi_6)~} & \meter \\
   \lstick{\ket{0}} & \gate{X} & \multigate{1}{~A(\theta_2, \phi_1)~} & \ghost{~A(\theta_3, \phi_1)~}        & \multigate{1}{~A(\theta_5, \phi_5)~} & \ghost{~A(\theta_6, \phi_6)~}        & \meter  \\
   \lstick{\ket{0}} & \qw      & \ghost{~A(\theta_2, \phi_1)~}        & \qw                                  & \ghost{~A(\theta_5, \phi_5)~}        & \qw                                  & \meter \\
}} \caption{Example of the ESP ansatz construction for $N= 4$ and $m=2$, as presented in \cite{Gard2020}, based on two-parameters exchange-conserving $A(\theta, \phi)$ gates, Eq.~(\ref{eq:two_param_xchange_gate}). The first three $\phi_i$ parameters are set to be equal as the number of optimal parameters has already be reached in the ansatz and further flexibility is not required} \label{fig:ESP_4_2}
\end{figure*}
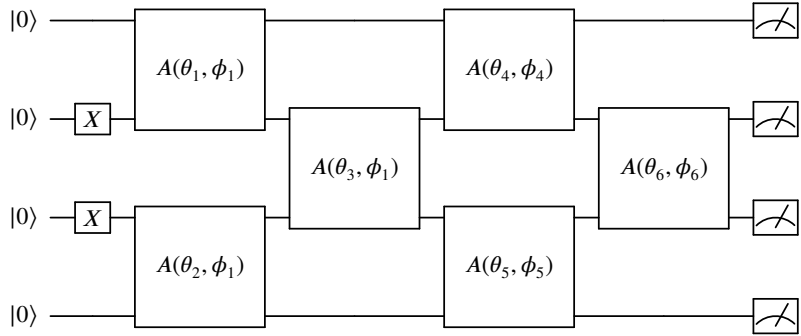

This ansatz can be systematically built by first applying X gates to a number of qubits equal to the number of electrons in the problem considered. For efficient circuit building, and unlike in Ref.~\cite{Barkoutsos2018}, Gard et al. \cite{Gard2020} advise that one must avoid placing these as neighboring qubits. Following this, one can apply A gates to each pair of $\ket{1}$ and  $\ket{0}$ zero qubits, and subsequently, apply additional A gates to span the complete register and until $\binom{N}{m}$ A gates have been placed ($N$ the number of orbitals or number of qubits, $m$ the number of electrons). We have reproduced an example initially presented in \cite{Gard2020} for $N=4$ and $m=2$ in Fig. \ref{fig:ESP_4_2}.

The optimal number of parameters to allow full parametrization of the Hilbert space, given particle-conserving symmetry can be derived from the number of complex coefficients required to span the subspace drawn by a given Hamiltonian, Gard et al. \cite{Gard2020} show that this number is $2\binom{N}{m} - 2$. Given there are $2\binom{N}{m}$ parameters in the ESP ansatz as described above, the optimal number can be obtained by setting two of the $\phi_i$ equal to a third one, as shown in Fig. \ref{fig:ESP_4_2}. They also point out that the method for constructing the ansatz is neither unique nor optimal in terms of the number of entangling gates and as such, further research may help extend the symmetry preserving ansatz (for instance by including ansatz options for fully-connected machines). Other symmetries can be imposed as follows:
\begin{itemize}
    \item Time reversal symmetry can be imposed by setting all $\phi_i$ to zero, resulting in the number of parameters scaling $\binom{N}{m}$
    \item Spin symmetries require to first order the qubit such that the first half of the register represents spin up orbitals, and the second half spin down orbitals (which requires some fermion to spin mapping modifications). The ladder structure described for the ESP ansatz is maintained, with a difference that any $A$ gate that links the first and second half of the register has parameters set to $0$ (or is removed), preventing swaps of electrons from spin-up to spin-down orbitals (and vice versa), and enforcing spin symmetry.
\end{itemize}

Based on this construction, the number of CNOTs scales at most as $3\binom{N}{m}$(without further optimization and compilation of the ansatz). This implies that it goes from linear if $m=1$, or $m=N-1$ to exponential if $m \sim N/2$. It can be slightly reduced using some of the symmetries mentioned above, however it remains significantly less efficient than UCC based ansatz in particular in terms of parameters, due to a larger span of the Hilbert space.

As such, we find that the cost of ESP (and other symmetry preserving ansatz) is too prohibitive to be implemented in cases where $m \sim N$. There could be scope for these ans{\"{a}}tze to outperform the likes of UCC type ans{\"{a}}tze when the problem can be treated accurately with $m$ being negligible or close to $N$. An interesting research question would be to study this ansatz in a plane wave basis where the number of basis functions (and therefore qubits) required for each electron to treat the system accurately can be extremely large \cite{Babbush2018}.

\subsubsection{The Hamiltonian Variational Ansatz and extensions} \label{sec:hva}

\paragraph{The Hamiltonian Variational Ansatz (HVA):} this ansatz was initially presented in Refs. \cite{Wecker2015, Wiersema2020}, and is inspired from the quantum approximate optimization algorithm (QAOA) ansatz \cite{Farhi2014} which draws from ideas developed for adiabatic quantum computation \cite{Farhi2000}. The first step in building this ansatz is to decompose the Hamiltonian into a sum of non-commuting operators:

\begin{equation}
    \hat{H} = \sum_a \hat{h}_a,
\end{equation}

where $[\hat{h}_i, \hat{h}_j] \neq 0$. These terms are used alternatively over $L$ layers to evolve an initial state $\ket{\psi_0}$.

\begin{equation}
    \ket{\psi_L} = \prod_{l=1}^L \left( \prod_a e^{-i \theta_{a, l} \hat{h}_a} \right) \ket{\psi_0}.
\end{equation}

The initial state $\ket{\psi_0}$ is picked as the ground state of any of the $\hat{h}_a$, except for the first one acting on it. There are a few points to note about this ansatz. The first one is that for a single repetition of the ansatz it has a number of variational parameters lower than the UCCSD ansatz, because there is only one parameter per commutative group in the Hamiltonian (we have seen in Sec. \ref{sec:Grouping}, that $N^3$ commutative group is a reasonable estimate for \textit{ab initio} molecular systems, though it could be much lower for lattice models). This means that layer repetitions is likely to be required to achieve the same level of expressibility, although exact scaling is not known. In Ref.~\cite{Wecker2015}, Wecker et al. suggest that the ansatz is trained layer-by-layer to facilitate optimization. The scaling in the number of entangling gates is similar to UCC based ansatz, as approximately the same number of operators are incorporated for each repeated layer. This also implies a similar depth scaling.

The HVA was shown however to be particularly well suited for models of many-body physics\cite{Wiersema2020}. Key points revealed in this study include the fact that the ansatz appears to exhibit strong resilience to barren plateaus, and that layer repetitions allow for over-parametrization of the ansatz and therefore smoothing out of the energy landscape to avoid local minima. It is clear that low Pauli-weight mapping (or lattice models) significantly improves the algorithm. The ansatz was also adapted to solve the Heisenberg antiferromagnetic model on the Kagome lattice in Refs. \cite{kattemolle2021variational, bosse2021probing}, both study showing promising results. In particular, Ref. \cite{kattemolle2021variational} shows that VQE approaches the true ground of the lattice exponentially as a function of circuit depth on a noiseless simulation of up to 20 sites. Bosse and Montanaro~\cite{bosse2021probing} find a similar result ton a simulation of 24 sites, though point out to the large number of variational parameters needed. Another adaptation of HVA to the Heisenberg model, accompanied by a scaling study is presented in Ref. \cite{Jattana2022}. Finally, a version of HVA for Fermi-Hubbard is studied by Stanisic \textit{et al.} \cite{Stanisic2021}, in which the ansatz is created by first applying a number of Givens rotations (rotation in a two dimensional subspace of the Hilbert space, which can be used as partial excitation exchanges between two qubits), allowing to efficiently prepare the ground state of the non-interacting Fermi-Hubbard model \cite{Jiang2018}, then followed by time evolution following the same logic as HVA. The paper shows that through this ansatz construction, few parameters are needed to reproduce key qualitative features of strongly-correlated electronic systems. 

\paragraph{Fourier-transform HVA \cite{Babbush2018}:} This ansatz was initially developed as a linear scaling method to variationally model the ground state of jellium (uniform electron gas). In this context, it was developed using a plane wave basis, allowing the one-body hamiltonian of jellium to be diagonal \cite{Babbush2018}. A slightly more general formulation is presented in Ref.~\cite{Choquette2021}. The method relies on the fermionic fast Fourier transform (FFFT) \cite{Verstraete2009, Ferris2014, Jiang2018} which was initially constructed to diagonalize dynamics in many-body systems, and which has been shown to have a depth scaling of $\mathcal{O}(N)$ if implemented on a planar lattice of qubits \cite{Babbush2018}. To construct this ansatz one must first separate the Hamiltonian into kinetic and potential energy terms, $\hat{H} = \hat{T} + \hat{V}$ (as presented in Sec. \ref{sec:Hamiltonian_representation}). The ansatz is then constructed in a manner similar to HVA, but applying FFFT to the kinetic energy term: 

\begin{equation}
    U(\boldsymbol{\theta}) = \prod_{l=1}^L \mathrm{FFFT}^{\dagger}\left( \prod_a e^{-i \theta_{a, l}^{(T)} \hat{T}_a} \right) \mathrm{FFFT} \left( \prod_b e^{-i \theta_{b, l}^{(V)} \hat{V}_b} \right) 
\end{equation}
with $\hat{T} = \sum_a \hat{T}_a$ and $\hat{V} = \sum_b \hat{V}_b$. In Ref.~\cite{Babbush2018}, Babbush et al. recommend to use $\ket{\psi_0} = \mathrm{FFFT} \ket{0}$ as initial state, as this makes $\ket{\psi_0}$ an eigenstate of $\hat{T}$ in the plane wave basis. They also suggest absorbing the component of the ansatz corresponding to $\hat{T}$ in the Hamiltonian as it can be done efficiently. It is worth noting that given use of a plane wave basis, for jellium, the operators of $\hat{V}$ are mapped to diagonal $Z$ operators using Jordan-Wigner, which provides a very convenient and compact ansatz scaling $\mathcal{O}(LN)$ on planar qubit lattices, and which can also be optimized layerwise \cite{Wecker2015, Babbush2018}. To the best of our knowledge, this ansatz has not been benchmarked or tested on alternative systems.

\paragraph{Symmetry breaking HVA \cite{Vogt2020, Choquette2021}:} The salient feature of this type of ans{\"{a}}tze is that it incorporates unitaries into the HVA to purposefully break symmetries, to allow for a better optimization path. Both the quantum-optimal-control-inspired ans{\"{a}}tze (QOCA) \cite{Choquette2021}, and the Variational Extended Hamiltonian Ansatz (VEVA) \cite{Vogt2020} are nearly identical and were initially published at the same time. 
The version presented in Ref.~\cite{Choquette2021} is motivated by optimal control theory \cite{Werschnik2007, Alessandro2007}. Quantum optimal control (QOC) is concerned about the efficient design of control pulses to manipulate a quantum system (for a recent review of QOC, we refer reader to M{\"{u}}ller \textit{et al.} \cite{Muller2021}). QOC theory relies on a control Hamiltonian which is composed of a set of drive terms $\{\hat{H}_b\}$, and a drift term $\hat{H}_0$, such that $\hat{H}^{(ctrl)}(t) = \hat{H}_0 + \sum_b c_b(t) \hat{H}_b$ (where $\{c_b(t)\}$ are time-dependent parameters). To build QOCA or VEHA, one must first create a Hamiltonian incorporating the structure of the control Hamiltonian:

\begin{equation}
    \hat{H}_{tot}(t) = \hat{H} + \sum_b c_b(t) \hat{H}_b^{(ctrl)}
\end{equation}
where $\hat{H}$ is the problem Hamiltonian. The ansatz is constructed from this Hamiltonian in a manner similar to HVA, thereby mimicking the time evolution of QOC. It can therefore be constructed as:

\begin{equation}
    U(\boldsymbol{\theta}, \boldsymbol{\phi}) = \prod_{l=1}^L \left( \prod_a e^{-i \theta_{a, l} \hat{H}_a} \right) \left( \prod_b e^{-i \phi_{b, l} \hat{H}^{(ctrl)}_b} \right) 
\end{equation}
where $\hat{H} = \sum_a \hat{H}_a$ and $\hat{H}^{(crtl)}_b$ are a set of drive terms selected for the purpose of the problem studied ($\boldsymbol{\phi}$ corresponds to the parameters $\{c_b(t)\}$). Following the description in Ref.~\cite{Choquette2021}, the problem Hamiltonian is used to explore the symmetry subspace of the Hilbert space, and the drive terms are used to provide shortcuts through the subspaces to increase efficiency. It is recommended in Ref.~\cite{Choquette2021} that the drive terms do not commute with $\hat{H}$. Choquette et al. suggest that this terms could be selected adaptively (akin to adaptive ans{\"{a}}tze described in Sec. \ref{sec:adaptive_ansatz}), but also provide a method to identify the most relevant drive operators. 

This method is applied to the Fermi-Hubbard model and compared to FT-HVA and HVA in Ref.~\cite{Choquette2021}. Results show excellent fidelities achieved with a reasonable number of entangling gates up to 12 qubits (which could also suggest some degree of resilience to barren plateaus). The method is also shown to perform well on a 12-qubit version of $\mathrm{H_2O}$, with fewer numbers of entangling gates compared to UCCSD, albeit a higher number of parameters \cite{Choquette2021}. The method is also tested on the Hubbard model in Ref.~\cite{Vogt2020}, with the stability of the ansatz being very much dependent on the problem parameters. Vogt et al. \cite{Vogt2020} extend the idea of the symmetry breaking HVA by adding classical mean-field parametrization to the ansatz (a method named Variational Mean Field Hamiltonian Ansatz, or VMFHA), which performs better where the VEHA fails. 

\subsection{Adaptative structure ans{\"{a}}tze}  \label{sec:adaptive_ansatz}

Adaptative structure ans{\"{a}}tze aim at creating a circuit structure tailored to the problem studied. As such, they all function similarly, in which the ansatz is grown iteratively throughout the optimization process by adding new operators at each step based on their contribution to the overall energy (or other metric). We distinguish three types of adaptive ansatz, those that grow the quantum circuit iteratively (Sec. \ref{sec:adapt-vqe}), those that incorporate (at least part of) the ansatz into the Hamiltonian by 'dressing' it (Sec. \ref{sec:Hamiltonian_dressing}), and those that variationally update the gate structure of the entire ansatz (Sec. \ref{sec:gate_structure_optimization}). At this point, it is challenging to build comparative metrics for adaptive ans\"atze reliably, and as such we do not provide a comparative table as we have done in other sections. Instead, we direct readers to the qualitative discussion at the end of this section. 

\subsubsection{Iterative ansatz growth methods (ADAPT-VQE and extentions)} \label{sec:adapt-vqe}

\paragraph{Fermionic-ADAPT-VQE:} The idea behind ADAPT-VQE \cite{Grimsley2019} is to progressively build the ansatz by sequentially incorporating into it the operators that contribute most to lowering the VQE energy towards the ground energy. Starting from an initial state (usually the Hartree-Fock wave function, and given a pool of operators $\{\hat{A}_i \}$ the ADAPT-VQE ansatz evolves as follows, with the iteration number as superscripts:
\begin{align} \label{eq:adapt-vqe}
    &\ket{\psi^{(0)}} = \ket{\psi_{HF}} \nonumber \\
    &\ket{\psi^{(1)}} = e^{\theta_1\hat{A}_1}\ket{\psi_{HF}} \nonumber \\
    &\ket{\psi^{(2)}} = e^{\theta_2\hat{A}_2}e^{\theta_1\hat{A}_1}\ket{\psi_{HF}} \nonumber \\
    &... \nonumber \\
    &\ket{\psi^{(k)}} = \prod^1_{i=k} e^{\theta_i\hat{A}_i}\ket{\psi_{HF}}.
\end{align}

The operators in the pool are similar to those produced as part of the UCC ansatz (see Eq. \ref{eq:ucc_operators}), however rather than specifying a number of possible excitation, one can include any one-, two body- or even higher body operator that is believed to be particularly relevant for the system considered: $\hat{A}_m \in \{ (\hat{\tau}_p^q - (\hat{\tau}_{p}^{q})^{\dagger}), (\hat{\tau}_{pq}^{rs} - (\hat{\tau}_{pq}^{rs})^{\dagger}), (\hat{\tau}_{pqr}^{stu} - (\hat{\tau}_{pqr}^{stu})^{\dagger}), ... \} $.

To decide on which operator to incorporate, Grimsley et al \cite{Grimsley2019} recommend choosing the one which results in the highest gradient of the energy functional with respect to the operator parameter. After initializing all operator parameters to $0$, to guarantee that they initially act as the identity operator, gradients can be simply computed as:

\begin{equation} \label{eq:adapt_vqe_grad}
    \frac{\partial E^{(k)}}{\partial \theta_i} = \bra{\psi^{(k)}} [ \hat{H}, \hat{A}_i] \ket{\psi^{(k)}}.
\end{equation}
Once an operator has been selected, and incorporated into the ansatz, a full VQE optimization is completed until parameters all parameters are at their optimum. This implies that at the next operator selection step, all existing $\theta_j$ are such that $\partial E^{(k - 1)}/\partial \theta_j = 0$, enforcing the results of Eq.~(\ref{eq:adapt_vqe_grad}) for all $\theta_i$ corresponding to new operators in following iterations.

The simplicity of the gradient computation at all iterations of ADAPT-VQE is key to its efficiency. However, assuming the case of a second-quantized electronic structure Hamiltonian (with $\mathcal{O}(N^4)$ terms), and given there are $\mathcal{O}(N^4)$ terms in a complete pool for up to double excitation operators, measurement scaling at each outer-loop iteration could naively reach $\mathcal{O}(N^8)$. This can be reduced to $\mathcal{O}(N^6)$ if one reformulates the energy gradients in terms of the three-body reduced density matrix \cite{Liu2021_adapt}. This method also allow predicting gradients from the two-body reduced density matrix of the previous iteration of ADAPT-VQE, reducing overall scaling of measurements back to $\mathcal{O}(N^4)$ albeit at a cost to accuracy.

Therefore, an important research question surrounding ADAPT-VQE is the optimal size of the operator pool, as it determines the number of iterations required, and the complexity of each step.
It is also important to note that the scaling of the ADAPT-VQE pool can also be adjusted the same way it is adjusted in certain variations of the UCC ansatz. For instance, one could restrict operator availability to those available in k-UpCCGSD \cite{Lee2019}, or proceed to only integrate double excitation operator on the quantum computer subroutine akin to the OO-UCC ansatz \cite{Kottmann2021_1}. Comparative performance of these respective approaches is yet to be assessed. 

Also contributing to the cost of ADAPT-VQE, a complete VQE convergence must be conducted between each incremental step. While this is largely tractable for early steps, as the ansatz remains shallow, it could rapidly become prohibitively expensive to implement. Grimsley {\it et al.} \cite{Grimsley2022} nonetheless showed that the build-in "recycling" of parameters (initialization using the optimal values from the previous step) allow for rapid convergence and avoidance of local optima. 
Grimsley \textit{et al.} \cite{Grimsley2019} show that using the ADAPT-VQE method outperforms UCCSD in terms of accuracy for a given number of ansatz parameters (which is a proxy for the number of operators as we need one parameter for each operator). Using $\mathrm{BeH_2}$ as an example, they show that ADAPT-VQE achieves an error of $10^{-3}$ kcal/mol for about 50-60 parameters (and still significantly less than 100 for $10^{-5}$ kcal/mol), while UCCSD requires nearly 300. They also tested against random ordering of the operator pool and showed that the gradient-based ordering described above performs significantly better.

It is difficult however to assess output at a comparable computational cost. While UCCSD only needs to be optimized once (over 300 parameters in $\mathrm{BeH_2}$ example), ADAPT-VQE needs to be optimized many times as the number of operators (or parameters) included (in addition to computation of $\mathcal{O}(N^4)$ gradients at each iteration). Therefore, it is important to assess the comparative cost of ADAPT-VQE in terms of total number of measurements required, in addition to the number of parameters required to achieve a given accuracy.
Large-scale numerical simulations may be required to assert the superiority of adaptive ansatz methods. Ref.~ \cite{Claudino2020} presents a numerical comparison of ADAPT-VQE and a UCCSD based VQE, corroborating the results presented in Ref.~\cite{Grimsley2019}, showing that the former can achieve excellent accuracy (on $\mathrm{H_2}$, $\mathrm{NaH}$, and $\mathrm{KH}$), however with on average significantly more total measurements than the latter. Due to the targeted expressibility of the ansatz produced, ADAPT-VQE has also been shown to be resilient to the barren plateau problem \cite{Grimsley2022} (other numerical evidence can be found in Refs. \cite{Skolik2021, Bilkis2021}). ADAPT-VQE could however suffer from exponential vanishing of pool operator gradients \cite{Grimsley2022}, though further research is needed to characterize and address this potential issue.  

ADAPT-VQE also inherits some advantageous properties of the UCC ans{\"{a}}tze such as the ease of management of particle number and time-reversal symmetries. Similar to UCC, however, it comes at the cost of a very large number of entangling gates required to implement the fermionic excitation operators into an ansatz, in particular when the qubit lattice has low connectivity (as discussed in the UCC section, see Sec. \ref{eq:ucca}). Qubit-ADAPT-VQE \cite{Tang2021} and Qubit Coupled Cluster (QCC) \cite{Ryabinkin2018} methods detailed below are methods to address this shortcoming. An alternative method called Energy Sorting VQE has also been proposed to not only compute the gradient of operator in the pool, but perform a full VQE optimization on each of them to rank them and include them in groups in the ansatz \cite{Fan2021}.
The principle of ADAPT-VQE has also been replicated in applications beyond electronic structure computation, in particular for periodic systems \cite{Liu2020}, and nuclear structure problems \cite{Romero2022}.

\paragraph{ADAPT-VQE for nuclear structure problems:}

While initially designed for electronic structure problems, ADAPT-VQE has been implemented for nuclear structures problems by Romero \textit{et al.} \cite{Romero2022}. The pool of fermionic operators can be replaced by one and two body operators for an $N$-nuclei wavefunction. The one body operators can be defined as $X_i$, $Y_i$, and $Z_i$, scaling as $3N$. For the two body operators, with $j < k$ the pool can be defined as: 
\begin{align} \label{eq:operator_pool_nuclear_adapt}
    &T_{+}^{jk} = \frac{1}{2}(X_jX_k - Y_jY_k), \quad T_{-}^{jk} = \frac{1}{2}(X_jY_k + Y_jX_k)\nonumber \\
    &U_{+}^{jk} = \frac{1}{2}(X_jX_k + Y_jY_k), \quad U_{-}^{jk} = \frac{1}{2}(Y_jX_k - X_jY_k)\nonumber \\
    &V_{+}^{jk} = X_jZ_k, \quad V_{-}^{jk} = Y_jZ_k, \quad V_{0}^{jk} = Z_jZ_k.
\end{align}

This results in $7N(N - 1)/2$ which results in a pool scaling $\mathcal{O}(N^2)$. This method is tested for both the Lipkin-Meshkov-Glick \cite{Lipkin1965} and the nuclear shell model \cite{Caurier2005, Heyde1994}, up to N=12 and demonstrates reasonable scaling in the number of outer-loop iteration required to reach good accuracy.  

\paragraph{qubit-ADAPT-VQE:}

The Qubit-ADAPT-VQE \cite{Tang2021} proposes an incremental selection of operators directly from the Pauli string exponentials produced from a given encoding rather than at the fermionic-operator level. Namely, this implies building an operator pool, with operators structured as 
\begin{equation} \label{eq:operator_pool_qubit_adapt}
    \hat{A}_a = i \bigotimes_i^N \hat{P}_i, \quad \hat{P}_i \in \{I, X, Y, Z\}^{\otimes N}.
\end{equation}
From this point, the ansatz is constructed as presented in Eq.~(\ref{eq:adapt-vqe}).

Further actions can be taken to restrict this operator pool. First restricting operations to those verifying time-reversal symmetry, removing from the pool in Eq.~(\ref{eq:operator_pool_qubit_adapt}) imaginary operators which have no impact on energy if $\hat{H}$ is time-reversal symmetric. The Pauli strings included in the pool are also selected only from those produced from the fermionic excitation mapping (a mapping that must happen anyway to produce the measurable Hamiltonian). Tang et al. \cite{Tang2021} also show analytically that a linear number of operator $(2N - 2)$ is required for the ansatz to be analytically exact and describe a method to select this pool of operators. They also suggest reducing the string Pauli weights by removing the Pauli $Z$ operators used for anti-symmetry of the wave function in the Jordan-Wigner mapping (Sec. \ref{sec:Encoding} and show that this does not impact energy estimate. This implies a maximum Pauli weight of 4 for any Pauli string used to construct the ansatz, allowing for significant depth contraction.

While it is clear that the Pauli weight reduction from removing Pauli-Z chains is not exclusive to this method we conjecture that it works in this instance because qubit-ADAPT-VQE uses one parameter for each Pauli string (instead of one per fermionic operator) and is, therefore, able to variationally learn the anti-commuting relationships required to reach ground-state energy (a view also supported in Ref.~\cite{Yordanov2021}).
Shkolnikov et al. \cite{Shkolnikov2021} also showed that $(2N - 2)$ is the minimum size of a complete pool comprised of Pauli strings. They also show that while minimal complete pools do not perform well on electronic structure Hamiltonian, these can be symmetry-adapted to successfully model ground state energy, achieving outer-loop measurement scaling of $\mathcal{O}(N^5)$.
Overall, this approach allows a significant reduction in the number of CNOT gates ($4$ per operator) required for the implementation of adaptive ansatz albeit at the cost of introducing a larger pool to sample from using the gradient-based selection rules, and as shown in Ref. \cite{Tang2021} more parameters to achieve a given accuracy. In particular, Mukherjee \textit{et al.} \cite{Mukherjee2022} show that qubit-ADAPT-VQE where operators are taken from a pool of Hamiltonian commutators produces a significantly more compact circuit that the HVA and the UCCSD ans{\"{a}}tze.

\paragraph{QEB-ADAPT-VQE:}

The qubit-excitation-based adaptative VQE \cite{Yordanov2021} (QEB-ADAPT-VQE) offers an interesting compromise between ADAPT-VQE and qubit-ADAPT-VQE, by having fewer CNOT gates than both, yet a similar number of parameters than ADAPT-VQE. It is based on a previous ansatz labeled Qubit Coupled Cluster Singles and Doubles (QCCSD) \cite{Xia2020} which uses the same qubit-excitations, up to second order, but without the adaptive aspect. It introduces two main changes:
\begin{itemize}
    \item Screening process includes a new step, whereby the top $k$ candidate operators in the pool (by gradient magnitude) are used for full VQE optimizations, with the one triggering the largest energy drop being selected for incorporation. \item It replaces the pool of operators with qubit-excitation operators \cite{Yordanov2020_QE}. These have been developed as a CNOT efficient alternative to the conventional implementation of fermionic excitation operators.
\end{itemize}

The premise for qubit-excitation operators is simply based on qubit creation and annihilation operators ($Q_i^{\dagger}, Q_i$), similar to the ones used in Jordan-Wigner, but without using the Pauli-Z chains maintaining anti-commutation relations (Ref. \cite{Yordanov2020_QE} shows how to re-incorporate anti-commutation properties, but this is not used in QEB-ADAPT-VQE). Therefore one can define
\begin{align}
    &Q_i^{\dagger} = \sigma_i^- = \frac{1}{2} (X_i - iY_i) \nonumber \\
    &Q_i = \sigma_i^+ = \frac{1}{2} (X_i + iY_i).
\end{align}
To construct operator pools from these, Ref. \cite{Yordanov2020_QE} proceeds by producing skew-Hermitian  operators (to later be exponentiated) of the type
\begin{align}
    &T_i^k= Q_k^{\dagger}Q_i - Q_i^{\dagger}Q_k \nonumber \\
    &T_{ij}^{kl}= Q_k^{\dagger}Q_l^{\dagger}Q_iQ_j - Q_i^{\dagger}Q_j^{\dagger}Q_kQ_l.
\end{align}
From there, we can derive the parametrized single and double qubit excitations:
\begin{align}
    &A_{ik}(\theta_a) = e^{T_i^k(\theta_a)}\nonumber \\
    &A_{ijkl}(\theta_a) = e^{T_{ij}^{kl}(\theta_a)}. \\
\end{align}
Operators from the pool are selected as in ADAPT-VQE, with the gradient computation rule identical to Eq.~(\ref{eq:adapt_vqe_grad}), and with the additional step mentioned above (although it is unclear from Ref.~\cite{Yordanov2021} whether this additional step has a significant impact on performance).
Overall, this method was shown to achieve a lower number of CNOT gates and parameter count than qubit-ADAPT-VQE. It also shows again that the inclusion of Pauli-Z chains into ansatz operators does not matter as long as there is enough variational flexibility for the correct phase to be recovered (which is likely not applicable using fermionic operators in the pool).

\paragraph{Entangler pool compression:} \label{sec:MI_assisted_VQE}

Research has also focused on addressing the wider question of operator pool optimization for adaptative ans{\"{a}}tze. An interesting method that achieves up to nearly two orders of magnitude reduction in the operator pool size is the Mutual Information (MI) assisted Adaptive VQE \cite{Zhang2021_CEP} (it is worth noting that the method described below can be applied to any adaptive ansatz described in this review).
The idea behind this method is to screen operators in the pool based on their mutual information \cite{Amico2008, Huang2005, Rissler2006}. Operators which have a higher correlation of information across the qubits it acts on for an approximate wave function are likely to contribute significantly more to the overall energy estimate. To implement this MI screening, one must first compute an approximation of the ground state wave function using the Density Matrix Renormalization Group (DMRG) method (see Sec. \ref{sec:vqe_vs_conventional}). From the DMRG wave function, the MI of each qubit pair must then be computed as
\begin{equation} \label{eq:mutual_information}
    I_{ij} = \frac{1}{2} \left( S(\rho_i) + S(\rho_j) - S(\rho_{ij} \right)) (1 - \delta_{ij}),
\end{equation}
where $\rho_{x}$ represents the reduced density matrix of system $x$, and $S(\rho)$ represents the von Neumann entropy of $\rho$, $\delta_{ij}$ is the Kronecker delta. The correlation strength for each operator $\tau_a$ in the pool can then be computed as
\begin{equation}
    C(\tau_a) = \frac{1}{q_{\tau_a}(q_{\tau_a} - 1)} \sum_{ij \in Q(\tau_a), i \neq j } I_{ij},
\end{equation}
where $q_{\tau_a}$ is the Pauli weight of the operator (or the number of qubit it acts on), and $Q(\tau_a)$ the set of qubit indices the operator acts on. Operators can then be ranked based on both their gradient (as in other adative methods) and their correlation strength). In Ref.~ \cite{Zhang2021_CEP}, Zhang \textit{et al.} suggest to cut off a certain percentage of operators with low correlation strength which they optimize numerically.

\subsubsection{Iterative Hamiltonian dressing (iterative Qubit Coupled Cluster (iQCC) and extensions)} \label{sec:Hamiltonian_dressing}
The QCC ansatz \cite{Ryabinkin2018} is not strictly speaking an adaptive ansatz in that its structure is not decided throughout the optimization process. However, because pre-processing of the relative importance of each of the operators incorporated in the ansatz is required to make it tractable, we thought it is more appropriate to include it in this section. It also makes it easier to present the iterative QCC, which is indeed an adaptive ansatz that makes use of the concept of Hamiltonian dressing to grow the Hamiltonian instead of the ansatz at each iteration.

\paragraph{The Qubit Coupled Cluster ansatz:} QCC was introduced as a means to bypass the detrimental non-local action arising in the UCC ansatz, and the large number of two-qubit gates that ensue. It is in essence very similar to the qubit-ADAPT-VQE, although with the addition of a Qubit Mean-Field wave function, and initially presented in Ref.~\cite{Ryabinkin2018} without adaptive looping and without prior selection of a restricted pool of Pauli strings.

The central premise of QCC is that, unlike UCC, it uses spin operators directly to construct the ansatz instead of fermionic excitation operators which are then transformed. The first step to construct this ansatz is to separate the wave function into mean field and correlated components, such that
\begin{equation}
    \ket{\psi(\boldsymbol{\tau}, \boldsymbol{\Omega})} = U(\boldsymbol{\tau}) \ket{\boldsymbol{\Omega}}.
\end{equation}
The mean field component $\ket{\boldsymbol{\Omega}}$ can be simply prepared through single qubit rotations on each individual qubit in the wave function.
\begin{align}
    &\ket{\boldsymbol{\Omega}} = \bigotimes_j^N \ket{\Omega}_j \nonumber \\
    & \ket{\Omega}_j = R(\theta_j, \phi_j) \ket{0}_j \nonumber \\
    &\ket{\Omega}_j =  \cos\frac{\theta_j}{2} \ket{0})j + e^{i\phi_j}\sin\frac{\theta_j}{2} \ket{1}_j,
\end{align}
with $\theta$, $\phi$ parameters for the rotation of each qubit. The correlation component consist in an entangling block unitary, built from amplitude parameters $\boldsymbol{\tau} = \{ \tau_a \}$, and a set of corresponding Pauli strings $\{ P_a\}$, as
\begin{align} \label{eq:qcc_generators}
    U(\boldsymbol{\tau}) = \prod_a^{\boldsymbol{P}}  e^{-i\tau_a \hat{P_a}/2},
\end{align}
with $\boldsymbol{P}$ the total number of Pauli strings used. With this we can define the VQE problem using the QCC ansatz:
\begin{align} \label{eq:cost_qcc}
    E_{QCC} = \min_{\boldsymbol{\tau}, \boldsymbol{\Omega}} \bra{\boldsymbol{\Omega}} U^{\dagger}(\boldsymbol{\tau}) \hat{H} U(\boldsymbol{\tau}) \ket{\boldsymbol{\Omega}}.
\end{align}

Because the number of possible Pauli strings candidates for this ansatz scales $\mathcal{O}(4^N)$, it is necessary to carefully select the appropriate operators to incorporate into the entangling block. A first proposal in Ref.~\cite{Ryabinkin2018} is to look at the (negative) contribution of using each operator individually once its amplitude parameter ($\tau_a$) is optimized. This rapidly becomes computationally prohibitive, and as such Ryanbinkin et al. \cite{Ryabinkin2018} also propose to rank operator contributions based on their first and second order gradients. In a manner that is similar to that presented in ADAPT-VQE, the first order derivative with respect to the amplitude parameter $\tau_a$ at $\tau_a = 0$ is given by
\begin{equation} \label{eq:qcc_grad}
    \frac{\partial E(\tau_a, \hat{P_a})}{\partial \tau_a} = \bra{\boldsymbol{\Omega}^*} \left(- \frac{i}{2}[ \hat{H}, \hat{P}_a] \right) \ket{\boldsymbol{\Omega}^*}
\end{equation}
where $\ket{\boldsymbol{\Omega}^*}$ is the optimized mean-field qubit wave function with respect to $\boldsymbol{\theta}$ and $\boldsymbol{\phi}$. This remains exponential in scaling, and further restrictions on the Pauli strings to be included in the pool must be implemented.
In particular, Ref.~\cite{Ryabinkin2020} introduces the concept of direct interaction set (DIS), which aims at excluding from the operator pool all those with zero gradients, and grouping the remaining $\mathcal{O}(4^N)$ operators into groups having identical gradients up to a sign. This results in a polynomial scaling number of groups ($\mathcal{O}(N^4)$) each composed of $2^{N-1}$ operators.

\paragraph{iQCC:} \label{sec:iqcc}

The iterative Qubit Coupled Cluster method \cite{Ryabinkin2020} is an improvement on the QCC method which aims to reduce the number of operators that can be included in the ansatz by incorporating operators into the Hamiltonian iteratively instead of increasing the ansatz as it is done for ADAPT-VQE and extensions. It works based on a slightly modified cost function, compared to Eq.~(\ref{eq:cost_qcc}):
\begin{align} \label{eq:cost_iqcc}
    E_{iQCC} = \min_{\boldsymbol{\tau}} \{ \min_{\boldsymbol{\Omega}} \bra{\boldsymbol{\Omega}}  \hat{H}_d(\boldsymbol{\tau})  \ket{\boldsymbol{\Omega}} \},
\end{align}
where $\hat{H}_d = U(\boldsymbol{\tau})^{\dagger} \hat{H}U(\boldsymbol{\tau})$ is called the \textit{dressed Hamiltonian}. The dressed Hamiltonian can be computed recursively, starting from $\hat{H}_d^{(0)} = \hat{H}$, and each $\hat{H}_d^{(k)}$ being dependent on the first $k$ amplitude in the vector $\boldsymbol{\tau}$, such that $\hat{H}_d^{(k)} =  \hat{H}_d^{(k)}(\boldsymbol{\tau}_{1 \rightarrow k})$:

\begin{align} \label{eq:dressed_hamiltonian}
    \hat{H}_d^{(k)} &= e^{i\tau_k \hat{P}_k / 2} \hat{H}_d^{(k-1)} e^{-i\tau_k \hat{P}_k / 2} \nonumber \\
    & = \hat{H}_d^{(k-1)} - \frac{i \sin{\tau_k}}{2} \left[\hat{H}_d^{(k-1)}, \hat{P}_k \right] \nonumber \\
    & + \frac{(1 - \cos{\tau_k})}{2} \hat{P}_k \left[\hat{H}_d^{(k-1)}, \hat{P}_k \right],
\end{align}
where the exponential terms $e^{i\tau_k \hat{P}_k / 2}$ are called the generators of the iQCC process, and akin to the elements of the QCC ansatz presented in Eq.~(\ref{eq:qcc_generators}). The process for completing the iQCC optimization process works as follows:
\begin{itemize}
    \item Inputs are the mean-field parameters of $\ket{\boldsymbol{\Omega}}$ optimized at the previous iteration, and the dressed Hamiltonian of the previous iteration  $\hat{H}_d^{(k-1)}$ (or $\hat{H}_d^{(0)}$ at the start).
    \item Using the optimized qubit mean-field wave function, and $\hat{H}_d^{(k-1)}$, use a screening procedure to select a number of generators $g$ with the largest gradient from an operator pool. The value for $g$ is at the discretion of the user (including $g=1$, although Ryabinkin et al. \cite{Ryabinkin2020} point out that increasing $g$ accelerates convergence).
    \item Using the selected generators, construct the appropriate ansatz and proceed to a VQE optimization on QPU of both the mean-field parameters and $g$ generator amplitudes $\tau$. Ryabinkin et al. \cite{Ryabinkin2020} suggest using random initializations for the amplitudes or set $\boldsymbol{\tau}= 0$, if random initialization does not start from a lower energy than at the previous iteration.
    \item Once parameters are optimized, incorporate the $g$ generators into the Hamiltonian by computing $\hat{H}_d^{(k)}$ using Eq.~(\ref{eq:dressed_hamiltonian}), and begin a new iteration unless convergence is reached.
\end{itemize}

Using this method, one can enforce a fixed size for the ansatz, at each iteration, determined uniquely by the constant $g$. This provides many advantages from a NISQ perspective as it allows to directly adjust the size of the ansatz in each iteration to the noise level acceptable (and which can be mitigated) for the device being used.
This comes however with two significant caveats. First, there is no guarantee that the method can converge to the exact solution, despite successfully demonstrating the method works for small systems \cite{Ryabinkin2020}. Secondly, the number of terms in the Hamiltonian grows in a somewhat prohibitive manner. It is estimated that the Hamiltonian grows by a factor of $\sim (3/2)^{g}$ at each iteration, which means that for a number of iterations $k$, and with $\mathcal{P}$ the number of terms in $\hat{H}_d^{(0)}$, the Hamiltonian scales $\mathcal{O}(\boldsymbol{P}(3/2)^{g k})$, or exponentially in the number of iterations (with an obvious upper bound of $\mathcal{O}(4^N)$).
This was improved upon in Ref.~\cite{Lang2020}, where Lang et al. show that using Truncation of the  Baker-Campbell-Hausdorff expansion, one can restrict the increase in the number of terms by a factor $\mathcal{O}(g^2)$ at each iteration, and hence an overall scaling of  $\mathcal{O}(g^{2k})$. They show, using an active space of $4$ electrons on $4$ molecular orbitals for $\mathrm{H_2O}$, with $k=1$, one can recover most of the electron correlation energy. Reaching chemical accuracy requires between $k=5$ and $k=10$ depending on the bond distance. That considered, numerical analysis of the growth in the number of terms in the Hamiltonian when using iQCC shows a plateau after a number of iterations, owing to the fact that many terms will have negligible weights, resulting in a low polynomial scaling \cite{Genin2022}. Finally, in Ref. \cite{Ryabinkin2021}, Ryabinkin et al. propose to use ex-post corrections on the energy obtained from iQCC based on perturbation theory, to reduce the number of iteration $k$ required for the implementation of the method.

Overall, despite the promising fixed-length ansatz, the viability of the method depends on further research regarding scalability and extension of the method to very large systems.The analysis presented by Genin {\it et al.}  \cite{Genin2022} suggests that that iQCC could compete with methods such as DFT in terms of scaling and accuracy using a minimum of 72 fully-connected and corrected qubits. 

\paragraph{ClusterVQE:} In Ref.~\cite{Zhang2021}, Zhang et al. present a method that borrows ideas from ADAPT-VQE, iQCC and Mutual Information (MI) assisted Adaptive VQE to optimize the use of quantum resources. The first step in this method is to create a cluster of qubits (across the qubit register of the QPU used) that have strong MI within, and low MI across, each cluster. MI was shown to be a strong metric to measure the correlation between these quantum clusters \cite{Rissler2006}, computed as presented in Eq.~(\ref{eq:mutual_information}). As described in Sec. \ref{sec:MI_assisted_VQE}, the MI can be initially approximated using a DMRG approximation of the ground state wave function. Once the MI for each qubit pair has been computed, a conventional clustering algorithm can be used to create a set of clusters that minimizes the MI between clusters (sum over MIs between inter-cluster qubits).

The idea behind ClusterVQE is to separate the pool of operators (akin to those found in other adaptive methods) into entanglers that act within a cluster of qubits (noted $U_{c_i}$, $i$ labeling a specific cluster) and those acting on two different clusters, coupling them (noted $U_{c_i, c_j}$). The former are used to grow the ansatz, similarly to qubit-ADAPT-VQE (see Sec. \ref{sec:adapt-vqe}), the latter are used to dress the Hamiltonian similarly to iQCC (see Sec. \ref{sec:iqcc}). The ClusterVQE optimization problem can then be written as
\begin{equation}
    E = \min_{\boldsymbol{\theta, \phi}}\bra{\psi_{HF}} \prod_i U_{c_i}^{\dagger}(\theta_i) \hat{H}^d(\boldsymbol{\phi}) \prod_i U_{c_i}(\theta_i) \ket{\psi_{HF}},
\end{equation}
with, the dressed Hamiltonian
\begin{equation}
    \hat{H}^d(\boldsymbol{\phi}) = \prod_{i\neq j}U_{c_{ij}}^{\dagger}(\phi_{ij}) \hat{H} \prod_{i\neq j}U_{c_{ij}}(\phi_{ij}).
\end{equation}

The optimization process works similar to the methods mentioned above, where a pool of entanglers is first created (Zhang et al. \cite{Zhang2021} recommend selecting qubit-operators from the pool generated by single and double fermionic excitations) and ranked based on their gradients, computed as in Eq.~(\ref{eq:adapt_vqe_grad}). Entanglers that couple qubit clusters are used to dress the Hamiltonian, others are used to grow the ansatz to be implemented on the quantum computer. Once top entanglers have been selected and integrated, a VQE optimization is conducted. The wave function obtained is then used to update the MI (which can be computed from the one and two-body reduced density matrices) and a new loop begins.

A first observation to make on the ClusterVQE is that it allows separating the ansatz into smaller, shallower depth quantum circuits corresponding to each qubit cluster \cite{Zhang2021}. The quantum circuit for each cluster can be optimized separately as their gradients are independent of the rest of the ansatz, allowing for implementation of ClusterVQE on smaller quantum computers. It is worth noting however that doing so means that the number of measurements required must be multiplied by the number of qubit-clusters, though these can be parallelized. It also means that the number of terms in the Hamiltonian is lower than the iQCC method as only certain entanglers are integrated into the dressed Hamiltonian.

Finally, Zhang et al. \cite{Zhang2021} show numerically that ClusterVQE can achieve similar accuracy as iQCC but with significantly lower numbers of Hamiltonian terms. The shallower circuits also offer some noise resilience compared to VQE and qubit-ADAPT-VQE, though iQCC remains the most resilient due to its fixed ansatz depth.
A few caveats are worth mentioning, however. Despite promising features, the dressed Hamiltonian still grows exponentially in the number of cross-qubit-cluster entanglers; while these have been shown to be few for small systems (e.g. $\mathrm{LiH}$ in Ref. \cite{Zhang2021}) there is at this stage no way to tell how this method scales on large, highly entangled systems. Also noted by Zhang et al., a potential bottleneck for the method is the inter-MI clustering process which could become too costly to compute accurately enough.

\subsubsection{Ansatz structure optimization} \label{sec:gate_structure_optimization}

Several methods have been developed to iteratively update the structure of the ansatz throughout the optimization process. Unlike the adaptive ans\"atze presented in the previous section, these methods do not propose to build the circuit layer by layer but start from an initial guess of the ansatz structure to progressively learn an optimal one. We outline a few of these methods at a high level below. 

\paragraph{RotoSelect \cite{ostaszewskiStructureOptimizationParameterized2021}:} this method is built as an extension to the RotoSolve optimizer \cite{Nakaji2021, ostaszewskiStructureOptimizationParameterized2021}, which we present in more details in Sec.~\ref{sec:analytical_opt}. Rotosolve uses the fact that the expectation value of the Hamiltonian is a $2\pi$ periodic sine-type function of a given parameter to find the optimal value. This can be done by sampling three different values of the parameter and finding the minimum using trigonometric relationships. RotoSelect extends that idea, by optimizing not only the parameter but also the Hilbert space dimension of the rotation gate. In other words, the algorithm finds which orientation, $X, Y$, or $Z$, allows for the largest reduction in energy for a given rotation gate and replaces the orientation of the gate accordingly. This is associated with an additional sampling cost on the initial iterations of the VQE, but has been shown in Ref.~\cite{ostaszewskiStructureOptimizationParameterized2021} to significant improve convergence rate over RotoSolve and other gradient-based optimizers. These results were tested on HEA, and could be different on the likes of UCC-based ans\"atze.     

\paragraph{Variable Ansatz (VAns) \cite{Bilkis2021}:} The VAns is another algorithm that proposes adapting the structure of the ansatz as the optimization proceeds. Unlike RotoSelect, mentioned above, it not only changes the axes of rotation gates but also incorporates and removes gates throughout the process. The first step in this method is to define a dictionary of parametrized gates. From an initial circuit configuration (which can be flexibly designed), pre-defined insertion rules stochastically select gates from the dictionary to add them to the circuit. Additional simplification rules are used to identify gates with lower impact on the resulting energy and remove them from the ansatz. The ansatz is primarily designed for use on Quantum Machine Learning tasks which cannot benefit from a problem tailored ansatz (as it is the case for VQE), but is also shown to outperform HEA on the Ising and Heisenberg models, and on $\mathrm{H_2}$ and $\mathrm{H_4}$ \cite{Bilkis2021}. The key advantage of this approach is the large flexibility it provides in terms of designing rules, allowing for efficient design and tailoring to any learning task, while offering some degree of resilience to the barren plateau problem. The cost of implementing the method is unclear however, and while it reduces the burden on quantum resources, it could result in a large pre-factor cost for its implementation, requiring further research. 

\paragraph{Machine learning of ansatz structure and Quantum Architecture Search (QAS):} These comprise a series of methods aiming at learning an optimal ansatz structure using machine learning tools. While it would probably deserve a review in its own right, we outline here some key publications regarding this family of approaches. The Evolutionary VQE (EVQE) \cite{Rattew2019} uses evolutionary programming to dynamically generate and optimize the ansatz and is shown to produce ansatz which are nearly twenty times shallower (with twelve times fewer CNOT gates) than UCCSD. Rattew et al. \cite{Rattew2019} also show numerically a significant ability of the EVQE to learn and thereby reduce the impact of inherent quantum noise, all while providing resilience to the barren plateau problem. Similarly, the Multi-objective Genetic VQE (MoG-VQE) \cite{Chivilikhin2020} uses a genetic algorithm to learn the circuit topology of a series of blocks constituting the ansatz. This results in blocks of gates resembling the HEA structure, but only spanning a limited number of qubits. The method achieves a significant reduction in two-qubit gates compared to HEA (showing that one can reach chemical accuracy on a 12-qubit version of $\mathrm{LiH}$ with only 12 CNOTs \cite{Chivilikhin2020}). 

QAS was initially developed in Ref.~\cite{Du2020} as a means to improve both trainability and noise resilience of variational quantum algorithms. Similar to the methods described above, QAS modifies the optimization problem by incorporating the structure of the quantum circuit in the cost function. In Ref.~\cite{Zhang2020_QAS}, Zhang et al. extend this idea by creating the Differentiable QAS (DQAS), which is based on the differentiable neural architecture search (DARTS) \cite{Liu2018}. The idea is to relax the discrete search space of network architectures onto a continuous and differentiable domain, accelerating QAS \cite{Zhang2020_QAS}. Additional work on QAS includes a version based on Neural Predictors \cite{Zhang2021_QAS}, on Deep Reinforcement Learning \cite{Kuo2021}, and on meta-learning \cite{Chen2021_QAS}. QAS can also be used to improve noise resilience of an ans\"atze \cite{Wang2021_QAS}.
Overall, while very promising, machine learning based ansatz all suffer from a common pitfall: the cost of training machine learning models to optimize ansatz for large-scale systems remains largely untested. Studying the training cost and cost scaling for these methods would be an interesting research avenue to warrant their applicability on larger systems than has been analyzed so far. 

\subsection{Discussion on ans\"atz used in the VQE context} \label{sec:ansatz_discussion}

There has been a very large amount of research published on VQE ans{\"{a}}tze over recent years, we summarize in this subsection the key findings from our research. The k-UpCCGSD ansatz \cite{Lee2019} (and extensions) have been numerically shown to achieve excellent accuracy while offering linear scaling, thereby arguably offering the best trade-off in cost to accuracy among ans\"atze proposed at the time of writing. Adaptive ans{\"{a}}tze could place themselves as a reliable alternative subject to further studies on their expected computational cost.
Layered HEA are likely not suitable for large systems. As they are not physically motivated, they require a large number of parameters and need to span a significant proportion of the Hilbert space to guarantee a good ground state approximation can be reached. This also makes HEA particularly prone to barren plateaus \cite{Holmes2021}. There are good arguments however to suggest that VQE may be resilient to barren plateaus if the right design choices are made in the ansatz (except for Noise-Induced Barren Plateau \cite{Wang2020}). In particular, these are; local encodings for the Hamiltonian \cite{Cerezo2021_BP, Uvarov2020}, an ansatz that is problem tailored (e.g. UCC), or is constructed adaptively during the optimization (e.g. ADAPT-VQE \cite{Grimsley2019} and others), using specific initialization techniques (e.g. Ref.~\cite{Grant2019, Patti2021}, and use of a local mapping for the Hamiltonian (e.g. Bravyi-Kitaev).

Qubit-based ans{\"{a}}tze \cite{Ryabinkin2018, Ryabinkin2020, Lang2020} (or qubit-excitations based ans{\"{a}}tze) could be more appropriate for the NISQ era than fermion based (or hardware efficient) ans{\"{a}}tze. That is because they offer the option to parametrize the ansatz with respect to each Pauli string used in the construction. Given enough variational flexibility, this allows forgoing Pauli-Z chains used for maintaining the antisymmetric relationship of fermionic operators under the Jordan-Wigner mapping. As such, each operator is acting on very few qubits (up to four in the Jordan-Wigner mapping, instead of $\mathcal{O}(N)$ when using a fermionic ansatz), reducing the reliance on costly two-qubit gates and allowing depth reduction. These however tend to require a larger number of parameters and operators, and as such may be more cumbersome to optimize and have slightly worse computational scaling.

Adaptive ans{\"{a}}tze have been a significant focus of recent research. While they have brought several critical breakthroughs, the scaling of the number of measurements required remains unclear. We find that the main advantages of adaptive ans{\"{a}}tze are the reduction in number of parameters (though as far as we have seen, less than an order of magnitude), and numerically demonstrated resilience to barren plateau. This however comes at a significant computational cost of gradient estimations (scaling with the size of the Hamiltonian) and repeated VQE optimizations (shallow at the beginning of the process, but not at the end). The relative benefit of adaptive ansatz should be studied in more detail.

Some research has been conducted on the impact of initialization on the performance of different ans\"tze, in particular, Choquette \textit{et al.} \cite{Choquette2021} numerically show that a computational basis initialization for HEA has a minor impact on target state fidelity compared to Hartree-Fock initialization. Of course, UCC-based ans\"atze are motivated as an evolution of a reference state, and, in general, apply to a qubit register initialized as the Hartree Fock wavefunction \cite{Peruzzo2014}. As a side note, Murta and Fern\'andez-Rossier \cite{Murta2021} propose a quantum routine to model the Gutzwiller wavefunction \cite{Gutzwiller1963} on a quantum device. This wavefunction is a physically-motivated model which can be applied to a variety of lattice systems \cite{Hubbard1963}. While not providing a full parametrized ansatz, it was shown to provide a better approximation than a mean-field method for the ground state of the Fermi-Hubbard model, and could therefore be used as an initialization strategy for lattice model VQE implementations \cite{Murta2021}. Further research on the preparation of the Gutzwiller wavefunction is presented in Ref. \cite{Seki2022}.

%% file: 07_optimization_main.tex
\section{Optimization strategies} \label{sec:Optimization}

The VQE is in essence an optimization problem, it aims at heuristically constructing an approximation of an electronic wavefunction through iterative learning of ansatz parameters. For the algorithm to be viable, it must be that it can learn a good enough approximation of the solution within a tractable number of learning steps. It was already demonstrated that optimization of the variational quantum ans{\"{a}}tze is NP-hard~\cite{Bittel2021}, meaning that there exist at least some problems in which finding an exact solution for the VQE problem is intractable. As such, efficient optimization strategies that provide a well-approximated solution within an acceptable number of iterations are essential for any variational algorithms to be put into practice. Compared to the conventional numerical optimization problem, however, optimizing the expectation value of a variational quantum ansatz faces additional challenges: 
\begin{itemize}
    \item Sampling noise and gate noise on NISQ devices disturb the landscape of the objective function. Such noise can be detrimental to the convergence of optimization~\cite{gentini_noise-resilient_2020, Wang2020}, and could limit the scope for quantum advantage~\cite{Franca2021}. 
    \item While the precision of conventional numerical optimization is generally not considered a problem, the precision of the measured expectation value is limited by the sample shot number. The cost of optimization is heavily dependent on the precision required for optimization. 
    \item Related to the point above, the landscape of the expectation value of variational ansatz may cause the vanishing of gradient very easily as a result of the barren plateau problem~\cite{McClean2018} (See Sec. \ref{sec:barren_plateau} for further details).
\end{itemize}

On the flip side, studies from recent years show the landscape of expectation value has some analytical properties that are useful to extract information, such as evaluating gradients directly on quantum devices~\cite{nakanishi_sequential_2020,ostaszewskiStructureOptimizationParameterized2021, Romero2019,schuld_evaluating_2019}. In addition, the ansatz' landscape can be efficiently approximated to accelerate the convergence~\cite{koczor_quantum_2020,parrish_jacobi_2019}. Utilizing such prior knowledge helps to develop efficient optimization strategies for variational quantum algorithms.   
This section presents the most relevant optimizers used in the context of the VQE and recent studies of optimization strategies, focusing on the strategies that target fast convergence rates. Some optimizers are developed to reduce the shot numbers \cite{Arrasmith2020,Kubler2020adaptiveoptimizer}, for which specific applications are discussed in  section \ref{sec:measurements_strategies}. We also compare these proposed algorithms and discuss the current challenge facing the optimization of variational quantum algorithms.  

\subsection{Background and notation}  

The objective function of a variational algorithm is constructed conventionally based on the measurement outcome. Denote $\mathbf{O}(\boldsymbol{\theta}) = (\hat{O}_1(\boldsymbol{\theta}^{(1)}),\hat{O}_2^{(2)}(\boldsymbol{\theta}^{(2)}),\dots,\hat{O}_a(\boldsymbol{\theta}^{(a)}))$ are the observables used to compose the objective function and $a$ is the number of observables. The objective function is given by 
\begin{equation}
    \mathcal{L}(\boldsymbol{\theta}) = C(\mathbf{O}(\boldsymbol{\theta}))
\end{equation}
where $C$ is a function that maps the observed expectation value to the objective function, and usually have the simple linear form 
\begin{equation}
    C(\mathbf{X}) = \sum_i c_i X_i
\end{equation}
where $c_i$ is a constant defined by the problem as the coefficient of each measurement expectation value, $X_i$ is the $i$-th component of $\mathbf{X}$. Note that such linear form preserves the analytical properties and it is essential for using the analytical methods to directly calculate the gradient or implement analytical gradient free optimization strategy~\cite{CostFunction2020Cao,nakanishi_sequential_2020,ostaszewskiStructureOptimizationParameterized2021,koczor_quantum_2020,schuld_evaluating_2019,Romero2019}.  

A measurement expectation value is given by
\begin{equation}
    \braket{\hat{O}_k(\boldsymbol{\theta}^{(k)})} = \bra{\psi_0}U^{(k)\dagger}(\boldsymbol{\theta}^{(k)}) \hat{M}_{k} U^{(k)}(\boldsymbol{\theta}^{(k)})\ket{\psi_0},
\end{equation}
where $\ket{\psi_0}$ is the initial state on the quantum computer. 
$\hat{M}^{(k)}$ is a Hermitian measurement operator, usually chosen to be the tensor product of Pauli operators to match the physical measurement implementation on quantum hardware.  $U_k(\boldsymbol{\theta}^{(k)})$ is the variational ansatz defined as
\begin{equation}
    U^{(k)}(\boldsymbol{\theta}^{(k)}) = \prod_j U^{(k)}_j(\theta^{(k)}_j)
\end{equation}
and each $U_j$ is a quantum gate, which is generalized as
\begin{align}
U^{(k)}_{j}(\boldsymbol{\theta}^{(k)}_j) = \exp(i \theta^{(k)}_{j} P^{(k)}_j)
\end{align}
where $P_j^{(k)}$ is a Hermitian matrix, usually is a tensor product of Pauli operators. 

It is sometimes convenient to utilize the superoperator formalism and consider the noise into the optimization process, the expectation value can be written as
\begin{equation}
    \braket{\hat{O}_k(\boldsymbol{\theta}^{(k)})} = \operatorname{Tr}[\hat{M}_k\Phi^{(k)}(\boldsymbol{\theta}^{(k)})\rho_0],
\end{equation}
where $\rho_0$ is the initial state density operator and $\Phi^{(k)}(\boldsymbol{\theta}^{(k)})$ denote the transformation matrix, which is given by 
\begin{equation}
    \Phi^{(k)}(\boldsymbol{\theta}^{(k)}) = \prod_j \Phi^{(k)}_j(\theta^{(k)}_j).
\end{equation}

For devices that support single shot readout, each sample the quantum device would yield a bit string $\mathbf{s}$. For each string $\mathbf{s}$ a measurement value could be calculated with $M_i(\mathbf{s})$, and the expectation value is the average of each $M_k(\mathbf{s})$.
\begin{equation}
    \braket{\hat{O}_k(\boldsymbol{\theta}^{(k)})} = \sum_j \mathrm{Prob(\mathbf{s}}(\boldsymbol{\theta}^{(k)}))=b_jM_k(b_j)
\end{equation}
where $b_j \in B$, $B$ covers all possible single shot bit string (all binary numbers from $0$ to $2^{n-1}$) of the measurement outcome.

Due to the physical implementation from the quantum hardware, not all quantum computing systems support single-shot readout. Some systems can only yield expectation value by averaging the signal from the readout. For example, some NMR systems use an ensemble of molecules to implement quantum computing and cannot read the state of each single molecule~ \cite{ebel_dispersive_2021,lu_nmr_2015}. In practice, the quantum hardware system may directly yield an expectation value, and the measurement approaches vary from different physical systems.  

In the following discussion, the upper label "$(k)$" is simplified when there is no ambiguity.

\subsection{Gradient evaluation}

A significant amount of numerical optimization methods requires the knowledge of the gradient of the objective function at a given parameter value. Therefore the evaluation of gradient is rather important and we would cover this topic in a separate subsection. 

\subsubsection{Stochastic approximation methods }\label{sec:stochastic_approximation}

Stochastic approximation (SA) is a family of methods used to reconstruct the properties of the expectation value of a function that depends on some random variables. Instead of measuring the expectation values directly, SA recovers the properties of the expectation value with random sampling. SA has been widely used for big data and machine learning applications when the objective function is too costly to evaluate directly. In the context of variational quantum algorithms, the objective function is some function of the expectation value of the quantum state; evaluating the expectation value of a variational ansatz is expensive. Therefore SA naturally fits into the context of variational quantum algorithms and has been used to approximate the gradient of the expectation values.      

\paragraph{Finite difference stochastic approximation (FDSA)}

One of the simplest methods to approximate the gradient is evaluating two function points and using its difference as the approximated gradient. Finite difference stochastic approximation (FDSA)~\cite{kiefer_stochastic_1952} is given by
\begin{equation}
    ({g}(\boldsymbol{\theta}_{t}))_{j}
        =
        {\frac {    \mathcal{L}(\boldsymbol{\theta}+c_{t}\mathbf{e}_{j})    -    \mathcal{L}(\boldsymbol{\theta}-c_{t}\mathbf{e}_{j})    }
        {2c_{t}}},
\end{equation}
where $({g}(\boldsymbol{\theta}_{t}))_{j}$ is $j$-th component of the gradient at point $\boldsymbol{\theta}_t$. $\mathbf{e}_j$ is the unit vector of the $j$-th dimension, $c_t$ is the displacement value that generally decrease with iteration step number $t$ \cite{kiefer_stochastic_1952}.

\paragraph{Simultaneous perturbation stochastic approximation (SPSA)}

SPSA algorithm approximates the gradient with only two measurements of the objective function. Instead of choosing the measurement points symmetrically, SPSA uses a small random vector to perturb the objective function. While FDSA calculates the gradient at one direction of the variable, SPSA calculates a direction composed of multiple variables.
\begin{equation}
    ({g}(\boldsymbol{\theta}_{t}))_{j}
        =
        {\frac {    \mathcal{L}(\boldsymbol{\theta}+c_{t}\mathbf{\Delta}_{j})    -    \mathcal{L}(\boldsymbol{\theta}-c_{t}\mathbf{\Delta}_{j})    }
        {2c_{t}(\Delta_t)_j}},
\end{equation}
where $\Delta_t$ is a random perturbation vector. The SPSA algorithm needs to measure two points per iteration, which is friendly to variational quantum algorithms. Also it can be applied to noisy optimization~\cite{s_performance_2012,morison_spsa_2003,wang_mixed_2018}. 

The SPSA methods can approximate the preconditioned gradient for second order optimization methods~\cite{Spall1997AcceleratedMeasurements}. Explanation of second order optimization methods are covered in Sec. \ref{sec:second_order_optimizers}. Consider the preconditioned gradient $g^\prime(\boldsymbol{\theta}) = F^{-1} g(\boldsymbol{\theta})$ where $F$ is some metric with information from second order derivatives of function $f(\boldsymbol{\theta})$. Denote $\Tilde{F}$ as the approximated value for $F$, we have

\begin{equation}
    \Tilde{F}^{(k)} = \frac{\delta f}{2\epsilon^2}\frac{\Delta^{(k)}_1\Delta^{(k)T}_2+\Delta^{(k)}_2\Delta^{(k)T}_1}{2}
\end{equation}

where 
\begin{align}
    \delta f^{(k)} &= f(\boldsymbol{\theta}^{(k)}+\epsilon\Delta_1^{(k)} +\epsilon\Delta_2^{(k)}) \\ 
    &- f(\boldsymbol{\theta}^{(k)}+\epsilon\Delta_1^{(k)}) \\
    &- f(\boldsymbol{\theta}^{(k)}-\epsilon \Delta_1^{(k)} + \epsilon \Delta_2^{(k)}) \\ 
    &+ f(\boldsymbol{\theta}^{(k)}-\epsilon \Delta_1^{(k)}) \\ 
\end{align}

The approximated value is estimated from all previously evaluated values with an exponentially smooth estimator 
 
 \begin{equation}
 \bar{F}^{(k)} = \frac{k}{k+1} \bar{F}^{(k-1)} + \frac{1}{k+1} \Tilde{F}^{(k)}
 \end{equation}


\subsubsection{Analytical gradient calculation}

The gradient of measurement observables can be evaluated on quantum computers directly by utilizing the analytical property of the ansatz. For ans{\"{a}}tze made from a sequence of parametrized quantum gates, the partial derivative is
\begin{align}\label{eq:exactDerivative}
\frac{\partial \braket{\hat{O}_k(\boldsymbol{\theta})}}{\partial \theta_j} &= 
2 \operatorname{Im} (\langle \phi_0| V^{j\dagger}_{k}(\boldsymbol{\theta}) \hat{M}_k U(\boldsymbol{\theta})|\phi_0\rangle),
\end{align}
where the operator $V^{j}_{k}(\mathbf{t})$ is defined as inserting $P^j_k$ between the $(j-1)$-th term and $j$-th term, which comes from the derivative of the $j$-th term:
\begin{align}
V^{j}_{k}(\boldsymbol{\theta}) = & e^{i \theta_{N_k} P^{N_k}_k} \cdots e^{i \theta_{j} P^{j}_{k}} P^j_{k} e^{i \theta_{j-1} P^{j-1}_{k}} \cdots  e^{i \theta^{1} P^{1}_{k}}.
\end{align}

The difficulty of evaluating the gradient directly is that $V^{j\dagger}_{k}(\boldsymbol{\theta}) \hat{M}_k U(\boldsymbol{\theta})$ is usually not Hermitian, therefore there is no known method that can directly convert it to a quantum circuit. Two methods to analytically measure the gradient have been developed, and it is worth noticing that these two methods are measuring the gradient with the exact same mechanism, but with direct and indirect measurements~\cite{mitarai_methodology_2019}. Here, direct measurement means the observed quantity is encoded into a single quantum observable, while indirect measurement calculates the quantity with multiple quantum observables. 


\paragraph{Direct analytical gradient measurement} 

The imaginary part of $\langle \phi_0| V^{j\dagger}_{k}(\boldsymbol{\theta}) \hat{M}_k U(\boldsymbol{\theta}) |\phi_0\rangle$ can be directly evaluate by introducing an ancillary qubit~\cite{Romero2019}. This is done by first prepare the ancillary qubit into the $1/(\sqrt{2})(\ket{0}+\ket{1})$ state, and prepare the rest of quantum register into $\ket{\phi(\boldsymbol{\theta})}$ state. Now the system has the state
\begin{align}
    \ket{\psi} = \frac{\sqrt{2}}{2} (\ket{0}+\ket{1})\otimes \ket{\psi(\boldsymbol{\theta})}.
\end{align}
We can then apply the $P_{j}$ gate controlled by the ancillary qubit and apply the remaining unitary. Finally we apply the measurement observable controlled by the ancillary qubit. This gives the new state
\begin{align}
    \ket{\psi} =\frac{(\ket{0}\otimes U(\boldsymbol{\theta})\ket{\psi_0}+\ket{1}\otimes \hat{M}_k V_k^j(\boldsymbol{\theta})\ket{\phi_0})}{\sqrt{2}}.
\end{align}
Now we apply the Hadamard gate to the ancillary qubit, which gives
\begin{align}
\ket{\psi} = \frac{\ket{0}\otimes(U\ket{\phi_0}+ \hat{M}_k V^j_k(\boldsymbol{\theta})\ket{\phi_0}) + \ket{1}\otimes(U\ket{\phi_0} - \hat{M}_k V^j_k(\boldsymbol{\theta})\ket{\phi_0})}{2}.
\end{align}
The imaginary part of $\langle \phi_0| V^{j\dagger}_{k}(\boldsymbol{\theta}) \hat{O}_i  U(\boldsymbol{\theta})|\phi_0\rangle$ is now encoded in the ancillary qubit in the $Y$-basis.
\begin{center}
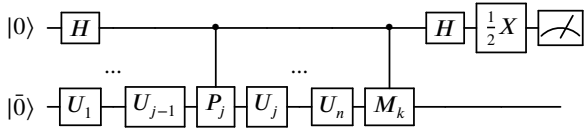
\begin{figure}[hbt]
\[\Qcircuit @C=0.5em @R=.7em {
\lstick{\ket{0}}&\gate{H}&\qw&\qw&\ctrl{2}&\qw&\qw&\qw&\ctrl{2}&\gate{H}& \gate{\frac{1}{2}X} &\meter\\
&&...&&&&...&\\
\lstick{\ket{\bar{0}}}&\gate{U_1}&\qw&\gate{U_{j-1}}&\gate{P_{j}}&\gate{U_{j}}&\qw&\gate{U_{n}}&\gate{M_{k}}&\qw&\qw&\qw\\
}\]
\label{fig:gradient_measurement}
\caption{Quantum circuit that evaluates $\operatorname{Im} (\langle \phi_0| V^{j\dagger}_{k}(\boldsymbol{\theta}) \hat{M}_k U(\boldsymbol{\theta})|\phi_0\rangle)$.}
\end{figure}
\end{center}

\paragraph{Indirect analytical gradient measurement} 

To illustrate this method first we make some modifications to our notations. We now absorb the common term of $U$ and $V$ into the state and measurement operator, and define
\begin{equation}
\begin{split}
    W_{k,1}(\boldsymbol{\theta}) &= \exp(i \theta_{j-1} P^{j-1}_{k}) \cdots  \exp(i \theta^{1} P^{1}_{k}) \\
    W_{k,2}(\boldsymbol{\theta}) &= \exp(i \theta_{n} P^{n}_k) \cdots \exp(i \theta_{j} P^{j}_{k}) \\
    \ket{\phi(\boldsymbol{\theta})} &= W_{k,1}(\boldsymbol{\theta})\ket{\phi_0} \\
    Q_k(\boldsymbol{\theta}) &=  W_{k,2}^\dag \hat{M}_i  W_{k,2}.
\end{split}
\end{equation}
Then we have 
\begin{equation}
    \frac{\partial \braket{\hat{O}_k(\boldsymbol{\theta})}}{\partial \theta_j} = 2 \operatorname{Im} (\langle \phi(\boldsymbol{\theta})| Q_k(\boldsymbol{\theta}) P_{k} |\phi(\boldsymbol{\theta})\rangle).
\end{equation}
We use following construct which is related to $\bra{\phi(\boldsymbol{\theta})}P_kQ_k(\boldsymbol{\theta})\ket{\phi(\boldsymbol{\theta})}$:

\begin{equation}
    \langle \phi(\boldsymbol{\theta})| (\unit \mp i P_{k})Q_k(\boldsymbol{\theta})(\unit \pm i P_{k} |\phi(\boldsymbol{\theta})\rangle  = \mp  \operatorname{Im} (\langle \phi(\boldsymbol{\theta})| P_{k}Q_k(\boldsymbol{\theta}) |\phi(\boldsymbol{\theta})\rangle) + C,
\end{equation}
where $C = \bra{\phi(\boldsymbol{\theta})}    Q_k(\boldsymbol{\theta})  \ket{\phi(\boldsymbol{\theta})} +  \bra{\phi(\boldsymbol{\theta})} P_k Q_k(\boldsymbol{\theta})P_k   \ket{\phi(\boldsymbol{\theta})})$ is a constant value.

For single qubit gate given by $U_j(\theta_j) = \operatorname{exp}(i\theta_jP_k), P_k \in {X,Y,Z}$ where ${X,Y,Z}$ are Pauli matrices and with exactly two eigenvalues, we have~\cite{schuld_evaluating_2019}
\begin{align}\label{eq:optimization:shift_rule_principal}
    \unit \pm i P_{k} = \operatorname{exp}(\mp\frac{\pi}{4} i P_k),
\end{align}
therefore
\begin{align}
    &\langle \phi(\boldsymbol{\theta})| (\unit \mp i P_{k})Q_k(\boldsymbol{\theta})(\unit \pm i P_{k} |\phi(\boldsymbol{\theta})\rangle \\=& \bra{\phi(\boldsymbol{\theta} \pm \frac{\pi}{2}\mathbf{e}_j)}    Q_k(\boldsymbol{\theta} \pm \frac{\pi}{2}\mathbf{e}_j)  \ket{\phi(\boldsymbol{\theta} \pm \frac{\pi}{2}\mathbf{e}_j)},
\end{align}
and 
\begin{align}
    \frac{\partial \braket{\hat{O}_k(\boldsymbol{\theta})}}{\partial \theta_j} = \braket{\hat{O}_k(\boldsymbol{\theta}+\frac{\pi}{2}\mathbf{e}_j)} - \braket{\hat{O}_k(\boldsymbol{\theta}-\frac{\pi}{2}\mathbf{e}_j)}.
\end{align}
The parameter shift rule can be generalized to a general parameter shift rule and give gradient $g_{gPSR}$ to obtain higher-order derivatives~\cite{schuld_evaluating_2019,Hubregtsen2022}.
\begin{equation}
    g_{gPSR}(\theta,\gamma_1,\gamma_2) := r[L(\theta + \gamma_1) + L(\theta + \gamma_2)]
\end{equation}

where $2r$ is the difference of the eigenvalue of the gate generator $G$. For single-qubit gates with Pauli operators as the generator, $r=1$. 

\begin{equation}
    g_{gPSR}(\theta,\gamma_1,\gamma_2) = \frac{sin(2r\gamma_1)+sin(2r\gamma_2)}{2}g^{(1)}(\theta) - \frac{cos(2r\gamma_1)-cos(2r\gamma_2)}{4r}g^{(2)}(\theta),
\end{equation}
and because of the sinusoidal property of the expectation value, the $k$-th order derivative has a constant multiplier different to the $k-2$-th order derivative.
\begin{equation}
 g^{(k+2)} = -\frac{1}{4r} g^{(k)}.
\end{equation}

So far the analytical gradient methods only apply to parametrized single-qubit gates. The relation presented in Eq. \ref{eq:optimization:shift_rule_principal} holds only when $P_k$ has exactly two different eigenvalues. To use the parameter shift rule for arbitrary two-qubit gates, one can decompose the multi-qubit gate into a sequence of single-qubit gate and product of the same Pauli matrices, which always has two different eigenvalues~\cite{crooks_gradients_2019}. When the operators has 3 different eigenvalues, the gradient can be evaluated with a modified 4-value shift rule~\cite{izmaylov_analytic_2021}. For other operators, the value $P$ can be polynomially expanded into a linear combination of low-rank (2 or 3 eigenvalues) operators~\cite{izmaylov_analytic_2021}. Alternatively, the objective function can be decomposed into trigonometric polynomials, and the gradient can be evaluated with trigonometric interpolation methods~\cite{GeneralGradientsWierichs2022,OptimalityCircuits2021Theis}.

It is worth mentioning that for VQE, the parameter shift rule can be implemented before encoding the fermionic excitation to the qubits, which reduces the required measurement amount~\cite{Kottmann2021_3}.

\subsection{Gradient-based searching strategy}

Gradient-based optimization strategies utilize information from cost function derivatives. First-order optimizers utilize only the first-order derivatives of the cost function, and second-order optimizers utilize both first-order and second-order derivatives, at the cost of computing the second-order derivatives. In this section, we discuss some of the popular gradient-based optimizers.

\subsubsection{First order optimizers}

First-order optimizers for variational ansatz are mostly borrowed from the deep learning communities. These methods have been widely used for the early studies of variational quantum algorithms.

\paragraph{Simple gradient descent}

Simple gradient descent is the simplest first-order method searching for local minimum with gradient~\cite{Lemarechal12cauchyand,Courant1943Variational}. The algorithm takes a step towards the opposite direction of the gradient, and the step size is calculated based on the absolute value of the gradient and a meta-parameter $\eta$ usually referred to as the learning rate. Almost all the gradient-based methods are developed based on the idea of simple gradient descent. 

\vspace{0.5cm}

\begin{algorithm}[H]
\SetAlgoNoLine%
\SetKw{KwVariable}{~} 
\KwVariable{$\eta$: Learning rate.}\\
 \While{Not converged}{
  Calculate gradient $g_t = g(\theta_t)$\;
  $\theta_{t+1} = \theta_{t} - \eta g_{t}$
 }
\caption{Simple gradient descent}
\end{algorithm}

Adaptive optimizers well developed from the deep learning community are adapted to the ansatz parameter optimization. 

\paragraph{RMSProp}
RMSProp is an adaptive learning rate optimization method. The RMSProp divides the gradient by the weighted moving average of the root mean square of gradient value, which enhances the direction of the gradient and reduces the significance of the gradient magnitude.

\vspace{0.5cm}

\begin{algorithm}[H]
\SetAlgoNoLine%
\SetKw{KwVariable}{~} 
\KwVariable{$\gamma$: Moving average parameter.}\\
\KwVariable{$\eta$: Learning rate.}\\
 \While{Not converged}{
  Calculate gradient $g_t = g(\theta_t)$\;
  
  $E[g^2]_t = \gamma E[g^2]_{t-1} + (1 - \gamma) g^2_t$\;
  
  $\theta_{t+1} = \theta_{t} - \dfrac{\eta}{\sqrt{E[g^2]_t + \epsilon}} g_{t}$
 }
\caption{RMSProp optimizer}
\end{algorithm}

\vspace{0.5cm}


\paragraph{Adam optimizer}

Adaptive moment (Adam) optimizer \cite{kingma_adam_2017} is a widely used optimizer developed from the deep learning community to solve the stochastic gradient-based optimization problem. The Adam optimizer performs efficient stochastic optimization with only first-order gradient. Adam optimizers utilize adaptive moment estimation as an extra on RMSProp. Instead of descent to the gradient direction, it jumps towards the momentum direction, the weighted moving average of the gradient.

\vspace{0.5cm}

\begin{algorithm}[H]
\SetAlgoNoLine%
\SetKw{KwVariable}{~} 
\KwVariable{$\beta_{1}$: Moving average parameter for past gradients.}\\
\KwVariable{$\beta_{2}$: Moving average parameter for past squared gradients.}\\
 \While{Not converged}{  
  Calculate gradient $g_t = g(\theta_t)$
  
  $m_t = \beta_1 m_{t-1} + (1 - \beta_1) g_t$ \; 
  $v_t = \beta_2 v_{t-1} + (1 - \beta_2) g_t^2$ \;

$m^\prime_t = \dfrac{m_t}{1 - \beta^t_1}$ \; 
$v^\prime_t = \dfrac{v_t}{1 - \beta^t_2}$ \;
$\theta_{t+1} = \theta_{t} - \dfrac{\eta}{\sqrt{v^\prime_t} + \epsilon} m^\prime_t$ \;
 }
 \caption{Adam optimizer}
\end{algorithm}

\vspace{0.5cm}


Since the Adam optimizer utilizes moving average for both momentum and magnitude estimations, it can be applied to scenarios with noisy gradients.

\subsubsection{Second order optimizers} \label{sec:second_order_optimizers}

Second-order optimization techniques make use of the second-order derivative of the objective function to determine the descent direction. 

\paragraph{Broyden–Fletcher–Goldfarb–Shanno (BFGS) algorithm} 

The BFGS algorithm~\cite{broyden_convergence_1970,fletcher_new_1970,goldfarb_family_1970,shanno_conditioning_1970} is a gradient-based iterative approach to solve standard nonlinear optimization problems. The direction of each step is obtained by preconditioning the gradient with curvature information. Preconditioning the gradient in the context of optimization means it modifies the gradient before taking the descending step. The BFGS algorithm modifies the gradient based on the curvature information obtained by the Hessian. In practice, the Hessian is difficult to calculate precisely, even for conventional problems with large parameter space. A different variant of the BFGS method has been developed to approximate the Hessian, such as Limited-memory BFGS (L-BFGS) \cite{byrdt_limited_nodate}. 

\vspace{0.5cm}

\begin{algorithm}[H]
\SetKw{KwVariable}{~}
\KwVariable{$B_{k}$: Approximated Hessian at step $k$.} \\
\KwVariable{$\mathcal{L}(\theta)$: Objective function}\\
\KwVariable{$\alpha_k$: stepsize at step $k$ }\\
\KwVariable{$\mathbf{p}_k$: Descent direction}\\

\SetAlgoNoLine%

Calculate $B_0$ from initial guess $x_0$\;

 \While{Not converge}{
 Obtain $\mathbf{p}_{k}$ by solving $B_{k}\mathbf {p} _{k}=-\nabla \mathcal{L}(\mathbf {\theta} _{k})$\;
 Perform line search for $\alpha _{k}=\argmin \mathcal{L}(\mathbf {\theta} _{k}+\alpha \mathbf {p} _{k})$ \;
 Set $ \mathbf {s} _{k}=\alpha _{k}\mathbf {p} _{k}$ and update $ \mathbf {\theta} _{k+1}=\mathbf {\theta} _{k}+\mathbf {s} _{k}$\;
Set $\mathbf {y} _{k}={\nabla \mathcal{L}(\mathbf {\theta} _{k+1})-\nabla \mathcal{L}(\mathbf {\theta} _{k})}$\;
Set $ B_{k+1}=B_{k}+{\frac {\mathbf {y} _{k}\mathbf {y} _{k}^{\mathrm {T} }}{\mathbf {y} _{k}^{\mathrm {T} }\mathbf {s} _{k}}}-{\frac {B_{k}\mathbf {s} _{k}\mathbf {s} _{k}^{\mathrm {T} }B_{k}^{\mathrm {T} }}{\mathbf {s} _{k}^{\mathrm {T} }B_{k}\mathbf {s} _{k}}}$\;
 }
 \caption{BFGS algorithm}
\end{algorithm}

\vspace{0.5cm}

Although the Hessian of a variational ansatz landscape can be calculated with analytical method \cite{huembeli_characterizing_2021}, it is still difficult to calculate the full Hessian. Therefore this method is only used in the very early age of simulating conventional algorithms or as a baseline of comparison for lately developed optimizers. The SPSA algorithm from section \ref{sec:stochastic_approximation} can be used to approximate a Hessian.

\paragraph{Quantum natural gradient}

Natural gradient descent is a well-studied second-order optimization technique for numerical optimization. Natural gradient descent uses the steepest descent direction for the information geometry instead of taking each step with the gradient from parameter space~\cite{amari_natural_1998,martens_new_2020,wierichs_avoiding_2020}. A similar principle can be applied to optimizing the parameter of quantum variational ansatz, which is then referred to as quantum natural gradient descent. The idea of taking steps in another manifold can also be extended beyond information geometry~\cite{Wiersema2022OptimizingGradientFlow}.

Before diving into the details of quantum natural gradients, it is worth noting that the pure state quantum natural gradient descent is, in fact, equivalent to imaginary time evolution~\cite{stokes_quantum_2020}.
The imaginary time evolution is defined by
\begin{equation}
    \ket{\psi(\tau)} = A(\tau)e^{-H\tau}\ket{\psi(0)}.
\end{equation}
The state at $\tau \rightarrow \infty$ is the ground state of $H$. Instead of evolving the state directly, variational imaginary time evolution calculates the evolution direction with McLachlan's variational principle and simulates the evolution by variating the ansatz parameters~\cite{McArdle2019}.

The idea of natural gradient descent can be explained as follows. Consider the objective function $\mathcal{L}(\theta)$ is a likelihood function $\mathcal{L}(\theta|x)$ where $x$ denote the measurement outcome of the distribution. For a general optimization, the likelihood function is given by the definition of the problem. While the traditional gradient descent algorithm takes one step towards the gradient direction in the parameter space, such direction may not necessarily be the direction that most significantly changes the difference of the distribution of $\mathcal{L}(\theta|x)$. From the statistical point of view, we would like each step to be taken in the direction that maximizes the distribution difference of the cost function. The natural gradient is then invented to present the gradient of the difference of the distribution function, and the corresponding geometry space is called information geometry.

The likelihood function's distribution difference from different $\theta$ can be characterized by the Kullback–Leibler (KL) divergence \cite{Kullback1951}. The KL divergence defines a distance measure between $\mathcal{L}(\theta|x)$, and when parameter $\theta$ is mapped to the likelihood function, it creates a Riemannian manifold with a metric $F$, where for small vector $d\theta$ in the parameter space, the distance $ds$ on the manifold is 
\begin{equation}
    ds^2 = d\theta^T F d\theta.
\end{equation}
Here $F$ is the Fisher information metric, which is the Hessian of the KL divergence. The Riemannian manifold describes the information geometry of the problem. The natural gradient descent instead takes the gradient from the information geometry and maximizes the KL divergence for each step. The natural gradient is defined as 
\begin{equation}
    \Tilde{g} = F^{-1}g  =F^{-1} \grad \mathcal{L}(\theta).
\end{equation}

So far our discussion about natural gradient descent are with real space. To implement natural gradient descent for variational quantum algorithms, we can use the Fubini-Study metric as the Fisher Information metric in Hilbert space~\cite{safranek_simple_2018,liu_quantum_2020}, which is given by \cite{wierichs_avoiding_2020,yamamoto_natural_2019,stokes_quantum_2020,Straaten2020}
\begin{equation}
  \begin{split}
    \left(F\right)_{ij} &\coloneqq \operatorname{Re}({\braket{\partial_{i}\psi|\partial_{j}\psi})-\braket{\partial_{i}\psi|\psi}}\braket{\psi|\partial_{j}\psi},
  \end{split}
\end{equation}
where $\ket{\partial_i \psi}\coloneqq \frac{\partial U(\boldsymbol{\theta})}{\partial \theta_i}\ket{\psi_0}$. These quantities can be evaluated with similar techniques as the Hadamard test. More details of Hadamard test has been included in appendix \ref{sec:hadamard-test}.

There are a few challenges for the quantum natural gradient. The first is the quantum natural gradient is defined on pure states, while in practice we need to consider the noise and non-unitary evolution. To solve this problem, the Fubini-Study metric can be generalized to a non-unitary circuit \cite{koczor_quantum_2020}.
\begin{equation} 
	(F)_{ij} = \frac{1}{2} \Tr[\rho(\boldsymbol{\theta})   (L_i L_j + L_j L_i) ],
\end{equation}
where $L$ is the symmetric difference defined as 
\begin{equation} 
	\partial_k \rho(\boldsymbol{\theta})  =: \frac{\partial \rho(\boldsymbol{\theta})}{\partial \theta_k} =  \frac{1}{2}(L_k \rho(\boldsymbol{\theta}) +\rho(\boldsymbol{\theta}) L_k).
\end{equation}
Suppose the density matrix of mixed state has fidelity $f_{id} = 1-\epsilon$, which means the eigenvalue $\lambda_1 = f_{id}$. Such density matrix is given by
\begin{equation}
    \rho_\epsilon = (1-\epsilon)\ket{\psi_1}{\bra{\psi_1}} + \epsilon \sum_{k=2}^d \lambda_k \ket{\psi_k}\bra{\psi_k}.
\end{equation}
The Fisher information matrix on NISQ device can be approximated efficiently as
\begin{equation}
    (F)_{ij} = 2\operatorname{Tr}[\frac{(\partial_i\rho_\epsilon)(\partial_j\rho_\epsilon)}{f_{id}}] + \mathcal{O}(\frac{1-f_{id}}{d}).
\end{equation}
Term $\frac{1}{f_{id}}$ is ignored since it is only a scale factor. $\operatorname{Tr}[(\partial_i\rho_\epsilon)(\partial_j\rho_\epsilon)]$ is a Riemannian metric tensor and can be evaluated with SWAP tests.

Secondly, the quantum natural gradient can be ill-conditioned and might lead to unreasonably large updates due to very small eigenvalues of $F$. In practice, the learning rate for the initial steps needs to be chosen very small or one needs to use \textit{Tikhonov} regularization to add a small constant to the diagonal of $F$ before the inversion. Another method to get a stable gradient is by doing a “half-inversion”, given by \cite{haug_optimal_2021}
\begin{equation}
    \Tilde{g} =F^{-\alpha} \grad \mathcal{L}(\theta)
\end{equation}

where $\alpha$ is considered a regularization parameter. When $\alpha=0$, no precondition is applied to the original gradient, and when $\alpha=1$ the formula gets back to the original natural gradient. In practice, $\alpha$ is often chosen to be $0.5$. 

The third challenge comes from the cost of constructing the Fisher information metric. To fully construct the Fisher information metric with $p$ parameters, $p^2$ different expectation values needs to be estimated from the quantum computer. For a large $p$ the cost of construct the Fisher information metric can be significant. Gacon \textit{et al.} \cite{Gacon2021simultaneous} proposed that quantum natural gradient can also be approximated with the SPSA methods previously discussed in Sec. \ref{sec:stochastic_approximation}. Given the Fisher information metric $F(\boldsymbol{\theta}_1,\boldsymbol{\theta}_2) = |\braket{\psi_0|U^T(\boldsymbol{\theta_1})U(\boldsymbol{\theta_2})|\psi_0}|$, the estimated gradient $\Tilde{g}$ is given by

\begin{equation}
    \Tilde{g}^{(k)} = \frac{\delta F}{2\epsilon^2}\frac{\Delta^{(k)}_1\Delta^{(k)T}_2+\Delta^{(k)}_2\Delta^{(k)T}_1}{2}
\end{equation}

where 
\begin{align}
    \delta F^{(k)} &= F(\boldsymbol{\theta}^{(k)},\boldsymbol{\theta}^{(k)}+\epsilon\Delta_1^{(k)} +\epsilon\Delta_2^{(k)}) \\ 
    &- F(\boldsymbol{\theta}^{(k)},\boldsymbol{\theta}^{(k)}+\epsilon\Delta_1^{(k)}) \\
    &- F(\boldsymbol{\theta}^{(k)},\boldsymbol{\theta}^{(k)}-\epsilon \Delta_1^{(k)} + \epsilon \Delta_2^{(k)}) \\ 
    &+ F(\boldsymbol{\theta}^{(k)},\boldsymbol{\theta}^{(k)}-\epsilon \Delta_1^{(k)}), \\ 
\end{align}

and the exponentially smooth estimator 
 
 \begin{equation}
 \bar{F}^{(k)} = \frac{k}{k+1} \bar{F}^{(k-1)} + \frac{1}{k+1} \Tilde{F}^{(k)}.
 \end{equation}

The simulation results from ~\cite{Gacon2021simultaneous} show that the stochastic approximated natural gradient does not perform as well as the original natural gradient, however, it still improves the convergence speed compared to the standard gradient descent.   

\subsection{Gradient-free searching strategy}

The gradient-free search strategy is the other large category of optimization strategy. Because of the analytical property of the quantum ansatz, a few gradient-free searching strategies are developed for optimizing the variational ansatz. 

\subsubsection{Gradient-free optimizers}

\paragraph{Nelder-Mead algorithm}

The Nelder-Mead algorithm~\cite{nelder_simplex_1965} is a gradient-free heuristic search method based on a simplex. A simplex $\mathcal{S}$ in $\mathbb{R}^k$ is defined as the convex hull of $k+1$ vertices in $\mathbb{R}^k$ . It can be considered a multidimensional version of triangles. For example, in $\mathbb{R}^2$ a simplex is a triangle, and in $\mathbb{R}^3$ a simplex is a tetrahedron. 

The algorithm first generates a random simplex and then transforms the simplex $\mathcal{S}$ and decreases the function values at each vertex iteratively. In each iteration, the objective function value at one or more test points is measured. Then a new simplex is updated by replacing the vertices with one of the test points.

\vspace{0.5cm}

\begin{algorithm}[H]
\SetAlgoNoLine%
\SetKw{KwVariable}{~} 
\KwVariable{$\alpha_r$: Reflection coefficient.}\\
\KwVariable{$\alpha_e$: Expansion coefficient.}\\
\KwVariable{$\alpha_c$: Contraction coefficient.}\\
\KwVariable{$\alpha_s$: Shrink coefficient.}\\
 \While{Not converge}{
 
  Ordering the existing test points, so that $\mathcal{L}(\boldsymbol{\theta}_{0}) \leq \mathcal{L}(\boldsymbol{\theta}_{1}) ... \leq \mathcal{L}(\boldsymbol{\theta}_{n+1})$
 
  Calculate $\mathbf {\theta} _{o}$, the centroid of all points except $\mathbf {\theta} _{n+1}$ \;

  Calculate reflection point $\boldsymbol{\theta}_{r}=\boldsymbol{\theta}_{o}+\alpha_r (\boldsymbol{\theta}_{o}-\boldsymbol{\theta}_{n+1})$ with $\alpha_r >0$.
  
  \If{The reflection point the not the best estimation, but better than the second worst point $\theta_n$}{
   replace the worst point $ \mathbf {\theta} _{n+1}$ with the reflected point $\mathbf {\theta} _{r}$\;
   continue\;
   }
   
  \If{Reflected point $\boldsymbol{\theta}_{r}$ is the best point}{
    Calculate expansion point $\boldsymbol{\theta}_{e}=\boldsymbol{\theta}_{o}+\alpha_e (\mathbf {\theta} _{r}-\mathbf {\theta} _{o})$ with $\alpha_e >1$
    
   replace the worst point $ \mathbf {\theta} _{n+1}$ with the best point from expansion point $\boldsymbol{\theta}_{e}$ or reflected point $\mathbf {\theta} _{r}$\;
  }
  
  Calculate contraction point $ \mathbf {\theta} _{c}=\mathbf {\theta} _{o}+\alpha_c (\mathbf {\theta} _{n+1}-\mathbf {\theta} _{o})$ with $ 0<\alpha_c \leq 0.5$.

  \If{Contraction point  $ \mathbf {\theta} _{c}$ is better than the worst point $ \mathbf {\theta} _{n+1}$}{
    Replacing the worst point $ \mathbf {\theta} _{n+1}$ with the contracted point $ \mathbf {\theta} _{c}$ \;
    Continue \;
  }
 
  \tcp{Shrink}
  Replace all points except the best point $ \mathbf {\theta} _{1}$ with
    $ \mathbf {\theta} _{i}=\mathbf {\theta} _{1}+\alpha_s (\mathbf {\theta} _{i}-\mathbf {\theta} _{1})$
 }
\caption{Nelder-Mead algorithm}
\end{algorithm}

\vspace{0.5cm}

The Nelder-Mead algorithm is gradient-free, and for each iteration, only very limited points need to be measured. Therefore it is very friendly to variational quantum algorithms.  A result from \cite{Lavrijsen2020ClassicalDevices,Pellow-Jarman2021} shows the Nelder-Mead algorithm performs well in a low-noise environment, but not as well when a noise model is included.  

\paragraph{Powell's conjugate direction algorithm}

Powell's conjugate direction algorithm is a gradient-free optimization method. Powell's algorithm takes a starting point and two non-parallel vectors as inputs. The algorithm first search bidirectionally through the direction parallel to the first vector, and uses the minimal point as the direction, and search through the next direction parallel to the second vector. A conjugate direction can be defined as the displacement between the initial starting point and the optimal point of the second iteration. The algorithm repeats this process until it converges. The linear directional search can be implemented with Golden-section search or Brent's method~\cite{Powell1964}. 

\begin{algorithm}[H]
\SetAlgoNoLine%

\SetKw{KwVariable}{~}
\KwVariable{$x_k$: Searched minimal point. $x_0$ is the starting point.} \\
\KwVariable{$\mathcal{L}(\theta)$: Objective function}\\
\KwVariable{$\alpha_k$: Searching displacement }\\
\KwVariable{$\mathbf{d}_k$: Searching direction. $\mathbf{d}_1$ and $\mathbf{d}_2$} are the starting direction.\\
 \While{Not converged}{  
  Search for $\alpha_k$ that minimize $\mathcal{L}(\mathbf{x}_{k-1}+\alpha_k \mathbf{d}_k)$ \\
  Set $x_k = x_{k-1} + \alpha_k d_k$.\\
  Remove the initial vectors by assigning $\mathbf{d}_j = \mathbf{d}_{j+1}$
  Set $\mathbf{d}_N = \mathbf{x}_N - \mathbf{x}_0 $ 
  Search for $\alpha_N$ that minimize $\mathcal{L}(\mathbf{x}_{N}+\alpha_N \mathbf{d}_N)$ \\
  Set $\mathbf{x}_0 = \mathbf{x}_0 + \alpha_N \mathbf{d}_N$

 }
 \caption{Powell's algorithm}
\end{algorithm}

\subsubsection{Analytical optimization} \label{sec:analytical_opt}

In this section, we review optimization methods that take advantage of the analytical property of the objective function landscape. Consider the ansatz circuit as a CPTP mapping as a product of individual quantum super-operators
\begin{equation}\label{fullansatz}
\Phi(\boldsymbol{\theta}) = \Phi_k(\theta_k) \dots \Phi_2(\theta_2) \Phi_1(\theta_1).
\end{equation}	
Here $\Phi_k(\theta_k)$ are $k$-th parameterised quantum gates where $\Phi_k(\theta_k) \rho := U_k \rho U_{k}^{\dagger} $
and $U_k=\exp(- i \frac{\theta_k}{2} P_k)$. Here $P_k$ are tensor products of single-qubit
Pauli operators as $P_k \in \{I, X, Y ,Z\}^{\otimes N}$. Each superoperator can be expanded around $\boldsymbol{\theta}_0$ as
\begin{equation}
\Phi_k(\theta_0 + \theta) = 
a(\theta) \Phi_{ak}
+ b(\theta) \Phi_{bk}
+ c(\theta) \Phi_{ck},
\end{equation}
where $a(\theta), b(\theta) = 1\pm\cos(\theta)$ and $c(\theta) = \frac{1}{2}\sin(\theta)$.

Now we only allow $n$ parameters to be variables and fix all the others, the whole ansatz can be expanded into 
\begin{equation}\label{eq:optimization:ansatz_expansion}
\Phi(\boldsymbol{\theta}_0 {+} \boldsymbol{\theta})
= 
\prod_{k=1}^\nu
[a(\theta_k) \Phi_{ak}
+ b(\theta_k) \Phi_{bk}
+ c(\theta_k) \Phi_{ck}],
\end{equation}	
where $\boldsymbol{\theta}$ is the displacement around the reference point $\mathbf{\theta_0}$.

\paragraph{Sequential optimization with sinusoidal fitting (Rotosolve)}

A variety of quantum variational algorithms are dealing with linear objective functions. For example, the energy value for VQE is a linear objective function. For these problems, the objective function landscape of every single variable is a sinusoidal function. The full information of such function can be obtained by simply measuring three different points of the objective function. The sequential optimization method \cite{Vidal2018, nakanishi_sequential_2020,ostaszewskiStructureOptimizationParameterized2021} utilizes this feature and calculates the minima of the sinusoidal function directly. By iteratively finding the minima of every single parameter while fixing the other parameters, a greedy method can be derived to efficiently optimize the parameters of the ansatz.

From Eq.~(\ref{eq:optimization:ansatz_expansion}) we can simplify $a(\theta), b(\theta)$ and $b(\theta)$ into a simpler form if we only allow one single parameter to change. When all the parameters are independent, the objective function can be transformed as 
\begin{align}
\mathcal{L}(\boldsymbol{\theta}_0 {+} \theta \cdot e_i) = \mathcal{L}_i(\theta)_{\boldsymbol{\theta}_0} = & \operatorname{Tr}[M \Phi_i(\boldsymbol{\theta}_0 {+} \theta \cdot e_i)]\\
= & \operatorname{Tr}[M(a(\theta) \Phi_{a}
+ b(\theta) \Phi_{b}
+ c(\theta) \Phi_{c})]\\
= & A \sin(\theta + \phi) + C.
\end{align}
The parameter can be evaluated analytically with following equation when the hermitian generator has exactly two eigenvalues.
\begin{equation}
\begin{split}
    \theta^* = \phi - \tfrac{\pi}{2} - \arctantwo \big( \phi_1 , ~ \phi_2\big) + 2\pi k,
\end{split}
\end{equation}
where 
\begin{equation}
    \phi_1 = 2\mathcal{L}_i(\theta)_{\boldsymbol{\theta}_0} - \mathcal{L}_i(\theta+\frac{\pi}{2})_{\boldsymbol{\theta}_0} - \mathcal{L}_i(\theta-\frac{\pi}{2})_{\boldsymbol{\theta}_0} 
\end{equation}
\begin{equation}
    \phi_2 =\mathcal{L}_i(\theta+\frac{\pi}{2})_{\boldsymbol{\theta}_0} - \mathcal{L}_i(\theta-\frac{\pi}{2})_{\boldsymbol{\theta}_0}.
\end{equation}

All the single qubit gates have Pauli operators as their generators, which satisfy the above equations. The cost function can also sum sinusoidal functions with different period~\cite{GeneralGradientsWierichs2022,Vidal2018CalculusQuantumCircuit} when the generator has more than two distinct eigenvalues.

When several gates share the same parameter, the objective function can be transformed into similar form but with different oscillation period:
\begin{align}
\mathcal{L}(\boldsymbol{\theta}_0 {+} \theta \cdot e_i)
= & A \sin( k \theta + \phi) + C,
\end{align}
where $k$ is the number of appearances of $\theta_i$ in the ansatz. The same approach can be applied to find the analytical solution.

It is important to note that the conditions for validity of this optimization method implies that it cannot be applied to some ans{\"{a}}tze (for instance in the case where a parameterized controlled unitary is used) In general terms, Rotosolve and its subsequent extension require that all parametrized gate subject to optimization have  $2\pi$-periodicity and full-rank for the Hermitian matrix generating the rotation. The work presented by Wierisch \textit{et al.} \cite{Wierichs2022}, relying on alternative general parameter-shift rules, demonstrates a generalization of Rotosolve which allows for it to be used on all quantum gates with arbitrary frequencies.

\paragraph{Analytical Free-Axis Selection with fixed rotation angles (Fraxis)}

The analytical method we mentioned so far has fixed rotation axis in the ansatz, but with a flexible rotation angle. The Analytical Free-Axis Selection (Fraxis) method implemented an alternative version where angles are fixed but the axis can be flexible \cite{Watanabe2021WatanabeOptimizingSelection}. The method is shown to provide faster optimization than Rotosolve in some numerical tests. Wada \textit{et al. } \cite{Wada2021SimulatingCircuits} further improved on this idea by optimizing the angle and the axis at the same time for time evolution simulations. 

The Fraxis method considers each single qubit gate in the ansatz as 
\begin{equation}
    U_k(\theta_k) = e^{-i\frac{\theta_k}{2}\hat{n}_k\cdot \Vec{P}}
\end{equation}

where $U_k(\theta_k)$ is the $k$-th single qubit rotation, $\theta_k$ is the rotation angle, and is the $\hat{n}_k$ is the direction vector characterize the rotation direction of the hermitian generator, given by  

\begin{equation}
    \hat{n}_k\cdot\Vec{P} =  n_{k,x}X + n_{k,y}Y + n_{k,z}Z \mathrm{~for~} \hat{n}_k \in \mathcal{R}^3, |\hat{n}_k| = 1.   
\end{equation}

Here $X$,$Y$,$Z$ are Pauli matrices.

With Lagrange multiplier method, the optimal $\hat{n}_k$ is found to satisfy following linear equations: 

\begin{equation}
    \sin^2(\frac{\theta_k}{2}\mathbf{R} - 2\lambda^*\hat{n}_{k^*}) = -\alpha_{\theta_k}\mathbf{b}
\end{equation}

where 

\begin{equation}
    \mathbf{b} = (tr(M[\rho,X]),tr(M[\rho,Y]),tr(M[\rho,Z]))^T \label{eq:fraxis}
\end{equation}

and

\begin{eqnarray}
    r_x &\equiv& \mbox{tr}\left(MX\rho X\right),\\
    r_y &\equiv& \mbox{tr}\left(MY\rho Y\right),\\
    r_z &\equiv& \mbox{tr}\left(MZ\rho Z\right),\\
    r_{(x+y)} &\equiv& \mbox{tr}\left(M \left(\frac{X+Y}{\sqrt{2}}\right) \rho  \left(\frac{X+Y}{\sqrt{2}}\right)\right),\\
    r_{(x+z)} &\equiv& \mbox{tr}\left(M \left(\frac{X+Z}{\sqrt{2}}\right) \rho  \left(\frac{X+Z}{\sqrt{2}}\right)\right),\\
    r_{(y+z)} &\equiv& \mbox{tr}\left(M \left(\frac{Y+Z}{\sqrt{2}}\right) \rho  \left(\frac{Y+Z}{\sqrt{2}}\right)\right).
\end{eqnarray}

\begin{equation}\label{eq:R-def}
    \mathbf{R} \equiv 
    \begin{pmatrix}
    2r_x  & 2r_{(x+y)} - r_x - r_y & 2r_{(x+z)} - r_x - r_z\\
    2r_{(x+y)} - r_x - r_y & 2r_y & 2r_{(y+z)} - r_y - r_z \\
    2r_{(x+z)} - r_x - r_z & 2r_{(y+z)} - r_y - r_z & 2r_z
    \end{pmatrix}.
\end{equation}

The optimum value $\hat{n}_k^*$ can be solved by measuring all elements in $\mathbf{R}$ and solving the linear equation \ref{eq:fraxis}.   When $\theta_k = \pi$, the method can be further simplified and $\hat{n}_k^*$ becomes the eigenvector of $\mathbf{R}$.  

\paragraph{Quantum analytical descent}

The full landscape of the expectation value could be too expensive to construct. However, it can be approximated. Previous works from Sung et.al \cite{sung_using_2020} approximate the local region with a predefined model, such as a quadratic polynomial.  The quantum analytical descent \cite{koczor_quantum_2020} approximates the landscape properly and utilizes a similar idea of sequential optimization with sinusoidal fitting.  Instead of fitting the sinusoidal function for each parameter iteratively, the entire landscape of the linear objective function can be approximated efficiently with a quadratic number of parameters and quadratic number of measurements. 

The expansion of the entire ansatz has $3^p$ terms, which cannot be efficiently evaluated. The full expansion of the ansatz can be approximated into 
\begin{align} \label{full-circuit}
\Phi(\boldsymbol{\theta}) = A(\boldsymbol{\theta}) \Phi^{(A)} &+ \sum_{k=1}^p
[B_k(\boldsymbol{\theta}) \Phi^{(B)}_k + C_k(\boldsymbol{\theta}) \Phi^{(C)}_k] \\
&+\sum_{l>k}^p [ D_{kl}(\boldsymbol{\theta}) \Phi^{(D)}_{kl}] + O(\sin^3 \epsilon). \nonumber
\end{align}
Here $A, B_k, C_k, D_{kl} : \mathbb{R}^p \mapsto \mathbb{R}$ are products of simple univariate trigonometric functions. And the loss function can be constructed as
\begin{align} \label{full-energy}
\mathcal{L}(\boldsymbol{\theta}) = A(\boldsymbol{\theta}) \mathcal{L}^{(A)} &+ \sum_{k=1}^p
[B_k(\boldsymbol{\theta}) \mathcal{L}^{(B)}_k + C_k(\boldsymbol{\theta}) \mathcal{L}^{(C)}_k] \\
&+\sum_{l>k}^p [ D_{kl}(\boldsymbol{\theta}) \mathcal{L}^{(D)}_{kl}] + O(\sin^3 \epsilon). \nonumber
\end{align}

Here $\mathcal{L}^{(A)},\mathcal{L}^{(B)}_k,\mathcal{L}^{(C)}_k, \mathcal{L}^{(D)}_{kl} \in \mathbb{R}$ can be calculated from the hardware by measuring expectation value at $1 + 2p^2 -2 p$ different points where $p$ is the number of parameters. 

After the objective function landscape has been approximated, the author uses natural gradient descent to find the minimum conventionally. The natural gradient descent requires the Fubini-Study metric tensor to find the natural gradient, which can be approximated as 

\begin{equation}
[F_Q]_{pq} =
F_{BB} F_{BB}(\boldsymbol{\theta})
+ F_{AB} F_{AB}(\boldsymbol{\theta}) + \dots  O(\sin^2\epsilon),
\end{equation}
where
\begin{align}
	F_{BB}(\boldsymbol{\theta}) :=& 2\frac{\partial  B_p(\boldsymbol{\theta})}{\partial \theta_p}  \frac{\partial  B_q(\boldsymbol{\theta})}{\partial \theta_q}\\
	F_{AB}(\boldsymbol{\theta}) :=&  2\frac{\partial B_p(\boldsymbol{\theta}) }{\partial \theta_p}  \frac{\partial A(\boldsymbol{\theta}) }{\partial \theta_q} 
	+ 2 \frac{\partial A(\boldsymbol{\theta}) }{\partial \theta_p}  \frac{\partial B_n(\boldsymbol{\theta}) }{\partial \theta_q}.
\end{align}

Once the local minima of the approximated landscape have been found, the algorithm updates the best-guessed parameters in this iteration to be the local minima. Then repeat the approximation and conventional optimization steps until the algorithm converges.


\paragraph{Jacobi diagonalization and Anderson acceleration}

An analytical method inspired by Jacobi diagonalization and Anderson acceleration has been introduced in Ref. \cite{parrish_jacobi_2019}. This method is an improved version of the sequential sinusoidal fitting. Instead of fitting the landscape one dimension at a time, the landscape can be accurately reconstructed by only allowing a small subset of parameters to vary. This technique is similar to the Jacobi diagonalization algorithm for large matrices by iteratively optimizing a random subset of parameters~\cite{golub_eigenvalue_2000}. The Anderson/Pulay DIIS sequence acceleration is then introduced to produce a better estimation for each iteration. 

This algorithm first constructs the analytical landscape with a few parameters and then iteratively chooses a random subset of parameters to fit the analytical landscape for the optimal value. This iteration approach is similar to the Jacobian diagonalization method. 

Parrish \textit{et al.} \cite{parrish_jacobi_2019} then introduce the DIIS (Direct inversion of the iterative subspace) method to better estimate and improve the convergence speed. Suppose the optimized parameter after $i$-th iteration is $\theta^i$, the error of the optimized parameter is $\epsilon^i$, which is the difference between the optimal value and the optimized parameter $\theta^i$. The DIIS algorithm gives a better estimation:
\begin{equation}
    \theta^{i\prime} = \sum_i c_i\theta^i,
\end{equation}
where $c_i$ is a real coefficient that minimizes the square of the 2-norm of $\epsilon$,
\begin{equation}
    O(c_i) = \sum_{ij}c_ic_j\epsilon^i \cdot \epsilon^j,
\end{equation}
and subject to the normalization condition 
\begin{equation}
   \sum_i c_i = 1.
\end{equation}
The DIIS algorithm utilizes the previous steps' historical value and extrapolates a better estimation for the next steps. Since the error $\epsilon^i$ is not accessible for each step, it can be approximated by 
\begin{equation}
    \epsilon^i \approx \delta \theta^i = \theta^i - \theta^{i-1} ~~\mathrm{(Anderson~style)},
\end{equation}
or 
\begin{equation}
    \epsilon^i \approx g(\theta^i) ~~\mathrm{(Pulay~style)},
\end{equation}
where $g(\theta^i)$ is the gradient of the cost function at $\theta^i$. Here the Anderson style and Pulay style are different approaches to approximate the error value.

\subsection{Engineering cost function} \label{sec:cost_function}

\paragraph{Collective optimization}\cite{zhang_collective_2020}
In practice, when using VQE to solve eigenenergies of molecules, there are different Hamiltonians with varied bond lengths that can be studied. Since the two different solutions should be close when the Hamiltonian is only different in a small amount, all of the optimal solutions of the series of Hamiltonian should also be chained together in the parameter space, forming a snake-like shape. The entire optimization process can be considered optimizing the arrangement of the whole snake instead of optimizing points separately. 

The collective optimization algorithm redefines the cost function into 
\begin{equation}
\mathcal{L(\boldsymbol{\theta}(\lambda))} = \int_{\lambda_0}^{\lambda_1} [L(\boldsymbol{\theta}(\lambda))+E(\boldsymbol{\theta}(\lambda))], 
\end{equation}
where $\boldsymbol{\theta}(\lambda)$ is the optimized ansatz parameter for Hamiltonian of bond distance $\lambda$, the $E$ term is the energy of the molecules, while the $L$ term defines the internal energy of the snake.
\begin{equation}
    L(\boldsymbol{\theta}(\lambda)) = \alpha \lvert\frac{\partial\boldsymbol{\theta}(\lambda)}{\partial\lambda}\rvert^2 + \beta \lvert\frac{\partial^2\boldsymbol{\theta}(\lambda)}{\partial^2\lambda}\rvert^2.
\end{equation}
The first term refers to the energy that depends on the snake's length, and the second term refers to the curvature of the snake. $\alpha$ and $\beta$ are meta parameters. 

In practice the snake is discretized into a sequence of parameters at different bond distance $r_i = (\theta_i(\lambda_1),\theta_i(\lambda_2),\dots,\theta_i(\lambda_K))$ where $i$ is the $i$-th component for each $\theta_i(\lambda_k)$. Then the $r_i$ can be solved iteratively by
\begin{equation}
    r_i^t=(\eta\mathbf{A}+\unit)^{-1}(r_i^{t-1}-\eta\frac{dE(r^{t-1})}{dr_i}),
\end{equation}
where $\eta$ is the learning rate, $E(r)=\sum_{k=1}^K E(\theta(\lambda_j))$ and $A$ is a pentadiagonal banded matrix with $A_{i-2,i}=A_{i,i-2}=\beta$, $A_{i-1,i}=A_{i,i-1}=-\alpha-4\beta$, $A_{i,i}=2\alpha+6\beta$.

The simulation result shows the collective optimization methods help to pull the parameters from local minimums.

\paragraph{Conditional Value-at-Risk as objective function}

Conditional Value-at-Risk (CVaR) is a measure that takes into account only the tail of the probability distribution. The CVaR of a random variable $X$ for a confidence level $\alpha \in (0, 1]$ is defined as
\begin{equation}
    \mathrm{CVaR}_\alpha(X) = E[X|X \leq F^{-1}_X(\alpha)],
\end{equation}
where $F_X$ is the cumulative density function of X. To illustrate the idea, consider the random variable $X$ has been sampled $N$ times. The CVaR with confidence $\alpha$ can be calculated by selecting $\alpha N$ samples with the lowest value and evaluating the average. When $\alpha=1$, CVaR equals the expectation value. For variational quantum algorithms, consider the result of each sampling to be a random variable. Then the CVaR can be used as an objective function and replace the expectation value~\cite{barkoutsos_improving_2020}. 
\begin{equation}
    \mathcal{L}(\theta) = \mathrm{CVaR}_\alpha(X(\theta)),
\end{equation}
where $X$ is the random variable for each sample result.

The CVaR could give a reasonable benefit. Suppose $\ket{\psi_0}$ is the ground state, and $\ket{\psi_1}$,$\ket{\psi_2}$,$\ket{\psi_3}$ are first, second and third excited state. Define $\ket{\psi_A} = (\ket{\psi_0} + \ket{\psi_3})\sqrt{2}$ and $\ket{\psi_B} = (\ket{\psi_1} + \ket{\psi_2})\sqrt{2}$. Suppose the energy level are equally separated, then $\bra{\psi_A}H\ket{\psi_A} = \bra{\psi_B}H\ket{\psi_B}$, therefore the optimizer will encounter a gradient plateau. However for our purpose, we would like to obtain the ground state, therefore $\ket{A}$ is better than $\ket{B}$ in practice. The CVaR could help with this problem by emphasizing the distribution $\ket{A}$. The CVaR emphasizes the best-observed samples and leads to a smooth objective function without introducing local minimum~\cite{barkoutsos_improving_2020}. Also, the implementation of CVaR is relatively straightforward. Since CVaR throws away some of the samples, the accuracy of the estimation decreases. In order to get the same accuracy, the sampling number needs to be increased. The same amount of samples should be involved in the calculation. 

\paragraph{Symmetry preserving cost function adjustments:} One means to maintain electron number conservation in the output wavefunction of VQE is to impose a constraint on the cost function \cite{mccleanTheoryVariationalHybrid2015, Ryabinkin2019}. Namely, one can include a penalty term corresponding to violation of symmetries. The VQE cost function can therefore be re-written as
\begin{equation}
 E(\theta, \mu) = \bra{\psi(\theta)} \hat{H} \ket{\psi(\theta)} + \sum_i \mu \left[ \bra{\psi(\theta)} \hat{O}_i \ket{\psi(\theta)} - O_i \right]^2,
\end{equation}
where $\hat{O}_i$ represents the symmetry operators that can be independently measured, for instance, the electron number operator, or the spin operator (square of total spin). $O_i$ is the target expectation value for each of these operators, and $\mu$ is a Lagrangian parameter to determine the strength of the constraints. This method, referred to as Constrained VQE in Ref.~\cite{Ryabinkin2019}, was shown to also eliminate 'kinks' appearing in the VQE implementation that had been shown to appear in Ref.~\cite{Kandala2017}. It is worth pointing out (as done in Refs.~ \cite{mccleanTheoryVariationalHybrid2015, Ryabinkin2019}) that the Pauli strings used for operators representing the electron number or the total spin already need to be computed for the Hamiltonian thereby the method does not require additional Quantum costs with respect to the computation of the energy function. One caveat to this is that, as is the case for any constrained optimization problem, the optimization landscape becomes more complex with the addition of constraints. This, in addition to the management of the hyper-parameter $\mu$, could result in additional optimization costs and risks of local minima. The use of this method, and further considerations regarding optimal application have been discussed in Ref. \cite{KuroiwaPenaltyEigensolver2021}.

\subsection{Discussion}

In this Section, we have reviewed some of the latest optimization strategies adapted to optimize variational quantum ans\"atze. Some methods are adapted from the traditional numerical optimization, while some new methods are developed to provide some essential features specifically for variational quantum ansatz optimization. 

\begin{table}[h]
\caption{Comparison of optimization strategies mentioned in this section. $C_M$ denotes the number of different measurement expectation values per iteration that need to be evaluated from the quantum computer. $C_C$ denotes the complexity of the classical algorithm for each iteration. Since the gradient can be evaluated or approximated with different methods, one can use $g^{(1)}$ to denote the cost of evaluating first order gradients and $g^{(2)}$ to denote the cost of evaluation second order gradients. $S$ denote the required sample shot number. $k$ denote the number of Hamiltonians with different bond distance being optimized simultaneously. $p$ denote the number of parameters in the ansatz.}
    \begin{tabularx}{\linewidth}{X l l l l l}
        \toprule
        Strategies             & Type                  & Meta parameters                                                                                                      & $C_M$                          & $C_C$            & References                                                                                                           \\ \midrule
        Simple gradient descent         & First order & $\eta$:Learning rate.                                                                                                & $S g^{(1)}$         & $\mathcal{O}(p)$ & \cite{Pellow-Jarman2021,wierichs_avoiding_2020}                                                                                                                    \\ \hline
        RMSProp                         & First order  & \begin{tabular}{@{}l@{}}$\gamma$: Moving average parameter. \\ $\eta$: Learning rate.\end{tabular}
                                                                                                                & $S g^{(1)}$         & $\mathcal{O}(p)$ & \cite{ExponentiallyNetworks2021}                                                                                                                    \\ \hline
        Adam                            & First order  & \begin{tabular}{@{}l@{}}$\beta_{1},\beta_{2}$: Moving average parameters. \\ $\eta$: Learning rate.\end{tabular}                                                                                                                 & $S g^{(1)}$         & $\mathcal{O}(p)$ & \cite{kingma_adam_2017}                                                                                              \\ \hline
        BFGS                            & Second order  & $\eta$:Learning rate.                                                                                                                 & $S (g^{(1)} + g^{(2)})$ & $\mathcal{O}(p^3)$ & \cite{Lavrijsen2020ClassicalDevices,Pellow-Jarman2021,wierichs_avoiding_2020} \\ \hline
        Quantum natural gradient decent & Second order & $\eta$:Learning rate.                                                                                                                 & $S (g^{(1)} + g^{(2)})$ & $\mathcal{O}(p^3)$ & \cite{amari_natural_1998,martens_new_2020,wierichs_avoiding_2020}                                                    \\ \hline
        Nelder-Mead                     & Gradient free         & \begin{tabular}{@{}l@{}}$\alpha_r$: Reflection coefficient.\\ $\alpha_e$:  Expansion coefficient.\\$\alpha_c$:  Contraction coefficient. \\$\alpha_s$:    Shrink coefficient. \\\end{tabular}                                                                                                                 & $S$                              & $\mathcal{O}(p)$ & \cite{Lavrijsen2020ClassicalDevices,Pellow-Jarman2021}                                                                                           \\ \hline
        Powell                          & Gradient free         & \begin{tabular}{@{}l@{}}$\alpha_k$:   Searching displacement. \\ $\mathbf{d}_k$:Searching direction.\end{tabular}                                                                                                                 & \begin{tabular}{@{}l@{}}Depends on linear \\  search method.\end{tabular}                              & $\mathcal{O}(1)$ & \cite{Pellow-Jarman2021}                                                                                                                    \\ \hline
        Rotosolve                       & Gradient free         & None                                                                                                                 & $\mathcal{O}(3n)$              & $\mathcal{O}(1)$ & \cite{nakanishi_sequential_2020,ostaszewskiStructureOptimizationParameterized2021}                                   \\ \hline
        Fraxis                       & Gradient free         & None                                                                                                                 & $\mathcal{O}(6n)$              & $\mathcal{O}(1)$ & \cite{Watanabe2021WatanabeOptimizingSelection,Wada2021SimulatingCircuits}                                   \\ \hline
        Quantum analytical descent      & Gradient free         & $\eta$:Learning rate.                                                                                                                 & $S(1+2p^2-2p)$                              & NP & \cite{koczor_quantum_2020}                                                                                           \\ \hline
        Anderson acceleration           & Addon         & None                                                                                                                 & No extra costs               & $\mathcal{O}(p)$ & \cite{golub_eigenvalue_2000}                                                                                         \\ \hline
        Collective optimization         & Addon                 & \begin{tabular}{@{}l@{}}$\alpha$:    Coefficient for snake length \\ $\beta$:Coefficient for snake curvature.\end{tabular} & No extra costs              & $\mathcal{O}(k)$ & \cite{zhang_collective_2020}                                                                                         \\ \hline
        CVaR                            & Addon                 & $\alpha$:Confident level                                                                                             & Multiply factor $1/\alpha$                              & $\mathcal{O}(1)$ & \cite{barkoutsos_improving_2020}                                                                                     \\
        \bottomrule
    \end{tabularx}
\label{table:optimizer_comparison}
\end{table}

The first key feature to consider is the speedup of the convergence for variational quantum ansatz. From the convergence speed per iteration, the analytical method is much faster than the gradient descent strategies. However, the analytical methods would require taking more measurement points to finish a single iteration, which is unfair to directly compare the convergence speed per iteration between optimization strategies. Numerical studies have nonetheless shown in multiple occasions that Rotosolve (and by association its extensions) indeed reach convergence significantly faster than other methods \cite{nakanishi_sequential_2020,ostaszewskiStructureOptimizationParameterized2021,koczor_quantum_2020}. 

Several studies have been performed to compare the relative strength of convergence of different optimizers. Mihlikov \textit{et al.} \cite{Mihlikov2022} show as part of a case study on Hydrogen that SPSA presents a clear advantage on the Nelder-Mead and the Powell optimizers. Bonet-Monroig \textit{et al.} \cite{BonetMonroig2021} test four different optimizers on a variety of small molecular systems and find that SPSA performs slightly better than all others. Beyond these studies, several methods allow for improved convergence on almost any optimizer: the covariance functions between the Hamiltonian and operator can also be used to increase the convergence speed when the optimization is almost converged~\cite{BoydCoVar2022}. Stochastic methods are also suggested as add-ons to reduce the impact of hardware noise, accelerating convergence speed~\cite{AcceleratingProcessesMueller2021,StochasticApplications2022Gidi,PattiMarkovChain2021}.  

Another key feature is the resilience of the barren plateau problem. The natural-gradient-based strategy is considered resilient to barren plateau, and its absolute value of gradient has a lower bound $1/(2^{2N+1})$~\cite{haug_capacity_2021,haug_optimal_2021}.  The analytical method is gradient-free and directly jumps into an optimal or approximately optimal position in each iteration, which can prevent entering a barren plateau through optimization dependent on a learning rate \cite{Vidal2018, nakanishi_sequential_2020,ostaszewskiStructureOptimizationParameterized2021,koczor_quantum_2020}. However, this is at the cost of increasing the measurement points and conventional computation. 

From the discussion above, no optimization strategy outperforms other strategies in every aspect; the convergence needs to trade with the implementation complexity and measurement points. In fact, multiple strategies can be applied together to improve the overall performance of the optimizer. In practice, the sequential analytical fitting method has been mainly used when a non-traditional optimizer optimizes the variational ansatz. Quantum natural gradient-based methods are more popular in simulation-based studies. 

A final comment is worth raising on the optimization of VQE ans{\"{a}}tze: it does not need to be done entirely using samples from observable obtained on the quantum computer. In particular Okada \textit{et al.} \cite{Okada2022} have shown that one can use efficient circuit simulation to optimize local order parameters and subsequently use quantum devices post-optimization to measure the global quantities. 

%% file: 08_error-mit.tex
\section{Error mitigation for VQE} \label{sec:error-mit}

When the noise of a quantum computer is below a certain threshold and a sufficient number of qubits are available, quantum error correction schemes can be applied to suppress the noise to arbitrarily small levels~\cite{nielsenQuantumComputationQuantum2010}. However, hardware demonstrations of simple quantum error correcting codes have been limited and have only demonstrated fault-tolerant universal quantum computation with limited error-correcting capabilities
\cite{andersenRepeatedQuantumError2020a,barendsSuperconductingQuantumCircuits2014,campagne-ibarcqQuantumErrorCorrection2020,niggQuantumComputationsTopologically2014,ofekExtendingLifetimeQuantum2016,waldherrQuantumErrorCorrection2014,Postler_2022,Krinner_2022}. Error correction also brings different types of overheads, including large amounts of extra ancilla qubits, fast decoding and communication between quantum and conventional devices~\cite{nielsenQuantumComputationQuantum2010}. For example, assuming an optimistic error rate threshold of $\varepsilon = 10^{-3}$, the required number of physical qubits to start exploring interesting quantum chemistry problems could be of the order of $10^6$~\cite{reiherElucidatingReactionMechanisms2017}.

As an alternative, a series of techniques for mitigating the effects of noise in quantum algorithms running on NISQ hardware have been developed. These techniques have been shown to achieve a reduction in the noise levels of expectation value estimates, without requiring the large resources involved in error correction. Such methods will be critical for early implementations of the VQE algorithm in order to achieve the required precision for Quantum Chemistry computation. Although error mitigation can also benefit other algorithms such as the Quantum Approximate Optimization Algorithm~\cite{Farhi2014}, we focus our discussion on the relevance of these methods for VQE. In this section, we cover the most relevant error mitigation techniques for VQE algorithms. For a more complete review of error mitigation techniques, we refer readers to~\cite{endoHybridQuantumClassicalAlgorithms2021}.

\subsection{Symmetry verification}
\label{sec:mit-symmetry-verification}

The computation of molecular Hamiltonians often comes with symmetry constraints. Formally, this results in the ground state of the Hamiltonian $\ket{\psi_0}$ being the eigenstate of the corresponding symmetry operator $\hat{S}$. One example is the particle number operator $\hat{N} = \sum_i \hat{a}^{\dagger}_i \hat{a}_i$, which is usually a symmetry of the system, and we are often sure that the corresponding ground state should have a certain number of particles. Symmetries can be used to `taper off qubits' (Sec.~\ref{sec:tappering_qubits}) or as properties to design the ansatz (Sec.~\ref{sec:symmetry_preserving_methods}). Here, we discuss how symmetry can be used to mitigate errors on a quantum state $\rho$ prepared by an ansatz.

The general idea is that we need to find a way to measure the expectation value $\langle \hat{S} \rangle$ of the symmetry operator $\hat{S}$. Then we can filter out the experiments $\rho$ having the wrong number $\langle \hat{S} \rangle$, i.e. we post-select the experiment outcome based on $\langle \hat{S} \rangle$. Such a verification process has many error-mitigating benefits. The first is that verification filters out errors that violate the symmetry, which applies regardless of the error rates of the quantum computer, although the failure rate is higher when the error rates increase~\cite{mcardleErrorMitigatedDigitalQuantum2019}. The potential for filtering noise on measurements has been demonstrated experimentally in a VQE algorithm \cite{Sagastizabal2019}. At the same time, the measurement and the post-selection on the result can project the state into a subspace that preserves the symmetry, and hence increase the overlap of the state $\rho$ prepared using the ansatz with the symmetry preserving subspace \cite{bonet-monroigLowcostErrorMitigation2018}. Therefore, the symmetry verification method should be considered together with other symmetry preserving methods when preparing the target state in a VQE algorithm \cite{Sagastizabal2019}.

Here we briefly mention two specific techniques to measure $\hat{S}$.
The first is called the \textit{final symmetry verification},
in which we only verify that symmetries are respected at the end of the computation, i.e. after the ansatz circuit. This usually comes at no additional cost as the Pauli terms included in $\hat{S}$ are often measured as part of the Hamiltonian.
The second is named \textit{bulk symmetry-verification}, in which the symmetry verification step may be carried
out during the computation, i.e. inside the ansatz circuit, without disrupting the computation.

\begin{figure}[ht]
    \centering
    $$
    \Qcircuit @C=1.0em @R=0.8em @!R { \\
    \lstick{\text{qubit }1} & \multigate{1}{U(\boldsymbol{\theta})} & \ctrl{2} & \qw & \qw & \qw \\ 
    \lstick{\text{qubit }2} & \ghost{U(\boldsymbol{\theta})} & \qw & \ctrl{1} & \qw & \qw \\ 
    \lstick{\text{ancilla qubit}} & \qw & \targ & \targ & \qw & \meter \\ }
    $$
    \caption{Particle number parity verification circuit, adapted from \citet{mcardleErrorMitigatedDigitalQuantum2019}.
    Here the measurement outputs of the ancilla qubit give the qubit number parity, which is equivalent to the electron number parity in the Jordan-Wigner mapping.
    }
    \label{fig:sym-verify-particle-number}
\end{figure}
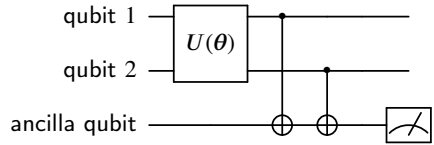

\paragraph{Final symmetry verification:}
Verifying the symmetry at the end of the computation is relatively straightforward. For example, when using a Jordan-Wigner mapping, the electron number parity and the spin number parity of the state $\rho$ produced by the ansatz can be directly measured at the end of the computation. We give an example of such measurement in Fig.~\ref{fig:sym-verify-particle-number}. On the other hand, if the symmetry operator is more complex but can still be decomposed into a weighted sum of Pauli operators, it could then be measured using the same technique in VQE, even reusing data obtained already in measuring the energy of the state $\rho$.

The cost of this method can be characterized by \cite{caiMultiexponentialErrorExtrapolation2021}:
\begin{equation}\label{eq:symmetry-verification-success-fraction}
    C=\frac{1}{\operatorname{Tr}[\Pi_{s}\rho]},
\end{equation}
where $\Pi_{s}$ projects into the eigenspace of the desired eigenvalue $s$. This is because
only the $\mathrm{Tr}[\Pi_{s}\rho]$ fraction of total circuit executions pass the symmetry verification test.

\begin{figure}
    \centering
    \subfloat[]{
        \label{fig:em-sym-ancilla-bulk-1}
        \includegraphics[width=0.45\linewidth]{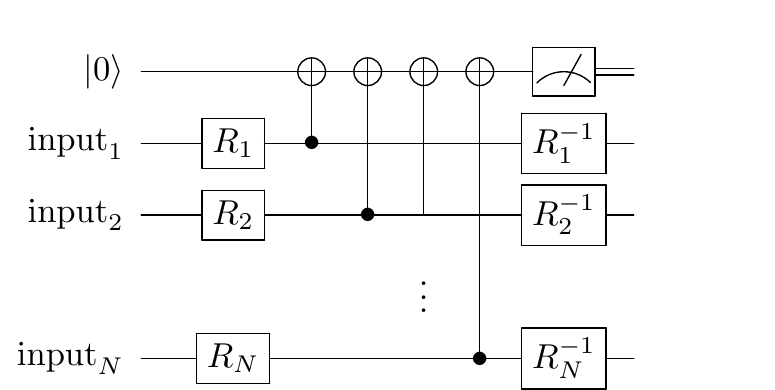}
    }
    \subfloat[]{
        \label{fig:em-sym-ancilla-bulk-2}
        \includegraphics[width=0.45\linewidth]{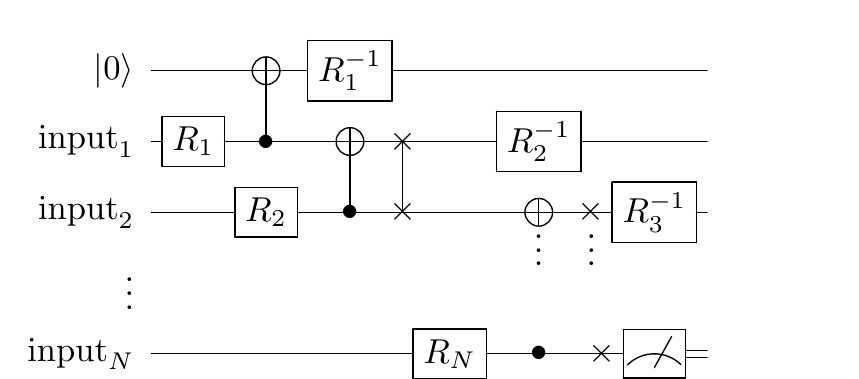}
    }\\
    \subfloat[]{
        \label{fig:em-sym-ancilla-bulk-mid-1}
        \includegraphics[width=0.45\linewidth]{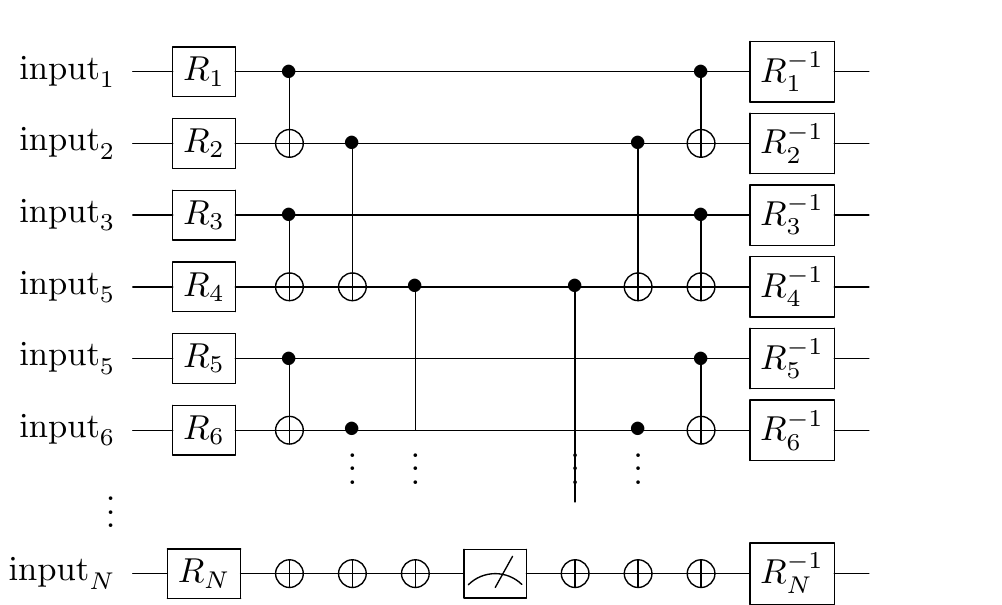}
    }
    \subfloat[]{
        \label{fig:em-sym-ancilla-bulk-mid-2}
        \raisebox{10pt}{  
            \includegraphics[width=0.45\linewidth]{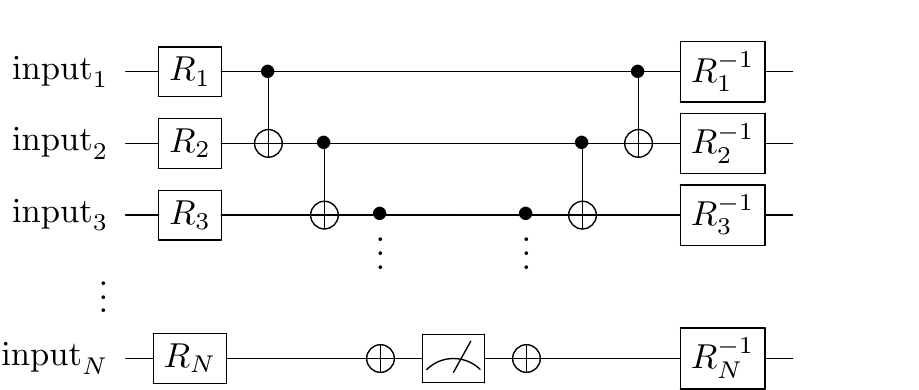}
        }
    }
    \caption{Four different circuits that achieve the same bulk symmetry verification 
    (Sec.~\ref{sec:mit-symmetry-verification}) with different hardware requirements (adapted from \citet{bonet-monroigLowcostErrorMitigation2018}).
    In \protect\subref{fig:em-sym-ancilla-bulk-1} and \protect\subref{fig:em-sym-ancilla-bulk-2}, 
    an ancilla qubit initialized in state $\ket{0}$ is required for measuring the symmetry.
    In \protect\subref{fig:em-sym-ancilla-bulk-mid-1} and \protect\subref{fig:em-sym-ancilla-bulk-mid-2},
    the quantum computer should be able to perform a mid-circuit measurement.
    In all four circuits, the rotation gates $\{R_i\}$ perform a basis transformation such that
    the symmetry operator is equivalent to the qubit parity operator in the transformed basis.
    }
    \label{fig:mit-bulk-sym}
\end{figure}

\paragraph{Bulk symmetry verification:}
Here we need to measure the observable $\hat{S}$ without disrupting the computation. At the moment, we can only achieve this when $\hat{S}$ can be decomposed into tensor products of one-qubit Pauli matrices
\begin{equation}
    \hat{S} = \hat{s}_1 \otimes \hat{s}_2 \otimes \cdots,
\end{equation}
where $\hat{s}_i\in \{I, X, Y, Z\}$. In \citet{bonet-monroigLowcostErrorMitigation2018}, four different but equivalent circuits have been proposed using a similar construction, each having different advantages depending on the characteristics of the quantum computer. In general, a basis transformation is used to map $\hat{s}_i$ to the computational basis, using the pre-rotation gates already mentioned in Sec. \ref{sec:pauli_grouping}. Information on all $\hat{s}_i$ in the computation basis is copied using CNOT gates to a qubit, which is then measured to extract the $\langle \hat{S}\rangle$ for filtering. We present all four circuits in Fig.~\ref{fig:mit-bulk-sym}.

\paragraph{Symmetry verification by post-processing:}
Verifying the symmetry $\hat{S}$ of a quantum state $\rho$ projects the state into $\rho_s$, which lies within the eigenspace of the symmetry operator with the desired valued $s$. Subsequently, an observable $\hat{O}$ is measured on the projected state,
giving a measured outcome $\mathrm{Tr}[\hat{O} \rho_s]$.
When both $\hat{S}$ and $\hat{O}$ are members of the Pauli group, it was shown \cite{bonet-monroigLowcostErrorMitigation2018} that 
\begin{equation}
    \operatorname{Tr}\left[\hat{O} \rho_{s}\right] =
    \frac{\operatorname{Tr}[\hat{O} \rho]+s \operatorname{Tr}[\hat{O} \hat{S} \rho]}{1+s \operatorname{Tr}[\hat{S} \rho]}.
\end{equation}
Therefore, the effect of symmetry verification, in this case, can be achieved with measurement outcomes of observables $\hat{O}$, $\hat{S}$, and $\hat{O}\hat{S}$. All three observables can be obtained as part of the measurements that in any case must be performed as part of the VQE optimization process.
Therefore, we can achieve symmetry verification without measuring the symmetry operator $\hat{S}$ and post-selecting the measured number. This adds one more measured observable ($\hat{O}\hat{S}$) per Pauli observable, and one more
measured observable $\hat{S}$ overall. This kind of symmetry verification is identical to a specific form of the quantum subspace expansion~\cite{bonet-monroigLowcostErrorMitigation2018}, where the excitation operators include the identity operator and the symmetry operator $\hat{S}$.

\subsection{Extrapolation based methods}
\label{sec:mit-extra}

Extrapolation methods are based on the simple idea that the measurement result $\langle \hat{O} \rangle$ on a quantum computer is affected by the strength of the noise on the device, i.e. $\langle \hat{O} \rangle= \langle \hat{O} \rangle( \varepsilon )$, where $\varepsilon$ characterizes the noise strength of the device. Therefore, using the data $\{ \langle \hat{O} \rangle (\varepsilon _{i})\}_{i\in \mathbb{N}}$ for the expectation value under several noise strength $ \{\varepsilon _{i}\}$, we can construct a mathematical model $ \langle \hat{O} \rangle ( \varepsilon |\alpha )$ describing $ \langle \hat{O} \rangle ( \varepsilon )$, where $\alpha =\{\alpha _{k}\}$ are the set of parameters parameterizing the model used for extrapolation. We can then use the same mathematical model to predict the error-free expectation value $ \langle \hat{O} \rangle_{\mathrm{exact}} \equiv \langle \hat{O} \rangle (0) \approx \langle \hat{O} \rangle ( \varepsilon =0|\alpha )$. We denote by $\varepsilon_{0}$ the smallest noise strength of the quantum computer that can be achieved by hardware engineering, i.e. $\varepsilon _{i} \geqslant \varepsilon _{0}$ for all $i\in \mathbb{N}$. Of course, on a noiseless quantum computer we would have $\varepsilon_{0}=0$, which is not a realistic assumption. Hence, $ \langle \hat{O} \rangle ( \varepsilon _{0})$ denotes the measurement outcome that would be achieved on the quantum computer without any error mitigation technique.

This method brings the additional overhead of the extra circuit executions required to obtain $\{ \langle \hat{O} \rangle ( \varepsilon _{i})\}$ at different noise levels. It also potentially add noise coming from the adjustment of the noise strength $\varepsilon$.

Care should be taken regarding implementing this method as part of a VQE algorithm: one may model the overall energy $E$ as $E=E(\varepsilon|\alpha)$, but we argue here that it is more practical to model each Pauli observable separately. That is, each Pauli observable measured separately should be modeled with an independent $ \langle \hat{O} \rangle (\varepsilon|\alpha)$.
The reason is that the difference in noise from measuring different Pauli observables is a result of the different circuits required to rotate the state in the desired measurement basis. While this is likely insignificant for naive measurements of all Pauli observables, it becomes very relevant when grouping strategies relying on general commutativity of the Pauli observables are introduced (see Sec.~\ref{sec:pauli_grouping}), since the basis rotation circuit depth scales as $\mathcal{O}(N^2)$ \cite{Gokhale2019_long}. 
It is worth pointing out that the parallel implementation for the VQE (as discussed in Sec. \ref{sec:parallelization}) also benefits from model fitting for each different hardware on which measurements are performed,
especially when the noise characteristics vary across the hardware used.
The effectiveness of extrapolation methods therefore depends on the ability to characterize and systematically increase the noise strength parameter $\varepsilon $ (Sec. \ref{sec:mit-increase_noise}), and on the soundness of the mathematical model $\langle \hat{O} \rangle( \varepsilon |\alpha )$ (see Sec. \ref{sec:mit-noise_model}).

\subsubsection{Method to systematically increase the noise}\label{sec:mit-increase_noise}

\paragraph{Re-scaling method}
This method, proposed originally by Temme \textit{et al.} \cite{temmeErrorMitigationShortDepth2017}, can achieve an effective scaling of the noise strength by rescaling the Hamiltonian $H_{\mathrm{QPU}}$ (here we are referring to the Hamiltonian driving the action of the quantum circuit on the initial state, rather than the Hamiltonian of the quantum chemistry problem we are trying to solve). The dynamics on a quantum computer
can be described by the master equation
\begin{equation}
    \partial _{t} \rho (t)=-i[\hat{H}_{\mathrm{QPU}}(t),\rho (t)]+\varepsilon _{0}\mathcal{L} [\rho (t)],
\end{equation}
where the term $\varepsilon _{0}\mathcal{L} [\rho (t)]$ represents the noise. An initial quantum state $\rho(0)$ is evolved under
the Hamiltonian $\hat{H}_{\mathrm{QPU}}(t)$ for a period of time $T$ to the state $\rho_{\varepsilon_0}$ where $\varepsilon_0$ characterize the noise strength of the quantum computer. Another copy of the same initial state is evolved under a different Hamiltonian $\hat{H}_{\mathrm{QPU}}(t/c)/c$ for a longer period of $cT$, resulted in another state $\rho_{c\varepsilon_0}$. Then, assuming the noise $\mathcal{L}$ is independent of time $t$ and of the Hamiltonian, one has
\begin{equation}
    \rho _{c\varepsilon_0} -\rho _{\varepsilon_0} =c \varepsilon _{0}\mathcal{L}.
\end{equation}
Therefore one can achieve an effective rescaling of the expectation value, resulting in $ \langle \hat{O} \rangle (\varepsilon=c \varepsilon _{0} )- \langle \hat{O} \rangle (\varepsilon=\varepsilon_0)=c \varepsilon _{0}\operatorname{Tr}[\hat{O}\mathcal{L}]$.
This scaling modifies the Hamiltonian on quantum computers directly and therefore requires precise control of the device, which has been demonstrated by the hardware team in \citet{kandalaErrorMitigationExtends2019}.
However, the recently proposed OpenPulse specification~\cite{mckayQiskitBackendSpecifications2018} allows a more
platform-independent way to control the underlying hardware which
has been used to implement this error boosting method  \cite{garmonBenchmarkingNoiseExtrapolation2020}.

\paragraph{Pauli Twirling}
Li and Benjamin \cite{liEfficientVariationalQuantum2017} propose that one can boost the physical error rate to any desired value by randomly applying Pauli gates before and after a Clifford entangling two-qubit gate, and then randomly generating additional Pauli gates after the twirled two-qubit gates. 

\paragraph{Inserting CNOT gates}
\label{sec:mit-cnot-boosting}
\newcommand{\ncnot}{\ensuremath{J_\mathrm{exact}}}
Refs.~\cite{dumitrescuCloudQuantumComputing2018, heResourceEfficientZero2020} showed that when every CNOT gate is replaced by $j$ consecutive CNOT gates ($j$ should be an odd number), the measurement outcome $\langle \hat{O} \rangle(j)$ depends on $j$ by
\begin{equation}
    \label{eq:mit-insert-cnot-r}
    \langle \hat{O} \rangle(j)=\langle \hat{O} \rangle_{\mathrm{exact}} +kj+\mathcal{O}\left(\left( j\sum _{i=1}^{\ncnot} \varepsilon _{i}\right)^{2}\right) ,
\end{equation}
where $\langle \hat{O} \rangle_{\mathrm{exact}}$ is the exact measurement outcome when there is no noise, $\varepsilon _{i}$ characterizes the noise strength of the $i$-th CNOT gate, $k$ is the parameter characterizing the noise strength, and $\ncnot$ is the number of CNOT gates in the original circuit. In this model, it is assumed that the noise is dominated by the two-qubit depolarizing noise on the CNOT gates, and all other forms of error are considered negligible. The extrapolation method can therefore be used for extrapolating the case of $j=0$.

This method becomes problematic when there are a large number of CNOT gates in the original circuit, each of which needs to be replaced by $j$ CNOT gates. Eq.~(\ref{eq:mit-insert-cnot-r}) means that errors in higher orders of $j$  quickly overshadow lower-order contributions used for modeling the error. 

As a solution, \citet{heResourceEfficientZero2020} provides an approach, which is appropriate for large numbers of CNOT gates. Instead of replacing each CNOT with $r$ CNOT gates, the proposal is to replace the $i$-th CNOT gate with $j_{i}$ CNOT gates, where $(j_i - 1)/2 \sim \mathrm{Poisson}( v)$, a Poisson distribution with mean $v$. That is, the number of additional CNOT gates is sampled each time from the Poisson distribution. As a result, Eq.~(\ref{eq:mit-insert-cnot-r}) is slightly changed to
\begin{equation}
    \label{eq:mit-insert-cnot-v}
    \langle \hat{O} \rangle( v) =\langle \hat{O} \rangle_{\mathrm{exact}} +(1+2v) k +\mathcal{O}\left(\left((1+2v)\sum_{i=1}^{\ncnot}\right)^{2}\right) .
\end{equation}
Using this adjustment, the circuit is run for different values of $v$ to extrapolate to the case of $v=-1/2$. Errors in the quantum computer can be mitigated by extrapolating from measurement outcomes $\langle \hat{O} \rangle$ corresponding to different values of $j$ (Eq.~\ref{eq:mit-insert-cnot-r}) or $v$ (Eq.~\ref{eq:mit-insert-cnot-v}) using a linear model (Sec.~\ref{sec:mit-poly-model}). In particular, $v$ can be any positive number chosen such that $1+2v$ is close to $1$, making the second method more noise friendly than the first method, where $j$ can only take integer values.

\paragraph{Unitary folding:}
This method is similar to the CNOT insertion method mentioned above. Instead of inserting CNOT gates,
\citet{giurgica-tironDigitalZeroNoise2020} proposes that one can insert only gates from the original circuit to boost the error rate. In this method, named \textit{unitary folding}, one assumes a unitary circuit $U$ is made from several layers $\{U_i\}_{i=1}^d$ of quantum gates
\begin{equation}
    U = U_d \cdots U_2 U_1,
\end{equation}
where $d$ is the number of layers. To boost the error, $U$ is transformed into an equivalent circuit
\begin{equation}
    U_{\mathrm{folded}} =U(U^{\dagger } U)^{L_{1}} U_{l_{L_{2}}}^{\dagger } U_{l_{L_{2} -1}}^{\dagger } \cdots U_{l_{1}}^{\dagger } U_{l_{1}} \cdots U_{l_{L_{2} -1}} U_{l_{L_{2}}} ,
\end{equation}
where $U$ has been \textit{folded} $2L_2+1$ times, and an additional subset $\{l_1,l_2,\cdots, l_{L_2}\}$ ($L_2<d$) of the $d$ layers is chosen to boost the error even further. There is no particular rule as to how this subset should be chosen. However, these additional gates give us additional freedom to control the noise strength. One can define $\lambda$ as the ratio of layers in the folded circuit over the original circuit
\begin{equation}
    \lambda = \frac{(2L_1+1)d + 2L_2}{d} = 2L_1+1 + \frac{2L_2}{d}.
\end{equation}

Form this point, $\lambda$ can be systematically increased with a step size of $2/d$, giving granular control of this noise parameter. Assuming a depolarizing noise model on all qubits, the measurement output of the folded circuit has the form
\begin{equation}
    \langle \hat{O} \rangle(\lambda) = \langle \hat{O} \rangle_{\mathrm{exact}} + b \varepsilon^\lambda,
\end{equation}
where $b$ and $\varepsilon$ are coefficients that are determined from extrapolating from the measured data. Therefore, extrapolation to the case of $\lambda=0$ can give the desired error-free measurement outcome.

\paragraph{Multi-parameter noise model:}
It is, in general, more realistic to characterize the noise of a quantum computer by multiple parameters, such as the  times $T_1$ and $T_2$ (see Sec.~\ref{sec:mit-noise-relaxation}), which are usually used as figure of merits for quantum computers. In this case, the noise parameter $\varepsilon$ becomes a vector
$\varepsilon = (\varepsilon^{(1)},\cdots, \varepsilon^{(l)},)$ of $l$ different noise parameters.
\citet{ottenRecoveringNoisefreeQuantum2019} discusses an interesting scenario where $T_1$ and $T_2$ are
estimated and used to build a model that recovers the population in state $\ket{1}$.
Although we have not seen a method that systematically controls errors on multiple noise parameters,
we note that all modeling methods discussed below may be generalized to this multi-parameter case.

\subsubsection{Modeling the noise and extrapolating from measured data} \label{sec:mit-noise_model}

Once a suitable technique has been developed to artificially increase the noise in the device, one can turn towards finding a model to extrapolate the data to obtain a better approximation of the true expectation value.  A general formulation of the extrapolation can be made: given the model $\langle \hat{O} \rangle(\varepsilon|\alpha)$ and several measurement outcomes $\langle \hat{O} \rangle(\varepsilon_i)$, an optimal coefficient $\alpha^*$ can be found by optimizing function $f(\alpha)$, which measures the squared error between prediction $\langle \hat{O} \rangle(\varepsilon_i|\alpha)$ and the measurement outcome $\langle \hat{O} \rangle(\varepsilon_i)$
\begin{equation}
    \label{eq:mit-model-square-loss}
    f(\alpha) = \sum_i |\langle \hat{O} \rangle(\varepsilon_i) - \langle \hat{O} \rangle(\varepsilon_i|\alpha)|^2.
\end{equation}
This method is very common in data analysis, and is often used in regression techniques (for further details, see Ref. \cite{goodfellowDeepLearning2016}). 

\paragraph{Linear/Polynomial model:} \label{sec:mit-poly-model}
In this model $\langle \hat{O} \rangle( \varepsilon |\alpha )$ can be generalized to
\begin{equation}
    \label{eq:mit-linear-1}
    \langle \hat{O} \rangle( \varepsilon |\alpha ) -\langle \hat{O} \rangle_{\mathrm{exact}} =\sum ^{K}_{k=0} \alpha _{k} \varepsilon ^{k} + \mathcal{O}(\varepsilon^{K+1}),
\end{equation}
which was shown to be valid in \citet{temmeErrorMitigationShortDepth2017}, when assuming that the noise on quantum computer is small enough, such that the residual noise $\langle \hat{O} \rangle( \varepsilon ) -\langle \hat{O} \rangle( \varepsilon |\alpha ) =\mathcal{O}( \varepsilon ^{K+1})$.
This function was also shown to characterize noise correctly when inserting additional identity gates into the quantum circuits, see Sec.~\ref{sec:mit-cnot-boosting}.

Having obtained $\langle \hat{O} \rangle(\varepsilon _{i} )_{i=0}^{L}$, where $L+1$ is the number of measurement outcomes we obtained for $L+1$ different values of $\varepsilon _{i}$, we mention two methods to extrapolate an approximation to the exact measurement outcome $\langle \hat{O} \rangle_{\mathrm{exact}}$. The first method is called Richardson extrapolation~\cite{liEfficientVariationalQuantum2017,temmeErrorMitigationShortDepth2017}, and can be applied when $L=K$. Here, one rewrites Eq.~(\ref{eq:mit-linear-1}) as
\begin{equation}
    \label{eq:Richardson_eq1}
    \langle \hat{O} \rangle(c_{i} )=\langle \hat{O} \rangle_{\mathrm{exact}} +\sum _{k=0}^{K} \alpha _{k}( c_{i} \varepsilon _{0})^{k} +\mathcal{O} (c_{i}^{K+1} ),
\end{equation}
where $c_{i} =\varepsilon _{i} /\varepsilon _{0}$ is the ratio of the noise strength of different cases relative to the reference value $\varepsilon _{0}$. Next, one seeks a series of combination coefficients $\beta _{i}$ such that
\begin{equation}
    \begin{split}
        \label{eq:mit-richardson-1}
        \sum _{i} \beta _{i} \langle \hat{O} \rangle( c_{i}) & =\langle \hat{O} \rangle_{\mathrm{exact}} ,
    \end{split}
\end{equation}
for any $\varepsilon_{0}$, ignoring the higher order terms $\mathcal{O} (\varepsilon_0^{N+1} )$ in Eq.~\ref{eq:Richardson_eq1}. That is, one considers Eq.~(\ref{eq:mit-richardson-1}) as an equivalence of two polynomials in $\varepsilon _{0}$, leading to the conditions
\begin{equation}
    \begin{split}
        \sum _{i} \beta _{i} & =1,\\
        \sum _{i} \beta _{i} c_{i}^{k} & =0,\ k=1,2,\cdots K,
    \end{split}
\end{equation}
which admit the solution
\begin{equation}
    \begin{split}\label{eq:mit-poly-beta}
        \beta _{i} & =\prod _{k\neq i}\frac{c_{k}}{c_{k} - c_{i}} .
    \end{split}
\end{equation}
It is worth noting that more explicit formulae for the constants $c_i$ can be obtained assuming depolarizing noise in the CNOT-insertion based error boosting method~\cite{heResourceEfficientZero2020}, see Sec.~\ref{sec:mit-cnot-boosting}.

The second method is a simple extension of the linear regression technique commonly found in machine learning \cite{hastie2015statistical}, which we already mentioned above (Eq.~\ref{eq:mit-model-square-loss}). Here, the optimization can be furthered simplified. One first rewrites Eq.~(\ref{eq:mit-linear-1}) into a matrix form by defining
\begin{equation}
    \begin{split}
        \label{eq:mit-poly-regression}
        A & =\begin{pmatrix}
            1      & \varepsilon _{0} & \cdots & \varepsilon _{0}^{K} \\
            1      & \varepsilon _{1} & \cdots & \varepsilon _{1}^{K} \\
            \vdots & \vdots           &        & \vdots               \\
            1      & \varepsilon _{L} & \cdots & \varepsilon _{L}^{K}
        \end{pmatrix} ,\\
        X & =( \langle \hat{O} \rangle_{exact} ,\alpha _{1} ,\cdots ,\alpha _{K})^{T} ,\\
        Y & =( \langle \hat{O} \rangle( \varepsilon _{0}) ,\langle \hat{O} \rangle( \varepsilon _{1}) ,\cdots ,\langle \hat{O} \rangle( \varepsilon _{L}))^{T} ,\\
        f( X) & =( AX-Y)^{T}( AX-Y) ,
    \end{split}
\end{equation}
where the cost function $f$ measures the squared difference between the measurement outcome $Y$ and the prediction $AX$ (i.e. $\langle \hat{O} \rangle( \varepsilon _{i} |\alpha )$). The minima of $f$ (which contains the desired $\langle \hat{O} \rangle_{\mathrm{exact}}$) can be obtained using the equation \cite{hastie2015statistical}
\begin{equation}
    \begin{split}
        X^{*} & =\argmin_{X} \ f( X) =\left( A^{T} A\right)^{-1} A^{T} Y,
    \end{split}
\end{equation}
when $A^{T} A$ is not singular. Note that one often controls the ratio $c_{i} =\varepsilon _{i} /\varepsilon _{0}$, and the method works in the same manner by replacing $\varepsilon _{i}$ with $c_{i}$ in Eq.~(\ref{eq:mit-poly-regression}).

Here we compare the two methods. The Richardson extrapolation coincides with the second minimization method when $K=L$. This can be seen by noticing that $A$ has the form of a Vandermonde matrix and its determinant when $K=L$ is $\prod _{0\leqslant i< j\leqslant K}( \varepsilon _{j} -\varepsilon _{i})$, which is always invertible when the $\varepsilon _{i}$ have distinct values. However, although it may be less interesting to consider the case when $K>L$ when there are insufficient measurement samples to fix the free parameters in our model Eq.~(\ref{eq:mit-linear-1}), we argue here that the $K<L$ case is interesting; this uses a polynomial of order $K$, which is smaller than $L$, the number of measurement samples. The argument is as follows: although when $K=L$ we may obtain a solution $\alpha $ such that each measurement outcome $M( \varepsilon _{i})$ could be exactly predicted by our model $M( \varepsilon |\alpha )$, a model with higher-order polynomials is more sensitive to statistical fluctuations \cite{giurgica-tironDigitalZeroNoise2020} and generally does not perform well for predicting new values. Similar arguments could be found in the machine learning literature under the topic of \textit{over-fitting }\cite{goodfellowDeepLearning2016}. In this case, the second method is also compatible with techniques addressing the over-fitting problem, such as the Lasso regression\cite{hastie2015statistical}.

Finally, it is very important to note that extrapolation based on this model increases the cost of the VQE algorithm, possibly prohibitively. For each Pauli observable evaluation, the required circuit evaluation is multiplied by the size of data used to fit the model, and the variance of the observable is also increased by a factor $\gamma$. When the Richardson extrapolation is used, or when the matrix $A$ in Eq.~(\ref{eq:mit-poly-regression}) is a square and non-singular matrix, $\gamma$ can be derived analytically \cite{suguru2019a,temmeErrorMitigationShortDepth2017} as
\begin{equation}\label{eq:mit-poly-variance-amplification}
    \gamma = \sum_i \beta_i^2,
\end{equation}
with $\beta_i$ as in Eq.~(\ref{eq:mit-poly-beta}).

\paragraph{Exponential model:} \label{sec:mit-exp-model}
For a deep circuit, it is natural to expect that the error dominates the quantum computation exponentially fast.
\citet{endoPracticalQuantumError2018} demonstrated this in a simple example,
where a quantum circuit consisting of $J$ noisy gates is executed on a noisy quantum computer. It assumes that the noise
on the $i$-th identity gate $\mathcal{E}_i$ can be modeled by a simple error model of the form
\begin{equation}
    \mathcal{E}_i(\rho) = (1-\varepsilon) \rho + \varepsilon \mathcal{\mathcal{E}'}_{i} (\rho).
\end{equation}
An example is the depolarizing noise, where $\mathcal{E'}_i(\rho) \propto \unit$. With this assumption,
the overall noise effect $\prod_i \mathcal{E}_i$, when $J$ is large, can be approximated by
\begin{equation}
    \prod_i \mathcal{E}_i \approx e^{-J\varepsilon}  \sum_{k=0}^{J} \frac{(J\varepsilon)^k}{k!} \mathcal{X}_k,
\end{equation}
where $\mathcal{X}_k$ contains contributions of $\mathcal{E'}$ applied $k$ times towards the product. In order to fit an exponential noise extrapolation model, one can re-write $\langle \hat{O} \rangle( \varepsilon |\alpha )$ as
\begin{equation}
    \label{eq:mit-exp-1}
    \langle \hat{O} \rangle( \varepsilon |\alpha ) -\langle \hat{O} \rangle_{\mathrm{exact}} =A\exp( -J \varepsilon ).
\end{equation}
\citet{endoPracticalQuantumError2018} gives an explicit formula for $\langle \hat{O} \rangle( 0)$, computable with prior knowledge of $\langle \hat{O} \rangle( \varepsilon _{0})$ and $\langle \hat{O} \rangle( k\varepsilon _{0})$,
\begin{equation}
    \label{eq:mit-exp-2}
    \langle \hat{O} \rangle(0) = \frac{k e^{J \varepsilon}\langle \hat{O}\rangle(\varepsilon)-e^{J k \varepsilon}\langle \hat{O}\rangle(k \varepsilon)}{k-1}.
\end{equation}
Furthermore, one can combine the exponential model with the polynomial model into a poly-exponential extrapolation \cite{giurgica-tironDigitalZeroNoise2020}, where the noise model is
\begin{equation}
    \langle \hat{O} \rangle( \varepsilon |\alpha ) = A + B \exp(\mathrm{poly}(\varepsilon)).
\end{equation}
Here the degree of the polynomial in $\varepsilon$ needs to be chosen by experience, and the coefficients can be obtained with the minimization method mentioned at the beginning of this section. Giurgica-Tiron \textit{et al.} \cite{giurgica-tironDigitalZeroNoise2020} demonstrate that this method successfully mitigates errors resulting from quantum circuit execution, however, they do not compare with other extrapolation methods in terms of cost and benefits. 

Extrapolation based on this model similarly increases the cost of the VQE  algorithm.  For each Pauli string evaluation, the number of circuit repetitions is multiplied by the size of data, and as above, the variance of the observable is also increased by a factor $\gamma$.  For a simple exponential model in Eq.~(\ref{eq:mit-exp-2}), $\gamma$ can be derived analytically as \cite{suguru2019a}
\begin{equation}
    \gamma = \frac{k^{2} \exp(2\varepsilon J)+\exp(2 k \varepsilon J )}{(k-1)^{2}},
\end{equation}
assuming the variance of $\langle \hat{O} \rangle( \varepsilon _{0})$ and $\langle \hat{O} \rangle( k\varepsilon _{0})$ are equal.
One can note that the variance in this exponential model tends to be significantly larger than for the polynomial model. 
However, it has been shown~\cite{endoPracticalQuantumError2018} that if enough sampling is affordable, this exponential model gives more accurate prediction of the exact measurement outcome than the polynomial model.
The polynomial model can be seen as an approximation to this exponential model when the noise
$\varepsilon$ is not very high.

\subsection{Probabilistic error cancellation}
\label{sec:mit_pec}
In Probabilistic error cancellation (PEC) \cite{temmeErrorMitigationShortDepth2017,endoPracticalQuantumError2018},
quantum circuits are actively modified with the overall goal of inverting the impact of noise on a quantum computer. Mathematically, the estimated ideal measurement value $\langle \hat{O} \rangle$ is obtained as
\begin{equation}
    \label{eq:mit-pec-1}
    \langle \hat{O} \rangle_{\mathrm{PEC}} =\sum _{i} q_{i} \langle \hat{O} \rangle_{\mathrm{PEC},i} \ ( q_{i} \in \mathbb{R}),
\end{equation}
where each measurement value $\langle \hat{O} \rangle_{\mathrm{PEC},i}$ corresponds to a different, slightly modified, quantum circuit. Specifically,
given a quantum circuit $U=\prod_k U_k$,
one finds weights $q_{i_k}$ and operations on quantum computer $B_{i_k}$ such that
\begin{equation}
    U_k = \sum_{i_k} q_{i_k} B_{i_k},
\end{equation}
and hence overall
\begin{equation}\label{eq:mit-pec-cor-all-decomp}
U = \prod_k \left(\sum_{i_k} q_{i_k} B_{i_k}\right) 
= \sum_{i_1,i_2,\cdots} \prod_k q_{i_k} B_{i_k}.
\end{equation}
Here the weights $\prod_k q_{i_k}$ are used as the coefficient $q_i$ in Eq.~(\ref{eq:mit-pec-1}),
where $i$ is an abbreviation of the many indices $i_1,i_2,\cdots$.
We provide an example correcting single-qubit gates under the depolarizing noise in Appendix~\ref{sec:mit-pec-toy-example}.

One expects the terms in the summation to explode exponentially
with respect to the depth of the circuit.
Therefore, it helps to implement Eq.~(\ref{eq:mit-pec-1}) probabilistically.
To do this, one rewrites Eq.~(\ref{eq:mit-pec-1}) as
\begin{equation}
    \label{eq:mit-pec-5}
    \begin{split}
        \langle \hat{O} \rangle_{\mathrm{PEC}} & =\gamma _{\mathrm{PEC}}\sum _{i}\mathrm{sgn}( q_{i}) P_{\mathrm{PEC}} \langle \hat{O} \rangle_{\mathrm{PEC} ,i} ,\\
        \gamma _{\mathrm{PEC}} & =\sum _{i} |q_{i} |,\ P_{\mathrm{PEC}} =\frac{| q_{i}| }{\gamma _{\mathrm{PEC}}} .
    \end{split}
\end{equation}
Here $\mathrm{sgn}( q_{i})$ is the sign of $q_{i}$, and $P_{\mathrm{PEC}}$ is a proper probability distribution normalized by the constant $\gamma _{\mathrm{PEC}}$. Eq.~(\ref{eq:mit-pec-5}) provides a probabilistic implementation of the summation in Eq.~(\ref{eq:mit-pec-1}). That is, we could sample from all the modified circuits according to the distribution $P_\mathrm{PEC}$ and execute them. The overall measurement $\langle \hat{O} \rangle_{\mathrm{PEC}}$ can be estimated by adding measurement result $\langle \hat{O} \rangle_{\mathrm{PEC} ,i}$ from each circuit weighted by the factor $\gamma _{\mathrm{PEC}}\mathrm{sgn}( q_{i}) P_{\mathrm{PEC}}$. When the number of modified circuits is large, it is more practical to estimate \ $\langle \hat{O} \rangle_{\mathrm{PEC}}$ using this probabilistic implementation. For this reason, this mitigation technique is called \textit{probabilities error cancellation}.
In the literature \cite{endoPracticalQuantumError2018, temmeErrorMitigationShortDepth2017} pairs of weights $q_{i_k}$ and operations $B_{i_k}$ are considered as a \textit{quasi-probability representation} of $U_k$, $q_{i_k}$ or $q_i$. These are referred to as the \textit{quasi-probability weights}, and incidentally the mitigation method is called the \textit{quasi-probability method}.

Before discussing the quasi-probability representation and its construction, we first discuss its cost. Although the number of noisy circuits in a quasi-probability representation of the ideal circuit $U$ grows exponentially with respect to the depth of the circuit (Eq.~\ref{eq:mit-pec-cor-all-decomp}),
with the probabilistic implementation the cost of PEC only depends on how fast the probabilistic sampling converges,
i.e. the variance of $\langle \hat{O} \rangle_{\mathrm{PEC}}$. Assuming that the measurement outcomes of different circuits $\langle \hat{O} \rangle_{\mathrm{PEC,i}}$ are independent and that their variance can be bounded by $1$
(which applies to the Pauli observable of interest in the VQE algorithm), the variance of $\langle \hat{O} \rangle_{\mathrm{PEC}}$ could be bounded by $\gamma_{\mathrm{PEC}}^2$.

Now we detail how the modified circuits are constructed in the general case \cite{endoPracticalQuantumError2018,strikisLearningbasedQuantumError2020,Takagi2021OptimalResource}. There are several approaches to this problem, and here we mention two important methods. The first constructs modified circuits using tomography data from quantum computers, the second constructs modified circuits based on data from a quantum computer and a given ansatz structure.

\paragraph{Tomography based method  \cite{endoPracticalQuantumError2018}:} Given a circuit $U$ that consists of unitary gates $U=\prod _{k} U_{k}$, we assume that the effect of noise on each unitary gate can be characterized (by tomography) as $\mathcal{U}_{k,\mathrm{noisy}} =\mathcal{N}_{k} \circ \mathcal{U}_{k}$, and that is there is no correlation between errors on quantum gates acting on different qubits (spatial correlation) or different time steps (temporal correlation).
Here $\mathcal{U}_k$ represents the quantum channel corresponding to the unitary gate $U_k$, and is typeset in an italic font to emphasize it being a quantum channel.
From there, one must find pairs of quasi-probability weights $q_{i,k}$ and quantum operations $\mathcal{B}_{i,k}$ (which are not necessarily unitary gates) such that
\begin{equation}
    \label{eq:mit-pec-6_3}
    \mathcal{U}_{k} =\sum _{i} q_{i,k} \mathcal{B}_{i,k} \mathcal{U}_{k,\ \text{noisy}},
\end{equation}
and overall the quantum channel $\mathcal{U}$ corresponding to the unitary gate $U$ is expressed as
\begin{equation}
    \label{eq:mit-pec-7}
    \mathcal{U} =\sum_{k,i} q_{i,k} \mathcal{B}_{i,k} \mathcal{U}_{k,\ \text{noisy}} .
\end{equation}
To find such $\mathcal{U}$, one obtains the gate-wise noise model $\mathcal{N}_k$ for each $k$ using tomography. Then, a different formalism to describe the quantum process $\mathcal{N}_k$ named Pauli transfer matrix formalism is used to obtain the (non-physical) channel $\mathcal{N}^{-1}_k$, described by a linear summation of several basis operations $\{\mathcal{B}'_{i}\}$. We refer the readers to Ref.~\cite{endoPracticalQuantumError2018} for implementation details.

\paragraph{Learning based method:}

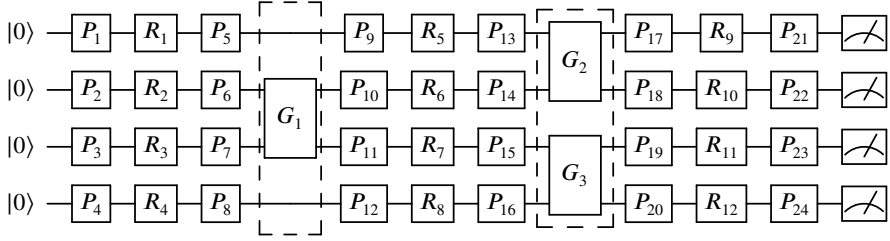
\begin{figure}[ht]
    \centering
    \input{\ProjectRoot/figs/errormit/mit-learning-PEC-setup.tex}
    \caption{Setup of the learning based probabilistic error cancellation method 
    (adapted from \citet{strikisLearningbasedQuantumError2020}).
    A quantum circuit is partitioned into single qubit gates $\mathbf{R}=(R_1,R_2,\cdots)$ followed by Clifford multi-qubit gates $\mathbf{G}=(G_1,G_2,\cdots)$. Additional, single qubit Pauli gates $\mathbf{P}=(P_1,P_2,\cdots )$ are inserted in between the single qubit gates and Clifford gates to mitigate errors.}
    \label{fig:mit-learning-PEC-setup}
\end{figure}

This method, proposed by Strikis \textit{et al.} \cite{strikisLearningbasedQuantumError2020} assumes that single-qubit errors are negligible, and can be applied to quantum circuits having a specific layer-wise structure.
In each layer of these quantum circuits, single qubit gates are followed by Clifford multi-qubit gates. Pauli gates are inserted in between single qubits gates and Clifford multi-qubit gates to mitigate errors (see
Fig.~\ref{fig:mit-learning-PEC-setup}). Specifically, we denote all the Clifford multi-qubit gates collectively by $\mathbf{G}$ (called \textit{frame gates} in the original paper), all single qubits in the unmodified circuit collectively by $\mathbf{R}$, and all the additional inserted Pauli gates by $\mathbf{P}$. To mitigate the error, different combinations of single qubit Pauli gates are inserted, labeled as $\mathbf{P}=(P_1,P_2,\cdots )$, resulting in different quantum circuits. Each of these gives a measurement value labeled by $\langle \hat{O} \rangle(\mathbf{R},\mathbf{P})$. The overall mitigation result $\langle \hat{O} \rangle^{\mathrm{mit}}( \mathbf{R}, q)$ is
\begin{equation}
    \label{eq:mit-pec-l}
    \langle \hat{O} \rangle^{\mathrm{mit}}( \mathbf{R},q) =\sum _{\mathbf{P}} q( \mathbf{P}) \langle \hat{O} \rangle( \mathbf{R},\mathbf{P}),
\end{equation}
where $q:\mathbf{P}\mapsto q( \mathbf{P}) \in \mathbb{R}$ associates a specific weight to a specific combination $\mathbf{P}$. Eq.~(\ref{eq:mit-pec-l}) resembles the quasi-probability formula Eq.~(\ref{eq:mit-insert-cnot-r}) closely, except for the structure of the circuit, and the procedure to modify the circuits. To find the appropriate weights $q(\mathbf{R})$, Strikis \textit{et al.} \cite{strikisLearningbasedQuantumError2020} define a loss function
\begin{equation}
    \label{eq:mit-pec-lcost}
    \mathrm{Loss}( q) =\frac{1}{|\mathbb{T} |}\sum _{\mathbf{R}\in \mathbb{T}} |\langle \hat{O} \rangle^{\mathrm{mit,qpu}}( \mathbf{R},q) -\langle \hat{O} \rangle^{\mathrm{mit,sim}}( \mathbf{R}) |^2,
\end{equation}
where the single qubit gates in the original circuit are substituted by single qubit gates in the set $\mathbb{T}$, called the training set. The substituted circuits are simulated to obtain an exact simulation result $\langle \hat{O} \rangle^{\mathrm{mit,sim}}(R)$. Pauli gates are inserted in the manner specified in Fig.~\ref{fig:mit-learning-PEC-setup} and executed on a quantum computer to obtain the noisy result $\langle \hat{O} \rangle^{\mathrm{qpu,noisy}}( \mathbf{R},q)=\sum_\mathbf{P} q(\mathbf{P}) \langle \hat{O} \rangle(\mathbf{R},\mathbf{P})$.

Overall, one aims to find the weights $q$ such that the loss function in Eq.~(\ref{eq:mit-pec-lcost}) is minimized. Strikis \textit{et al.} \cite{strikisLearningbasedQuantumError2020} show that if the training set $\mathbb{T}$ is chosen as the set of all Clifford circuits, and if one assumes that there exists a $q^{*}$ such that $\mathrm{Loss}\left( q^{*}\right) =0$, then
\begin{equation}
    \langle \hat{O} \rangle^{\mathrm{mit,qpu}}\left( \mathbf{R},q^{*}\right) =\langle \hat{O} \rangle^{\mathrm{exact}}\left( \mathbf{R}\right),
\end{equation}
for any set of single qubit gates $\mathbf{R}$. Here it is also assumed that the single qubit errors (i.e. errors for $\mathbf{R}$) are negligible. The quantity $\langle \hat{O} \rangle^{\mathrm{exact}}\left( \mathbf{R}\right)$ is the exact result of the original circuit without error.


Although Strikis \textit{et al.} \cite{strikisLearningbasedQuantumError2020} has not discussed the optimization of the loss function $\mathrm{Loss}(q)$, it discussed in detail the approach to sampling the training set $\mathbb{T}$ and the set of all Pauli operators $\mathbb{P}$ for estimating the loss function, which might be a concern since both sets grow exponentially with respect to the number of qubits.


\subsection{Exponential error suppression}
\label{sec:mit-ees}
This method was independently proposed in \citet{koczorExponentialErrorSuppression2021,hugginsVirtualDistillationQuantum2021}, and further developed by \citet{obrienErrorMitigationVerified2021, huoDualstatePurificationPractical2021, cai2021resourceefficient}. It is called \textit{exponential suppression by derangement} in \citet{koczorDominantEigenvectorNoisy2021}, and \textit{virtual distillation} in \citet{hugginsVirtualDistillationQuantum2021}.
For fairness, we refer to this method as exponential error suppression to highlight its ability to suppress errors exponentially. Using different circuits, all variants of this method compute the value

\begin{equation}
    \label{eq:ees-computed-rhon}
    \frac{\operatorname{Tr}\left[ \rho ^{\nEES} \hat{O}\right]}{\operatorname{Tr}\left[ \rho ^{\nEES}\right]},
\end{equation}
where a quantum state $\rho$ is prepared by a circuit, e.g. the ansatz circuit of VQE. Here, $\hat{O}$ can be any observable, and $\nEES$ is a free parameter that can be adjusted. The effect of calculating Eq.~(\ref{eq:ees-computed-rhon})
instead of directly $\operatorname{Tr}[\rho O]$ is that the dominant eigenvector in $\rho$ is be amplified exponentially with respect to $\nEES$. 
More precisely, from the spectral decomposition of~$\rho$
\begin{equation}
    \label{eq:ees-spectral-decomposition}
    \rho = \sum_{i} p_i \ket{\psi_i}\bra{\psi_i},
\end{equation}
where we adopt the convention that $p_i\geqslant p_{i+1}$, one can show \cite{koczorExponentialErrorSuppression2021} that
\begin{equation}
\begin{split}
    \frac{\mathrm{Tr}\left(\rho ^{\nEES} \hat{O}\right)}{\mathrm{Tr}\left( \rho ^{\nEES}\right)} 
    &= \langle \psi _{1} |\hat{O}|\psi _{1} \rangle + \mathcal{O}(Q^\nEES),
\end{split}
\end{equation}
where $Q\equiv \left(p_1^{-1} -1\right)p_2<1$.

Therefore, if one wants the difference between the contribution from the dominant eigenvector $\bra{\psi_1} \hat{O} \ket{\psi_1}$ 
and the overall calculated value to be only $\epsilon$, one needs $\nEES$ to be\cite{koczorExponentialErrorSuppression2021}:
\begin{equation}
    \nEES = \operatorname{ceil}(\ln{\frac{1}{\epsilon}} + \frac{\ln(2\,p^{-1}_2)}{\ln Q^{-1}})
\end{equation}
where $\operatorname{ceil}(x)$ gives the ceiling of $x$, i.e. the smallest integer that is greater than or equal to $x$.
The calculation of the numerator in the Eq.~(\ref{eq:ees-computed-rhon}) on a quantum computer can be achieved
with the circuit outlined in Fig.~\ref{fig:ees-controlled-D-implement}, or by several alternative methods detailed in the Appendix~\ref{sec:derangement-details}.

\begin{figure*}
    \subfloat[Example of exponential error suppression circuit for $\nEES=3$]{
        \label{fig:ees-general-circ}
        \makebox[0.5\linewidth]{
        \input{\ProjectRoot/figs/errormit/ees-general-circ}
        }
    }
    \subfloat[Example of derangement circuit on $\nEES$ copies of $\rho$]{
        \label{fig:ees-VD-swap-circ}
        \makebox[0.5\linewidth]{
        \resizebox{0.2\linewidth}{!}{
        \input{\ProjectRoot/figs/errormit/ees-VD-swap-circ}
        }
        }
    }
    \caption{General circuits to compute the numerator in Eq.~(\ref{eq:ees-computed-rhon})
    for exponential error suppression 
    (adopted from \citet{koczorExponentialErrorSuppression2021}).
    Here for simplicity we only considered the case when $\nEES=3$,
    but the circuit could be easily generalized to arbitrary $\nEES\geq 2$.
    $D_\nEES$ is the circuit which performs a derangement on all indices $\{i_1,\cdots,i_\nEES\}$. An example of $D_\nEES$
    is shown to the right. It swaps all nearest neighbor qubits, as well as the first and the last qubit. Details on properties of $D_\nEES$ could be found in the appendix~\ref{sec:derangement-details}.
    }
    \label{fig:ees-controlled-D-implement}
\end{figure*}
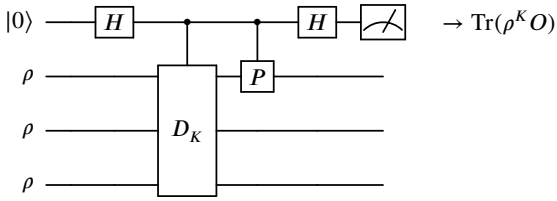
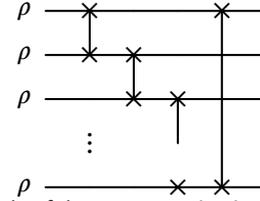

To apply this method to the VQE algorithm, the mixed state $\rho$ can be prepared by the ansatz circuit $U(\boldsymbol{\theta})$
in a noisy quantum computer, and then each Pauli observable $P$ in the decomposed Hamiltonian is measured
on $\rho$ with this suppression technique. 
If the energy has been minimized, the dominant state $\ket{\psi_1}$ in $\rho$ prepared by the ansatz circuit
has energy exponentially close to the true ground state energy of the Hamiltonian.
Therefore, exponential error suppression purifies the mixed state and helps to prepare a pure eigenstate of Hamiltonian when combined with VQE.

Meanwhile, the variance of the measured quantity, when $\hat{O}$ is again also a unitary
operator, is comparable to the variance of measuring $\hat{O}$ directly. More precisely,
the number of samples $S$ required to achieve a small error $\epsilon_s^2$ in variance of estimating either the numerator or the denominator is of the order~\cite{koczorExponentialErrorSuppression2021}
\begin{equation}\label{eq:ees-variance}
    \mathcal{O}[\epsilon ^{-2(1+2f)}_{s}],
\end{equation}
where $f=\ln(p_1^{-1})/\ln(Q^{-1}))>0$, which converges to the limiting case without error mitigation method ($\mathcal{O}(\epsilon^{-2})$) as $p_1$ goes to $1$ and $p_2$ goes to $0$.

Next, we note that $\ket{\psi_1}$ is not necessarily the ideal error-free computation result $U(\boldsymbol{\theta})\ket{0}$.
Although this does not affect the VQE algorithm itself as mentioned above, 
it affects some VQE-based algorithms. 
For example, the excited state VQE algorithms (see Sec. \ref{sec:excited_states}) that measure the fidelity between 
$\ket{\psi_1}$ and other quantum states, are affected by the fact that
the inverse unitary $U(\boldsymbol{\theta})^\dagger$ no longer prepares the adjoint $\bra{\psi_1}$ from $\bra{0}$.
Although there is no known efficient solution to this problem, the infidelity between the ideal error-free computation result and $\ket{\psi_1}$, called the coherent mismatch, has been studied extensively in \citet{koczorDominantEigenvectorNoisy2021},
where it is shown that the error from coherent mismatch can be guaranteed to be smaller than the
error caused by the limitation in creating more copies (i.e. the limitation in $\nEES$).

Finally, it is important to mention that there are several methods with different advantages to compute the quantities in Eq.~(\ref{eq:ees-computed-rhon}), which we provide a high-level summary in the Appendix~\ref{sec:derangement-details}. Some of the methods requires several copies of the same state $\rho$. 
In these methods, each copy $\rho^{(i)}$ is slightly different from $\rho$:

\begin{equation}
    \rho^\nEES \mapsto \rho^{(1)} \rho^{(2)} \cdots \rho^{(\nEES)},\;\rho^{(i)} \approx \rho.
\end{equation}

However, as long as the dominant state (the state with the largest amplitude) in the spectral decomposition of $\rho^{(i)}$ remains $\ket{\psi_i}$, which is a reasonable assumption, the conclusions made above remain valid with minor modifications (see \citet{koczorExponentialErrorSuppression2021}).

\subsection{Measurement readout error mitigation}
\label{sec:mit-readout}

State preparation and measurement (SPAM) errors are also important in NISQ quantum computation. It is often important to counter their influence with simple mitigation measures in order to achieve good accuracy on VQE experiments
\cite{kandalaErrorMitigationExtends2019,Kandala2017},
which is particularly important for Pauli operators with large weight. In particular, the choice of encoding (see Sec.~\ref{sec:Encoding}) has a direct implication for the level of readout errors affecting the VQE algorithm. For instance, the operators in the Jordan-Wigner mapping can have up to $\mathcal{O}(n)$ Pauli weight (non-identity operators), while the Bravyi-Kitaev mapping has up to $\mathcal{O}(log(n))$ Pauli weight. This in turn means that less of the qubits are measured on each operator, thereby lowering the risk of readout error \cite{Huggins2021} (although once the operators are grouped, this advantage of low Pauli weight encoding disappears as the entire register must be measured). Similarly, grouping methods based on commutative groups of Pauli strings (see Sec.~\ref{sec:pauli_grouping}) transform the measurement basis such that multiple operators can be jointly measured by simply measuring one qubit at a time and therefore reduces measurement error probability (but could result in a much larger impact in case of error).

To mitigate this type of noise, typically one first performs basic calibration experiments on the quantum computer to obtain a measurement calibration matrix $T$ (introduced below) which maps the probability of each state to be measured as each other, and use this information to mitigate SPAM errors on quantum computer outputs.

\subsubsection{Measurement calibration matrix}

The measurement calibration matrix is defined element-wise as
\begin{equation}
    T_{i,j} = \text{Probability}\left(\text{measured state }i|\text{prepared in state }j\right),
\end{equation}
where $i,j\in\{0,1,\cdots 2^N-1\}$, are all the computational basis states on the $N$ qubits which we want to measure. $T$ models measurement errors as a conventional Markov process, which can be justified using a specific quantum measurement mode~\cite{gellerRigorousMeasurementError2020}.
$T$ could be measured on a quantum computer by first preparing the state in the corresponding basis $j$ using $X$ gates, and measuring the resulting quantum computation in the computational basis. However, since the size of $T$ grows
exponentially with respect to the number of measured qubits, it can rapidly become too expensive to obtain. In this case, an approximation based on the assumption that crosstalk between different qubits during the measurement process is negligible helps alleviate the problem. Here, one approximates $T$ by
\begin{equation}
    \label{eq:mit-readout-t-tensor-prod-approx}
    T \approx T_1 \otimes T_2 \cdots \otimes T_N,
\end{equation}
where $T_k$ is the measurement calibration matrix on the $k$-th qubit among the $N$ qubits to measure. $\{T_k\}_{i=1}^N$ could be obtained in a similar method, but it costs only $\mathcal{O}(N)$ number of measurements to obtain instead of $\mathcal{O}(2^N)$ for $T$. $T_k$ is usually routinely reported by major quantum computing platforms such as IBM Quantum~\cite{IBMQ}. Although more elaborate schemes to efficiently estimate $T$ while taking account of crosstalk effects on neighboring qubits have been studied in \citet{gellerEfficientCorrectionMultiqubit2020}, they do not resolve the issue of exponential growth in the size of $T$ matrix, and more research is needed to adapt this scheme for large $n$ problems. At the moment, the approximation in Eq.~(\ref{eq:mit-readout-t-tensor-prod-approx}) remains the best scalable approximation to $T$.

\subsubsection{Correcting measurement outcomes}

Having obtained $T$ periodically during the experiment, the measurement outcome is corrected using $T$ by the following procedure. One cannot simply apply the inverse of $T$ to the measurement outcomes, since $T$ may not be invertible, and the Moore–Penrose inverse of $T$ may produce unphysical quantities, as even though $T$ satisfies statistical properties (i.e. $T_{i,j}\geq 0$ and $\sum_i T_{i,j}=1$), the inverse may not. Therefore, one approaches the inverse problem by solving the following optimization problem \cite{Qiskit}
\begin{equation}
    x = \argmin_X |Y - T X|^2,
\end{equation}
subject to $\sum_i X_i = \sum_i Y_i$ and, $X_i \geqslant 0$,
where $Y$ is the vector of raw measurement outcome and $x$ is the vector of error mitigated measurement outcomes. In the $i$-th position of each vector is the number of occurrences of the measurement outcome in state $i$. The vector norm is defined as $|v| = \sqrt{v\cdot v}$. The condition ensures that the optimized value $x$ is physical.

Obviously, the size of the matrix $T$ and the vectors $X$ and $Y$ scales exponentially with $N$, the number of qubits measured, making it impractical for large $N$. To alleviate this problem, one can use the approximation scheme Eq.~(\ref{eq:mit-readout-t-tensor-prod-approx}), and store both matrices and vectors in a sparse data format for processing.
Meanwhile, correlated measurement errors between different qubits are ignored in this approximation, making it possible to derive a simple formula giving the mitigated measurement outcome from the noisy measurement outcome. This approach was first explored in \citet{Kandala2017}, and later generalized \cite{yeter-aydenizPracticalQuantumComputation2020,yeter-aydenizScalarQuantumField2019} into the formula
\begin{equation}
    \langle \hat{Z}_{i} \dotsc \hat{Z}_{j}\rangle 
    = \sum_{x} p(x)\prod _{k=i,\cdots ,j}\frac{(-1)^{x_{k}} -p_{k}^{-}}{1-p_{k}^{+}},
\end{equation}
where $\langle \hat{Z}_{i} \dotsc \hat{Z}_{j}\rangle$ is the mitigated measurement outcome of $\hat{Z}$ operators on qubits $i,\cdots,j$,
the summation $\sum_x$ is taken over all possible measurement outcomes $x$, $p(x)$ is the probability of each outcome $x$, $x_k$ is the $k$-th qubit component of $x$, $p_k^{\pm} = p_k(0|1)\pm p_k(1|0)$, $p_k(0|1)$/$p_k(1|0)$ is the probability of measuring $0$/$1$ when the state is $1$/$0$ on the $k$-th qubit and can be easily obtain from $T_k$ (Eq.~\ref{eq:mit-readout-t-tensor-prod-approx}).


It is worth noting that when the measurement calibration matrix $T$ is pathological; statistical uncertainties can be amplified and can result in oscillatory behaviour of the mitigated measurement outcome \cite{nachmanUnfoldingQuantumComputer2020}. Although an improved inversion method can alleviate this problem \cite{nachmanUnfoldingQuantumComputer2020}, when the readout errors are sufficiently small, the method mentioned above was shown experimentally to be good enough \cite{nachmanUnfoldingQuantumComputer2020}, see \citet{nachmanUnfoldingQuantumComputer2020} for details of the improved method.

\subsubsection{Exploiting state-dependent noise}

In certain quantum computer architectures, the design of the quantum computer makes certain states more prone to measurement readout errors than other states. For example, state $\ket{1}$ is often prepared as a state having higher energy than $\ket{0}$, which may have a stronger tendency to drop back to $\ket{0}$ than for a state in $\ket{0}$ to be excited to state $\ket{1}$. In other words, the measurement error can be biased towards certain states. \citet{tannuMitigatingMeasurementErrors2019} examined this phenomenon on several IBMQ quantum computers and discovered a strong correlation of the measurement readout error and the hamming weight of the measured state.
To decrease the readout error, \citet{tannuMitigatingMeasurementErrors2019} proposed splitting all the measurements into two parts, the standard part, and the inverted part. In the inverted part, additional $X$ gates are inserted before all the measurements to invert the qubits, whereas in the standard part no modification is made. The measurement outcomes of the two parts are merged together with simple post-processing to invert the measurement outcomes on the inverted part. Therefore, the effect of the measurement bias on the readout can be averaged in the merged measurement outcomes and can lead to an improvement of performance. However, it was noted that the improvement in performance depends highly on the nature of the output state, and if the output state is favoured by the measurement bias (such as the $\ket{0\cdots 0}$ state), this mitigation method may degrade the performance. \citet{tannuMitigatingMeasurementErrors2019} proposed another adaptive scheme to address this problem, which depends on our ability to predict the probability of certain states and its measurement bias relative to other states. The adaptive scheme therefore does not generalize well to experiments with more number of qubits. Therefore, we suggest accessing the performance of the non-adaptive approach on simplified quantum circuits where the exact quantum outcome can be calculated analytically before using this mitigation method.

\subsection{Other error mitigation methods}

There are a wide range of other error mitigation methods, which we mention briefly in what follows: 

\begin{itemize}
\item Combining different error mitigation methods: \citet{caiMultiexponentialErrorExtrapolation2021} proposes combining exponential extrapolation techniques with quasi-probability and symmetry verification methods, showing by simulation that a lowered estimation bias and a reduced sampling cost could be achieved through mixing different methods. \citet{mari2021extending} combines the quasi-probability method (Sec.~\ref{sec:mit_pec}) with extrapolation methods (Sec.~\ref{sec:mit-extra}) to offer more advantages, including avoiding the need for a full gate-set tomography on the quantum computer, reducing the noise level to a "virtual" noise level that is below the hardware noise level, and potentially reducing the sampling cost of the quasi-probability method.

\item Detecting errors in computation: \citet{urbanekQuantumErrorDetection2019} uses a simple $4$-qubit error detecting code to filter out experiments where one detects an error on a quantum computer. Although it shows certain error mitigation capabilities, the limited set of quantum gates that can be executed with the error detecting code makes it difficult to apply this method to general VQE algorithms.
Also, \citet{mezher2021mitigating} uses different simplified circuits of the original VQE ansatz, constructed with Clifford gates, to detect a degradation in the quality of the output of the original circuit.

\item Combining with reduced density matrix methods: \citet{Rubin2018} proposes using the physical constraint of reduced density matrices (RDMs) for error mitigation. This method has been experimentally tested in \citet{mccaskeyQuantumChemistryBenchmark2019} for $2$-RDMs where the McWeeny purification formula \cite{mcweenypurificationformula} is applied to constrain the RDMs as valid physical states.

\item Modeling the noise: If the noise on the quantum computer can be described by an effective model, we can use data from the device to estimate the strength of noise in it and correct the measurement outcome based on this estimation. An example is discussed in the Appendix A of ~\citet{Tilly2021}, where a simple probabilistic error model is used. In a similar spirit, \citet{vovroshSimpleMitigationGlobal2021} argues that a global depolarization model provides a good effective noise model on the quantum computer, which is used for error mitigation in the experiment to archive significant error reduction in the quantum algorithm. However, it remains unclear whether this method is scalable for algorithms requiring more qubits and deeper quantum circuits.

\item Incorporating machine learning: we may take the advancement of machine learning technologies and apply them for mitigating errors. For example, one can learn the mapping from the noisy observables $\langle \hat{O} \rangle_{\mathrm{noisy}}$ from the quantum computer to the exact one: $f:\langle \hat{O} \rangle _{\mathrm{noisy}} \mapsto \langle \hat{O} \rangle _{\mathrm{exact}}$. Although for a general quantum state $\ket{\psi }$ it is exponentially difficult to compute the expectation value $\langle \psi |\hat{O} |\psi \rangle $ on a conventional computer, such computation is efficient when $\ket{\psi }$ is prepared by certain class of quantum circuits: $\ket{\psi '} =U_{\mathrm{simulable}}\ket{0}$. Therefore, one can use machine learning to learn a function $f'$ mapping from $\langle \psi ' |\hat{O} |\psi'\rangle_{\mathrm{noisy}}$ to $\bra{\psi '}\hat{O}\ket{\psi '}_{\mathrm{exact}}$, and use same function $f'$ mitigating errors on the expectation value on any state $\bra{\psi }\hat{O}\ket{\psi }$, with assumption that $f'$ is similar to $f$. This method has been explored experimentally for Clifford circuits \cite{czarnikErrorMitigationClifford2021} and for quantum circuits in the fermionic linear optics \cite{montanaroErrorMitigationTraining2021}. This approach has been further combined with the extrapolation based method mentioned in Sec.~\ref{sec:mit-extra}~\cite{loweUnifiedApproachDatadriven2020a}. Further, this idea has taken been further to develop artificial neural networks that predict the noise, or the correction needed, on given quantum circuits \cite{kimQuantumErrorMitigation2020, zlokapaDeepLearningModel2020}.

\item Noise aware circuit design: Here, instead of trying to mitigate the noise from a given quantum circuit, it is asked whether one could design an ansatz that is better adapted to the noise on the quantum computer. For example, certain quantum circuits are naturally more resilient to quantum noise \cite{kim2017noiseresilient}. Then, machine learning can help produce more noise-resilient circuits for a specific task such as computing state overlap \cite{cincioMachineLearningNoiseResilient2021}. Recently, several works have been proposed applying machine learning approaches to design better ans\"atze (see Sec.~\ref{sec:Ansatz} for a related discussion on ansatz design), and the resilience towards noise can be naturally incorporated in this design process \cite{Wang2021_QAS,Du2020}.

\item Noise resilience from enhanced sampling: \citet{Wang2021MinimizeEstimationRuntime} discusses a method to enhance the rate of information gain when sampling in quantum computers for estimating expectation values. It was found that by incorporating a well-calibrated noise model into their method, deeper quantum circuits can be utilized to obtain more accurate results in less time \cite{katabarwa2021reducing}. It remains to be seen whether this result holds for experiments involving more qubits.

\item Canceling noise using second derivatives: \citet{ito2021universal} gives a different interpretation of noise on a quantum computer as a fluctuation of parameters in the cost function. This interpretation is valid for certain stochastic noise channels which include the depolarization noise channel. Based on this interpretation, the second derivatives of the cost function can be used to cancel noise in the cost function.


\end{itemize}

\subsection{Impact of error mitigation on the scaling of VQE}

Error mitigation procedures bring extra overheads to VQE algorithms and therefore affect the cost scaling. In this section, we describe how error mitigation procedures affect the scaling of VQE algorithms and discuss potential directions to alleviate the error mitigation overhead.

Firstly, error mitigation procedures
tend to increase the cost of estimating expectation values of each Pauli term in the Hamiltonian. This can be quantified by a systematic increase of the variance of expectation values when more repeated experiments are required to ensure the statistical error of estimating the expectation value is smaller than a fixed threshold. 
Specific analysis of the increase in variance has been done for most error mitigation methods, which are
summarized in the Table~\ref{tab:mit-costs}.

However, this analysis lacks a fair comparison between the cost of different error mitigation methods. Such a comparison is difficult due to the fact that the cost depends on the nature of noise on the quantum computer, the Hamiltonian considered, and sometimes on the desired accuracy of the method itself.
Recently, Takagi et~al. \citep{RyujiFundamentalLimits2021} attempt to set a general
lower bound on the increase in variance in a broad class of error mitigation methods by showing that error mitigation methods tend to increase the difficulty of distinguishing different quantum states. It shows that
\begin{equation}
    \gamma ^{2} \varpropto \frac{1-2b_{\max}}{K\sqrt{NQ} }\left(\frac{1}{1-\varepsilon }\right)^{L/2},
\end{equation}
where $\gamma^2$ is the amplification ratio of the variance, $b_{\max}$ is the maximum error of the error mitigation method, also called the bias of the error mitigation method, $N$ is the number of qubits, $Q$ and $K$ are constants that characterize the error mitigation method considered, $L$ is the number of layers applied when using a layer-wise ansatz, and $\varepsilon$ characterizes the noise present in the quantum computer.
Hence, most error mitigation methods inevitably increase the variance exponentially
with respect to both the number of qubits or the number of repeated layers in the ansatz circuit for VQE.
This is expected, since, in comparison with the scalable error correction technology, where one actively interacts with
the quantum system in order to suppress the errors in it,
error mitigation methods are mostly ``passive'', allowing errors to accumulate in the quantum system.

Secondly, in the NISQ era, where the error correction technology is not applicable due to hardware limitations,
there is substantial interest in studying and improving the efficiency of error mitigation methods. For example,
the result of \citet{RyujiFundamentalLimits2021} indicates that the variance increase due to the error mitigation method
may be offset by compromising on the accuracy of mitigation 
(see theorems 4 and 5 in \citet{RyujiFundamentalLimits2021}). 
\citet{piveteau2021quasiprobability} also provides an example where one trades off the accuracy of the
mitigation method for a smaller variance (i.e. overhead) in the probabilistic error cancellation method 
(see Sec.~\ref{sec:mit_pec}). Another potential direction concerns the assumption made in 
\citet{RyujiFundamentalLimits2021} that the error mitigation method should apply to all possible
quantum states. In a VQE algorithm this may not be necessary, since only the ground state is important, or more generally the eigenstates of a Hamiltonian. It remains to be researched whether the variance increase resulting from error mitigation methods can be adapted to specific subspaces of the whole Hilbert space in order
to decrease its overhead.

Thirdly, it is important to note that it is unclear whether error mitigation is necessary during the initial training stage of the VQE algorithm. It is possible that error mitigation could be applied only when the optimization algorithm in VQE starts to converge. Choosing when to apply error mitigation methods can significantly reduce their cost, and therefore more research is required on this aspect. 

Finally, we want to call for an investigation of how different error mitigation methods compare with each other. The increase in variance often depends on the mitigation parameter of each method, which often affects the accuracy of the method as well. \citet{cai2021practical} makes a fair comparison of different error mitigation methods based on the same noise model. There, it is shown that the symmetry verification method has a better extraction rate than several error mitigation methods analyzed (including the probabilistic error cancellation method and the exponential error suppression method mentioned in this section), which means that it can extract all components of the error mitigated state out of the noisy state. However, the size of the error mitigated state is measured by the \textit{fidelity boost} metric in that paper, and is different for different error mitigation methods. Obviously,
symmetry verification's fidelity boost depends on the number of symmetry elements available in the Hamiltonian.
In general, although Cai \cite{cai2021practical} does not cover a few important error mitigation methods, notably the Richardson extrapolation methods, we believe that Cai provides an important step towards understanding the dependency of the computational cost on the noise in the computer, and a rigorous comparison of different error mitigation methods.

\begin{table}[ht]
    \centering
    \caption{
    The extra cost brought by different error mitigation methods is measured by a relative increase in the number of shots required to achieve certain statistical precision (relative shots amplification). 
    Assuming a shot number of $S_0$ is required to achieve a statistical error of $\epsilon$, with the introduction of error mitigation methods an increased number of shots $\lambda S_0$ is required to achieve the same precision. Most of the increases in the methods presented in the table are due to the variances of observables being amplified by the error mitigation methods.
    An exception is the symmetry verification method where the extra cost comes from the probability that verification fails.
    Note that for the exponential error suppression method, the estimation can only be done approximately
and depends on the factor $f$ (which in turn depends on the amount of noise within the circuit), defined in Sec.~\ref{sec:mit-ees}.
    Although there is an exponential number of circuits that need
    to be evaluated in the probabilistic error cancellation method, in practice it only
    requires randomly sampling from these circuits, and the shot amplification is therefore only determined by its
    variance amplification factor $\gamma_{\mathrm{PEC}}$. 
    For the readout error mitigation method we expect the measurement calibration matrix $T$ to increase the variance of the final observable, although an analytical result is currently not available.
    }
    \label{tab:mit-costs}
    \begin{tabularx}{\textwidth}{llll}
        \toprule
        Model & Relative shots amplification & Relevant section.
        \\ \midrule
        Symmetry Verification \cite{mcardleErrorMitigatedDigitalQuantum2019,Sagastizabal2019,bonet-monroigLowcostErrorMitigation2018} & $1/\mathrm{Tr}[\Pi_{s}\rho]$ 
        & Sec.~\ref{sec:mit-symmetry-verification}, Eq.~(\ref{eq:symmetry-verification-success-fraction}).
        \\  
        Richardson extrapolation \cite{temmeErrorMitigationShortDepth2017,liEfficientVariationalQuantum2017} & $\sum_i \beta_k^2$
        & Sec.~\ref{sec:mit-poly-model}, Eq.~(\ref{eq:mit-poly-variance-amplification}).
        \\  
        Exponential extrapolation \cite{endoPracticalQuantumError2018} & $ \frac{k^{2} \exp(2\varepsilon J)+\exp(2 k \varepsilon J )}{(k-1)^{2}}$
        & Sec.~\ref{sec:mit-exp-model}, Eq.~(\ref{eq:mit-exp-2}).
        \\  
        Probabilistic error cancellation \cite{temmeErrorMitigationShortDepth2017}& $\gamma_{PEC}^2$ & Sec.~\ref{sec:mit_pec}, Eq.~(\ref{eq:mit-pec-5}).
        \\  
        Exponential error suppression \cite{koczorExponentialErrorSuppression2021,hugginsVirtualDistillationQuantum2021} & $\mathrm{(approx.)}\,\mathcal{O}(\frac{1}{\epsilon^{4f}})$ 
        & Sec.~\ref{sec:mit-ees}, Eq.~(\ref{eq:ees-variance}).
        \\  
        Readout error mitigation & Analytical expression not available. & Sec.~\ref{sec:mit-readout}
        \\ 
        \bottomrule
    \end{tabularx}
\end{table}

\subsection{Noise robustness of VQE algorithms}

It has been argued that the VQE algorithm is naturally robust to certain types of noise \cite{mccleanTheoryVariationalHybrid2015}.
However, the reality is more complicated. For example, a systematic error such as a systematic deviation resulting from an over-rotation error can be mitigated in VQE by the optimization process. In general, if a coherent error results in a unitary $\tilde{U}_{\mathrm{ansatz}}( \boldsymbol{\theta} )$ that is different from the expected $U_{\mathrm{ansatz}}( \boldsymbol{\theta} )$, the optimization algorithm might be able to either correct it with a different angle $\boldsymbol{\theta} _{1}$, satisfying $\tilde{U}_{\mathrm{ansatz}}( \boldsymbol{\theta} _{1}) =U_{\mathrm{ansatz}}( \boldsymbol{\theta} )$~\cite{yuanTheoryVariationalQuantum2019}, or the correction might not be needed as long as $\tilde{U}_{\mathrm{ansatz}}( \boldsymbol{\theta} )$ produced the desired ground state for VQE. However, this robustness also ties the results from a VQE to specific hardware, since the optimized angles can no-longer be adapted to another quantum computer.
Furthermore, errors on the hardware affect the landscape of the cost function, potentially creating more local minima and therefore jeopardizing the optimization procedure. This has been experimentally investigated for the variational quantum factoring algorithm \cite{karamlouAnalyzingPerformanceVariational2021} (see Sec.~\ref{sec:barren_plateau} for relevant discussions) which is a close relative of the VQE algorithm discussed here. Finally, the variation in noise on different quantum computers poses a challenge to the parallelization of the VQE algorithm, see Sec. \ref{sec:parallelization}.


%% file: figs/errormit/mit-learning-PEC-setup.tex
$
\Qcircuit @C=1.0em @R=0.8em @!R { \\
\lstick{\ket{0}} & \gate{P_1} & \gate{R_1} & \gate{P_5} & \qw & \gate{P_9} & \gate{R_5} & \gate{P_{13}} & \multigate{1}{G_2} & \gate{P_{17}} & \gate{R_9} & \gate{P_{21}} & \meter \\
\lstick{\ket{0}} & \gate{P_2} & \gate{R_2} & \gate{P_6} & \multigate{1}{G_1} & \gate{P_{10}} & \gate{R_6} & \gate{P_{14}} & \ghost{G_2} & \gate{P_{18}} & \gate{R_{10}} & \gate{P_{22}} & \meter \\
\lstick{\ket{0}} & \gate{P_3} & \gate{R_3} & \gate{P_7} & \ghost{G_1} & \gate{P_{11}} & \gate{R_7} & \gate{P_{15}} & \multigate{1}{G_3} & \gate{P_{19}} & \gate{R_{11}} & \gate{P_{23}} & \meter \\
\lstick{\ket{0}} & \gate{P_4} & \gate{R_4} & \gate{P_8} & \qw & \gate{P_{12}} & \gate{R_8} & \gate{P_{16}} & \ghost{G_3} & \gate{P_{20}} & \gate{R_{12}} & \gate{P_{24}} & \meter 
\gategroup{2}{5}{5}{5}{2.5em}{--}
\gategroup{2}{9}{5}{9}{1em}{--}
\\
}
$

%% file: figs/errormit/ees-general-circ.tex
$
\Qcircuit @C=1.0em @R=0.8em @!R { \\
\lstick{\ket{0}} & \qw & \gate{H} & \ctrl{1} & \ctrl{1} & \gate{H} & \meter & \rstick{\rightarrow\operatorname{Tr}(\rho^\nEES{} O)} \\
\lstick{\rho} & \qw & \qw & \multigate{2}{D_\nEES} & \gate{P} & \qw & \qw \\
\lstick{\rho} & \qw & \qw & \ghost{D_\nEES} & \qw & \qw & \qw \\
\lstick{\rho} & \qw & \qw & \ghost{D_\nEES} & \qw & \qw & \qw \\
}
$

%% file: figs/errormit/ees-VD-swap-circ.tex
$
\Qcircuit @C=1.5em @R=1.5em @!R { \\
\lstick{\rho} & \qswap & \qw & \qw & \qswap\qwx[4] & \qw \\
\lstick{\rho} & \qswap\qwx & \qswap & \qw & \qw & \qw \\
\lstick{\rho} & \qw & \qswap\qwx & \qswap\qwx[1] & \qw & \qw \\
& \vdots & & & \\
\lstick{\rho} & \qw & \qw & \qswap & \qswap & \qw \\
}
$

%% file: 09_VQE_extensions.tex
\section{Beyond the ground state of isolated molecules: Extensions of VQE} \label{sec:Extensions_of_VQE}

Up until this point, we have provided details of the VQE in the context of finding the ground state of an isolated system. However, in reality, a molecule is generally coupled with a wider environment, as well as the physics being strongly influenced by its electronic excitations. In this section, we briefly review some of the modifications to VQE which enlarge the scope of applicability, including accessing beyond ground state properties of the system, as well as the use of VQE as a sub-component of other algorithms to access multi-resolution descriptions of larger systems.

\subsection{Excited states VQE} \label{sec:excited_states}

The computation of excited states is key to many processes in quantum chemistry and materials science, governing the dominant optical, transport and reactive properties \cite{Karim2018, ZeinalipourYazdi2018}. However, it is in general a significantly more challenging task than ground state computation, owing to the state generally being further away from a mean-field description, as well as less straightforward optimization to avoid the variational collapse to the ground state. Conventional correlated quantum chemical approaches \cite{Matsika2018} include Equation of Motion (EOM) coupled-cluster \cite{Stanton1993}, linear response theory \cite{Monkhorst2009}, as well as multi-reference approaches for stronger correlation \cite{Jeziorski1981, Lyakh2011}. Quantum computing methods can be broadly divided into two main types of methods, those that rely on computing excited states within a subspace, and fully variational methods relying on modification of the VQE cost function. We briefly review the core aspects of some of these approaches below.

\paragraph{Quantum Subspace Expansion:} The quantum subspace expansion relies on finding an approximate Hamiltonian that spans a subspace of the full Hilbert space, but whose dimension is small and grows as only a low-order polynomial of the system size. The matrix elements of these Hamiltonians are sampled on quantum computers, but can then be tractably diagonalized on classical resources, with the higher-lying eigenvalues of these subspace Hamiltonians approximating true eigenvalues of the system. In practice, this approach starts with a ground state VQE calculation. From this ground state, it is then necessary to add additional states in order to define the span of a subspace into which the Hamiltonian can be computed. For reliable excited states, it is necessary to ensure that this space spans the dominant low-energy excitations of interest, as the whole spectrum will not be reproduced by construction. There are different approaches to choose these low energy states to span these relevant excitations, including approaches based on Krylov (or Lanczos) subspaces \cite{Motta2019,yeter-aydenizPracticalQuantumComputation2020,sunQuantumComputationFiniteTemperature2021}, and low-rank excitations of the ground state motivated by an equation-of-motion formalism \cite{McClean2017,Ollitrault2020}. These approaches can also be used to yield improved ground state estimates \cite{Motta2019,Parrish2019}.

In the quantum subspace expansion based around the equation-of-motion expansion, the resulting Hamiltonian (and overlap) matrix between these states can be found via high-order reduced density matrices evaluated from the ground state, as initially proposed in Ref.~\cite{McClean2017}, and subsequently implemented on a quantum device \cite{collessComputationMolecularSpectra2018}. The advantage of these methods is that they do not require particularly deep circuits to evaluate the relevant matrix elements of this subspace Hamiltonian. However, the quantum subspace expansion approaches can be quite sensitive to noise, while high-order density matrices can be expensive to sample and accumulate. Furthermore, noisy (yet unbiased) matrix elements can lead to systematic biases in eigenvalues \cite{Blunt2018, Epperly2021}.

\paragraph{Variational approaches:} An alternative approach relies on directly optimizing an ansatz for specific excited states, using a modified cost function, which affords a fully variational flexibility, while maintaining orthogonality to lower-energy states. These have the advantage of not suffering from the limitations and biases of subspace expansion methods, but usually come at a higher cost in terms of quantum resources, and a restriction to a specific ansatz chosen. The simplest approach is to simply enforce symmetry constraints on the ansatz to a different symmetry sector to the ground state, in which case orthogonality to the ground state is guaranteed for the lowest-lying excitations in each symmetry \cite{Ryabinkin2019} (for details about this method, please refer back to Sec. \ref{sec:cost_function}). This is however restricted to only specific excited states and limited by the symmetry of the system studied. Another approach which was proposed early on in the development of variational quantum algorithms (initially suggested for quantum computation in Ref. \cite{mccleanTheoryVariationalHybrid2015}), is to use the folded spectrum Hamiltonian \cite{Wang1994}: $\hat{H}^{\prime} = (\hat{H} - \gamma \unit)^2 $, for which the ground state is now the eigenstate of $\hat{H}$ which has an eigenvalue closest to $\gamma$. It was applied by Liu \textit{et al.} \cite{Liu2021_MBL} as a mean to probe many-body localization on a quantum computer. This method however implies squaring the Hamiltonian, which can result in a significant increase in measurements required if the operator is dense, and requires prior knowledge of the eigenspectrum (which is somewhat less of a problem in the case of vibrational spectroscopy than in the case of electronic structure computation \cite{Sawaya2021}). 

The subspace search VQE (SSVQE) \cite{nakanishiSubspacesearchVariationalQuantum2019} leverages the fact that a unitary transformation between states cannot change the orthogonality of the states it is applied to. Therefore by preparing different orthogonal input states and training a VQE ansatz to minimize the energy of all these states at the same time (for instance by modifying the VQE cost function to include the sum of expectation values of the Hamiltonian with respect to each of the states, or by creating a mixed state using ancilla qubits), one can simultaneously learn the ground state and any number of subsequent excited states. It is likely however that this simultaneous optimization of the ansatz becomes increasingly more constrained with the number of excited states desired.

Higgott \textit{et al.} \cite{higgottVariationalQuantumComputation2019} proposed using a deflated Hamiltonian to iteratively compute successive excited states (oftentimes referred to as Variational Quantum Deflation, VQD). The algorithm works by first computing the ground state with VQE. Once discovered, the cost function is modified to add a penalty term, which corresponds to the overlap between the ground state and a new trial wavefunction. This new trial wavefunction is then trained to minimize both the expectation value of the Hamiltonian and maximize the overlap with the lower energy states. This process can be repeated iteratively for any number of excited states. The key challenge of this method is the computation of the overlap term which may require quantum cost that could be significant for a NISQ device (i.e. a large number of SWAP gates), or possibly subject to additional noise (by implementing as a circuit the complex conjugate of the ansatz used to prepare previous excited states, a method also applied in Ref. \cite{Lee2019}), though improvements have been proposed. For instance, Jones \textit{et al.} \cite{Jones2019} propose to compute the overlap term with a low depth SWAP test, and uses variational time evolution \cite{McArdle2019}. Chan \textit{et al.} \cite{Chan2021} extend this excited state method by merging it with ADAPT-VQE \cite{Grimsley2019} (see Sec. \ref{sec:adapt-vqe}). Kottmann \textit{et al.} \cite{Kottmann2021_3} independently also proposed an adaptation of VQD to an adaptive method which benefits from efficiency gained from gradient evaluation process presented in the same work (see Sec. \ref{sec:Optimization}). Wakaura and Suksomo \cite{Wakaura2021} propose an adaptation of the VQE cost function to minimize the norm of the tangent vector to the energy rather than just the energy, dubbed Tangent-Vector VQE (TVVQE). While this can be used for ground state energies, it is also combined with VQD to compute excited states. While the method is shown to provide improved accuracy compared to a UCC based VQE on simple models (Hubbard, $\mathrm{H_2}$, $\mathrm{LiH}$), it is reported to require a run time on average five times longer than VQE \cite{Wakaura2021}. 

The discriminative VQE (DVQE) \cite{Tilly2020} is a further alternative method, and relies on training of a generative adversarial network to enforce orthogonality between the ground state and a trial excited state. Generative Adversarial Networks (GANs) are machine learning tools, which are composed of two neural networks competing against each other: a generator, which is trained to produce a specific data pattern (e.g. an image), and a discriminator, which is trained to distinguish between true instances of this data pattern, and generated instances. When the training is successful, the generator learns to generate data patterns, which are indistinguishable from true ones. This concept was ported to quantum computing with the Quantum GAN \cite{Lloyd2018, Benedetti2019}, a method, which can be used to learn an approximation of an unknown pure state. The DVQE proposed in \cite{Tilly2020} inverts the logic of the QGAN, forcing generator and discriminator to collaborate for the generator to generate a state, which is as easy to distinguish as possible from the ground state: an orthogonal state. In order to ensure that the generated state is the first excited state, one must at the same time minimize the expectation value of the Hamiltonian. Subsequent excited states can be found by repeating the procedure iteratively. The scaling of the depth required for the discriminator remains unknown and could become an impediment for the method.

An alternative approach is the Variance VQE method \cite{zhang2020variance_minimization}, which replaces the usual cost function of VQE by minimizing the variance of a Hamiltonian with respect to a state, rather than its expectation value. The idea behind this method is that the variance of the expectation value of a Hamiltonian must be equal to zero if the state used to perform the measurement is an eigenstate of that Hamiltonian (on the zero-energy variance principle, we direct readers to Ref.~\cite{Bartlett1935, Siringo2005, Umrigar2005, Khemani2016, Pollmann2016, Vicentini2019}). Because all eigenstates have zero-energy variance, a simple approach will not guarantee convergence to a low-energy state. This problem is addressed in Ref.~ \cite{zhang2020variance_minimization} by combining both energy and variance minimization in order to allow for computation of low lying excited states. Zhang \textit{et al.} \cite{zhangAdaptiveVariationalQuantum2021} propose an adaptative variant of this method to computed highly excited states of Hamiltonians. The ansatz is grown by choosing operators from a pool of Pauli operators, akin to the methods listed in Sec. \ref{sec:adaptive_ansatz}.

\paragraph{Dynamical correlation functions}
Equilibrium dynamical correlation functions are the key quantities governing the linear response behavior of quantum systems, encoding the information of the excitation spectrum over all energy scales. These functions can either be represented in the time or frequency (energy) domain, with a key dynamic response function being the single-particle Green's function, describing the charged excitation spectrum of the system. Any method to systematically calculate individual excited state energies (e.g. via a quantum subspace expansion) and the relevant transition amplitudes coupling them to the ground state, can in principle compute these dynamical correlation functions via its spectral representation \cite{Endo2019GF,runggerDynamicalMeanField2020,zhuCalculatingGreenFunction2021,Jamet2021,Jamet2022}. However,  other VQE approaches exist which directly target these correlation functions in either the (imaginary) time or frequency domains \cite{Endo2019GF,sunQuantumComputationFiniteTemperature2021, Wecker2015_Solving}. These include VQE-based variational approaches to directly solve the linear equations for the response of a system at a given frequency \cite{xuVariationalAlgorithmsLinear2021}, which can be cast as a modified cost function, with a similarly parameterized VQE ansatz \cite{Chen2021,caiQuantumComputationMolecular2020, Tong2021}. While these approaches can describe the correlation functions to high accuracy over the whole energy range without restricting to a low-energy subspace, their challenge arises chiefly from the substantially more difficult optimization problem for the ansatz, originating from the larger condition number of the cost function, as well as the necessity for Hadamard tests to compute the transition amplitudes between the excited (or response) and ground states (for more details, see Appendix.~\ref{sec:hadamard-test}).


\subsection{VQE as a solver of correlated subspaces in multiscale methods} \label{sec:multiscale}

The VQE has been applied as a sub-routine to resolve the low-energy electronic structure in a number of existing approaches, thereby adapting many hybrid methods of conventional quantum chemistry methods to exploit quantum computing. These include a number of `quantum embedding' methods, where the full space of the problem is partitioned, with each solved at a differing level of theory. In these, it is generally the strongly correlated low-energy partition of orbitals that are amenable to use within a VQE solver which are  then, in various ways, coupled back to the rest of the system (potentially self-consistently) at a lower level of theory on a classical device. These multi-resolution methods can substantially extend the scope and applicability of the VQE, under additional constraints arising from this choice of partitioning and coupling of the spaces. We have provided below three examples of embedding methods adapted in this way:

\paragraph{Complete active space approaches:} 
The simplest and most widespread approach in quantum chemistry for isolating and treating a correlated set of low-energy degrees of freedom at a higher level of theory are the Complete Active Space (CAS) approaches. In these, a subset of high-energy occupied and low-energy unoccupied Hartree--Fock orbitals are considered to span the dominant strongly correlated quantum fluctuations, and treated with an accurate correlated treatment within this subspace (often full configuration interaction, see Sec.~\ref{sec:full_configuration_interaction}). This subspace Hamiltonian includes the presence of a Coulomb and exchange mean-field potential from the remaining electrons outside this space. In this way, the active space electrons are fully correlated within that manifold, leading to the Complete Active Space Configuration Interaction (CAS-CI) approach \cite{Jensen2017, Levine2021}. Furthermore, the CAS-CI can be variationally optimized, by updating the choice of molecular orbitals defining the low-energy CAS space via single-particle unitary rotations among the entire set of orbitals in the system. 
This method is generally referred to as Complete Active Space Self-Consistent Field (CASSCF) \cite{Roos1980, Sun2017}, or the related Multi-Configurational Self-Consistent Field (MCSCF) where the active space is not solved at the level of full configuration interaction. The CASSCF wavefunction can therefore be written as follows:
\begin{align}
\label{eq:QCAS_WF_1}
    |\Psi_{\mathrm{CASSCF}} \rangle = \ket{\textbf{R}, \textbf{c}} = e^{-\textbf{R}} \sum_{\mu} c_{\mu} \ket{\mu},
\end{align}
where $\textbf{R}$ parameterizes the single-particle anti-unitary operator defining the rotation of the active space, $\ket{\mu}$ the complete set of Slater determinants in the active space, and $\textbf{c}$ defines the coefficients of the configurations indexed by $\mu$. In implementation on a quantum device, the rotation operator defining the active space, $\textbf{R}$, can be optimized on a classical device, while the parameterized description of the active space wavefunction can be sought via the VQE. These approaches constitute the bedrock for simulation of molecular systems with strong correlation, in particular in systems with competing spin states, excited states, systems at bond-breaking geometries, and inorganic chemistry \cite{RetaManeru2014,Li2015,olsen11}. These CAS-based approaches were initially proposed in combination with VQE as a solver for the active space in Ref.~\cite{reiherElucidatingReactionMechanisms2017} and were subsequently successfully demonstrated practically on quantum computers in Refs.~\cite{Takeshita2020, Yalouz2021, Tilly2021}, including self-consistent optimization of the active space. 

It should be noted that in order to achieve this optimization of the active space, the two-body reduced density matrix of the active space is required, which can have ramifications on the number of measurements required by the VQE \cite{Tilly2021}. However, in strongly correlated quantum chemistry, it is generally also important to include a description of the correlation within the orbitals external to the active space, generally via low-order perturbation theory, resulting in methods such as complete active space second-order perturbation theory (CASPT2) \cite{Abe2008}. These however require computation of the 3-body reduced density matrix (and potentially higher) in order to couple the active space correlations to this perturbative treatment and are therefore considered a daunting proposition for VQE. There is also a wider range of extensions to the CASCI approach, including extensions to embedding with density functional theory (DFT) description of the environment, which has also been explored by Rossmannek et al. \cite{Rossmannek2021} within a VQE description of a correlated active space. Shade \textit{et al.} \cite{Schade2022} also extend these ideas to the reduced density matrix function theory (RDMFT) and demonstrate an implementation of their method to a Hubbard-like system on a quantum device.

\paragraph{Density matrix embedding theory (DMET):} Similar to the active space methods mentioned above, DMET~\cite{Knizia2013, Wouters2016} aims at embedding an accurately correlated subspace in a mean-field environment. In contrast to CAS-CI, this `active space' is chosen through locality criteria, starting from a local fragment space and augmenting it with the minimal number of additional orbitals (denoted the bath space) to ensure that the active space recovers the Hartree--Fock description, and explicitly captures quantum entanglement between the fragment and its environment. In this way, the DMET approach can be considered as having a similar ambition to dynamical mean-field theory \cite{Georges1996}, but cast as a static wavefunction theory (see below). In order to optimize the mean-field state of the system, the one-body reduced density matrix is matched between the individual fragment spaces between the correlated and mean-field descriptions.

Integration of DMET with a VQE for the correlated subspace solver has been the subject of several publications \cite{Rubin2016, Yamazaki2018, Ma2020, Mineh2021, Li2021}, and has been implemented on quantum computers with proof of principles for relevant applications such as protein-ligand interactions for drug design \cite{Kirsopp2021} (with an alternative method based on perturbation theory proposed in Ref.~\cite{Malone2021}). Energy weighted DMET (EwDMET) which builds on DMET to improve its description of dynamical fluctuations for small fragment sizes (thereby moving systematically towards a DMFT description described below) \cite{Fertitta2018, Fertitta2019, Sriluckshmy2021} was also tested and implemented on a quantum device \cite{Tilly2021}, allowing quantum phase transitions to be captured which were out of the scope of DMET. A wide range of possible alternative formulations exist for embedding correlated subspaces in (static) mean-field environments - especially when that subspace is only weakly coupled to the environment, and the explicit entanglement between the subspace and the environment can be neglected.

\paragraph{Dynamical Mean-Field Theory (DMFT):} DMFT again relies on a similar embedding of a (local) correlated subspace in a mean-field environment. However, this environment allows for local quantum fluctuations in its description, thereby including the effects of correlation in the local propagation of particles through the environment. This effect is captured by a local self-energy, which is the self-consistent quantum object in DMFT \cite{Georges1996}. This necessitates a formalism built around the single-particle Green's function (a specific dynamical correlation function), which is the object which must be sampled within DMFT on a quantum device. At the heart of DMFT is a mapping from the system of interest to an impurity model, which describes a local correlated fragment coupled to a wider non-interacting set of degrees of freedom, denoted the `bath'. This impurity model can be represented in a Hamiltonian formulation, from which the single-particle Green's function must be sampled, with various approaches to solve for this Green's function known as `impurity solvers'. The techniques presented earlier in this section for can be used to sample this Green's function in either a time or frequency domain at each iteration in the self-consistent loop. 
The use of quantum computers as an impurity solver was proposed initially in Ref. \cite{Bauer2016} in the time domain, but frequency domain solvers have often been more amenable to the low-depth NISQ era. These were explored in the context of DMFT impurity solvers in physical realizations of correlated material systems via VQE-type parameterized algorithms in Refs.~\cite{Endo2019GF,runggerDynamicalMeanField2020, Kreula2016,Jamet2021}. An alternative method to compute the Green's function over the whole energy range is based on the quantum subspace expansion \cite{Jamet2022}.

Overall, embedding methods using the VQE as a high accuracy and scalable solver to describe the correlations within a subspace self-consistently coupled to a wider environment are a promising avenue to extend the applicability of quantum computation towards practical applications. In general, they allow for recovery of significant parts of the electron correlation energy, while avoiding treatment of the full system, thereby reducing qubits number in exchange for additional classical resources in defining the embedding, as well as a self-consistent loop. It is worth noting that the possibilities for embedding the VQE and more general quantum algorithms within wider multi-method and multi-resolution hybrid schemes extends far beyond just the quantum embedding methodologies presented above, and are likely to be of central importance in the utility of quantum algorithms in molecular modeling in all contexts in the future.

%% file: acknowledgements.tex
\section*{Acknowledgements} \label{sec:acknowledgements}

J.Tilly is supported by an industrial CASE (iCASE) studentship, funded by and UK EPSRC [EP/R513143/1], in collaboration with University College London and Odyssey Therapeutics.
H.C. is supported through a Teaching Fellowship from UCL.
D.P.  is supported by an industrial CASE (iCASE) studentship, funded by and UK EPSRC [EP/T517793/1], in collaboration with University College London and Odyssey Therapeutics.
Y.L. is supported by the National Natural Science Foundation of China (Grant No. 11875050 and 12088101) and NSAF (Grant No. U1930403).
E.G. is supported by the UK EPSRC [EP/P510270/1].  
L.W. is an honorary research fellow at the Department of Computer Science, University College London.
I.R. acknowledges the support of the UK government department for Business, Energy and Industrial Strategy through the UK National Quantum Technologies Programme.
G.H.B. gratefully acknowledges support from the Royal Society via a University Research Fellowship, as well as funding from the European Research Council (ERC) under the European Union\textquotesingle s Horizon 2020 research and innovation programme (Grant Agreement No. 759063).

\section*{Declaration of competing interest}

The authors declare that they have no known competing financial interests or personal relationships that could have appeared to influence the work reported in this paper.

%% file: appendices.tex
\section{Qubit encodings and Fenwick trees} \label{sec:Fenwick_trees}

The qubit encondings that have been presented in Sec. \ref{sec:gen_encoding} can be translated into the language of graph theory and be visualized in a simple way as Fenwick trees \cite{Havlek2017}.
\begin{figure}
\centering
\begin{tikzpicture}
\node[shape=circle,fill=white,thick,draw=cyan] (7) at (0,0) {7};  \node[shape=circle,draw=black,fill=gray!5] (6) at (-1.5,-1) {6};
\node[shape=circle,draw=black,fill=white, thick] (5) at (0,-1) {5};
\node[shape=circle,fill=white,thick,draw=red] (4) at (0,-2) {4};
\node[shape=circle,fill=white,thick,draw=green] (3) at (1.5,-1) {3};
\node[shape=circle,draw=black,fill=gray!5] (2) at (1,-2) {2};
\node[shape=circle,draw=black,fill=gray!5] (1) at (2,-2) {1};
\node[shape=circle,draw=black,fill=gray!5] (0) at (2,-3) {0};

\path [-] (6) edge (7);
\path [-] (5) edge (7);
\path [-] (3) edge (7);
\path [-] (4) edge (5);
\path [-] (2) edge (3);
\path [-] (1) edge (3);
\path [-] (0) edge (1);
\end{tikzpicture}
\caption{Fenwick tree for the Bravyi-Kitaev encoding of 8 qubits showing the update set $U(5) = \{7\}$ (cyan), flip set $F(5) = \{4\}$, (red) and remainder set $R(5) = \{3\}$ (green) for qubit $5$, from which we can read off the Brayvi-Kiteav operators (See Sec. \ref{sec:bravyi-kitaev})}
\label{fig:BK8qubit-5}
\end{figure}
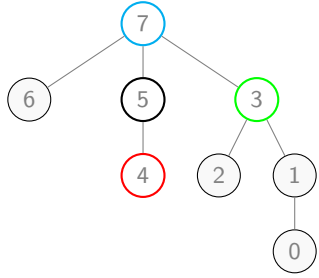
The tree corresponding to the Bravyi-Kitaev (like the one shown in Fig. (\ref{fig:BK8qubit-5}) and in Fig (\ref{fig:BK8qubit-2})) can be built recursively according to a procedure that mirrors the construction of the change of basis matrix of Eq. (\ref{eq:bravyikitaev2}): the tree corresponding to the single-qubit encoding is the trivial graph, and the tree corresponding to the $2^x$-qubit enconding is built from two copies of the tree corresponding to the $2^{x-1}$-qubit enconding, where vertex $k$ on the first tree corresponds to vertex $k + 2^{x-1}$ on the second, and the two trees are joined together by making vertex $2^{x-1} - 1$ a child of vertex $2^x  - 1$.
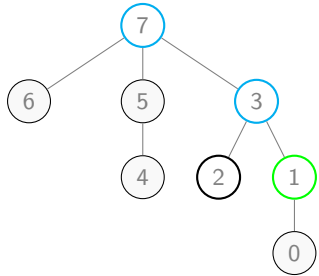
\begin{figure}
\centering
\begin{tikzpicture}
\node[shape=circle,fill=white,thick,draw=cyan] (7) at (0,0) {7};  \node[shape=circle,draw=black,fill=gray!5] (6) at (-1.5,-1) {6};
\node[shape=circle,draw=black,fill=gray!5] (5) at (0,-1) {5};
\node[shape=circle,draw=black,fill=gray!5] (4) at (0,-2) {4};
\node[shape=circle,fill=white,thick,draw=cyan] (3) at (1.5,-1) {3};
\node[shape=circle,draw=black,fill=white, thick] (2) at (1,-2) {2};
\node[shape=circle,fill=white,thick,draw=green] (1) at (2,-2) {1};
\node[shape=circle,draw=black,fill=gray!5] (0) at (2,-3) {0};

\path [-] (6) edge (7);
\path [-] (5) edge (7);
\path [-] (3) edge (7);
\path [-] (4) edge (5);
\path [-] (2) edge (3);
\path [-] (1) edge (3);
\path [-] (0) edge (1);
\end{tikzpicture}
\caption{The same Fenwick tree as in Fig. (\ref{fig:BK8qubit-5}), this time showing the update set $U(2) = \{3, 7\}$ (cyan) and remainder set $R(2) = P(2) = \{1\}$ (green) for qubit $2$ (because of how the Bravyi-Kitaev tree is constructed the leaves are the even vertices and so we see again that the flip set of an even qubit is always empty$F(2) = \emptyset$)}
\label{fig:BK8qubit-2}
\end{figure}

The definitions of the qubit sets given when describing the Bravyi-Kitaev encoding (see Sec. \ref{sec:bravyi-kitaev}) now translate into the following simple statements about Fenwick trees \cite{Havlek2017}:
\begin{itemize}
\item The update set of the $j$-th qubit $U(j)$ corresponds then to the set of ancestors of vertex $j$ on the tree.
\item The flip set $F(j)$ is the set of children of the $j$-th vertex.
\item The remainder set $R(j)$ is the set of children of the ancestors of vertex $j$ whose values is less than $j$.
\end{itemize}
This construction allows us to read off qubits sets for each qubit from the tree corresponding to our encoding (as shown in Fig. (\ref{fig:BK8qubit-5}) and in Fig (\ref{fig:BK8qubit-2}) for our usual examples of qubits 2 and 5) and hence the representation of the creation and annihilation operators using Eq. (\ref{eq:BKladder4}). In the case where the number of qubits $n$ is not a power of $2$ we construct the Fenwick tree for the next power of $2$ and determine the qubit sets, and then we discard from the qubit sets all the qubits greater or equal to $n$.

\begin{figure}
\centering
\begin{tikzpicture}
\node[shape=circle,fill=white,thick,draw=cyan] (7) at (0,0) {7};  \node[shape=circle,fill=white,thick,draw=cyan] (6) at (0, -1) {6};
\node[shape=circle,draw=black,fill=white,thick] (5) at (0, -2) {5};
\node[shape=circle,draw=red,fill=white,thick] (4) at (0, -3) {4};
\node[shape=circle,draw=black,fill=gray!5] (3) at (0, -4) {3};
\node[shape=circle,draw=black,fill=gray!5] (2) at (0, -5) {2};
\node[shape=circle,draw=black,fill=gray!5] (1) at (0, -6) {1};
\node[shape=circle,draw=black,fill=gray!5] (0) at (0, -7) {0};
\path [-] (7) edge (6);
\path [-] (6) edge (5);
\path [-] (5) edge (4);
\path [-] (4) edge (3);
\path [-] (3) edge (2);
\path [-] (2) edge (1);
\path [-] (1) edge (0);

\node[shape=circle,draw=black,fill=gray!5] (7) at (3,0) {7};  \node[shape=circle,draw=black,fill=gray!5] (6) at (3, -1) {6};
\node[shape=circle,draw=black,fill=white,thick] (5) at (3, -2) {5};
\node[shape=circle,draw=green,fill=white,thick] (4) at (3, -3) {4};
\node[shape=circle,draw=green,fill=white,thick] (3) at (3, -4) {3};
\node[shape=circle,draw=green,fill=white,thick] (2) at (3, -5) {2};
\node[shape=circle,draw=green,fill=white,thick] (1) at (3, -6) {1};
\node[shape=circle,draw=green,fill=white,thick] (0) at (3, -7) {0};
\end{tikzpicture}
\caption{Fenwick trees for the parity (left) and Jordan-Wigner (right) encoding of 8 qubits showing update $U(5)$ (cyan), flip $F(5)$ (red) and remainder sets $R(5)$ (green) for qubit 5}
\label{fig:JW8qubit-5}
\end{figure}
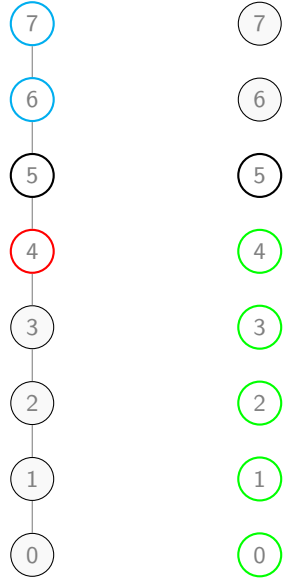

The parity encoding can also be represented this way, giving rise to the linear graph in Fig. \ref{fig:JW8qubit-5}. We can further generalize this to other encodings by considering disconnected Fenwick trees. We can then change our definition of $R(j)$ to include both the children of vertex $j$'s ancestors and the roots whose value is less than $j$. We then have that the Jordan-Wigner encoding corresponds to a totally disconnected graph (a graph with no edges). In the case of both encodings inspection of the corresponding Fenwick tree together with Eq. (\ref{eq:BKladder4}) recovers the same representations of $\hat{a}^\dagger_j$ and $\hat{a}_j$ operators as in Eq. (\ref{eq:parityladder2}) and Eq. (\ref{eq:JW2}). We can turn this argument upside down, and define new fermionic encodings starting from collections of Fenwick trees \cite{Havlek2017}.

\section{Hadamard test}\label{sec:hadamard-test}

\begin{figure} [ht]
  \centering
  \subfloat[$\mathrm{Re}\langle \psi | U | \psi \rangle$\label{fig:hadamard-real}]{
  \Qcircuit @C=1.0em @R=0.2em @!R { 
     \nghost{ |0\rangle  } & \lstick{ |0\rangle } & \gate{\Had} & \ctrl{1} & \gate{\Had} & \meter & \cw\\ 
     \nghost{ |\psi\rangle } & \lstick{ |\psi\rangle } & {/} \qw & \gate{U} & \qw & \qw & \qw \\ 
  }
  }
  \subfloat[$\mathrm{Im}\langle \psi |U|\psi\rangle$\label{fig:hadamard-imag}]{
  \Qcircuit @C=1.0em @R=0.2em @!R { 
     \nghost{ |0\rangle  } & \lstick{ |0\rangle } & \gate{\Had} & \gate{\mathrm{Rz}(-\frac{\pi}{2})} & \ctrl{1} & \gate{\Had} & \meter & \cw\\ 
     \nghost{ |\psi\rangle } & \lstick{ |\psi\rangle } & {/} \qw & \qw & \gate{U} & \qw & \qw & \qw \\ 
  }
  }
  \caption{
  Hadamard test circuits. The probability of measuring $0$ minus the probability of measuring
  $1$ in \protect\subref{fig:hadamard-real} and \protect\subref{fig:hadamard-imag} 
  gives respectively the real the imaginary part of the amplitude
  of $\langle \psi|U|\psi \rangle$.
  }
  \label{fig:hadamard-test}
\end{figure}
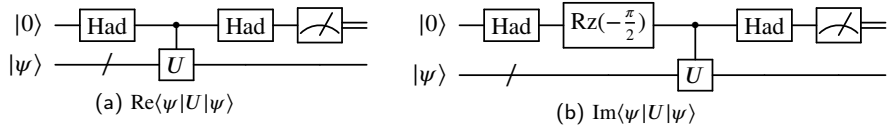

Here we briefly introduce a common quantum subroutine called the Hadamard test \cite{nielsenQuantumComputationQuantum2010}. Hadamard test is frequently used to compute the amplitude $\langle \psi |U|\psi \rangle $ (both its real and imaginary part) of an initial state $|\psi \rangle $ and a unitary gate $U$. This algorithm is summarized in the circuit diagram in Fig.~\ref{fig:hadamard-test}. Here we explain the circuit in detail. To measure the real part of the amplitude, the quantum computer is initialized in a product state $|\psi \rangle \otimes |0\rangle $ with one ancilla qubit. The Hadamard gate $\Had$ converts the ancilla from $|0\rangle $ to $( |0\rangle +|1\rangle ) /\sqrt{2}$. Then the controlled-$U$ gate with control qubit on the ancilla results in the state $( U|\psi \rangle \otimes |1\rangle +|\psi \rangle \otimes |0\rangle ) /\sqrt{2}$. The second Hadamard gate gives the state $(( I+U) |\psi \rangle \otimes |0\rangle +( I-U) |\psi \rangle \otimes |1\rangle ) /2$. In the final measurement, the probability of obtaining $0$ is $( 1+\mathrm{Re} \langle \psi |U|\psi \rangle ) /2$, and the probability of obtaining $1$ is $( 1-\mathrm{Re} \langle \psi |U|\psi \rangle ) /2$. Therefore, the difference between the two probabilities is $\mathrm{Re} \langle \psi |U|\psi \rangle $. Similarly calculation shows that in the second circuit (Fig.~\ref{fig:hadamard-imag}), the probability of obtaining 0 minus the probability of obtaining 1 is $\mathrm{Im} \langle \psi |U|\psi \rangle $.

\input{\ProjectRoot/08_error-mit_appendix.tex}

%% file: 08_error-mit_appendix.tex
\section{Error mitigation appendix}
\subsection{Common noise models} \label{sec:common_noise_models}
Here we mention a few noise models that may be common in the literature, as for most relevant concepts in quantum computing, readers can refer to Ref.~\cite{nielsenQuantumComputationQuantum2010} for further details.

\subsubsection{Relaxation rates}
\label{sec:mit-noise-relaxation}

We start with the $T_1$ and $T_2$ relaxation time of a quantum computer. They are commonly used as a figure of merit for the noise robustness of a quantum computer. $T_1$ and $T_2$ captures the error on a single qubit quantum state $\rho$ due to thermalization to an equilibrium state with the environment. Specifically, assume the initial quantum state $\rho$ is
\begin{equation}
    \label{eq:error-mit-noise-model-rho}
    \rho = \begin{pmatrix}
        a     & b   \\
        b^{*} & 1-a
    \end{pmatrix}.
\end{equation}
Then, after a period of time $t$, the quantum state is changed by the noise $\mathcal{N}$ into~\cite{nielsenQuantumComputationQuantum2010}

\begin{equation}
    \mathcal{N}(\rho) = \begin{pmatrix}
        ( a-a_{0}) e^{-t/T_{1}} +a_{0} & be^{-t/T_{2}}                     \\
        b^{*} e^{-t/T_{2}}             & ( a_{0} -a) e^{-t/T_{1}} +1-a_{0}
    \end{pmatrix},
\end{equation}
where $a_{0}$ represents the thermal equilibrium state, and usually $a_{0}=1$ on superconducting qubits.
The parameters $T_{1}$ and $T_2$ are also known as spin–lattice (or `longitudinal') and spin–spin (or `transverse') relaxation rates. They can be obtained experimentally are widely available metrics for most quantum computers today.

\subsubsection{Over-rotation}

This type of error happens when a physical pulse used to generate a particular gate are mis-calibrated
\cite{merrillProgressCompensatingPulse2012}. In the case or Pauli rotation gates $R_{i}(\theta)$, $i\in\{x,y,z\}$,
this error manifests a small deviation $\delta$ in the angle $\theta$, resulting in an actual gate of modified
rotation gate $R_i(\theta + \delta)$. The deviation may be systematic\cite{bultriniSimpleMitigationStrategy2020}, but may also be stochastic and usually varies differently from different quantum computer.

\subsubsection{Depolarizing noise model}

The depolarizing model $\mathcal{N}_{\text{depolar}}$ is a common noise model. On a quantum state $\rho$ of $n$ qubits, it changes $\rho$ according to

\begin{equation}
    \mathcal{N}_{\text{depolar}}( \rho ) =( 1-\varepsilon ) \rho +\varepsilon \frac{\unit{}}{2^{n}},
\end{equation}
where $\unit{}$ is the identity matrix on $n$ qubits, and $\varepsilon$ characterizes the strength of the depolarizing noise.

\subsubsection{Dephasing noise model}

The dephasing noise model $\mathcal{N}_{\mathrm{dephasing}}$ is related to the disappearance of off-diagonal part of a density matrix, hence closely related to the $T_2$ relaxation rate of a quantum computer. Its action on a single qubit state $\rho$ is described by

\begin{equation}
    \begin{aligned}
        \mathcal{N}_{\mathrm{dephasing}}(\rho ) & =( 1-\varepsilon ) \rho +\varepsilon Z\rho Z \\
                                                & =\left(\begin{array}{ c c }
                a                       & b( 1-2\varepsilon ) \\
                b^{*}( 1-2\varepsilon ) & 1-a
            \end{array}\right),
    \end{aligned}
\end{equation}
where $a$ and $b$ characterize the initial state of $\rho$ (Eq.~\ref{eq:error-mit-noise-model-rho}), $\varepsilon$ characterizes the strength of this noise model. It's clear that the depolarizing noise does not affect the diagonal part of a single qubit state, therefore it commutes with most symmetry operator and could not be detected by the symmetry verification technique mentioned in Sec.~\ref{sec:mit-symmetry-verification}.

\subsubsection{Damping error}

Here as suggested by the name, the damping error decreases the population of $\ket{1}$ in a single qubit state, and is closely related to the $T_1$ relaxation time.
This error model can be described by the Kraus operators
\begin{equation}
    K_{0} =\begin{pmatrix}
        1 & 0                     \\
        0 & \sqrt{1-\varepsilon }
    \end{pmatrix},\;
    K_{1} =\begin{pmatrix}
        0 & 0                   \\
        0 & \sqrt{\varepsilon }
    \end{pmatrix},
\end{equation}
where $\varepsilon$ characterizes the strength of this noise model. Its action on the initial state $\rho$ in Eq.~(\ref{eq:error-mit-noise-model-rho}) is

\begin{equation}
    \begin{split}
        & \mathcal{N}_{\mathrm{damping}}(\rho) = \sum_{i=0}^1 K_i \rho K_i^\dagger \\
        & = \begin{pmatrix}
            1+(a-1)(1-\varepsilon )    & b\sqrt{1-\varepsilon } \\
            b^{*}\sqrt{1-\varepsilon } & (1-a)(1-\varepsilon )
        \end{pmatrix}.
    \end{split}
\end{equation}

\subsection{Example for probabilistic error cancellation}
\label{sec:mit-pec-toy-example}

Here we present an illustrative example of using probabilistic error cancellation to correct the depolarizing error on a one-qubit quantum computer. The depolarizing noise $\mathcal{N}_{\mathrm{dep}}$ changes an initial state $\rho$ to

\begin{equation}
    \label{eq:mit-pec-2}
    \mathcal{N}_{\mathrm{dep}}( \rho ) =\left( 1-\frac{3}{4} p\right) \rho +\frac{p}{4}\sum _{\sigma \in \{X,Y,Z\}} \sigma \rho \sigma .
\end{equation}
Here $p\in \left[ 0,\frac{4}{3}\right]$ characterizes the strength of the depolarizing noise, and $\{X,Y,Z\}$ are the Pauli matrices. One can show that $\mathcal{N}^{-1}_{\mathrm{dep}}$ can be written as~\cite{temmeErrorMitigationShortDepth2017}

\begin{equation}
    \label{eq:mit-pec-3}
    \begin{split}
        \mathcal{N}^{-1}_{\mathrm{dep}}( \rho ) & =\gamma _{\mathrm{PEC} ,\ \mathrm{dep}}\left( q_{\mathrm{dep} ,1} \rho +q_{\mathrm{dep} ,2}\sum _{\sigma \in \{X,Y,Z\}} \sigma \rho \sigma \right) ,\\
        \gamma _{\mathrm{PEC} ,\ \mathrm{dep}} & =\frac{p+2}{2-2p} ,\\
        q_{\mathrm{dep} ,1} & =\frac{4-p}{2p+4} ,\\
        q_{\mathrm{dep} ,2} & =-\frac{p}{2p+4} .
    \end{split}
\end{equation}

Suppose the effective quantum channel of an ideal unitary $U$ on quantum computer could be characterized as $\mathcal{N}_{1} \circ U$, where the circle $\circ $means composition of quantum channels. One could define three additional quantum circuits, each formed by appending one of the three Pauli gates after the unitary $U$.
One obtains the measurement outcome from the four circuits $\{U,X\circ U,Y\circ U,Z\circ U\}$ containing the original one, and calculates according to the weights in Eq.~(\ref{eq:mit-pec-3})

\begin{equation}
    \begin{split}
        \label{eq:mit-pec-4}
        & \langle \hat{O} \rangle_{\mathrm{PEC} ,\ \mathrm{dep}}                                                                                                                                                \\
        = & \gamma _{\mathrm{PEC} ,\ \mathrm{dep}}\bigg(q_{\mathrm{dep} ,1}\underbrace{\operatorname{Tr}[\hat{O} \mathcal{N}_{1} \circ U( \rho )]}_{\text{original circuit}} + q_{\mathrm{dep} ,2}\sum _{\sigma \in \{X,Y,Z\}}\underbrace{\operatorname{Tr}[ \hat{O} \sigma \circ \mathcal{N}_{1} \circ U( \rho )]}_{\text{appended by a Pauli gate}}\bigg)               \\
        = & \operatorname{Tr}\Big[\hat{O}\underbrace{\mathcal{N}^{-1}_{1}(\mathcal{N}_{1} \circ U( \rho ))}_{\text{quasi-probability\ representation\ of\ } U\ \text{acting\ on} \ \rho }\Big] \\
        = & \operatorname{Tr}[ \hat{O} U( \rho ))].
    \end{split}
\end{equation}
Therefore, the noise effect of $\mathcal{N}_{1}$ can be canceled with the PEC method.

\subsection{Implementation of exponential error suppression}
\label{sec:derangement-details}

Here we will mention several methods to actually compute the value in Eq.~(\ref{eq:ees-computed-rhon}) using quantum computers. For simplicity, we mostly consider the computation of the numerator
\begin{equation}
    \label{eq:ees-numerator}
    \mathrm{Tr( \rho^{\nEES} \hat{O})},
\end{equation}
since the denominator is a special case of the numerator with $\hat{O}$ being the identity observable.

\subsubsection{Ancilla assisted method}
\label{sec:derangement-details_ancilla-assisted}
The first method needs the assistance of an extra ancilla qubits. Considering
$n$ copies of the same state $\rho^{\otimes \nEES}$. We want to measure the observable $O$ only on one system, but
in as an overlap of the original state, and a permuted state. Specifically, we want to compute the quantity

\begin{equation}
    \label{eq:ees-single1}
    \begin{aligned}
          & \langle \psi _{i_1} |\otimes \cdots \otimes \langle \psi _{i_n} |O^{( 1)} |\psi _{s ( i_{1})} \rangle \otimes \cdots \otimes |\psi _{s ( i_{\nEES})} \rangle \\
        = & \langle \psi _{i_{1}} |O|\psi _{i_{1}} \rangle \delta _{i_{1} ,s ( i_{1})} \cdots \delta _{i_{n} ,s ( i_{\nEES})},
    \end{aligned}
\end{equation}
where for simplicity we consider only one possibility
$|\psi _{i_1} \rangle\otimes \cdots \otimes |\psi _{i_\nEES}\rangle$ in the
statistical mixture $\rho^{\otimes \nEES}$,
$O^{(1)}$ is the same observable $O$ on any one of the $n$ copies of the same system,
$s$ is called the $\nEES$-cycle derangement in the Group theory. Explicitly, $s$ is a permutation in the indices $\{1,2,\cdots \nEES\}$, and can be written as mappings
\begin{equation}
    \label{eq:ees-VD-ncycle}
    \begin{aligned}
        s( i_{1})   & =i_{2}\\
        s( i_{2})   & =i_{3}   \\
                    & \cdots   \\
        s( i_{n-1}) & =i_{\nEES}   \\
        s( i_{n})   & =i_{1} ,
    \end{aligned}
\end{equation}
where all indices $\{i_1,\cdots,i_\nEES\}$ are distinct numbers. An example is
\begin{equation}
    \label{eq:ees-cycle-all}
    s(i) = i+1,\,(i< \nEES),\, s(\nEES) = 1.
\end{equation}

$s$ being a permutation of all indices (i.e. $\{i_1,\cdots,i_\nEES\}$ are distinct numbers above)
enforces the delta functions $\delta _{i_{1} ,s ( i_{1})} \cdots \delta _{i_{n} ,s ( i_{\nEES})}$
in Eq.~(\ref{eq:ees-single1}).
For the statistical mixture $\rho^{\otimes \nEES}$, we are effectively computing $\operatorname{Tr}[\rho^\nEES O]$ since

\begin{equation}
    \begin{aligned}
          & \sum _{i_{1} \cdots i_{n}} p_{i_{1}} \cdots p_{i_{\nEES}}                                                                                                                         \\
          & \times \langle \psi _{i_{1}} |\otimes \cdots \otimes \langle \psi _{i_{\nEES}} |O^{( 1)} |\psi _{\sigma ( i_{1})} \rangle \otimes \cdots \otimes |\psi _{\sigma ( i_{\nEES})} \rangle \\
        = & \sum _{i_{1}} p_{i_{1}}^{\nEES} \langle \psi _{i_{1}} |O|\psi _{i_{1}} \rangle =\operatorname{Tr}\left[ \rho ^{\nEES} O\right].
    \end{aligned}
\end{equation}

Given an explicit expression for $s$, it is easy to construct a circuit $D_\nEES$ which performs the unitary mapping
\begin{equation}
    |\psi _{i_1} \rangle\otimes \cdots \otimes |\psi _{i_\nEES}\rangle
    \mapsto
    |\psi _{s ( i_{1})} \rangle \otimes \cdots \otimes |\psi _{s ( i_{\nEES})} \rangle.
\end{equation}
For example, for the $s$ in Eq.~(\ref{eq:ees-cycle-all}), $D_m$ can be constructed by
performing CNOT operations on pairs of qubits $(1,2),\cdots,(\nEES-1,\nEES),(\nEES,1)$.
Having constructed $|\psi _{s ( i_{1})} \rangle \otimes \cdots \otimes |\psi _{s ( i_{\nEES})} \rangle$,
measuring the overlap in Eq.~(\ref{eq:ees-single1}) could be achieved easily with a modified
Hadamard test, which is illustrated in the circuit diagram in Fig.~\ref{fig:ees-general-circ}.

We note that several improvements of the derangement operation $D_\nEES$ exists in the literature~\cite{hugginsVirtualDistillationQuantum2021, czarnikQubitefficientExponentialSuppression2021}. In particular,
\citet{czarnikQubitefficientExponentialSuppression2021} utilize a deep circuit to achieve derangement on a single copy, avoiding creating $\nEES$ copies of $\rho$ and applying the derangement operation on all $\nEES$ copies.
Although $D_\nEES$ seems to be prone to errors on actual quantum computer due to its long-range nature,
as long as the error does not affect the orthogonal relations in Eq.~(\ref{eq:ees-single1}), the noisy $D_\nEES$
is still a valid derangement operation, and it is shown in \citet{koczorExponentialErrorSuppression2021}
that the error induced by $D_\nEES$ could be
mitigated effectively with extrapolation based mitigation techniques.

\subsubsection{Diagonalization method}
\label{sec:derangement-details_diagonalization}
In the second method, we reformulate the derangement operation
as a matrix $S$, and implement a unitary version of
$S$ and the observable after the derangement operation $OS$ in order to
estimate the denominator and the numerator of Eq.~(\ref{eq:ees-computed-rhon}).
Compared with the previous method, this method does not require additional
ancilla qubit and the long range controlled-$D_\nEES$ operation (Fig.~\ref{fig:ees-general-circ}). This method only performs local operations $B_i$ which connects the same set of qubits in each copy of the state $\rho$ in $\rho^{\otimes \nEES}$.
We will first describe a general method, and then specialize on a specific case ($\nEES=2$, $O$ acts only one qubit only)
to illustrate the potential of this method.

We can define a matrix $S$ which swaps quantum systems as defined by the derangement permutation
$s$ in Eq.~(\ref{eq:ees-VD-ncycle})
\begin{equation}
    \label{eq:ees-derangement-matrix-S}
    S | \psi _{i_1} \rangle \otimes \cdots \otimes | \psi _{i_\nEES} \rangle
    = |\psi _{s ( i_{1})} \rangle \otimes \cdots \otimes |\psi _{s ( i_{\nEES})} \rangle.
\end{equation}
It should be easy to verify the equalities (see for example Eq.~\ref{eq:ees-single1})
\begin{equation}
    \label{eq:ees-VD-trOSrho}
    \mathrm{Tr}(O^{(1)} S \rho^{\otimes \nEES}) = \mathrm{Tr}(O \rho^\nEES),\,
    \mathrm{Tr}(S \rho^{\otimes \nEES}) = \mathrm{Tr}(\rho^\nEES), 
\end{equation}
where the observable $O^{(1)}$ is the same observable $O$ on the first one of the $n$ copies of the system.
It should be noted that $S$ in Eq.~(\ref{eq:ees-VD-trOSrho}) is not intended to define a unitary operation,
since it acts on the density matrix $\rho$ as $S\rho$ instead of $S\rho S^\dagger$.
Although $S$ and $O^{(1)}S$ are not Hermitian in general, in many cases they may be diagonalized by unitary matrices.
That is, we could find unitary matrices $B_S$ and $B_{OS}$ such that
\begin{equation}
    \label{eq:ees-sym-diag-gates}
    B_S \Lambda_S B_S^\dagger = S,\, B_{OS} \Lambda_{OS} B_{OS}^\dagger = O^{(1)} S,
\end{equation}
where $\Lambda_S$ and $\Lambda_{OS}$ are diagonal matrices.
Then, in principle we may perform the unitaries $B_S$ and $B_{OS}$ on the state $\rho^{\otimes \nEES}$,
and measure the $\nEES$ copies of $\rho$ each in their computation basis to obtain the numerator and denominator of
Eq.~(\ref{eq:ees-computed-rhon}). Naively, it may seem difficult to calculate the diagonalization since both $S$ and $O^{(1)}S$ operate on the large Hilbert space of $\rho^{\otimes \nEES}$. However, since $S$ permutes the $\nEES$ copies of $\rho$,
$S$ can be easily factorized into tensor products
\begin{equation}
    S = S_1 \otimes S_2 \otimes \cdots \otimes S_N,
\end{equation}
where $N$ is the number of qubits of state $\rho$, $S_{i}$ permutes all the $i$-th qubit on each copy.
For example, when $\nEES=2$, $S_i$ interchanges the qubits in the same position on the two copies of $\rho$,
i.e.
\begin{equation}
    S_i = \begin{pmatrix}
        1 & 0 & 0 & 0 \\ 0& 0& 1& 0\\ 0& 1& 0& 0\\ 0& 0& 0& 1
    \end{pmatrix}.
\end{equation}

\begin{figure}[ht]
    \centering
    \includegraphics[width=0.50\linewidth]{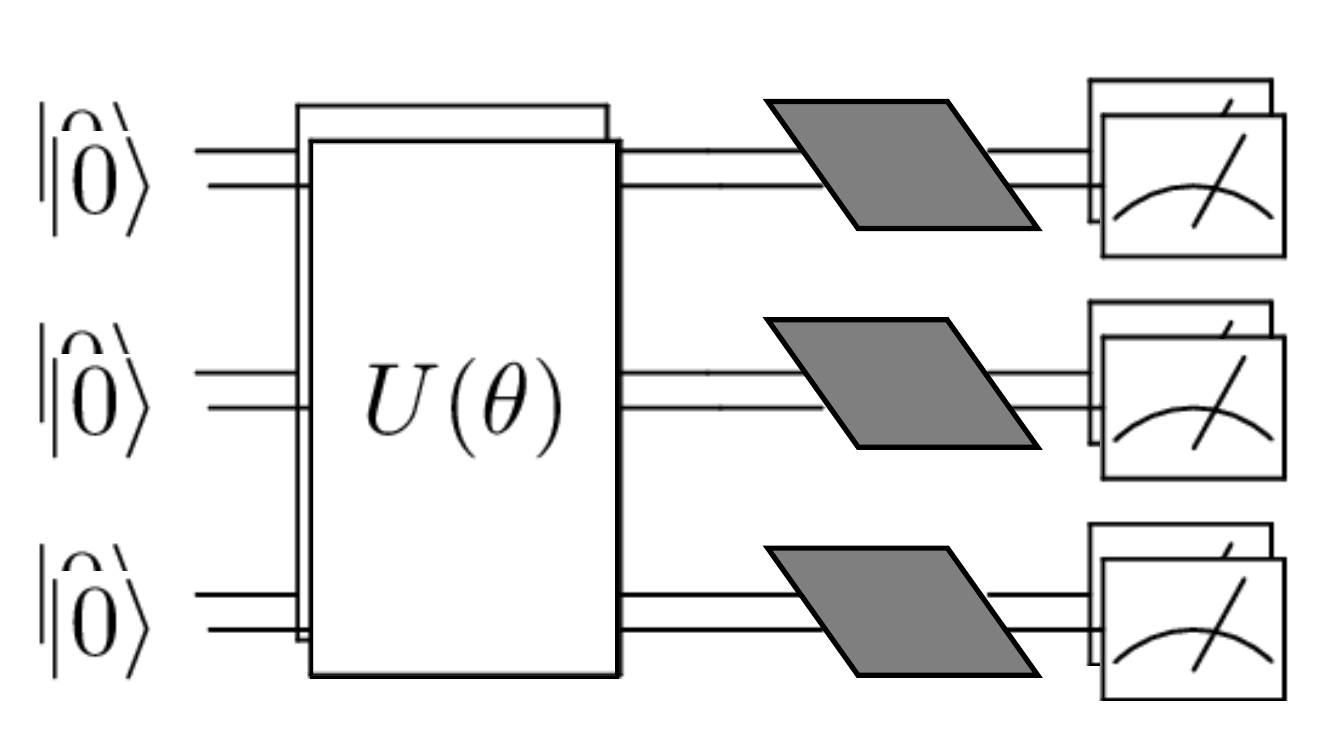}
    \caption{An illustration of the exponential error suppression mitigation implemented with the simultaneous diagonalization method mentioned in Sec.~\ref{sec:derangement-details}
    (adopted from \citet{hugginsVirtualDistillationQuantum2021}).
    We prepare two copies of the same quantum state using the same ansatz circuit $U(\boldsymbol{\theta})$,
    and then apply the diagonalizing gate on each pair of the qubit as specified by Eq.~(\ref{eq:ees-sym-diag-gates}),
    and measure the observables in the computation basis (Eq.~\ref{eq:ees-sym-diag-gates}) in the computation basis.
}
    \label{fig:ees-sym-diag}
\end{figure}

Therefore, for the qubits which are not related to the observable $O$, diagonalizing $S$ is simplified
into diagonalizing $S_i$, which would not be difficult assuming $\nEES$ is not large.
Meanwhile, in many cases of VQE, the observable $O$ may be limited to act only on a few qubits.
In this case, the diagonalization of $O^{(1)}S$ could be simplified into the diagonalization of $S_i$
above, and the diagonalization of
\begin{align}
    O \otimes_{i\in \mathbb{S}_O} S_i,
\end{align}
where $\mathbb{S}_O$ is the subset of qubits on which $O$ acts.
Furthermore, when $O$ is a one-qubit observable (e.g. $O=Z_1$), we can define the symmetrized
version of $O$ as
\begin{equation}
    O_{\mathrm{sym}} = \sum_{j=1}^\nEES O^{(j)}.
\end{equation}
where $O^{(j)}$ is the same observable $O$ on the $j$-th copy among $n$ copies of $\rho$.
Obviously $O_{\mathrm{sym}}$ and $S$ commutes. Additionally, $O_{\mathrm{sym}}$ and $S$
can be factorized into local unitaries acting on the same qubit in the $\rho$ among the $\nEES$ copies.
Hence, the two diagonalization unitaries coincide ($B_O = B_{OS}$) and can also be factorized
as tensor products of unitaries acting only on the same qubit in the $\rho$ among the $\nEES$ copies.
In particular,
we illustrate an explicit example when $\nEES=2$ of this method in Fig.~\ref{fig:ees-sym-diag}. Sec.~[II.A] of \citet{hugginsVirtualDistillationQuantum2021} also gives the explicit formula for $\Lambda_{O/OS}$ and diagonalizing
unitary $B_{O/OS}$ when $\nEES=2$.

Finally, using grouping methods (see Sec. \ref{sec:Grouping}), we can turn a complex multiple qubit measurement into a simple one-qubit measurement, making this implementation more appealing since it only needs
local qubit connections.

\subsubsection{Dual state purification method}
\label{sec:derangement-details_dual-state-pure}

\begin{figure}[ht]
    \centering
    \subfloat[]{
        \label{fig:dual-state-purification-a}
        \includegraphics[width=0.5\linewidth]{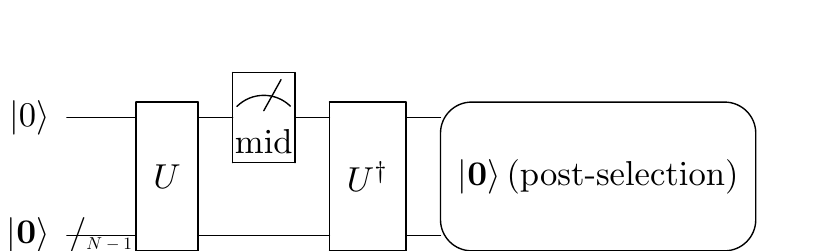}
    }
    \subfloat[]{
        \label{fig:dual-state-purification-b}
        \includegraphics[width=0.40\linewidth]{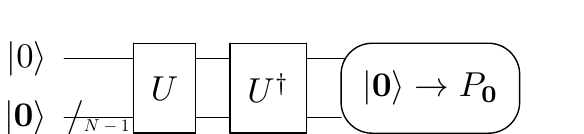}
    }
    \caption{Quantum circuit to measure Pauli $Z$ on the first qubit using the dual state purification method in exponential error supprresion (Appendix~\ref{sec:derangement-details_dual-state-pure}). 
    In \protect\subref{fig:dual-state-purification-a}, a mid-circuit measurement labelled by $\mathrm{mid}$ is carried out in addition to the final
    measurement on all qubits, and only experiments where all the final qubits are measured to be $0$ are considered.
    In \protect\subref{fig:dual-state-purification-b}, no mid-circuit
    measurement is carried out, and it should be noted that dual to the noise present in the quantum computer, the combined effect of applying $U$ and $U^\dagger$ is not identity in \protect\subref{fig:dual-state-purification-b}.
    }
    \label{fig:dual-state-purification}
\end{figure}

It was shown in \citet{huoDualstatePurificationPractical2021} that for the special case $\nEES=2$, exponential error suppression could be achieved without creating multiple copies of the state $\rho$, and without any controlled unitary gates. The disadvantage is that this method requires measuring all qubits as a final step, and it doubles the circuit depth. The core idea is to ``measure'' $\rho$ on the state $O\rho$, therefore giving the observable $\operatorname{Tr}(\rho O\rho) = \operatorname{Tr}(O\rho^2)$.
It should be noted that, although that same spirit can be applied to the general case where $m$ is an even number $\nEES=2, 4, etc$~\cite{cai2021resourceefficient}, only in the case of $\nEES=2$ can we achieve exponential error suppression without the over-head of controlled unitary gates.

The circuit diagram in Fig.~\ref{fig:dual-state-purification} shows an example of dual state purification method when the observable to be measured is Pauli-$Z$ on the first qubit (denoted by $Z_1$).
In the diagram, the circuit $U$ which prepares a state $\rho$ from $\ket{0}$ is applied twice:
$U$ in the first time and its inverse $U^\dagger$ in the second time.
In the first diagram (Fig.~\ref{fig:dual-state-purification-a}),
a mid-circuit measurement is carried out, and one also post-select on the experiments where the final measurement on all qubits gives the same outcome $0$. In this case, one denotes the probability that the final measurement obtaining $0$ in all qubits when the mid-circuit measurement gives an outcome of $b\in{0,1}$ as $P_{\mathbf{0}, b}$.
Also, in a separate experiment illustrated in Fig.~\ref{fig:dual-state-purification-b}, the mid-circuit measurement $\nEES$ is not carried out, and one
denotes the probability that the final measurement obtaining $0$ in all qubits as $P_{\mathbf{0}}$.
Then, the two circuits outcome can be combined to the following value, which closely approximate $\mathrm{Tr}(Z_1 \rho^2)/\mathrm{Tr}(\rho^2)$:
\begin{equation}
    \frac{P_{\mathbf{0}, 0} - P_{\mathbf{0}, 1}}{P_{\mathbf{0}}}.
\end{equation} 
This implementation can be adapted to measure any quantum observable $O$ by expanding $O$ into a linear sum of Pauli observables. Details could be found in \citet{huoDualstatePurificationPractical2021} 
and an explicit extension of the circuit in Fig.~\ref{fig:dual-state-purification} to arbitrary observable $O$ which squares to
the identity (e.g. Pauli observables) is available in the Appendix A of~\citet{cai2021resourceefficient}.

\subsubsection{Shadow tomography based method}
\label{sec:derangement-details_shadow-distillation}

This method has been proposed in \citet{Seif2022ShadowDistillation,Hu2022LogicalShadowTomography}, and uses the shadow tomography technique (see Sec.~\ref{sec:pauli_grouping-inference-methods}) to estimate the quantity
in Eq.~(\ref{eq:ees-computed-rhon}), or more simply the numerator  $\mathrm{Tr( \rho^{\nEES} \hat{O})}$ (Eq.~\ref{eq:ees-numerator}).
The method proposed in \citet{Hu2022LogicalShadowTomography} is more complicated and involves error correction codes, but shares the same spirit with \citet{Seif2022ShadowDistillation} in using shadow tomography to estimate the numerator $\mathrm{Tr( \rho^{\nEES} \hat{O})}$.
By applying a random unitary $U$ generated from a pool on $\rho$, and measuring it on the computational basis, we can obtain a classical shadow $\tilde{\rho}$ of $\rho$ (see Sec~\ref{sec:pauli_grouping-inference-methods} for details), and compute the numerator using the shadow $\tilde{\rho}$ as a surrogate for $\rho$.
An explicit formula is available in \citet{Seif2022ShadowDistillation} when the pool of random unitaries satisfy the 3-design property. 
One significant advantage of the shadow tomography based method over other methods presented above, is that shadow tomography does not require additional copies of the same state (Sec.~\ref{sec:derangement-details_ancilla-assisted} and \ref{sec:derangement-details_diagonalization})
or doubling of the depth of the circuit (Sec.~\ref{sec:derangement-details_dual-state-pure}).
On the other hand, the number of required random unitaries ($N_U$) to sample in order to
achieve an estimate of $\mathrm{Tr( \rho^{\nEES} \hat{O})}$ within certain precision, 
might scale exponentially with respect to the number of qubits $N$. A preliminary numerically study in \citet{Seif2022ShadowDistillation} suggest that $N_U\approx 2^{0.82N}$ for $\nEES=2$, which is better than the cost of performing state tomography on $\rho$,
but is out of reach for large scale quantum experiments.